\ifx\pdfoutput\undefined
\newcommand\texorpdfstring[2]{#1}
\else
\fi
\RequirePackage{silence}
\documentclass[prd,twocolumn,showpacs,showkeys,preprintnumbers,floatfix,nofootinbib,superscriptaddress,10pt,aps]{revtex4-1}
\usepackage{adjustbox}
\usepackage{amsfonts} 
\usepackage{amssymb} 
\usepackage{amsmath} 
\usepackage{morefloats}
\usepackage{graphicx} 
\usepackage{subfigure} 
\usepackage{array} 
\usepackage{dcolumn} 
\usepackage{bm} 
\let\bold=\bm 
\usepackage{latexsym} 
\usepackage{longtable} 
\usepackage{hyperref} 
\usepackage{bm}
\usepackage{slashed}
\usepackage{bbold}
\usepackage{color}
\usepackage{multirow}
\usepackage{rotating}
\usepackage{comment}

\graphicspath{{./figsX/}}

\usepackage{booktabs}   
\usepackage{multirow}   
\usepackage{xcolor}     
\usepackage{colortbl}   
\usepackage{longtable}

\usepackage[margin=1in]{geometry}
\usepackage [english]{babel}
\usepackage [autostyle, english = american]{csquotes}
\MakeOuterQuote{"}
\usepackage{graphicx}
\usepackage{subfigure}
\usepackage{xcolor}
\usepackage[toc,page]{appendix}
\usepackage{setspace}
\usepackage{amsmath, amssymb}
\usepackage{slashed}
\newcommand{\be}{\begin{eqnarray}}
\newcommand{\ee}{\end{eqnarray}}

\newcommand{\gf}{\gamma^5}
\newcommand{\ra}{\rangle}
\newcommand{\la}{\langle}
\newcommand{\ripmom}{${\rm RI}^\prime{\rm -MOM}$ }
\newcommand{\MSbar}{\overline{\rm MS} }

\def\ol{\overline}

\def\nn{\nonumber}
\def\dag{\dagger}

\makeatletter
\newcommand{\overleftrightsmallarrow}{\mathpalette{\overarrowsmall@\leftrightarrowfill@}}
\newcommand{\overrightsmallarrow}{\mathpalette{\overarrowsmall@\rightarrowfill@}}
\newcommand{\overleftsmallarrow}{\mathpalette{\overarrowsmall@\leftarrowfill@}}
\newcommand{\overarrowsmall@}[3]{%
  \vbox{%
    \ialign{%
      ##\crcr
      #1{\smaller@style{#2}}\crcr
      \noalign{\nointerlineskip}%
      $\m@th\hfil#2#3\hfil$\crcr
    }%
  }%
}
\def\smaller@style#1{%
  \ifx#1\displaystyle\scriptstyle\else
    \ifx#1\textstyle\scriptstyle\else
      \scriptscriptstyle
    \fi
  \fi
}
\makeatother
\newcommand{\olra}[1]{\overleftrightsmallarrow{#1}}

\providecommand{\abs}[1]{\lvert#1\rvert}
\providecommand{\matrixe}[3]{\langle#1\lvert#2\rvert#3\rangle}

\definecolor{green}{rgb}{0.1, 0.8, 0.1}

\newcolumntype{.}[1]{D{.}{.}{#1}}

\allowdisplaybreaks

\begin{document}


\title{The Momentum Fraction, Helicity and Transversity Isovector Moments of Nucleons from \texorpdfstring{$2+1$}{2+1}-flavor Lattice QCD}
%
%


\author{Santanu Mondal}
\email{santanu.sinp@gmail.com}
\affiliation{EcoDev Solutions \& Technologies Pvt. Ltd, Kolkata 700099, West Bengal, INDIA}

\author{Rajan Gupta}
\email{rajan@lanl.gov}
\affiliation{Los Alamos National Laboratory, Theoretical Division T-2, Los Alamos, NM 87545, USA}

\author{Sungwoo Park}
\email{sungwoo@sejong.ac.kr}
\affiliation{Department of Physics, Sejong University, Seoul 05006, South Korea}

\author{Jun-sik Yoo}
\email{junsik.yoo@outlook.com}
\affiliation{Department of Physics, Korea University, Seoul 02841, Korea}

\author{Tanmoy Bhattacharya}
\email{tanmoy@lanl.gov}
\affiliation{Los Alamos National Laboratory, Theoretical Division T-2, Los Alamos, NM 87545, USA}

\author{Boram Yoon}
\email{byoon@nvidia.com}
\affiliation{NVIDIA Corporation, Santa Clara, CA 95051, USA}

\author{B\'alint~Jo\'o}
\email{bjoo@nvidia.com}
\affiliation{NVIDIA Corporation, Santa Clara, CA 95051, USA}



\author{Frank~Winter}
\email{fwinter@jlab.org}
\affiliation{Jefferson Lab, 12000 Jefferson Avenue, Newport News, Virginia 23606, USA}

\collaboration{Nucleon Matrix Elements (NME) Collaboration}
\noaffiliation
\preprint{LA-UR-26-26706}
%
\date{\today}
\begin{abstract}
Results for  
the isovector momentum fraction, $\langle x \rangle_{u-d}$,
helicity moment, $\langle x \rangle_{\Delta u-\Delta d}$, 
and the transversity moment, 
$\langle x\rangle_{\delta u-\delta d}$, of the
nucleon are presented using high-statistics data on thirteen NME 
ensembles of gauge configurations generated by the JLab/W\&M/LANL/MIT/Marseille
collaborations using $2+1$-flavors of dynamical Wilson-clover
quarks. The much higher statistics facilitated better control
over all systematics compared to our previous lattice
calculation. The least controlled systematic---excited-state
contamination---is quantified by studying the variation of the results
as a function of three estimates of the mass gap of the first
excited state, obtained from two- and three-point correlation
functions. The final results are obtained using a simultaneous 
fit to extrapolate in the lattice spacing,
$a$, pion and kaon masses, $M_\pi$ and $M_K$, and the finite volume 
parameter, $M_\pi L$. The data show no significant finite-volume 
correction, and some dependence on the lattice spacing and the 
renormalization factors. The largest systematic uncertainty 
is due to possible remaining excited states contributions. Our final 
results, in the $\overline{\rm MS}$ scheme at 2~GeV, are 
$\langle x \rangle_{u-d} = 0.154(10)(9)$, 
$\langle x \rangle_{\Delta u-\Delta d} = 0.177(10)(15)$ and 
$\langle x \rangle_{\delta u-\delta d} = 0.197(12)(18)$, where the first error
is the overall statistical uncertainty and the second represents the various 
systematic uncertainties added in quadrature.   Results for the momentum fraction and 
helicity moment are consistent with phenomenological global fit values, 
while the transversity moment is a prediction. 

\end{abstract}
\maketitle
%
%
%
%
\section{Introduction}
\label{sec:intro}

Over the next decade, significant progress in the elucidation of the hadron 
structure in terms of quarks and gluons is expected from experiments 
at the Electron-Ion Collider (EIC) at BNL~\cite{Accardi:2012qut},
Jefferson Lab~\cite{Dudek:2012vr} and the Large Hadron Collider at
CERN that will extend the range of
Bjorken $x$ and the momentum transfer $Q^2$ in the collisions.  Phenomenological 
analyses of these experimental data (global fits), combined with 
higher order calculations of electroweak and strong corrections, 
will improve the precision of parton distribution
functions (PDFs)~\cite{Brock:1993sz,Ji:2020ect}, transverse momentum
dependent PDFs (TMDs)~\cite{Yoon:2017qzo}, and generalized parton
distributions (GPDs)~\cite{Diehl:2003ny}.  A limitation in their extraction is that these distributions are not 
measured directly in experiments~\cite{Cichy:2018mum,Karthik:2019},
and the phenomenological analyses rely on theoretical inputs that 
are based on models. The uncertainty can be greatly reduced 
by incorporating results from first principle lattice QCD 
calculations.  

\begin{figure*}[ht]  
\begin{subfigure}
\centering
\includegraphics[trim=100 200 150 100,angle=0,width=0.6\textwidth,clip=true]{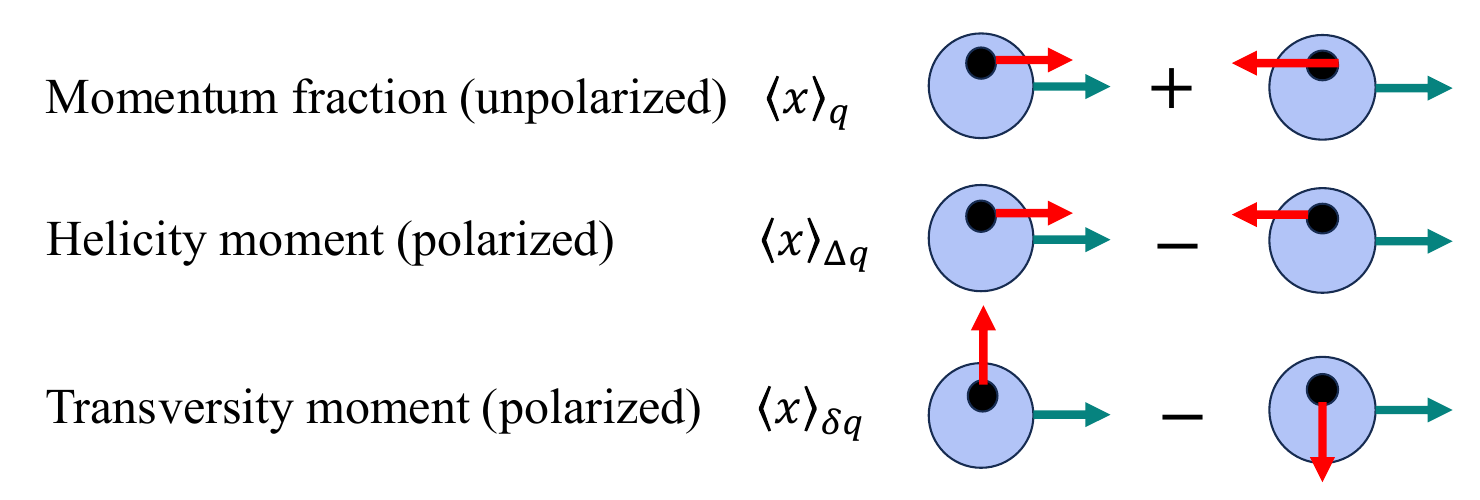}
\end{subfigure}
\vspace{-0.08in}
\caption{A pictorial description of the alignment of the direction of the spin of the quark (red arrow) 
with respect to the nucleon momentum (green arrow) for the three moments. }
\label{fig:Moments_diag}
\end{figure*}

Prospects of improvements through a 
cross-fertilization between experiments, phenomenology and 
lattice QCD  have been reviewed in Refs.~\cite{Lin:2017snn,Lin:2020rut}. Over the last decade, 
the precision of lattice QCD calculations has increased significantly due to a factor of almost 100 increase in 
computing power and advances in algorithms.  Today, there 
exist many quantities for which there is good agreement with
experimental results, and for some, the lattice results are the most
precise. Results 
considered robust are reviewed periodically by the Flavor Lattice Averaging Group (FLAG)~\cite{Aoki:2019cca,FlavourLatticeAveragingGroupFLAG:2021npn,FlavourLatticeAveragingGroupFLAG:2024oxs}. The 2024 FLAG report~\cite{FlavourLatticeAveragingGroupFLAG:2024oxs} includes results for the isovector momentum fraction, $\la x \ra_{u-d}$, the helicity $\la x \ra_{\Delta u-\Delta d}$ and transversity
moments $\la x \ra_{\delta u-\delta d}$, which currently have roughly 10\%  uncertainty in contrast to nucleon charges that have 1--5\% errors.  This 
is because the charges are obtained from the matrix elements of local quark 
bilinear operators, $\bar q(x) {\cal O}(x)q(x)$~\cite{Gupta:2018qil,Aoki:2019cca}, 
while the first moments are obtained from one derivative operators  described in Sec.~\ref{sec:moments}. The difference between the three  moments differ 
is in the alignment of the direction of the spin of the quark with respect 
 to the nucleon momentum as shown in Fig.~\ref{fig:Moments_diag}.\looseness-1

The calculation of the isovector $\la x \ra_{u-d}$, $\la
x \ra_{\Delta u-\Delta d}$ and $\la x \ra_{\delta u-\delta d}$ 
was done using thirteen NME ensembles  generated with 2+1-flavors 
of Wilson-clover fermions by the JLab/W\&M/LANL/MIT/Marseille
collaborations~\cite{Yoo:2026uul}.  
It represents four significant improvements over results 
presented in Ref.~\cite{Mondal:2020ela} based on a subset of the data on seven ensembles: 
\begin{itemize}
    \item Six additional ensembles, including one with physical pion mass, have been simulated.
    \item The statistics on two other ensembles have been increased.
    \item Data cover a larger range of lattice spacings, $ 0.053 < a <0.117$, and pion masses, $135 < M_\pi < 310$~MeV.
    \item The chiral fits now include dependence on both the light and strange quark masses.
\end{itemize}
  These improvements allowed a more controlled analysis of excited-state contributions (ESC) and more robust chiral-continuum-finite-volume fits to get results at the
physical point defined by $a = 0$, $M_\pi = 135$~MeV, $M_K = 494$~MeV
and the infinite volume characterized by $M_\pi L \to \infty$.

{\it {An important caution to the reader.  The IDs 
of the ensembles, the lattice spacing $a$, the strong coupling constant $\alpha_s$, and the values of $M_\pi$ and $M_K$ 
in physical units given in Tables~\ref{tab:ensembles},~\ref{tab:statistics} and ~\ref{tab:PSmasses} (and used throughout this paper) are different compared 
to those in Ref.~\cite{Mondal:2020ela} even though the raw data for the common ensembles are the same. This 
is because the values given here come from the recent comprehensive 
analysis of the lattice scale and spectrum on these ensembles published in Ref.~\cite{Yoo:2026uul}. These new results will be used to characterize these NME ensembles here and in future works. }}

Highlights of the analysis are: (i) investigation of the dependence of
the results on the spectra of possible, $a\ priori$ unresolved,
excited states that need to be included in the fits to remove excited-state
contamination (ESC). Concretely, the full analysis is carried out
using three probable estimates of the mass gap of the first excited
state in fits to the three-point correlation functions using Eq.~\eqref{eq:3pt}. 
The spread in the results is used to assign a systematic uncertainty to
account for possible remaining contributions from excited-states. (ii) Renormalization of the operators is carried out using two methods for handling discretization errors; (iii) the chiral fits are carried out including dependence on both the light and the strange quark masses.

Our final results for the isovector moments, in the $\MSbar$ scheme at 2~GeV, 
are given in Eq.~\eqref{eq:finalresults}. They are:   
$\langle x \rangle_{u -d} = 0.154(10)_{stat.} (9)_{sys.}$, 
$\langle x \rangle_{\Delta u- \Delta d} = 0.177(10)_{stat.} (15)_{sys.} $, and 
$\langle x \rangle_{\delta u- \delta d} =  0.197(12)_{stat.} (18)_{sys.}$. 
These estimates are in good agreement
with other lattice calculations and phenomenological global fit results as
summarized in Table~\ref{tab:Compare} in Sec.~\ref{sec:results}. 
The most extensive and precise
results from global fits are for the unpolarized moments of the
nucleons, the momentum fraction $\langle x \rangle_q$, which have 
smaller uncertainty than the lattice estimates. Global results for 
the helicity fraction, the polarized moment $\langle x \rangle_{\Delta
q}$, have a large spread and the lattice results are at the lower end.  Lattice 
results for the transversity $\langle x\rangle_{\delta q}$ are a
prediction as  sufficient experimental
data for their extraction do not exist~\cite{Lin:2017snn,Lin:2020rut}.

This paper is organized as follows: Section~\ref{sec:lattice} 
summarizes the lattice parameters and methodology. The definitions of the moments,  
the lattice operators used, and their connection to the matrix elements calculated  are given in Sec.~\ref{sec:moments}. The extraction 
of the matrix elements from the two- and three-point functions is described in
Sec.~\ref{sec:correlators}. The analysis of excited state
contributions in the extraction of the ground state matrix elements is presented in
Sec.~\ref{sec:ESC}.  The chiral-continuum-finite-volume (CCFV) extrapolation 
carried out to obtain results at the physical point is discussed in Sec.~\ref{sec:CCFV}, 
and the averaging over various analysis models to get the final values 
is described in Sec.~\ref{sec:Models}. 
A comparison with other lattice calculations
and global fit values is made in
Sec.~\ref{sec:results} and our conclusions are presented in
Sec.~\ref{sec:summary}. Four appendices give further details of the 
calculations. The figures in Appendix~\ref{sec:ratios} show the 
raw data for the ratio of the 3- to 2-point functions, 
$C_\mathcal{O}^{3\text{pt}}(\tau;t)/C^{2\text{pt}}(\tau)$, multiplied
by the appropriate factors given in Eqs.~\eqref{eq:me2momentT}  
to convert them into the moments. These figures also show the fits used to remove excited-state 
contamination. Results for the fit parameters and the unrenormalized 
moments are summarized in  Appendix~\ref{sec:rawfits}. The third  
Appendix~\ref{sec:renormalization} discusses the
calculation of the renormalization factors, $Z_{VD,AD,TD}$,  for the
three operators and their values from two methods 
used to control discretization errors in them. Figures in the last Appendix~\ref{sec:A_CCFV} 
compare the CCFV fits to the data. 

\section{Lattice Methodology}
\label{sec:lattice}

Our overall methodology for the lattice calculation has been described in 
Ref.~\cite{Mondal:2020ela} for 2+1-flavors of Wilson-clover fermions, 
and in Ref.~\cite{Mondal:2020cmt} where we presented results
for 2+1+1-flavors clover-on-HISQ formulation. This work extends the 
results in Ref.~\cite{Mondal:2020ela} to thirteen clover
ensembles (six new ensembles and increased 
statistics in two others) whose parameters are summarized in Table~\ref{tab:ensembles}.  
The source-sink separations $\tau$ used to remove excited-state contributions,
the number of configurations analyzed, and the 
high precision
(HP) and low precision (LP) measurements made to cost-effectively
increase statistics using the bias-corrected truncated-solver
method~\cite{Bali:2009hu,Blum:2012uh} are listed in
Table~\ref{tab:statistics}. Results for the pion and kaon masses, and
the scale setting flow parameters $w_0/a$ and $t_0/a^2$ are summarized in
Table~\ref{tab:PSmasses}. Results for the nucleon masses and excited state energies
from the various fits are given in Table~\ref{tab:massgap}. 

Compared to Ref.~\cite{Mondal:2020ela}, the data now cover a larger 
range of lattice spacings $0.053 \leq a \leq
0.117$ fm and lattice size $3.7 \leq M_\pi L \leq 6.2$.  The number of
values of the pion mass have been increased to four: $M_\pi \approx
290,\ 230,\ 180$ and 135~MeV.  With data on these 13 ensembles, we
note significant improvement in the control over the
chiral-continuum-finite-volume (CCFV) fits used to extrapolate the
data to the physical point defined by $ a \to 0$, $M_\pi \to 135$~MeV,
$M_K \to 494$~MeV, and $M_\pi L \to \infty$.  The value of the kaon mass has
considerable variation, lying between $475 < M_K < 575$~MeV, as shown in
Table~\ref{tab:PSmasses}.  We, therefore, carry out the chiral part of the CCFV
analysis keeping both $M_\pi$ and $M_K$ as independent parameters.
\looseness-1

The clover action in this unitary calculation is the same as used in
Ref.~\cite{Mondal:2020ela,Yoo:2026uul}. It includes one iteration
of stout smearing with weight $\rho = 0.125$ for the
staples~\cite{Morningstar:2003gk}, and a tadpole corrected tree-level
Sheikholeslami-Wohlert coefficient $c_{SW} =
1/u_0$~\cite{Sheikholeslami:1985ij}, where $u_0$ is the fourth root of
the plaquette expectation value. This determination of $c_{SW}$ is
very close to the nonperturbative value determined, a posteriori,
using the Schr\"odinger functional method~\cite{Luscher:1996ug}. 
This a consequence of the stout smearing. The update of
the gauge configurations by the JLab/W\&M/LANL/MIT/Marseille 
collaborations~\cite{Yoo:2026uul} 
was carried out using the rational hybrid Monte Carlo
(RHMC) algorithm~\cite{Duane:1987de}, as described in
Ref.~\cite{Yoon:2016jzj,Yoo:2026uul}.

The parameters used to construct the Gaussian smeared
sources for the quark propagators~\cite{Gusken:1988yi,Yoon:2016dij,Gupta:2018qil,Mondal:2020cmt}
are given in Table~\ref{tab:ensembles}. To construct the smeared
source, the gauge links were first smoothened using twenty hits of the
stout algorithm with $\rho=0.08$ with only the spatial
staples included~\cite{Morningstar:2003gk}. The root-mean-square size of the
Gaussian smearing, $\sqrt{\int dr \, r^4 S^\dag S /\int dr \, r^2
S^\dag S} $ with $S(r)$ the value of the smeared source at radial
distance $r$, is about 0.5~fm and found to reduce ESC.
The quark propagators from these smeared sources were generated by
inverting the Dirac operator using the multigrid
algorithm~\cite{Babich:2010qb,Clark:2009wm,10.5555/3014904.3014995}.
The spin and color indices of these propagators were then tied together 
at the source and sink points by the nucleon interpolating operator 
and at the operator insertion point to construct the two- and three-point
correlation functions.

The statistical analysis of the data was carried out as follows. We first averaged 
the $O(100)$  measurements of correlation functions made on a given 
configuration.  The fits to these configuration averaged 2- and 3-point 
functions were made using the full covariance matrix. The final values 
and errors were determined using a single-elimination jackknife procedure. Since the data after binning over measurements on each configuration showed no significant auto-correlations, therefore no augmentation of the jackknife errors was done.

The full analysis was done separately for the three strategies used 
to remove excited state contamination that are the same as used  in
Refs.~\cite{Mondal:2020cmt,Mondal:2020ela}. These data were 
renormalized using two methods to handle 
discretization errors arising in the intermediate RI'-MOM scheme as discussed in Appendix~\ref{sec:renormalization}.  In the chiral part of CCFV fit ansatz, we now include the dependence on both the light and strange quark mass, and the continuum extrapolation explores three possibilities for the leading correction, i.e., $O(a)$, $O(\alpha_s a)$ and $O(a^2)$. Variations in results from the various models used in the different steps in the analysis are used to estimate the systematic errors. The model averaging is done using the Akaika Information score~\cite{1100705}.

\begin{table*}[phtb]  
\centering
\setlength{\tabcolsep}{5pt}
\renewcommand{\arraystretch}{1.2} 
\begin{tabular}{|l||c|c|c|c|c|c|c|c| }
\hline
        &         &    &              &          &               &                           &         \\  
Ensemble& $\beta$ & $a$ & $L^3\times T$&$M_\pi L$ & $c_{\text{SW}}$  & Smearing                  & RMS radius  \\
ID      &         &(fm)&              &          &               & $\{\sigma, N_{\text{KG}}\}$  & $r_{\rm RMS}/a$  \\                 
\hline
\hline
$a117m310$   & 6.1 & 0.1165(21)   &$32^3\times96$ & 5.85  &   1.24931    &  \{5, 50\}    &      4.10(1)    \\ 
\hline                                                                                                          
$a087m290$   & 6.3 & 0.0874(16)  &$32^3\times64$ & 4.11  &   1.20537    &  \{7, 91\}    &      5.46(2)    \\ 
$a087m290L$  & 6.3 & 0.0874(16)  &$48^3\times128$& 6.16  &   1.20537    &  \{7, 91\}    &      5.49(3)    \\ 
$a087m230$*  & 6.3 & 0.0873(16)  &$48^3\times128$& 5.0   &   1.20537    &  \{7, 98\}    &      5.54(1)    \\ 
$a087m230X$* & 6.3 & 0.0865(16)  &$48^3\times128$& 5.0   &   1.20537    &  \{7, 98\}    &      5.56(1)    \\ 
                                                                 
$a086m180$   & 6.3 & 0.0859(16)  &$48^3\times96$ & 3.75  &   1.20537    &  \{7, 91\}    &      5.41(2)    \\ 
$a086m180L$* & 6.3 & 0.0859(16)  &$64^3\times128$& 5.08  &   1.20537    &  \{7, 91\}    &      5.49(3)    \\ 
\hline                                                                                                          
$a068m290$   & 6.5 & 0.0682(12)  &$48^3\times128$& 4.82  &   1.170082   &  \{9, 150\}   &      6.96(1)    \\ 
$a067m230$*  & 6.5 & 0.0673(12)  &$64^3\times192$& 5.2   &   1.170082   &  \{10,200\}   &      7.53(1)    \\ 
$a067m175$*  & 6.5 & 0.0670(12)  &$72^3\times192$& 4.26  &   1.170082   &  \{10,185\}   &      7.57(2)    \\ 
$a067m135$*  & 6.5 & 0.0667(12)  &$96^3\times192$& 4.4   &   1.170082   &  \{10,200\}   &      7.57(1)    \\ 
\hline                                                                                                         
$a053m295$*  & 6.7 & 0.0534(10)  &$64^3\times192$& 4.8   &   1.142727   &  \{14,365\}   &      10.09(1)   \\ 
$a053m230$*  & 6.7 & 0.0531(10)  &$72^3\times192$& 4.5   &   1.142727   &  \{14,392\}   &      10.18(1)   \\ 
\hline
\end{tabular}
\caption{Lattice parameters of the 2+1-flavor clover ensembles generated by the 
JLab/W\&M/LANL/MIT/Marseille collaborations and analyzed in this study. Listed
are the ensemble ID assigned, gauge coupling $\beta$, the lattice spacing $a$,
the lattice size $L^3\times T$, and the lattice size $L$ in units of
$M_\pi$.  Column six gives the Sheikholeslami-Wohlert
improvement coefficient $c_{\rm SW}$ in the clover action. The parameters $\{\sigma, N_{\text{KG}}\}$ 
in the Chroma convention~\cite{Edwards:2004sx}, 
used to construct Gaussian smeared
sources~\protect\cite{Gusken:1989ad}, are given in 
column seven where $N_{\text{KG}}$ is the number of applications of the
Klein-Gordon operator and $\sigma$ controls the width of the smearing.
The resulting root-mean-square radius of the smearing, defined as $r_{\rm RMS} = \sqrt{\int dr \, r^4 S^\dag S /\int dr \, r^2
S^\dag S} $ with $S(r)$ the value of the smeared source at radial
distance $r$, is given in the last column. The ensembles that are new or have increased
statistics compared to those presented in Ref.~\protect\cite{Mondal:2020ela}
are marked with an ${}^\ast$. }
\label{tab:ensembles}
\end{table*}

\begin{table*}[phtb]  
\centering
\setlength{\tabcolsep}{5pt}
\renewcommand{\arraystretch}{1.2} 
\begin{tabular}{|l||c|c|c|c| }
\hline
          &           &           &         &   \\                                               
Ensemble  & $\tau/a$  & $N_{conf}$ & $N_{HP}$ & $N_{LP}$  \\ 
ID        &           &           &         &          \\                                                  
\hline
\hline

$a117m310$   &$\{8,{\textcolor{red}{10,12,14}}\}$       & 2001  &  8,004 & 256,128   \\
\hline                                                                                 
$a087m290$   &$\{10,{\textcolor{red}{12,14,16}}\}$      & 1464  &  4,392 & 140,544   \\
$a087m290L$  &$\{8,10,12,{\textcolor{red}{14,16,18}}\}$ & 4501  & 18,004 & 576,128   \\
$a087m230$*  &$\{10,12,{\textcolor{red}{14,16,18}}\}$   & 2000  &  8,000 & 256,000   \\
$a087m230X$* &$\{10,12,{\textcolor{red}{14,16,18}}\}$   & 2006  &  8,024 & 256,768   \\
$a086m180$   &$\{8,10,{\textcolor{red}{12,14,16}}\}$    & 4015  & 16,060 & 513,920   \\
$a086m180L$* &$\{8,10,{\textcolor{red}{12,14,16}}\}$    & 2532  & 12,660 & 405,120   \\
\hline                                                                                 
$a068m290$   &$\{11,13,{\textcolor{red}{15,17,19}}\}$   & 4477  & 17,908 & 573,056   \\
$a067m230$*  & $\{13,15,{\textcolor{red}{17,19,21}}\}$  & 2000  & 12,000 & 192,000   \\
$a067m175$*  &$\{13,15,{\textcolor{red}{17,19,21}}\}$   & 3120  & 18,720 & 299,520   \\
$a067m135$*  &$\{13,{\textcolor{red}{15,17,19}},21\}$   & 2501  & 15,006 & 240,096   \\
\hline                                                                                 
$a053m295$*  &$\{18, 21, {\textcolor{red}{24,27, 30}}\}$& 2700  & 16,200 & 259,200   \\
$a053m230$*  &$\{{\textcolor{red}{18,21,24}},27,30\}$   & 2410  & 14,460 & 231,960   \\ 

\hline
\end{tabular}
\caption{
The second column gives the values of source-sink separation $\tau$ at which data were collected 
for removing ESC. Values in red were used in the final fits to remove ESC. The last
three columns give the number of configurations analyzed, and the
total number of high precision (HP) and low precision (LP)
measurements made.  The ensembles that are new or have increased
statistics compared to those presented in Ref.~\cite{Mondal:2020ela}
are marked with an ${}^\ast$.  Ensembles $a087m230X$ and $a087m230$
differ primarily in the value of $m_s$ as can be inferred from the data in
Table~\ref{tab:PSmasses}.  }
\label{tab:statistics}
\end{table*}

\begin{table*}[htbp]      
\centering
\setlength{\tabcolsep}{2pt}
\renewcommand{\arraystretch}{1.2} 
\begin{tabular}{|l|ccc|ccc|c|c|}
\hline
Ensemble            & $a m_l$    & $aM_\pi$   & $M_\pi$     &   $am_s$    & $a M_K$    & $M_K$     & $w_0/a$      & $t_0 /a^2$  \\
ID                  &            &           &  (MeV)     &             &            & (MeV)     &              &            \\
                    &            &           &            &             &            &           &              &            \\
\hline                                                                                                      
\hline                                                                                                      
$a117m310$          & $-0.2850$  & 0.1834(3) & 310(6) &  -0.2450    & 0.3052(5)  & 517(10) & 1.3880(29)  & 1.4953(44) \\
\hline                                                                                                                       
$a087m290$          & $-0.2390$  & 0.1279(4) & 288(8) &  -0.2050    & 0.2436(2)  & 550(6)  & 1.8709(45)  & 2.6566(59) \\
$a087m290L$         & $-0.2390$  & 0.1281(1) & 289(5) &  -0.2050    & 0.2436(3)  & 550(10) & 1.86770(75) & 2.6517(13) \\
$a087m230X$         & $-0.2406$  & 0.1001(2) & 233(4) &  -0.2050    & 0.2342(2)  & 585(11) & 1.86580(52) & 2.65879(73)\\
$a087m230$          & $-0.2415$  & 0.1033(1) & 228(4) &  -0.1974    & 0.2589(3)  & 534(10) & 1.89853(90) & 2.7079(14) \\
$a086m180$          & $-0.2416$  & 0.0781(2) & 179(3) &  -0.2416    & 0.2280(2)  & 523(6) & 1.9187(19)  & 2.7442(30) \\
$a086m180L$         & $-0.2416$  & 0.0781(1) & 179(3) &  -0.2416    & 0.2279(3)  & 523(10) & 1.91977(44) & 2.74588(76)\\
\hline                                                                                                                       
$a068m290$          & $-0.2070$  & 0.1001(1) & 290(5) &  -0.1750    & 0.2120(3)  & 614(11) & 2.4115(19)  & 4.3619(38) \\
$a067m230$          & $-0.2080$  & 0.0797(1) & 234(4) &  -0.1788    & 0.1939(3)  & 568(10) & 2.46372(95) & 4.4779(18) \\
$a067m175$          & $-0.2091$  & 0.0594(1) & 175(3) &  -0.1778    & 0.1924(2)  & 567(10) & 2.48351(68) & 4.5158(11) \\
$a067m135$          & $-0.2095$  & 0.0457(1) & 135(3) &  -0.1793    & 0.1851(2)  & 548(10) & 2.50490(61) & 4.5620(10) \\
\hline                                                                                                                       
$a053m295$          & $-0.1830$  & 0.0796(1) & 295(5) &  -0.1650    & 0.1496(3)  & 553(10) & 3.1304(18)  & 7.1233(41) \\
$a053m230$          & $-0.1843$  & 0.0609(1) & 226(4) &  -0.1640    & 0.1480(1)  & 550(10) & 3.1616(25)  & 7.1904(55) \\
\hline
\end{tabular}
\caption{Data for 
  the light and strange quarks masses defined as 
  $m \equiv 1/2\kappa -1/4$ with $\kappa$ the hopping parameter
  used in the simulations.  Also given are the measured pion and kaon
  masses, and the two flow parameters, $w_0$ and $t_0$ that are used to set the
  lattice scale. These data are the same as presented in~\cite{Yoo:2026uul}.}
\label{tab:PSmasses}
\end{table*}

\section{Moments and Matrix elements}
\label{sec:moments}

The first moments of spin independent (or
unpolarized), $q=q_\uparrow+q_\downarrow$, helicity (or polarized),
$\Delta q=q_\uparrow-q_\downarrow$, and transversity, 
$\delta q =q_\top+q_{\perp}$ distributions, are defined as
\be
\langle x \rangle_q &=& \int_0^1~x~[q(x)+\ol{q}(x)]~dx \,, \\
\langle x \rangle_{\Delta q} &=& \int_0^1~x~[\Delta q(x)+\Delta \ol{q}(x)]~dx \,, \\
\langle x \rangle_{\delta q} &=& \int_0^1~x~[\delta q(x)+\delta \ol{q}(x)]~dx \,,
\ee
where $q_{\uparrow(\downarrow)}$ corresponds to quarks with helicity
aligned (anti-aligned) with that of a longitudinally polarized target,
and $q_{\top(\perp)}$ corresponds to quarks with spin aligned
(anti-aligned) with that of a transversely polarized target. These
alignments are shown pictorially in Fig.~\ref{fig:Moments_diag}.

The three moments, at leading twist, are extracted from the forward matrix
elements of one-derivative vector, axial-vector and tensor operators calculated
within ground state nucleons at rest.  The complete
set of the relevant twist two operators are
\be
{\cal O}^{\mu \nu}_{V^a}&=&\ol{q} \gamma^{\{\mu}\olra{D}^{\nu\}} \tau^a q\nn \,, \\
{\cal O}^{\mu \nu}_{A^a}&=&\ol{q}\gamma^{\{\mu}  \olra{D}^{\nu\}} \gf \tau^a q\nn \,, \\
{\cal O}^{\mu \nu \rho}_{T^a}&=&\ol{q} \sigma^{[\mu\{\nu]} \olra{D}^{\rho\}} \tau^a q \,, 
\label{operators}
\ee 
where $q=\{u,d\}$ is the isodoublet of light quarks and  
$\sigma^{\mu\nu} = (\gamma^\mu\gamma^\nu - \gamma^\nu\gamma^\mu)/2$. 
The derivative $\olra{D}^{\nu}\equiv\frac{1}{2}(\overrightarrow{D}^\nu-\overleftarrow{D}^\nu)$ 
consists of four terms as defined in Ref.~\cite{Mondal:2020cmt}. 
Lorentz indices within $\{ ~\}$ in Eq.~\eqref{operators} are
symmetrized and within $[\, ]$ are antisymmetrized. It is also
implicit that, where relevant, the traceless part of the above
operators is taken.  Their renormalization is carried out
nonperturbatively in the regularization independent RI${}^\prime$-MOM
scheme as discussed in Appendix~\ref{sec:renormalization}. A more
detailed discussion of these twist-2 operators and their
renormalization can be found in Refs.~\cite{Gockeler:1995wg}
and~\cite{Harris:2019bih}.

In our setup, we take $\tau^a
= \tau^3$ and fix the spin of the nucleon state to be in the
``3'' direction. With these choices, the explicit operators calculated
are
\be
{\cal O}^{44}_{V^3} &=&  \ol{q} (\gamma^{4}\olra{D}^{4}  -\frac{1}{3}
{\bm \gamma} \cdot \olra{\bf D}) \tau^3 q \,,
\label{eq:finaloperatorV} \\
{\cal O}^{34}_{A^3} &=&\ol{q} \gamma^{\{3}\olra{D}^{4\}} \gf \tau^3 q \,,
\label{eq:finaloperatorA} \\
{\cal O}^{124}_{T^3} &=& \ol{q} \sigma^{[1\{2]}\olra{D}^{4\}} \tau^3 q \,. 
\label{eq:finaloperatorT}
\ee
The forward matrix elements ($ME$) of these operators within the
ground state of the nucleon with mass $M_N$ are related to the moments
as follows:
\be
\la 0 | {\cal O}^{44}_{V^3}| 0 \ra &=&  -  M_N\, \la x \ra_{u-d} \,, 
\label{eq:me2momentV} \\
\la 0 | {\cal O}^{34}_{A^3}| 0 \ra &=&  - \frac{i  M_N}{2} \, \la x \ra_{\Delta u-\Delta d} \,, 
\label{eq:me2momentA} \\
\la 0 | {\cal O}^{124}_{T^3}| 0 \ra &=& - \frac{i M_N}{2} \, \la x \ra_{\delta u-\delta d} \,. 
\label{eq:me2momentT}
\ee 
These moments are, by construction, dimensionless. 

\section{Correlation functions and Moments}
\label{sec:correlators}

To construct the two- and three-point correlation functions needed to calculate 
the matrix elements, the interpolating operator ${\mathcal N}$ used to
create/annihilate the nucleon state is 
\be 
{\mathcal N} = \epsilon^{abc} \Big[  q_1^{aT} (x) C \gf \frac{(1\pm \gamma_4)}{2}q_2^b(x) \Big] q^c_1(x) \,, 
\label{nucop}
\ee 
where $\{a,b,c\}$ are color indices,
$q_1,q_2 \in \{u,d\}$ and $C=\gamma_4 \gamma_2$ is the charge conjugation
matrix used in our calculation.  The nonrelativistic projection $(1\pm \gamma_4)/2 $
is inserted to improve the signal, with the plus and minus signs
applied, respectively, to the forward and backward propagation in Euclidean time~\cite{Gockeler:1995wg}. At zero momentum, this operator
couples only to the spin-$\frac12$ states.  The zero momentum
two-point and three-point nucleon correlation functions are defined as \looseness-1
\begin{flalign}
\bold{C}^{\rm 2pt}_{\alpha \beta} (\tau ) &= \sum_{\bm x}\la 0 | {\mathcal N}_\alpha (\tau, {\bm x}) \ol{{\mathcal N}}_\beta(0,{\bm 0})| 0\ra\\ 
\bold{C}^{\rm 3pt}_{\mathcal{O},\alpha \beta}
(\tau,t ) &= \sum_{{\bm x}',{\bm x}} \la 0 | {\mathcal N}_\alpha (\tau, {\bm x}) {\cal O} (t, {\bm x}') \ol{{\mathcal N}}_\beta(0,{\bm 0})| 0\ra 
\end{flalign}
where $\alpha$, $\beta$ are spin indices. The source is placed at time
slice 0, the sink is at $\tau$ and the one-derivative operators,
defined in Sec.~\ref{sec:moments}, are inserted at all intermediate time slices $t$.
Data have been collected for the values of source-sink separation $\tau$ specified in
Table~\ref{tab:ensembles}, and for each $\tau$ for operator insertion at all intermediate
times $0  < t < \tau$.

To isolate the various contributions and extract the ground state matrix elements 
defined in Eqs.~\ref{eq:me2momentV},~\ref{eq:me2momentA} and~\ref{eq:me2momentT}, 
projected $2$- and $3$-point functions are constructed as
\be
C^{\rm 2pt}&=& {\rm Tr} \big( {\cal P}_{\rm 2pt} \bold{C}^{\rm 2pt} \big)\\
C_\mathcal{O}^{\rm 3pt}&=& {\rm Tr} \big( {\cal P}_{\rm 3pt} \bold{C}^{\rm 3pt}_{\mathcal{O}} \big) \,.
\ee
The projector ${\cal P}_{\rm 2pt} = \frac{1}{2}\, (1 + \gamma_4)$ in the nucleon 
correlator gives the positive parity contribution for the nucleon propagating
in the forward direction.
For the connected $3$-point functions, ${\cal P}_{\rm 3pt}= \frac{1}{2}(1 + 
\gamma_4)(1+i \gf \gamma^3)$ is used. 

The size of the ESC can be visualized from the ratios
\begin{equation}
R_\mathcal{O}(\tau;t) = C_\mathcal{O}^{3\text{pt}}(\tau;t)/C^{2\text{pt}}(\tau)
\label{eq:Ratio}
\end{equation}
that give the ground state matrix element in the limits $t \to \infty$ and
$(\tau - t) \to \infty$. We show these ratios for the 8 "new" ensembles in
Figs.~\ref{fig:Ratio-mom-1}--\ref{fig:Ratio-transversity-2} in Appendix~\ref{sec:ratios}. 
We re-emphasize that the ground state matrix elements,  
$\matrixe{0}{\mathcal{O}}{0}$ are not obtained from the ratio in 
Eq.~\eqref{eq:Ratio} but from fits to
$C_\mathcal{O}^{3\text{pt}}(\tau;t)$ with input of the spectral quantities from
$C^{2\text{pt}}(\tau)$.

\begin{table*}[tbhp]   
\centering
\setlength{\tabcolsep}{2.5pt}
\renewcommand{\arraystretch}{1.3}
\begin{tabular}{|c|c|c|c|c|c|c|c|c|c| }
\hline
      &     &  &                     &                        &            &    &   &   \\
Ensemble    &$aM_N^{\{4\}}$ &$a M_N^{\{4^{N\pi}\}}$  
& $a \Delta M_1^{\{2\}}$ & $a \Delta M_1^{\{4\}}$ & $a \Delta M_1^{\{4^{N\pi}\}}$ 
& $a \Delta M_1^{\{2^{\rm free}\}}$
& $a \Delta M_1^{\{2^{\rm free}\}}$ 
& $a \Delta M_1^{\{2^{\rm free}\}}$ \\
ID          &                        &                        &            &     & 
& $\langle x \rangle_{u -d}$ 
&$\langle x \rangle_{\Delta u- \Delta d}$ 
&$\langle x \rangle_{\delta u- \delta d}$ \\
\hline
\hline
$a117m310$  & $0.6181(19)$ & $0.6167(14)$ & $0.457(28)$  & $0.381(41)$ &  $0.416(29)$  &  $0.359(35)$ &  $0.706(58)$  &  $0.64(11)$   \\ 
\hline                                                                                                                         
$a087m290$  & $0.4706(39)$ & $0.4706(24)$ & $0.349(91)$  & $0.264(75)$ &  $0.263(11)$ &  $0.521(64)$ &  $0.647(66)$  &  $0.510(66)$  \\ 

$a087m290L$ & $0.4668(12)$ & $0.4656(9) $ & $0.357(19)$  & $0.303(42)$ &  $0.249(27)$  &  $0.344(26)$ &  $0.400(53)$  &  $0.466(29)$  \\

$a087m230$ & $0.4478(12)$ & $0.4464(22) $ &  $0.351(15)$ & $0.357(31)$ &  $0.2071(52)$  &  $0.359(61)$ &  $0.300(61)$  &  $0.498(138)$  \\

$a087m230X$ & $0.4331(28)$ & $0.4279(29) $ & $0.307(29)$  & $0.250(34)$ & $0.160(18)$  &  $0.345(49)$ &  $0.458(68)$  &  $0.418(79)$  \\ 

$a086m180$  & $0.4163(23)$ & $0.4119(19)$ & $0.346(22)$  & $0.293(45)$ &  $0.195(19)$  &  $0.311(48)$ &  $0.343(96)$  &  $0.424(61)$  \\

$a086m180L$ & $0.4140(21)$ & $0.4061(33)$ & $0.327(16)$  & $0.294(18)$ &  $0.141(24)$  &  $0.303(50)$ &  $0.377(60)$  &  $0.294(66)$   \\ 

\hline                                            

$a068m290$  & $0.3719(11)$ & $0.3716(8)$  & $0.321(18)$  & $0.229(41)$ &  $0.217(24)$  & $0.311(16)$ &  $0.457(14)$  &  $0.431(16)$  \\

$a067m230$  & $0.3468(20)$ & $0.3455(15)$  & $0.301(22)$  & $0.233(33)$ & $0.172(19)$  & $0.295(47)$ &  $0.353(48)$  &  $0.297(56)$  \\ 

$a067m175$  & $0.3283(21)$ & $0.3251(15)$ & $0.264(16)$  & $0.212(45)$ & $0.141(11)$ &  $0.222(60)$ &  $0.264(95)$  & $0.435(50)$  \\ 
$a067m135$  & $0.3125(25)$ & $0.3110(18)$  & $0.278(11)$  & $0.141(17)$ & $0.125(8)$  & $0.289(36)$ &  $0.350(84)$  &  $0.455(44)$  \\ 
\hline
$a053m295$  & $0.2939(6)$ & $0.2931(6)$ & $0.284(16)$  & $0.260(19)$ & $0.224(24)$ &  $0.352(276)$ &  $0.473(110)$  & $0.300(37)$  \\
$a053m230$  & $0.2767(8)$ & $0.2741(12)$  & $0.274(13)$  & $0.281(19)$ & $0.142(19)$  & $0.212(28)$ &  $0.347(28)$  &  $0.326(38)$  \\ 


\hline
\end{tabular}
\caption{The nucleon mass from two types of 
four-state fits, $aM_N^{\{4\}}$ and $a M_N^{\{4^{N\pi}\}}$, is 
given in columns two and three.  The next six columns
give the values of the mass gap, $a \Delta M_1 \equiv a(M_1-M_0)$,
of the first excited state
obtained from different fits studied in this work. The notation 
is---$\{2\}$ ($\{4\}$) is a two-state (four-state) fit to the 
two-point functions while $\{4^{N\pi}\}$ is a four-state fit 
to the two-point functions with a narrow width prior for 
$a\Delta M_1$ corresponding to the energy of the 
non-interacting $N \pi$ state. The three $a \Delta M_1^{\{\rm free\}}$ are outputs for the mass gap from fits  
to the three-point functions used to
extract the three moments as explained in the text. Results for nucleon mass and the mass gaps in columns 2--6 are different from those presented in~\cite{Yoo:2026uul} as they are based on smaller statistics 
specified in Table~\ref{tab:statistics} and used for the moments.
}
\label{tab:massgap}
\end{table*}


\section{Controlling excited state contamination}
\label{sec:ESC}

A major challenge to precision results for nucleon matrix elements is removing 
the contribution of multihadron excited states to 
three-point functions. These occur because the
lattice nucleon interpolating operator ${\mathcal N}$, defined in
Eq.~\eqref{nucop}, couples to the nucleon and all its excitations with 
the same quantum numbers,  including multihadron states such as $N \pi$ and $N\pi\pi$. Previous lattice
calculations~\cite{Mondal:2020cmt,Bhattacharya:2013ehc,Bali:2014gha,
Bali:2012av} have shown that these ESC can be 
large even at the largest source-sink separations simulated, and in some cases the contributions of $N\pi$ states are enhanced. 
To remove these artifacts, we follow the same set of strategies  as 
described in Refs.~\cite{Mondal:2020cmt,Mondal:2020ela}: (i) reduce ESC by
using smeared sources in the generation of quark propagators, (ii) high statistics to get a good signal at larger $\tau$,  and (iii) 
fit the data at multiple source-sink separations $\tau$ using the
spectral decomposition of the correlation functions
(Eqs.~\eqref{eq:2pt} and ~\eqref{eq:3pt}) keeping as many excited
states as possible without overparameterizing the fits.  The goal is to 
develop a data-driven analysis, i.e., the goodness of the fit 
should discriminate between the various possible values of masses of 
the excited-states that contribute significantly. In this work,
we examine three strategies, labeled $\{4,3^*\}$, $\{4^{N\pi},3^*\}$ and
$\{4,2^{\rm free}\}$, that use different estimates of the first excited-state 
mass, yield results that differ by $1-2\sigma$ but are not distinguished by the $\chi^2/dof$ of the fits. In the end, we resort to physics arguments to choose a model for the final value or take a model average using the AIC score.

\subsection{Two-point Correlation Functions}
\label{sec:2pt}

The spectral decomposition of the zero-momentum two-point function,
${ C^{\rm 2pt}}$, truncated at four states, is given by
\begin{equation}
C^{\rm 2pt}(\tau) = \sum_{i=0}^3  |{\cal A}_i|^2e^{-M_i \tau} \,. 
\label{eq:2pt}
\end{equation}
We fit the data over the largest time range,
$\{\tau_{min}-\tau_{max}\}$, allowed by the statistics, i.e., by the
stability of the covariance matrix, to extract the masses $M_i$ and 
the amplitudes  ${\cal A}_i$ for the creation/annihilation of the 
four lowest states by the interpolating operator ${\mathcal N}$.  
A straightforward (called standard) fit using the empirical Bayesian 
technique described in Ref. [24] is labeled $\{4\}$, and in it 
wide priors are used only to stabilize the three
excited-state parameters. 

For zero-momentum correlators, excited states can be of the "radial" 
type, i.e. single hadron like $N(1440)$, or multihadron like 
$N({-\bm p}) \pi({\bm p})$ and $N({\bm 0}) \pi({-\bm p}) \pi({\bm p})$. 
Both, $N({-\bm 1}) \pi({\bm 1})$ and $N({\bm 0}) \pi({\bm 0}) \pi({\bm 0})$
are the lowest states in their respective towers of positive parity 
multihadron states with nucleon quantum numbers, and their (non-interacting) energy   
is roughly equal for all thirteen ensembles analyzed here. Most important, 
their non-interacting energy becomes smaller than the radial excitation $N(1440)$
(roughly the $M_1$ obtained from the
$\{4\}$ fit) for the $M_\pi \lesssim 200$~MeV ensembles.

Unconstrained fits to the two-point functions do not reveal multihadron 
states such as ${N\pi}$. This is because their amplitude 
is suppressed by the $1/V$ normalization factor  
for each additional particle in a multihadron system~\cite{Bar:2018xyi}. 
In our fits, when there 
is reason to believe that these states contribute, they are input 
into Eq.~\eqref{eq:2pt} via normally distributed priors with $M_1$ taken 
to be their non-interacting energy, i.e., the sum of the energies of the  
individual hadrons calculated in this work, 
and a relative width of $\approx 10\%$.~\footnote{The quantity minimized in the fits with 
priors is the augmented $\chi^2$ 
defined as the standard  $\chi^2$ using the covariance 
matrix plus the square of the
deviation of the parameter from the prior mean normalized by the prior
width. With this understanding, the  
quoted $\chi^2$/dof  should be read as the standard or the augmented $\chi^2$ depending on the fit, divided by the 
appropriate number of degrees of freedom. In cases 
where the prior is used only to stabilize the
fit, i.e., with a large width, the difference between the two is small.}.  Such a 4-state fit is 
labeled$\{4^{N\pi}\}$. With the 
current data, the $\{4\}$ and $\{4^{N\pi}\}$
strategies are not distinguished on the basis of the $\chi^2/$dof.
In fact, the full range of values for $M_1$ between the two estimates,
from $\{4\}$ and $\{4^{N\pi}\}$, are viable on the basis of
$\chi^2/$dof alone. The same is true of $M_2$ and $M_3$,
indicating a large flat region in the eight parameter space of the $\{A_i,M_i\}$. The determination of 
$\{{\cal A}_0,M_0\}$, is, however, robust.

Because of this lack of resolution of the excited states, i.e., 
a large region of possible values for the excited-state masses $M_i$,
we carry out the full analysis with three strategies that use
different estimates of $M_i$ and investigate the sensitivity of the
results on them.  The ground-state nucleon mass obtained from the
$\{4\}$ and $\{4^{N\pi}\}$ fits is denoted by the common symbol
$M_N \equiv M_0$ and the successive mass gaps by $\Delta M_i \equiv
M_i - M_{i-1}$. These are given in Table~\ref{tab:massgap}, and in Tables~\ref{tab:5strategy-fits-momfrac},
~\ref{tab:5strategy-fits-helfrac} and~\ref{tab:5strategy-fits-transvmom} in Appendix~\ref{sec:rawfits}. 

The need to consider multihadron states arises because their contribution to 
some 3-point functions is enhanced sufficiently to cancel 
the normalization factors of $1/V$. 
For example, the $N({-\bm 1}) \pi({\bm 1})$ state has been shown to 
contribute significantly in the pseudoscalar and the axial channels~\cite{Jang:2019vkm}, a manifestation of the pion-pole dominance hypothesis, whereas the  
$N({\bm 0}) \pi({\bm 0}) \pi({\bm 0})$ state is expected to be the 
relevant for the vector channel  based on the vector meson
dominance hypothesis. Note that in fits with a single interpolating operator (as opposed to a variational 
analysis) only the energies $E_i$ and not 
the identity of the states matter. This is why we do not 
distinguish between the contributions of the nearly degenerate 
$\{N\pi\}$ and $\{ N\pi\pi \}$ states, 
rather label their joint contribution by $\{{N\pi}\}$. On the other hand, in a 
variational analysis where interpolating operators for all states that contribute significantly need to be included, 
one would need to determine which contributes more or include both.

We also emphasize that even
though we use a Bayesian procedure for stabilizing the fits, the
errors are calculated using the jackknife method and are thus the
usual frequentist standard errors.

\subsection{Three-point Correlation Functions}
\label{sec:3pt}

The analysis of the three-point functions,
$C_\mathcal{O}^{3\text{pt}}$, with insertion of  operators with zero-momentum 
defined in
Eqs.~\eqref{eq:finaloperatorV},~\eqref{eq:finaloperatorA}
and~\eqref{eq:finaloperatorT}, is performed retaining up to three
states $|i\rangle$ in the spectral decomposition, i.e., 
\begin{equation}
  C_\mathcal{O}^{3\text{pt}}(\tau;t) = 
   \sum_{i,j=0}^2 \abs{\mathcal{A}_i} \abs{\mathcal{A}_j}\matrixe{i}{\mathcal{O}}{j} e^{-M_i t - M_j(\tau-t)}\,,
   \label{eq:3pt}
\end{equation}
and simplified to the forward matrix element by fixing the momentum at the
sink to zero. The full set of values 
of $\tau$ simulated are given in Table \ref{tab:ensembles} and $t$ varies between $0$ and $\tau-1$ in each case.
To remove the ESC and extract the desired ground-state
matrix element, $\matrixe{0}{\mathcal{O}}{0}$, a simultaneous
fit versus $t$ and $\tau$ is made to data using Eq.~\eqref{eq:3pt}. 

Unconstrained  $3$-state fits using Eq.~\eqref{eq:3pt} are not stable 
and the output $M_i$ have errors much larger than those in the $M_i$ determined from 
the 2-point functions. Furthermore, as already stated,  fits to
the two-point function show a large flat region with roughly 
the same $\chi^2/$dof in the space of the
$M_i$. Thus, external information on the $M_i$ is needed for robust 
fits to Eq.~\eqref{eq:3pt} as, {\it a priori}, it is not known 
which states make a significant contribution to a given  three-point 
function with the insertion of a given operator ${\mathcal O}$. 

Theoretically, candidate lowest mass  intermediate states are radial starting 
with $N(1440)$ and multihadron starting with $N ({\bm p}=1) \pi({\bm p}=-1)$ 
and $N (0) \pi(0) \pi(0)$).   
Furthermore, the ordering depends on $M_\pi$---for $M_\pi < 200$~MeV and 
the value of the lowest non-zero lattice momentum ($\approx 200$~MeV  
for all 13 ensembles),  $N ({\bm p}=1) \pi({\bm p}=-1)$ 
and $N (0) \pi(0) \pi(0)$ are the lightest positive
parity states that contribute to the forward matrix elements. Fortunately, 
the non-interacting energies for these two states are roughly 
equal, $M^{N\pi} \approx M^{N\pi\pi}$, 
and since Eq.~\eqref{eq:3pt} only cares about the mass, $M_i$, the contribution of the 
two states can be effectively combined. In the $\{4^{N\pi},3^*\}$ strategy 
described below, this joint contribution should 
be considered implicit even though we use the 
single label $N\pi$. In $\{4^{N\pi}\}$ fits to 
the $C^{\rm 2pt}$, their non-interacting energy $E_{N\pi} = E_\pi + E_N$ is 
input for $M_1$ using narrow width Bayesian priors.\looseness-1

To investigate the sensitivity of
$\matrixe{0}{\mathcal{O}}{0}$ to possible values of $M_i$,  
we carry out the full analysis for the following three strategies 
using the mnemonic $\{m,n\}$ to denote an $m$-state fit to the
two-point function and an $n$-state fit to the three-point function:\looseness-1
\begin{itemize}
\item
$\{4,3^*\}$: The values of ${\cal A}_0$ and $M_0, M_1, M_2$ are taken 
from a $\{4\}$ state fit to the two-point function using Eq.~\eqref{eq:2pt} and 
then a $\{3^\ast\}$ fit is made to the three-point function using
Eq.~\eqref{eq:3pt}. The notation $3^\ast$ denotes a three-state fit 
with the term involving 
$\matrixe{2}{\mathcal{O}}{2}$ neglected. Both fits are 
made within a single jackknife loop.
This is labeled the standard strategy---it assumes that the same set of
states are dominant in the two- and three-point functions.
\item
$\{4^{N\pi},3^*\}$: The values of  ${\cal A}_0$ and $M_0, M_1$ are taken from a 
four-state fit to the two-point function. The energy of a non-interacting $N
({\bm p}=1) \pi({\bm p}=-1)$ state (or $N (0) \pi(0) \pi(0)$ that has
roughly the same energy) is input using a narrow prior and designated as  the first excited state mass. The $M_1$ from the $\{4\}$ state fit 
is designated as the mass of the second excited state.  
This spectrum is then used in a
$\{3^\ast\}$ fit to the three-point function, i.e., this strategy 
assumes that the lowest of the theoretically
allowed tower of $N \pi$ (or $N \pi \pi $) states, $M_1^{4^{N\pi}}$,  
contributes significantly.
\item
$\{4,2^{\rm free}\}$: The only parameters taken from the $\{4\}$ state
fit are the ground state amplitude ${\cal A}_0$ and the mass $M_0$, whose 
determination is robust---the variation in $M_0$ between
$\{4\}$ and $\{4^{N\pi}\}$ is $\lesssim 1\%$ as shown in
Table~\ref{tab:massgap}. In this 
two-state fit to the three-point function, the mass of the first
excited state, $M_1$, is left as a free parameter, i.e., the most
important determinant of ESC, $M_1$, is obtained from the fit to the
three-point function itself, making it a data driven strategy.  
The relative limitation of the $\{4,2^{\rm free}\}$
strategy is that, with the current data, we can only make two-state
fits to the three-point functions, i.e., include only one 
excited state, and the output $M_1$ has large errors 
as shown in Table~\ref{tab:massgap}.
\end{itemize}

The data for the ratios for the three operators, $R_\mathcal{O}(\tau;t)$, 
are shown in Figs.~\ref{fig:Ratio-mom-1}--\ref{fig:Ratio-transversity-2} 
in Appendix~\ref{sec:ratios} for 
the eight new ensembles along with fits using the three
strategies. Plots for the other five, unchanged, ensembles  
were published in Ref.~\cite{Mondal:2020ela}. 

In choosing the set of points,
$\{t,\tau\}$, to include in the final fit, we balance between 
statistical and systematic errors. First, we neglect $t_{\rm skip}$
points next to the source and sink in the fits as these have the
largest ESC. Second, noting that the data at smaller $\tau$ have smaller
errors but larger ESC, we have mostly picked the 
largest three values of $\tau$ shown 
in red in Table~\ref{tab:ensembles}. Third, since errors in the data grow with
$\tau$, we partially compensate for the larger weight given to smaller
$\tau$ data (smaller errors) by choosing $t_{\rm skip}$ to be the same for all $\tau$,
i.e., by including increasingly more $t$ points at larger $\tau$,
the weight of the larger $\tau$ data points is increased.  Note that the $3^\ast$-fit 
in the two strategies $\{4,3^*\}$ and $\{4^{N\pi},3^*\}$ denotes that the 
$\langle 2 | {\mathcal O} | 2\rangle$ term is set to zero as it is
essentially undetermined and its inclusion results in an overparameterization
based on the AIC score~\cite{1100705}.

The data exhibit the following noteworthy features:\looseness-1
\begin{itemize}
\item
The fractional  errors (statistical) are less than $2\%$ for all three
operators and on all thirteen ensembles.  The only exceptions are the
data for $\tau =\{21, 30\}$ on the $\{a067m135,\ a053m230\}$  
ensembles, respectively. For this reason, data at these two $\tau$ are not 
included in the final fits used to remove ESC. 
\item
The errors grow, on average, by roughly a factor of two for every 
two units increase in $\tau/a$. This is consistent with the 
asymptotic growth factor, $e^{(M_N - 3M_\pi/2) \tau}$, expected for nucleon correlation
functions. There is also a small increase in this factor between
$\langle x \rangle_{u-d} \to \langle x \rangle_{\Delta u- \Delta
d} \to \langle x \rangle_{\delta u- \delta d} $.
\item
The data for all three operators show symmetry about $t=\tau/2$ as required by the spectral decomposition of zero-momentum correlation functions with the 
same nucleon interpolating operators at the source and the sink. The two exceptions 
are the two large error cases already mentioned in the first item above.
\item
In all cases (operators and ensembles), the convergence of the data
towards the $\tau \to \infty$ value is monotonic and from above. Thus,
for all three operators, any remaining ESC will result in the  moments being overestimated.
\end{itemize}
With the data showing the expected behavior, we make the three
fits, $\{4,3^*\}$, $\{4^{N\pi},3^*\}$ and $\{4,2^{\rm free}\}$, to
the largest three values of $\tau$ in all but the two cases mentioned above. We have
also checked that the results from fits keeping the largest four
values of $\tau$ overlap with these within $1\sigma$. 
Each panel in Figs.~\ref{fig:Ratio-mom-1}--\ref{fig:Ratio-transversity-2} 
shows the result of
the fit and the $\tau=\infty$ value (blue band). 
The three panels in each row have the same data but compare 
fits with the three ESC strategies.  The
ensemble ID, the value of the unrenormalized moment obtained using
Eqs.~\eqref{eq:me2momentV}, or~\eqref{eq:me2momentA},
or~\eqref{eq:me2momentT}, the $\chi^2/$dof of the fit, and the values
of $\tau$ for which the data were collected are also given in the figure
labels. The scale of the y-axis is chosen to be the same for all the
plots to facilitate comparison.  

The values of $\Delta M_1$ entering/determined by the various fits are
given in Table~\ref{tab:massgap}. They display the following
qualitative features:
\begin{itemize}
\item 
The values $a \Delta M_1^{\{4\}} \approx 0.6 aM_N^{\{4\}}$. This
suggests that the lowest excited state mass in the $\{4\}$-fit to the
two-point function is close to the $N(1440)$.
\item
$ a \Delta M_1^{\{4^{N\pi}\}}$ is significantly smaller than $a \Delta
M_1^{\{4\}}$ for the $M_\pi \le 230$~MeV ensembles as explained above.
\item
On seven ensembles, $a \Delta M_1^{\{2^{\rm free}\}}$ from fits to the
momentum-fraction data are consistent with $a \Delta M_1^{\{2\}}$. (
Table~\ref{tab:massgap} gives both $a \Delta M_1^{\{2\}}$ and
$a \Delta M_1^{\{4\}}$ to highlight the variation between a two- and
four-state fit.)  To check whether this rough agreement holds
for all ensembles, we made fits with a range of priors 
but did not find a
flat direction with respect to $a \Delta M_1^{\{2^{\rm
free}\}}$. Further understanding of $a \Delta M_1^{\{2^{\rm
free}\}}$ needs higher precision data that would allow a 3-state fit. 
\item
The $a \Delta M_1^{\{2^{\rm free}\}}$ for helicity and transversity
moments are roughly the same and much larger than  $a \Delta
M_1^{\{2\}}$ or even the $a \Delta M_1^{\{2^{\rm free}\}}$ for the momentum fraction---a contraindication for the  $\{4^{N\pi},3^*\}$ strategy.
\end{itemize}

The results for the unrenormalized moments obtained using
Eqs.~\eqref{eq:me2momentV},~\eqref{eq:me2momentA},
and~\eqref{eq:me2momentT} are given in Table~\ref{tab:ESC-fits}.  The parameters 
and the $\chi^2/$dof of the fits for the various strategies are given in
Tables~\ref{tab:5strategy-fits-momfrac},~\ref{tab:5strategy-fits-helfrac},
and~\ref{tab:5strategy-fits-transvmom} in Appendix~\ref{sec:rawfits}. 
In these tables, we reproduce
results with $\{4,2\}$ and $\{4^{N\pi},2\}$ strategies for the five unchanged 
ensembles  published in~\cite{Mondal:2020ela}. There we compared these 
to the $\{4,3^\ast\}$, $\{4^{N\pi},3^\ast\}$ and $\{4,2^{\rm free}\}$ 
results and found that the change on including the second excited state is
small, i.e., $\Delta M_1$ is the  dominant parameter to get right to
 control ESC.

We draw the following conclusions from the results presented in
Tables~\ref{tab:massgap}--\ref{tab:5strategy-fits-transvmom} and the
fits shown in
Figs.~\ref{fig:Ratio-mom-1}--\ref{fig:Ratio-transversity-2}:
\begin{itemize}
\item
The statistics on the $a067m135$ and $a053m230$ ensembles need to be
increased further to improve the CCFV fits. 
\item
The $\chi^2/$dof of most fits are reasonable. 
\item 
The $\{4,2^{\rm free}\}$ fits have reasonable $\chi^2/$dof but do not
prefer the small $\Delta M_1^{N\pi}$ given in
Table~\ref{tab:massgap}. The output $\Delta M_1^{\rm free}$ mostly lie significantly higher
than even $\Delta M_1^{\{2\}}$.
\item
The $\Delta M_1$ from a two-state fit is expected to be larger since
it is an effective  mass gap including the contributions of all 
excited states. This is illustrated by the difference between $\Delta
M_1^{\{2\}}$ and $\Delta M_1^{\{4\}}$.  Consequently,  we take the values
$\Delta M_1^{\{4^{N\pi}\}}$ and $\Delta M_1^{\{2^{\rm free}\}}$ to bracket
possible values of $\Delta M_1 $, and in each case 
$\Delta M_1^{\{4\}}$ lies in between.
\end{itemize}
Based on the above arguments, we will choose the $\{4,3^\ast\}$
results obtained after performing the CCFV fits for the final central
value.  We will also assign half the spread between the
$\{4^{N\pi},3^\ast\}$ and $\{4,2^{\rm free}\}$ values as a conservative 
estimate of the systematic uncertainty due to possible remaining ESC.


\begin{figure*}[htbp]   
\begin{subfigure}
    \centering
\includegraphics[trim=0 30 0 185,angle=0,width=0.96\textwidth]{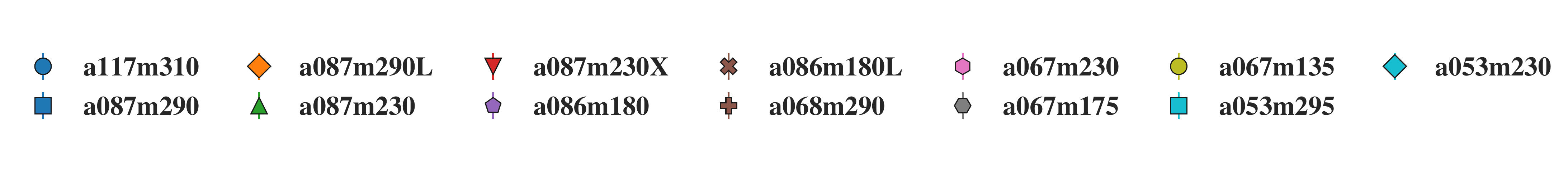}
\end{subfigure}
\vskip -0.2 cm
\begin{subfigure}
\centering
\includegraphics[angle=0,width=0.32\textwidth]{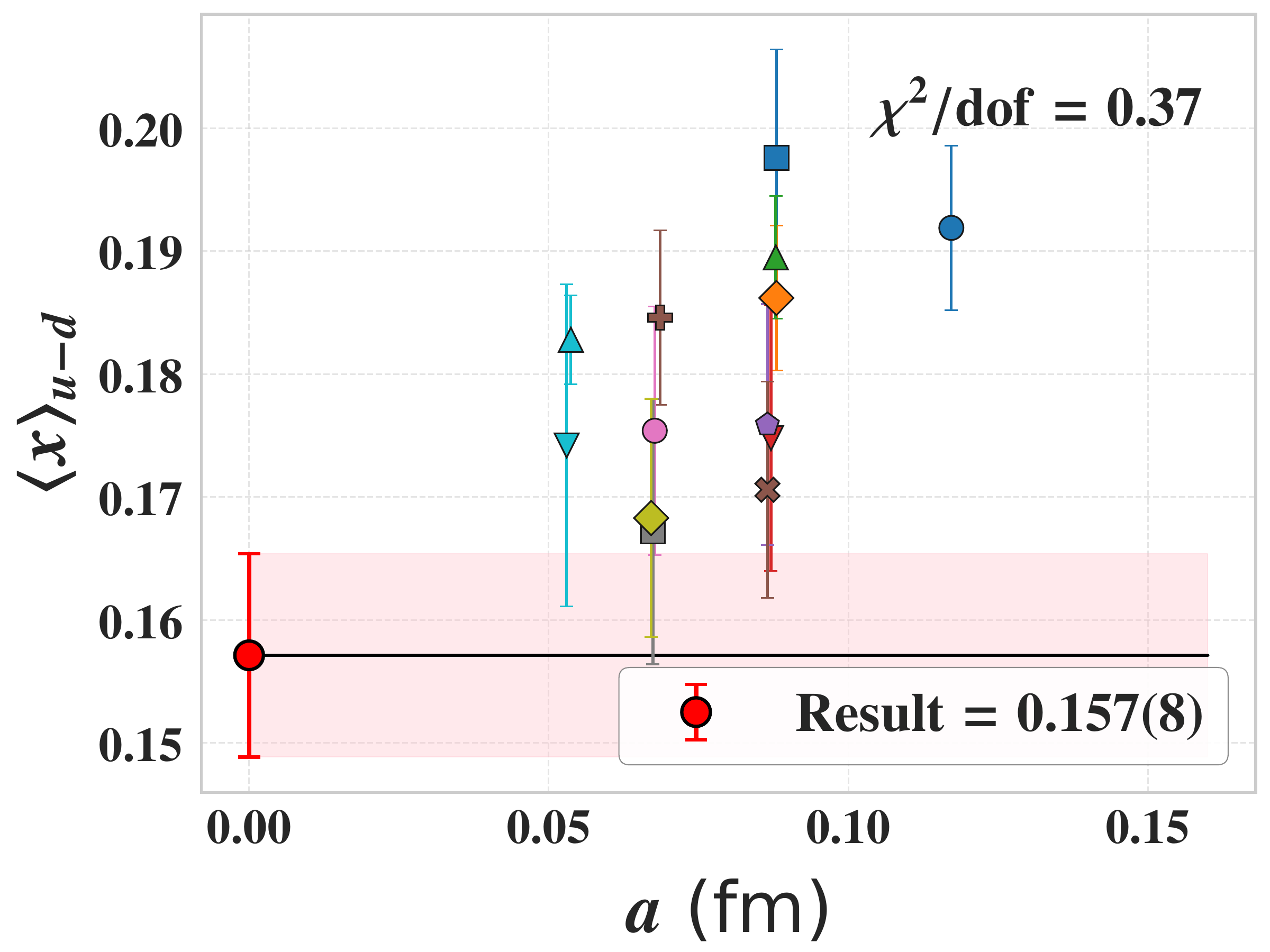}
\includegraphics[angle=0,width=0.32\textwidth]{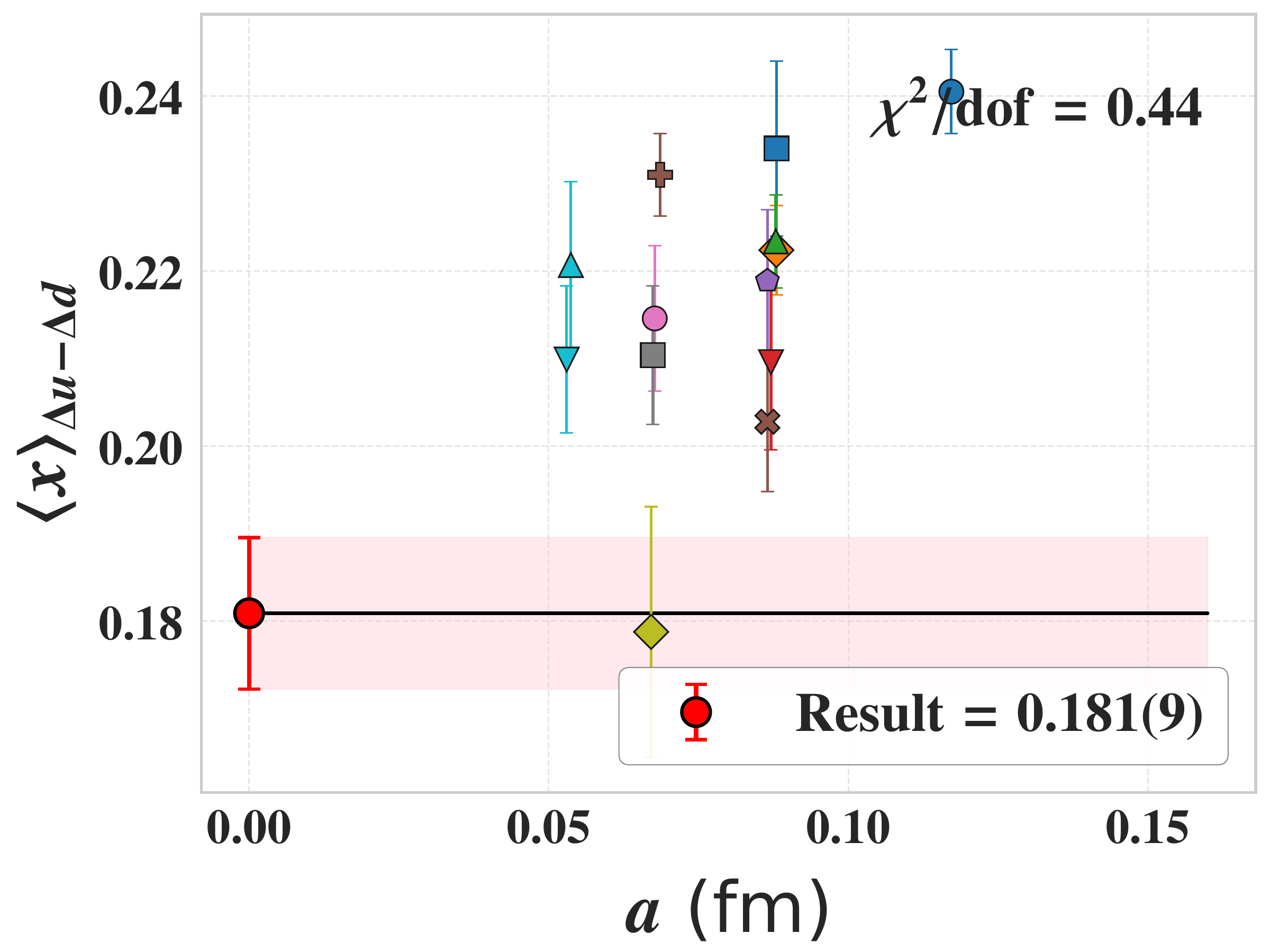}
\includegraphics[angle=0,width=0.32\textwidth]{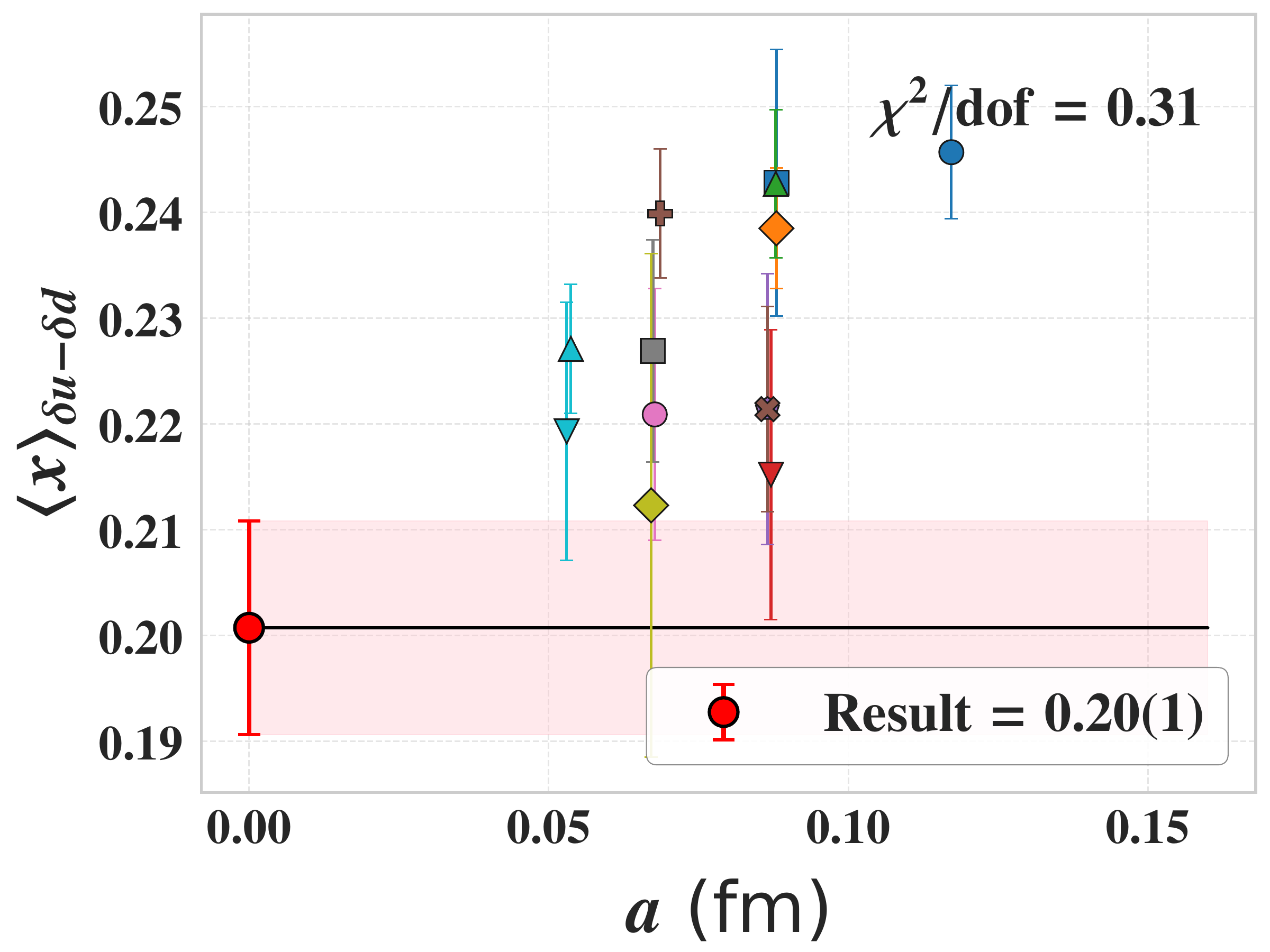}
\end{subfigure}

\begin{subfigure}
\centering
\includegraphics[angle=0,width=0.32\textwidth]{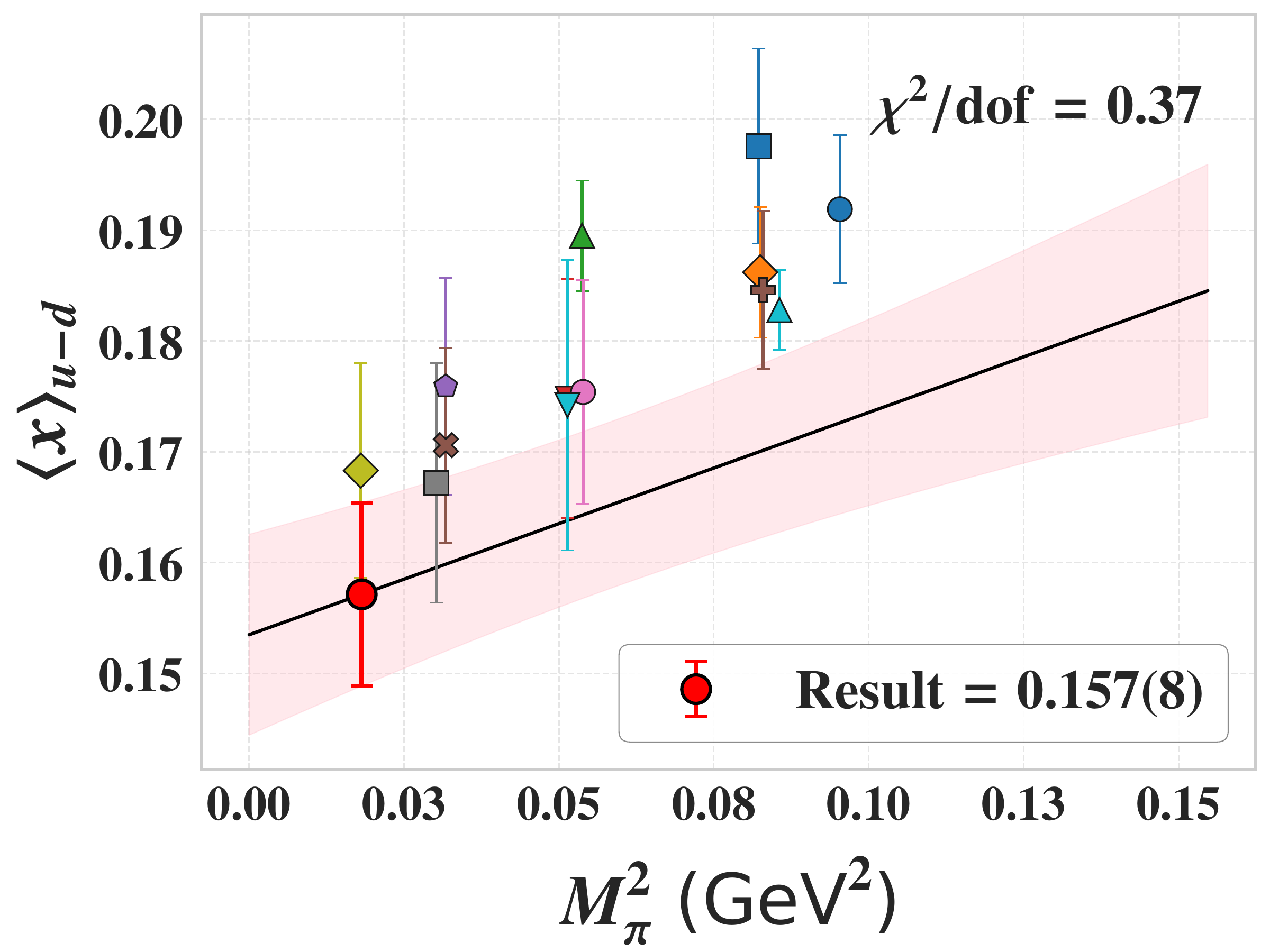}
\includegraphics[angle=0,width=0.32\textwidth]{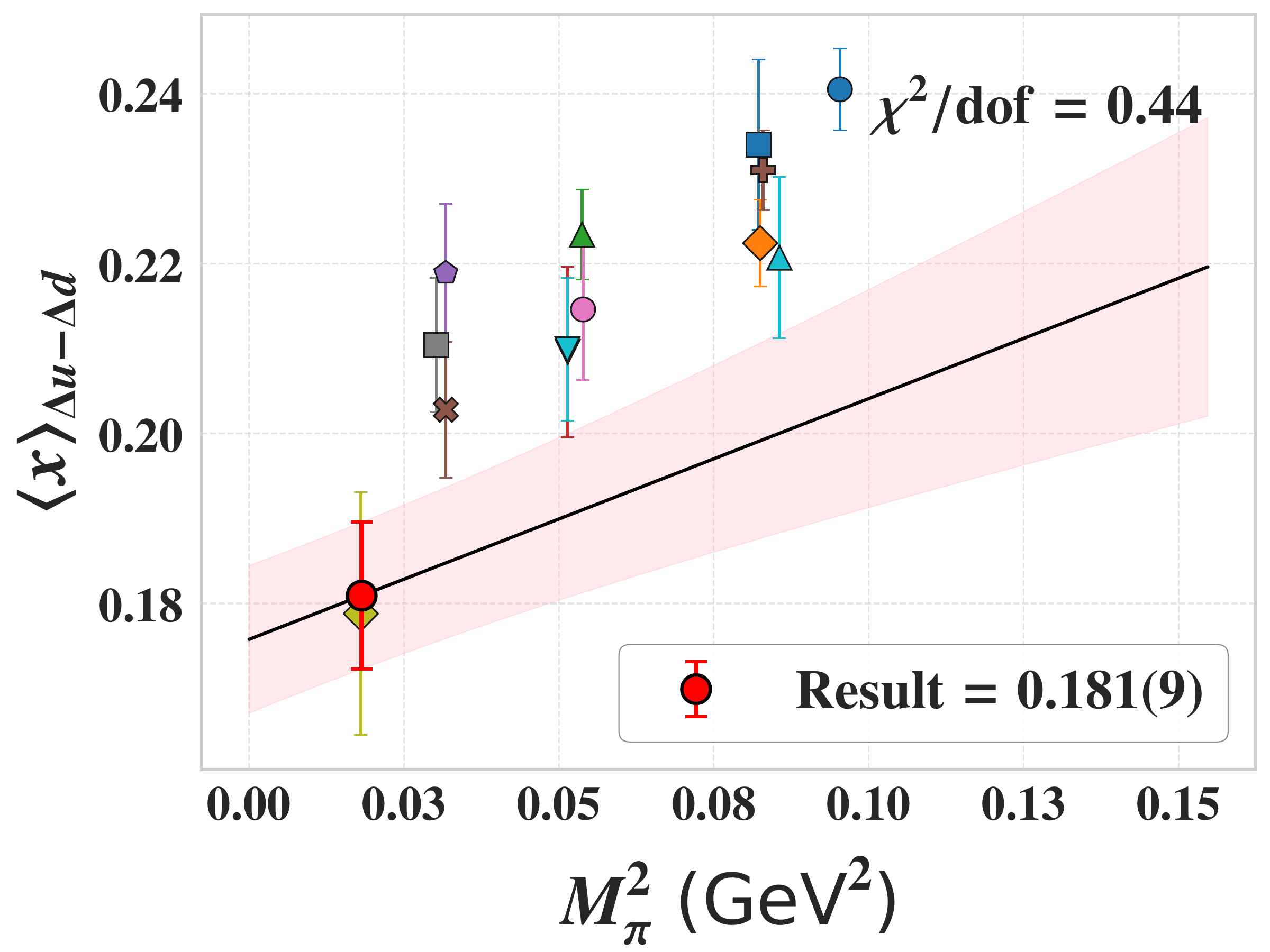}
\includegraphics[angle=0,width=0.32\textwidth]{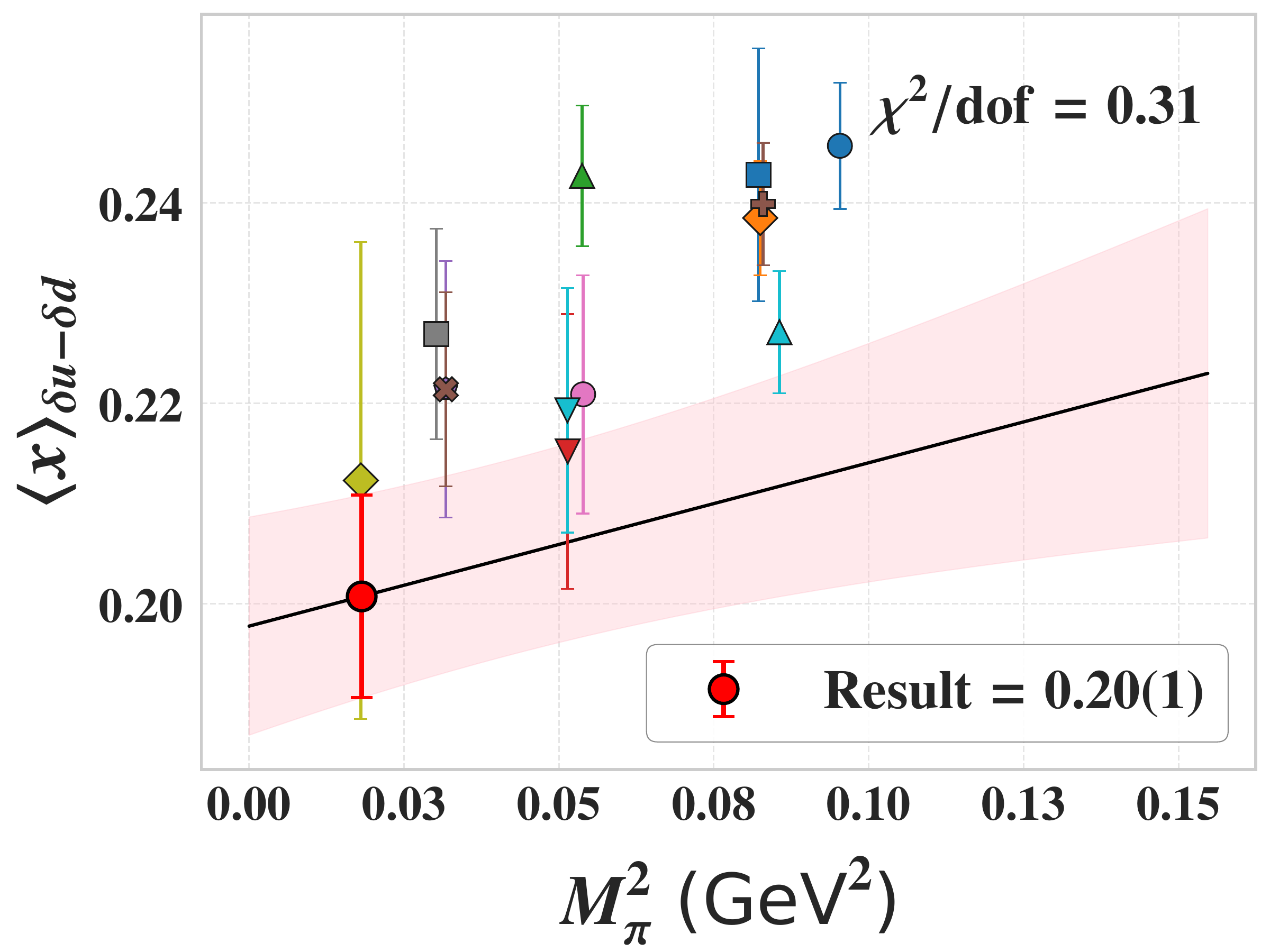}
\end{subfigure}
\begin{subfigure}
\centering
\includegraphics[angle=0,width=0.32\textwidth]{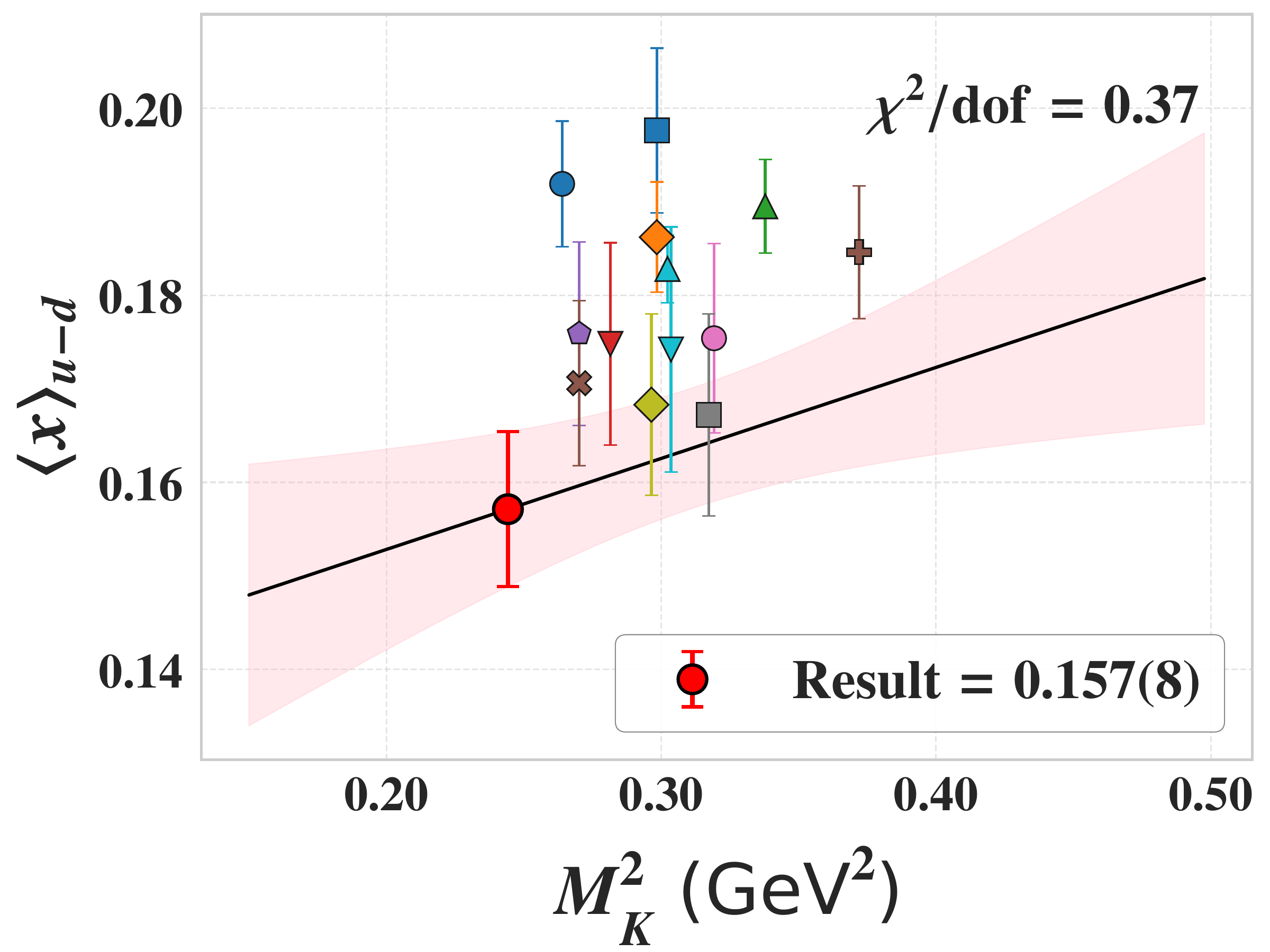}
\includegraphics[angle=0,width=0.32\textwidth]{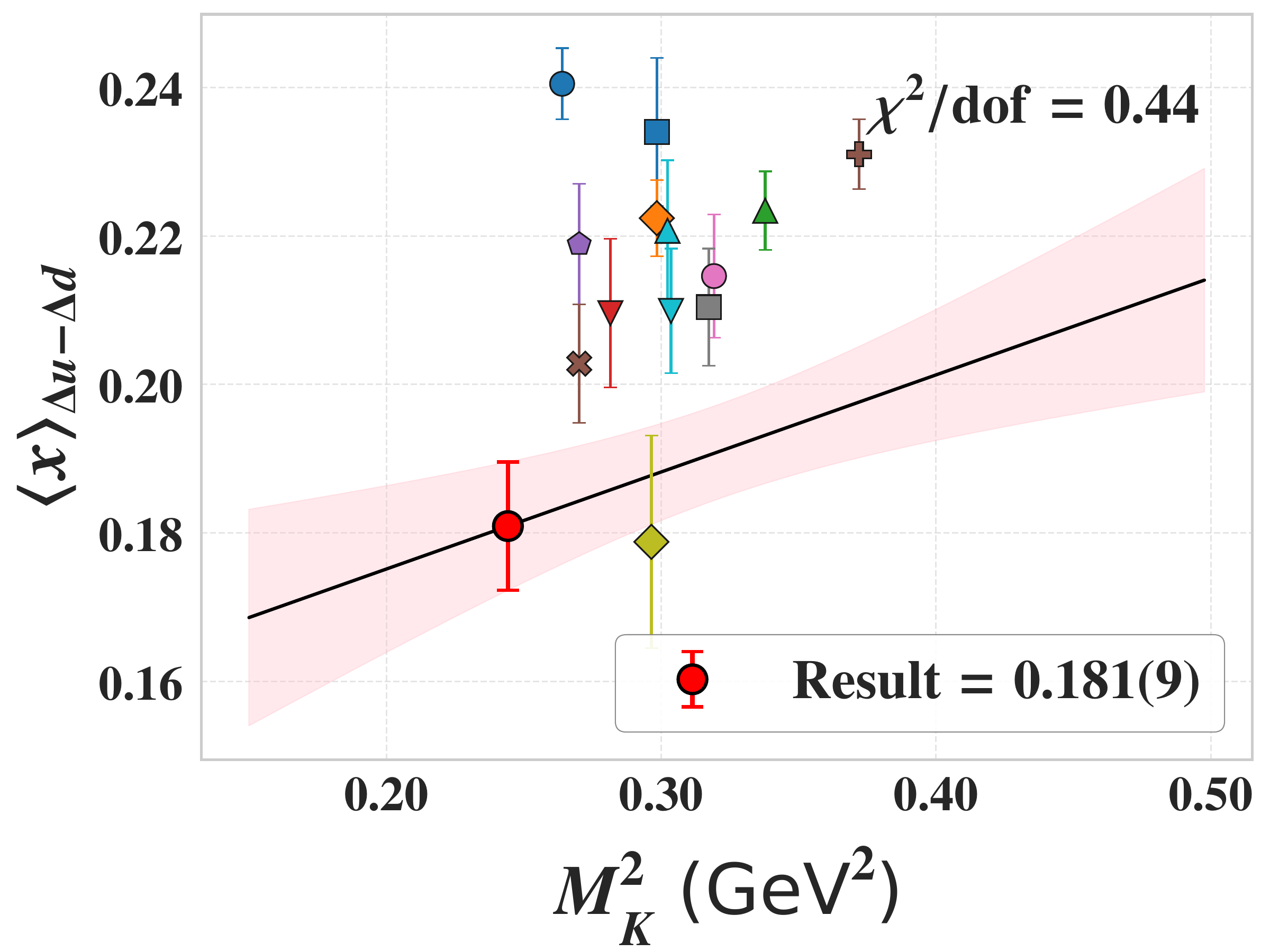}
\includegraphics[angle=0,width=0.32\textwidth]{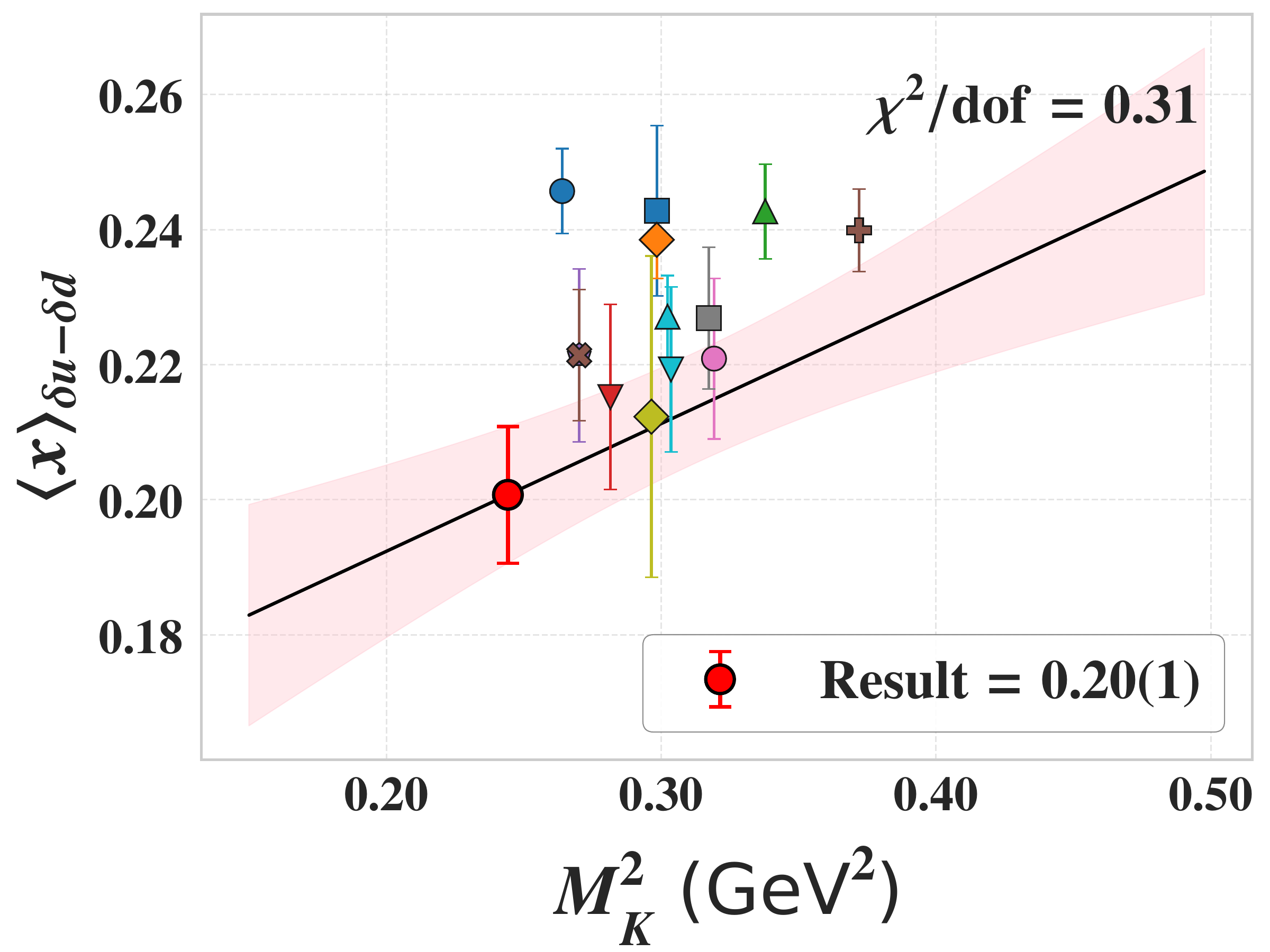}
\end{subfigure}

\begin{subfigure}
\centering
\includegraphics[angle=0,width=0.32\textwidth]{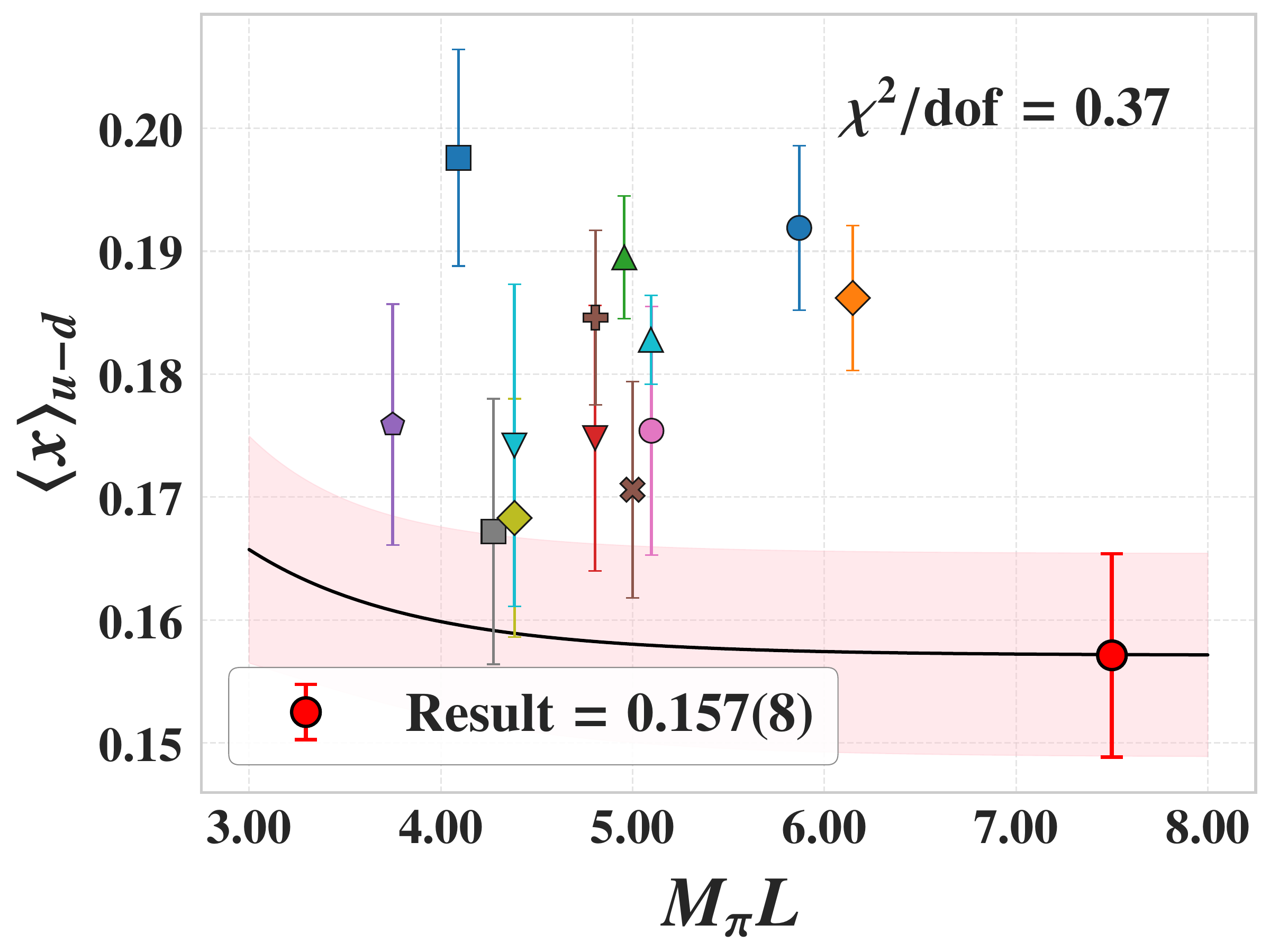}
\includegraphics[angle=0,width=0.32\textwidth]{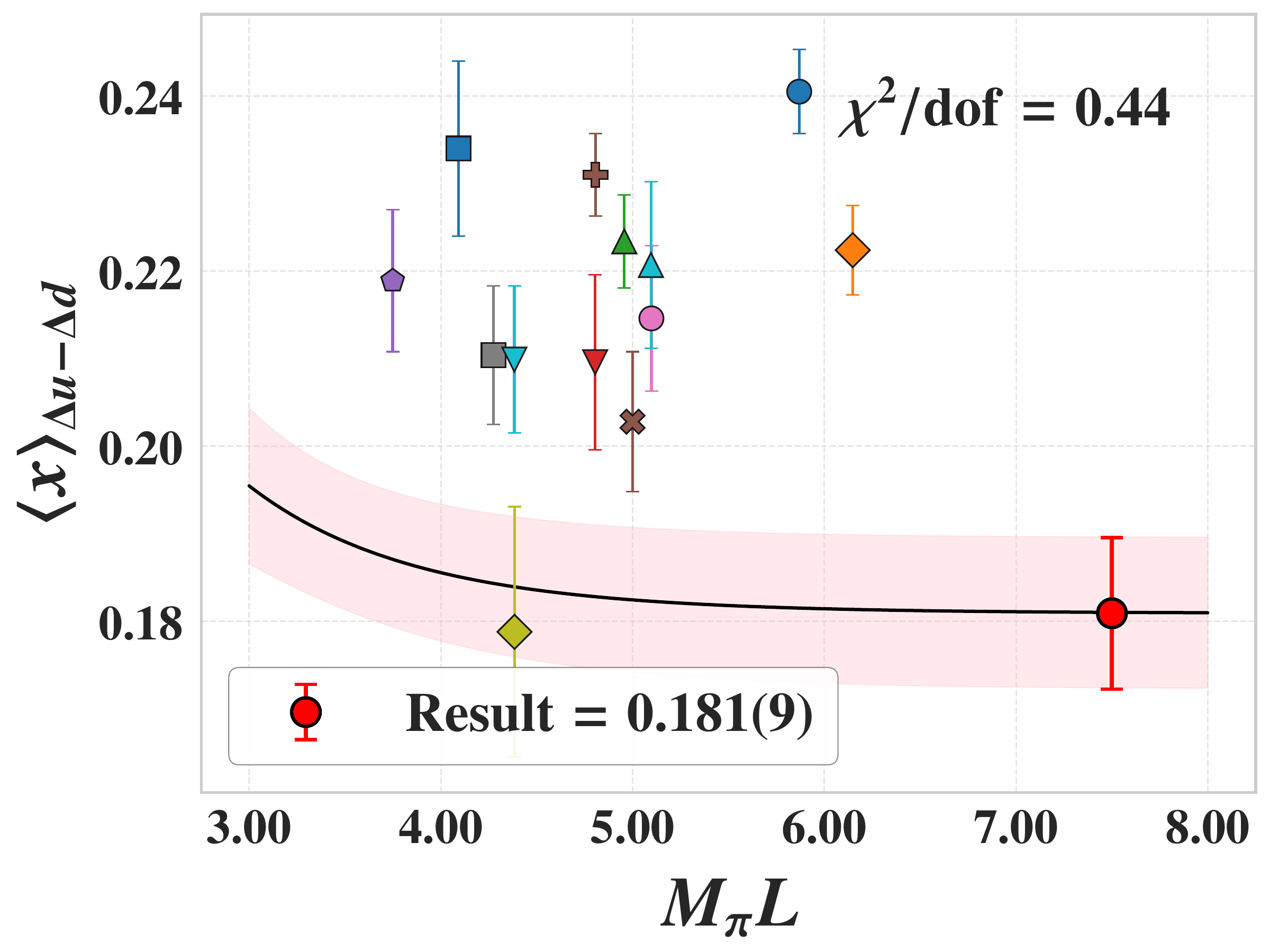}
\includegraphics[angle=0,width=0.32\textwidth]{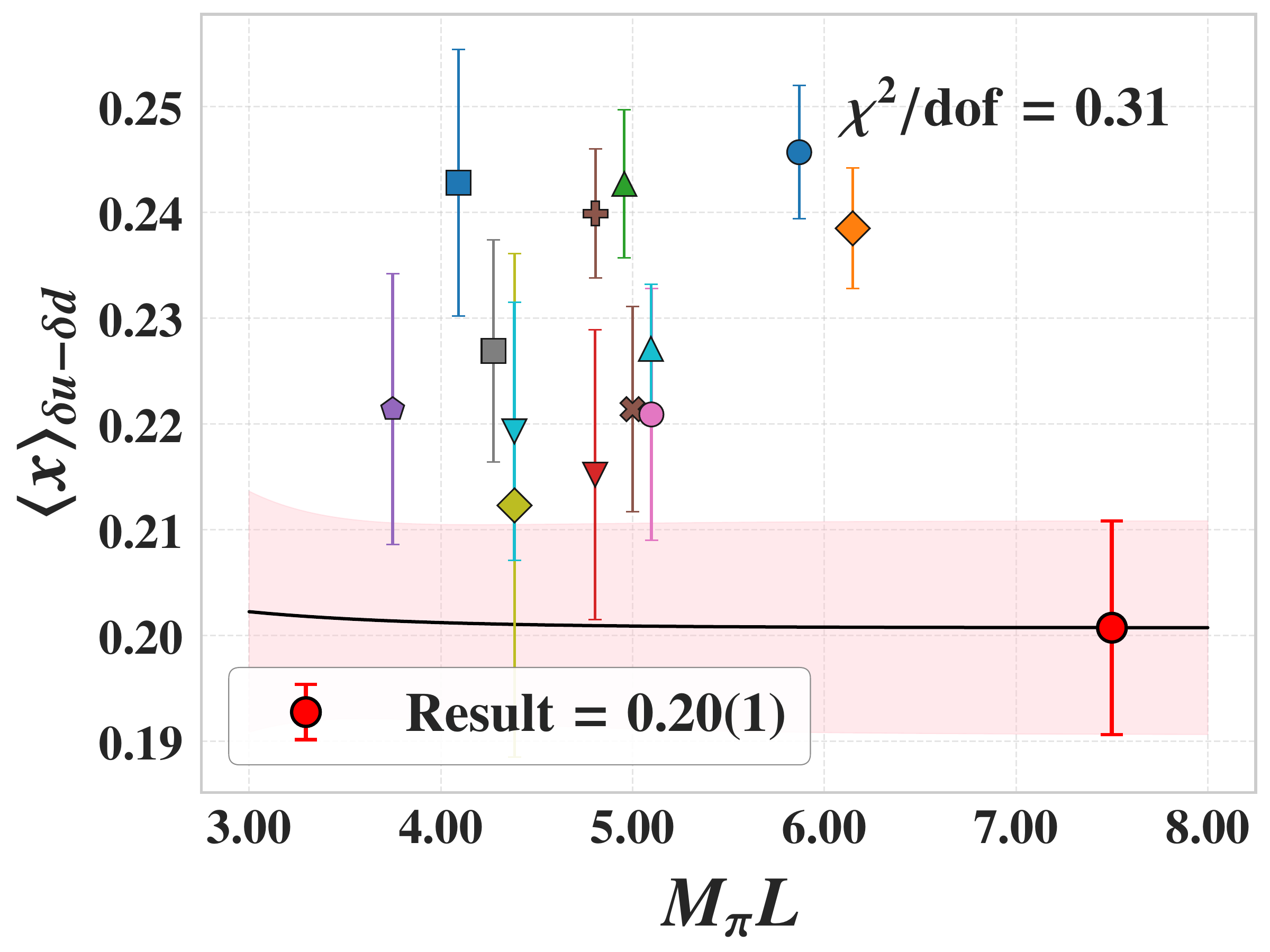}
\end{subfigure}
\vspace{-0.08in}
\caption{Data for the momentum fraction $\la x \ra_{u-d}$ (left panels), 
helicity moment $\la x \ra_{\Delta u- \Delta d}$ (middle panels), 
and transversity moment $\la x \ra_{\delta u- \delta d}$ (right panels) from
the thirteen ensembles using strategy $\{4,3^*\}$ to remove ESC, 
method A to renormalize moments  in the $\MSbar$ scheme at $\mu=2$ GeV, 
and with the discretization ansatz $f(a) = \alpha_s a$.   
The pink band shows the result of the CCFV fit, defined in Eq.~\protect\eqref{eq:CCFV}, 
versus $a$ (top row), versus $M_\pi^2$ (second row), 
versus $M_K^2$ (third row) and versus
$M_\pi L$ (bottom row) with the other three variables set to their
physical values in each case. In each panel, the data are not shifted to 
account for the corrections due to the other three variables, nevertheless, the 
correction can be read as roughly the difference between the data point and the result 
of the simultaneous fit (pink band with the 
central value given by the black line). 
The desired value at the physical point is shown by the symbol 
filled red circle. The listed $\chi^2$/dof is reasonable for all the fits. 
The labels for the data symbols are given in the very top panel. }
\label{fig:CCFV_explicit}
\end{figure*}

To get the renormalized values of the moments, the renormalization
factors, $Z_{VD},~Z_{AD}$, and $ Z_{TD}$, are calculated on the
lattice in the \ripmom scheme, converted to the $\MSbar$
scheme, and run to 2~GeV as discussed in Appendix~\ref{sec:renormalization}. 
The two methods used to control discretization errors in the
calculation of the $Z$'s are also described in
Appendix~\ref{sec:renormalization},  and the final values of
$Z_{VD},~Z_{AD}$, and $ Z_{TD}$ used in the analysis are given in
Table \ref{tab:Z-fac}.  The values of the three renormalized moments
from the thirteen ensembles and with the three ESC strategies are summarized
in Table~\ref{tab:renormalized-moments-A} for renormalization method A
and in Table~\ref{tab:renormalized-moments-B} for method B.
These data are used to perform the CCFV fits discussed next.

\begin{table*}[htbp]   
\centering
\begin{adjustbox}{width=1\textwidth}
  \def\hline{\noalign{\kern1pt\vrule\kern1pt}}
  \catcode`|\active
  \def|{\omit\hrule&}
\def\\{&\omit\hrule\cr}
\leavevmode{
\valign{&\hbox spread3pt{\hfil\strut\ignorespaces#\unskip\hfil}\cr\noalign{\vrule}%
\hline 
|moment &| strategy &| $a117m310$ & $a087m290$ & $a087m290L$ & $a087m230$ & $a087m230X$ & $a086m180$ & $a086m180L$ & $a068m290$ & $a067m230$ & $a067m175$ & $a067m135$ & $a053m295$ & $a053m230$ \\
\hline
\hline
|$\langle x\rangle_{u-d}$ &| $\{4^{N\pi},3^*\}$ &| $0.1828(49)$ & $0.1976(78)$ & $0.1799(49)$ & $0.1790(108)$ & $0.1591(190)$ & $0.1558(149)$ & $0.1579(264)$ & $0.1824(52)$ & $0.1668(90)$ & $0.1640(108)$ & $0.1672(108)$ & $0.1775(43)$ & $0.1657(100)$ \\
\hline
|$\langle x\rangle_{u-d}$ &| $\{4,3^*\}$ &| $0.1919(67)$ & $0.1976(88)$ & $0.1862(59)$ & $0.1895(50)$ & $0.1748(108)$ & $0.1759(98)$ & $0.1706(88)$ & $0.1846(71)$ & $0.1754(101)$ & $0.1672(108)$ & $0.1683(97)$ & $0.1828(36)$ & $0.1742(131)$ \\
\hline
|$\langle x\rangle_{u-d}$ &| $\{4,2^{\rm free}\}$ &| $0.1879(76)$ & $0.2111(60)$ & $0.1893(41)$ & $0.1895(78)$ & $0.1895(78)$ & $0.1780(129)$ & $0.1780(129)$ & $0.1955(45)$ & $0.1872(81)$ & $0.1651(203)$ & $0.1917(108)$ & $0.1849(44)$ & $0.1732(90)$ \\
\hline
|$\langle x\rangle_{\Delta u - \Delta d}$ &| $\{4^{N\pi},3^*\}$ &| $0.2415(64)$ & $0.2340(100)$ & $0.2192(43)$ & $0.2245(91)$ & $0.1990(141)$ & $0.2092(143)$ & $0.1888(248)$ & $0.2310(55)$ & $0.2091(83)$ & $0.2071(132)$ & $0.2060(143)$ & $0.2164(58)$ & $0.2002(104)$ \\
\hline
|$\langle x\rangle_{\Delta u - \Delta d}$ &| $\{4,3^*\}$ &| $0.2405(48)$ & $0.2340(100)$ & $0.2224(51)$ & $0.2234(53)$ & $0.2096(100)$ & $0.2189(81)$ & $0.2028(80)$ & $0.2310(47)$ & $0.2146(83)$ & $0.2104(79)$ & $0.1788(143)$ & $0.2207(95)$ & $0.2099(84)$ \\
\hline
|$\langle x\rangle_{\Delta u - \Delta d}$ &| $\{4,2^{\rm free}\}$ &| $0.2508(44)$ & $0.2530(63)$ & $0.2287(52)$ & $0.2149(111)$ & $0.2351(72)$ & $0.2318(82)$ & $0.2210(91)$ & $0.2442(42)$ & $0.2289(64)$ & $0.2115(176)$ & $0.2256(133)$ & $0.2272(231)$ & $0.2240(58)$ \\
\hline
|$\langle x\rangle_{\delta u - \delta d}$ &| $\{4^{N\pi},3^*\}$ &| $0.2499(82)$ & $0.2418(115)$ & $0.2363(57)$ & $0.2525(235)$ & $0.2196(223)$ & $0.2026(203)$ & $0.2303(182)$ & $0.2399(70)$ & $0.2141(108)$ & $0.2168(171)$ & $0.2134(216)$ & $0.2226(75)$ & $0.2048(152)$ \\
\hline
|$\langle x\rangle_{\delta u - \delta d}$ &| $\{4,3^*\}$ &| $0.2457(63)$ & $0.2428(126)$ & $0.2385(57)$ & $0.2427(70)$ & $0.2152(137)$ & $0.2214(128)$ & $0.2214(97)$ & $0.2399(61)$ & $0.2209(119)$ & $0.2269(105)$ & $0.2123(238)$ & $0.2271(61)$ & $0.2193(122)$ \\
\hline
|$\langle x\rangle_{\delta u - \delta d}$ &| $\{4,2^{\rm free}\}$ &| $0.2615(64)$ & $0.2646(77)$ & $0.2483(44)$ & $0.2503(89)$ & $0.2471(119)$ & $0.2402(109)$ & $0.2181(214)$ & $0.2581(49)$ & $0.2323(120)$ & $0.2506(84)$ & $0.2642(85)$ & $0.2304(68)$ & $0.2326(85)$ \\
\hline
}\unskip}
\end{adjustbox}
\caption{Renormalized moments for the three ESC strategies using the $Z$ factors obtained using Method A.}
\label{tab:renormalized-moments-A}
\end{table*}


\begin{table*}[htbp]   
\centering
\begin{adjustbox}{width=1\textwidth}
  \def\hline{\noalign{\kern1pt\vrule\kern1pt}}
  \catcode`|\active
  \def|{\omit\hrule&}
\def\\{&\omit\hrule\cr}
\leavevmode{
\valign{&\hbox spread3pt{\hfil\strut\ignorespaces#\unskip\hfil}\cr\noalign{\vrule}%
\hline 
|moment &| strategy &| $a117m310$ & $a087m290$ & $a087m290L$ & $a087m230$ & $a087m230X$ & $a086m180$ & $a086m180L$ & $a068m290$ & $a067m230$ & $a067m175$ & $a067m135$ & $a053m295$ & $a053m230$ \\
\hline
\hline
|$\langle x\rangle_{u-d}$ &| $\{4^{N\pi},3^*\}$ &| $0.1765(51)$ & $0.1928(77)$ & $0.1756(48)$ & $0.1737(106)$ & $0.1544(185)$ & $0.1505(145)$ & $0.1525(256)$ & $0.1789(56)$ & $0.1603(89)$ & $0.1545(102)$ & $0.1575(102)$ & $0.1713(56)$ & $0.1600(103)$ \\
\hline
|$\langle x\rangle_{u-d}$ &| $\{4,3^*\}$ &| $0.1852(68)$ & $0.1928(86)$ & $0.1817(57)$ & $0.1839(51)$ & $0.1697(105)$ & $0.1698(96)$ & $0.1648(86)$ & $0.1811(74)$ & $0.1685(100)$ & $0.1575(102)$ & $0.1585(92)$ & $0.1765(52)$ & $0.1682(132)$ \\
\hline
|$\langle x\rangle_{u-d}$ &| $\{4,2^{\rm free}\}$ &| $0.1813(76)$ & $0.2060(59)$ & $0.1847(41)$ & $0.1839(77)$ & $0.1839(77)$ & $0.1719(125)$ & $0.1719(125)$ & $0.1917(51)$ & $0.1799(82)$ & $0.1555(191)$ & $0.1805(103)$ & $0.1785(57)$ & $0.1672(94)$ \\
\hline
|$\langle x\rangle_{\Delta u - \Delta d}$ &| $\{4^{N\pi},3^*\}$ &| $0.2333(67)$ & $0.2269(98)$ & $0.2126(44)$ & $0.2175(91)$ & $0.1928(138)$ & $0.2026(140)$ & $0.1829(241)$ & $0.2247(60)$ & $0.2006(83)$ & $0.1959(127)$ & $0.1949(137)$ & $0.2068(76)$ & $0.1913(110)$ \\
\hline
|$\langle x\rangle_{\Delta u - \Delta d}$ &| $\{4,3^*\}$ &| $0.2323(53)$ & $0.2269(98)$ & $0.2156(52)$ & $0.2165(56)$ & $0.2031(99)$ & $0.2120(81)$ & $0.1964(80)$ & $0.2247(53)$ & $0.2058(83)$ & $0.1990(77)$ & $0.1691(136)$ & $0.2109(105)$ & $0.2006(95)$ \\
\hline
|$\langle x\rangle_{\Delta u - \Delta d}$ &| $\{4,2^{\rm free}\}$ &| $0.2423(50)$ & $0.2453(63)$ & $0.2218(52)$ & $0.2083(109)$ & $0.2279(74)$ & $0.2244(82)$ & $0.2140(91)$ & $0.2375(49)$ & $0.2195(67)$ & $0.2000(167)$ & $0.2134(127)$ & $0.2171(227)$ & $0.2140(78)$ \\
\hline
|$\langle x\rangle_{\delta u - \delta d}$ &| $\{4^{N\pi},3^*\}$ &| $0.2428(82)$ & $0.2375(117)$ & $0.2322(63)$ & $0.2472(232)$ & $0.2150(220)$ & $0.1976(199)$ & $0.2246(179)$ & $0.2367(74)$ & $0.2085(108)$ & $0.2079(165)$ & $0.2047(208)$ & $0.2154(101)$ & $0.1982(161)$ \\
\hline
|$\langle x\rangle_{\delta u - \delta d}$ &| $\{4,3^*\}$ &| $0.2387(65)$ & $0.2386(127)$ & $0.2343(63)$ & $0.2376(74)$ & $0.2107(137)$ & $0.2160(127)$ & $0.2160(98)$ & $0.2367(66)$ & $0.2151(119)$ & $0.2177(102)$ & $0.2036(229)$ & $0.2197(93)$ & $0.2122(137)$ \\
\hline
|$\langle x\rangle_{\delta u - \delta d}$ &| $\{4,2^{\rm free}\}$ &| $0.2540(66)$ & $0.2600(82)$ & $0.2440(52)$ & $0.2451(92)$ & $0.2419(119)$ & $0.2344(109)$ & $0.2128(210)$ & $0.2547(57)$ & $0.2261(120)$ & $0.2404(83)$ & $0.2534(84)$ & $0.2229(99)$ & $0.2251(111)$ \\
\hline
}\unskip}
\end{adjustbox}
\caption{Renormalized moments for the three ESC strategies using the $Z$ factors obtained using Method B.}
\label{tab:renormalized-moments-B}
\end{table*}


\section{Chiral, continuum and infinite volume extrapolation}
\label{sec:CCFV}

To obtain the final results at the  physical point defined by $M_\pi=135$~MeV, $M_K=494$~MeV, $M_\pi L \to
\infty$ and $a=0$, we make a simultaneous CCFV fit keeping only the
leading correction term in each variable. This truncated ansatz 
has five parameters, $c_i$, defined as follows
\begin{align}
 \la x \ra (M_\pi, M_K, a , L) &=& c_0   \nonumber \\
&+& c_1\, M_\pi^2 ~t_0 \nonumber \\
&+& c_2\,  (M_K^2 - M_\pi^2) ~ t_0 \nonumber \\
&+& c_3\, f(a) \nonumber \\
&+& c_4\, \frac{M_\pi^2 t_0~ e^{-M_\pi L}}{\sqrt{M_\pi L}}  \,.
\label{eq:CCFV}
\end{align}
Note that the flow scale $t_0$  used to make quantities dimensionless 
is also calculated in this study. 
Since our clover-on-clover formulation is only tadpole improved and the operators 
are not $O(a)$ improved, we explore three possibilities 
for the leading discretization correction factor $f(a)$: $a/\sqrt t_0$ or $\alpha_s~ a /\sqrt t_0$ or $a^2/t_0$. 
The chiral part includes the dependence on both $m_l$ and $m_s-m_l$ as the tuning of the strange quark mass 
on each of the thirteen ensembles is very approximate as shown  
by the values of $M_K$ in Table~\ref{tab:PSmasses}. 
 Having cast Eq.~\eqref{eq:CCFV} in terms of dimensionless variables, the only place where the value for the lattice scale $a$ on each ensemble 
 enters explicitly in the analysis is in the calculation of the renormalization constants discussed in Appendix~\ref{sec:renormalization}.

Figure~\ref{fig:CCFV_explicit} shows the data and the CCFV fit to the 13 data points for all three moments using 
strategy $\{4,3^*\}$ to remove ESC, renormalized  in the $\MSbar$
scheme at $\mu=2$ GeV using method A, and the discretization ansatz $\alpha_s a$. In each panel, the CCFV fit 
result (pink band) is plotted versus $a$, or 
$M_\pi^2$, or $M_K^2$ or $M_\pi L$  with the other 
three variables set to their physical values.
These plots show that the dependence on the 
lattice spacing, $a$, and finite 
volume parameter, $M_\pi L$, is small. The dependence on 
$M_\pi$ and $M_K$ is similar, which is not surprising 
since the Gell-Mann-Oakes-Renner relation, $M_{\rm PS}^2 \propto m_q$, is 
observed to work well for $m_q \le m_s$. Of these, the  variation of 
the data with $M_\pi^2$ is more significant as the range of variation of $M_\pi^2$ is 
larger. There is further dependence on $M_\pi^2$ in the $\{4^{N\pi},3^\ast\}$ 
analysis where the $\Delta M_1$ is $M_\pi$ dependent and the ES extrapolation 
becomes larger as $\Delta M_1$ decrease. This leads to a larger 
slope versus $M_\pi^2$ in the CCFV fits.   

We have performed a second extrapolation labeled CC to 
understand the finite-volume effects. In 
it we drop the finite volume correction term $\propto c_4$ 
and the two small volume
points $a087m290$ and $a086m180$ that also have the 
largest statistical errors. Comparisons between the predictions of the CCFV and 
CC fits with the data renormalized
using method A are shown in
Figs.~\ref{fig:momfrac_comp},~\ref{fig:helfrac_comp}
and~\ref{fig:trans_comp} in Appendix~\ref{sec:A_CCFV} 
for the three strategies, and in Table~\ref{tab:CCFVresults}. The only remarkable 
feature is that the CC fits 
do not capture the small-volume $a087m290$ and $a086m1890$ data that are excluded 
from the fit.  In the CCFV fit, the FV term $\propto c_4$  
is fixed mainly by these two data points.

All the results from the CCFV and CC fits for the various analysis models are
summarized in Tables~\ref{tab:CCFVresults}, ~\ref{tab:CCFV_13pt_fit} and~\ref{tab:CC_11pt_fits}.   
With these fits in hand, results for the moments at 
the physical point are obtained by setting 
$a=0$, $M_\pi L \to \infty$, $M_\pi=135$~MeV, $M_K=494$~MeV 
and  $\sqrt{t_0^{\rm phy}}=0.14474(57)$~fm taken 
from the FLAG 2024 report~\cite{FlavourLatticeAveragingGroupFLAG:2024oxs}.
%


\section{Model Averaging And Results}
\label{sec:Models}

The complete analysis for each moment is done for the following 36 choices (models) 
\begin{itemize}
    \item Three strategies, labeled $\{4^{N\pi},3^\ast\},\ \{4,3^\ast\}$ 
    and $ \{4,2^{\rm free}\}$, 
    to remove ESC as discussed in Sec.~\ref{sec:ESC}.  The data, the fits, and the results are 
    shown in Appendices~\ref{sec:ratios} and~\ref{sec:rawfits}.
    \item Two methods, labeled A and B, are used to calculate the renormalization constants as described in  Appendix~\ref{sec:renormalization}. Data for the  renormalized moments are given in Tables~\ref{tab:renormalized-moments-A} and~\ref{tab:renormalized-moments-B}. 
    \item Three ansatz for the discretization correction factor $f(a)$---$a/\sqrt t_0$ or $\alpha_s~ a /\sqrt t_0$ or $a^2/t_0$---are used. 
    \item The CCFV extrapolation is made using (i) the ansatz in Eq.~\eqref{eq:CCFV} 
    and (ii) a further truncated form labeled CC in which  $c_4=0$.
\end{itemize}

\begin{table*}[htbp]      
\centering
\setlength{\tabcolsep}{5pt}
\renewcommand{\arraystretch}{1.4}
\begin{tabular}{ |c|c|c|c|c|c|c|c| }
\hline
        &   Renorm  & \multicolumn{2}{|c|}{$\langle x \rangle_{u -d}$}   
                   & \multicolumn{2}{|c|}{$\langle x \rangle_{\Delta u- \Delta d}$} 
                   & \multicolumn{2}{|c|}{$\langle x \rangle_{\delta u- \delta d}$}  \\
strategy& Method               & CCFV         & CC   &  CCFV     & CC     &  CCFV     & CC   \\
\hline                                                                                                                                                                          
\hline                                                                                                                                                                          
$\{4^{N\pi},3^*\}$  & A   & $0.149(9)$  & $0.156(9)$  & $0.175(11)$ & $0.177(11)$ & $0.188(14)$ & $0.191(14)$ \\
$\{4^{N\pi},3^*\}$  & B   & $0.141(10)$ & $0.145(10)$ & $0.164(11)$ & $0.165(11)$ & $0.180(15)$ & $0.182(15)$ \\
$\{4^{N\pi},3^*\}$  & av. & $0.146(10)$ & $0.150(10)$ & $0.169(11)$ & $0.171(11)$ & $0.184(15)$ & $0.187(15)$ \\
\hline                                                                                                                                                                          
$\{4,3^*\}$         & A   & $0.158(9)$  & $0.160(9)$  & $0.182(10)$ & $0.183(9)$  & $0.202(11)$ & $0.202(11)$ \\
$\{4,3^*\}$         & B   & $0.151(10)$ & $0.153(10)$ & $0.171(10)$ & $0.172(10)$ & $0.193(12)$ & $0.193(12)$ \\
$\{4,3^*\}$         & av. & $0.154(10)$ & $0.157(10)$ & $0.177(10)$ & $0.178(10)$ & $0.197(12)$ & $0.197(12)$ \\
\hline                                                                                                                                                                          
$\{4,2^{\rm free}\}$ & A   & $0.167(9)$  & $0.173(9)$  & $0.201(10)$ & $0.207(9)$  & $0.221(9)$  & $0.225(9)$  \\
$\{4,2^{\rm free}\}$ & B   & $0.158(9)$  & $0.164(9)$  & $0.192(11)$ & $0.196(11)$ & $0.213(11)$ & $0.216(11)$ \\
$\{4,2^{\rm free}\}$ & av. & $0.162(9)$  & $0.169(9)$  & $0.196(11)$ & $0.201(11)$ & $0.217(11)$ & $0.221(11)$ \\
\hline
\end{tabular}
\caption{Summary of results for the three moments for the  
three strategies used to remove the excited state
contamination, the two methods (A and B) of renormalization 
discussed in appendix~\ref{sec:renormalization}, and the 
chiral-continuum-finite-volume (CCFV) and chiral-continuum (CC) fits. 
In each case, the estimates from the three ans\"atze used to model discretization 
corrections have been averaged over. Errors shown correspond to the 
maximum uncertainty across the averaged models.  
The value "av." for each case is the average of the results from renormalization methods A and B. 
The ``CC'' fits neglect finite-volume corrections and 
exclude data from the two smaller volume ensembles, $a087m290$ and 
$a086m180$. The full data for all the models 
are given in Tables~\protect\ref{tab:CCFV_13pt_fit} and~\protect\ref{tab:CC_11pt_fits}.}
\label{tab:CCFVresults}
\end{table*}
\begin{table*}[h]  
\setlength{\tabcolsep}{5pt}
\renewcommand{\arraystretch}{1.2}
\centering
\begin{tabular}{|c|c|c|c|c|c| }
\hline
Collaboration&Ref.&$\langle x \rangle_{u-d}^{{\rm \ol{MS}}}$&$\langle x \rangle_{\Delta
u-\Delta d}^{{\rm \ol{MS}}}$& $\langle x \rangle_{\delta u-\delta d}^{{\rm \ol{MS}}}$& Lattice Action \\
\hline
\hline
NME 26   & This &$0.154(10)(9)$&$0.177(10)(15)$&$0.197(12)(18)$& $N_f=2+1$\\
&Work&&&& Clover-on-clover\\
\hline
NME 20   &\cite{Mondal:2020ela} &$0.155(17)(20)$&$0.183(14)(20)$&$0.220(18)(20)$& $N_f=2+1$\\
&&&&& Clover-on-clover\\
\hline
Mainz 24&\cite{Djukanovic:2024krw}&$0.153(15)(10)$& $0.207(15)(06)$ & $0.195(17)(15)$ &$N_f=2+1$\\
        &                     & & & & Clover-on-clover  \\
\hline
Mainz 19&\cite{Harris:2019bih}&$0.180(25)_{\rm stat}$&$0.221(25)_{\rm stat}$&$0.212(32)_{\rm stat}$&$N_f=2+1$\\
        &                     &$(+14,-6)_{\rm sys}$  &$(+10,-0)_{\rm sys}$  &$(+20,-10)_{\rm sys}$ & Clover-on-clover  \\
\hline
$\chi$QCD 18&\cite{Yang:2018nqn} &$0.151(28)(29)$&        &          &$N_f=2+1$ \\
&&&&& Overlap on Domain Wall\\
\hline
FLAG  2024 & \cite{FlavourLatticeAveragingGroupFLAG:2024oxs} &$0.153(13)$&$0.200(13)$&$0.206(17)$& $N_f=2+1$\\
 Average& &&&& \\

\hline
\hline
ETM 22 &\cite{Alexandrou:2022dtc} &$0.126(32)$& &0.168(44) &$N_f=2+1+1$ \\
&&&&& Twisted Mass\\
\hline
PNDME~20 &\cite{Mondal:2020cmt} &$0.173(14)(07)$&$0.213(15)(22)$&$0.208(19)(24)$& $N_f=2+1+1$\\
&&&&& Clover-on-HISQ\\
\hline
%
FLAG 2024   & \cite{FlavourLatticeAveragingGroupFLAG:2024oxs} &$0.158(32)$&$0.213(27)$&$0.195(25)$& $N_f=2+1+1$\\
Average&&&&& \\
\hline
\hline
DIS23&\cite{Li:2023yda}&$0.143(5)$& & &\\
\hline
JAM21${}^\#$&\cite{Cocuzza:2021rfn}&$0.157(2)$& & &\\
\hline
CT18&\cite{Hou:2019efy}&$0.156(7)$& & &\\
\hline
JAM17${}^\dagger$&\protect\cite{Ethier:2017zbq,Lin:2017snn}& &0.241(26) & &\\
\hline
NNPDF3.1&\cite{Ball:2017nwa}&$0.152(3)$& & &\\
\hline
ABMP2016&\cite{Alekhin:2017kpj}&$0.167(4)$& & &\\
\hline
CJ15&\cite{Accardi:2016qay}&$0.152(2)$& & &\\
\hline
HERAPDF2.0&\cite{Abramowicz:2015mha}&$0.188(3)$& & &\\
\hline
CT14&\cite{Dulat:2015mca}&$0.158(4)$& & &\\
\hline
MMHT2014&\cite{Harland-Lang:2014zoa}&$0.151(4)$& & &\\
\hline
NNPDFpol1.1&\cite{Nocera:2014gqa}& &$0.195(14)$& &\\
\hline
DSSV08&\cite{deFlorian:2009vb,deFlorian:2008mr}& &$0.203(9)$& &\\
\hline
\end{tabular}
\caption{Our NME~26 results are compared with other lattice QCD
  calculations using $N_f=2+1$ (top) and $2+1+1$ (middle) 
  flavors of dynamical fermions that meet the FLAG criteria for 
  inclusion in the FLAG 
  averages~\protect\cite{FlavourLatticeAveragingGroupFLAG:2024oxs} given 
  in rows 6 and 9. (The FLAG 2024 report included results published 
  by April 30, 2024.)  Results from  phenomenological global fits are 
  given in the bottom part of the table to facilitate 
  comparison. All results are in the $\MSbar$ scheme at scale $2$~GeV. 
  For a  discussion and comparison of the lattice and global fit results up to
  2020, see Ref.~\protect\cite{Lin:2020rut} and also the 
  comparison in~\cite{Hou:2019efy} for $\langle x \rangle_{u-d}$. The
  JAM17${}^\dagger$ estimate for $\langle x \rangle_{\Delta u-\Delta d}$ 
  is obtained from~\cite{Lin:2017snn}, where, as part of the
  review, an analysis was carried out using the data in
  \cite{Ethier:2017zbq}. JAM21${}^\#$ value is taken from Ref.~\cite{Li:2023yda} that also 
  contains the DIS23 value. }
\label{tab:Compare}
\end{table*}

 \begin{figure*}[htbp]   
 \centering
 \begin{subfigure}
 \centering
 \includegraphics[height=9cm]{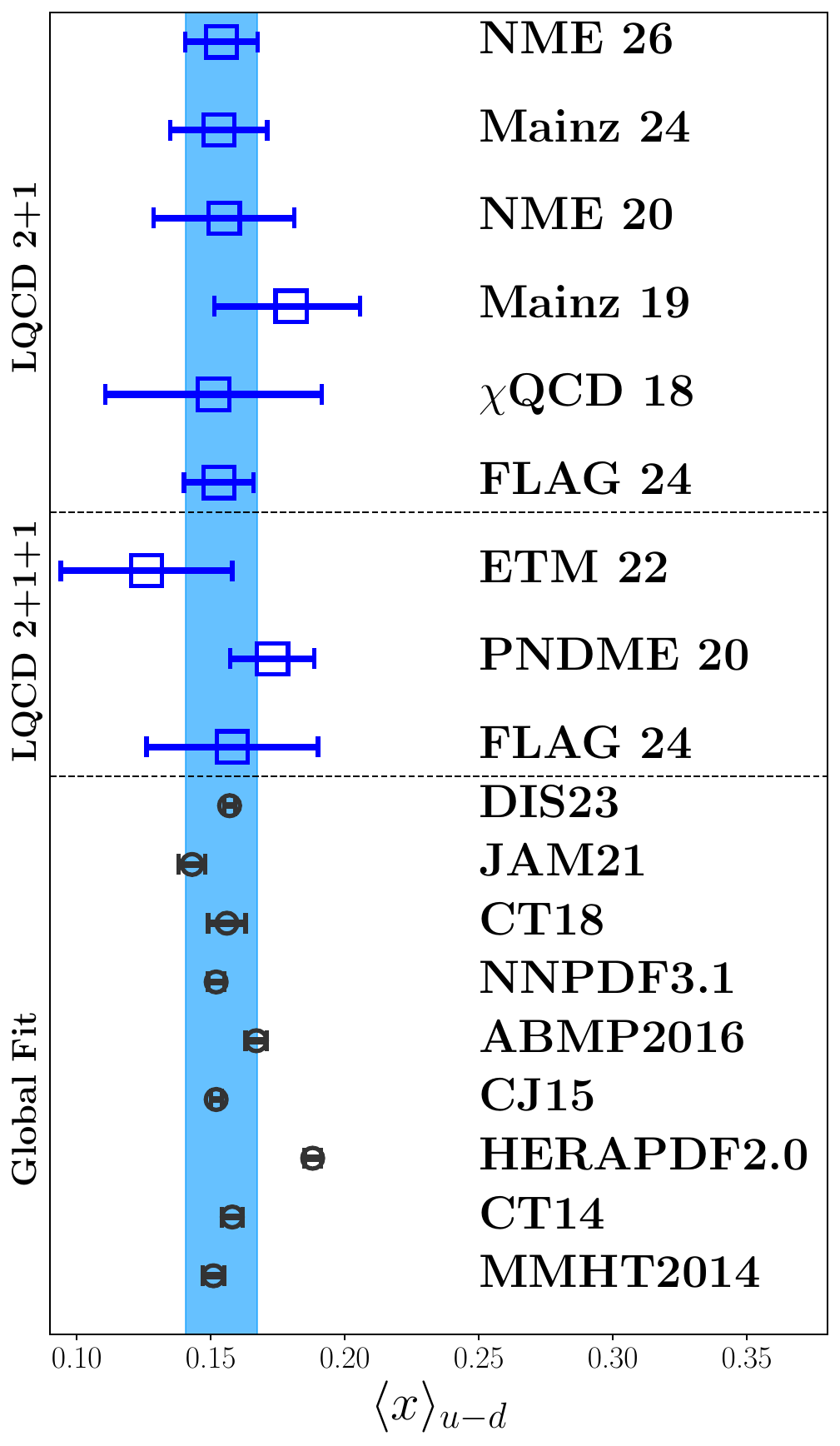}
 \includegraphics[height=9cm]{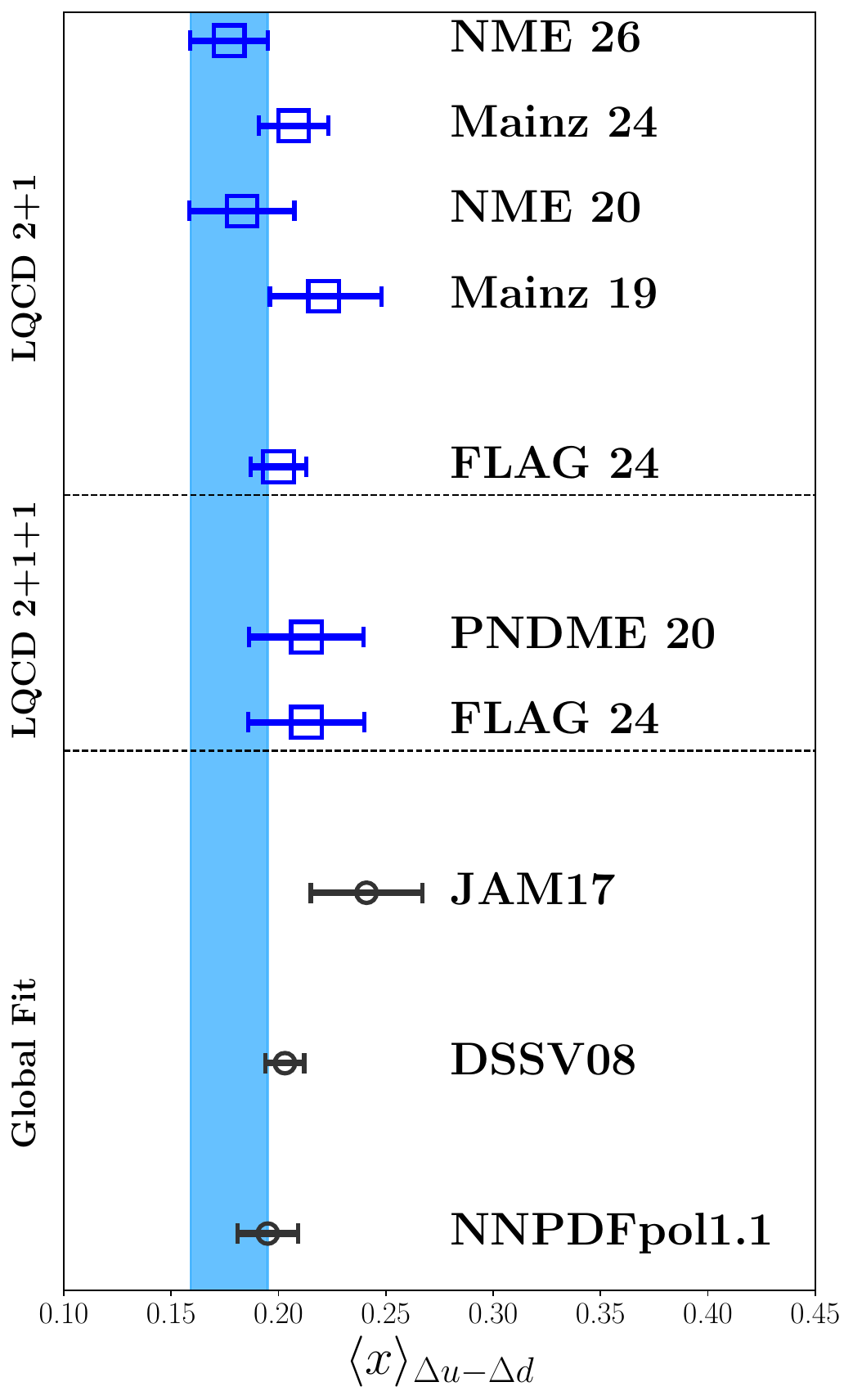}
 \includegraphics[height=9cm]{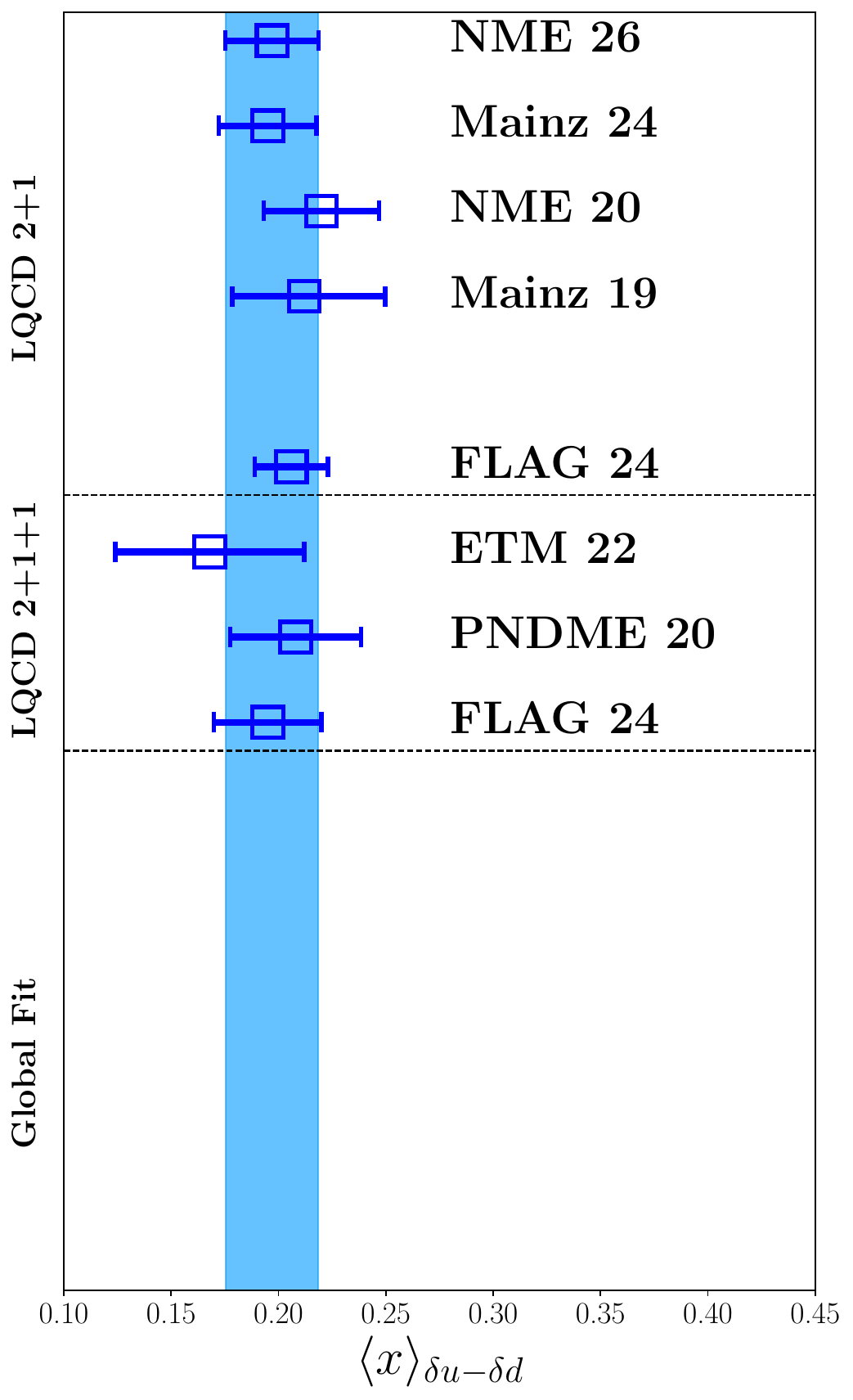}
 \end{subfigure}
 \caption{A comparison of results from lattice  QCD calculations using  
 $2+1$ and $2+1+1$-flavors of dynamical fermions with results of global 
 fits given at the bottom.  
 The left panel compares results for
the momentum fraction, $\la x \ra_{u-d}$, the middle for the helicity moment, $\la
x \ra_{\Delta u-\Delta d}$, and the
right for the transversity moment, $\la x \ra_{\delta u-\delta d}$. 
Our NME 26 result is also shown 
as the blue band to facilitate comparison.  These data are also given in
Table~\protect\ref{tab:Compare}.}
\label{fig:FLAG}
\end{figure*}


The results for all 36 models are summarized in Tables~\ref{tab:CCFV_13pt_fit} and~\ref{tab:CC_11pt_fits}. 
Rather than 
presenting an average over all these weighted by the AIC score 
as the final estimate, we use physics input to 
highlight that not all models are equally well motivated. Having chosen models 
for quoting the central value, we assign systematic errors 
using the spread in the data for each 
of the above four variants in the analysis as discussed below. Three of these 
can be inferred from the final 
compressed data given in Table~\ref{tab:CCFVresults}.

Based on the observation that the $\{4,2^{\rm free}\}$ fits
do not prefer the small $\Delta M_1^{\{4^{N\pi}\}}$, but are closer to
$\Delta M_1^{\{2\}}$ (momentum fraction) or even larger (helicity and
transversity), and lacking a physical motivation for a large contamination 
from the $N\pi$ intermediate state in these 
correlation functions, we take the $ \{4,3^\ast\}$ results 
as our best estimates.  
However, to account for not having
fully resolved the excited state contribution, we add a
systematic error to the final results, which is taken to be  half 
the difference between the 
$\{4^{N\pi},3^\ast\} $ and $ \{4,2^{\rm free}\}$ values.

Results with the two renormalization methods, summarized in
Table~\ref{tab:CCFVresults}, overlap---the differences ($\approx 0.005$) 
are a fraction of the errors from the rest of the
analysis. Also, note that these differences are much smaller than the
differences between the $Z$'s from the two methods given in
Table~\ref{tab:Z-fac}.  This is expected since the
differences between the two sets of $Z$'s are largely due to discretization errors that 
are removed on taking the continuum limit. The AIC scores for the two estimates are roughly equal and the two methods are 
equally well motivated, so the two values are averaged. A 
second systematic uncertainty is assigned as half the 
difference. 

The most likely discretization correction is $O(\alpha_s a)$ for 
the tadpole-improved action used, and the corresponding 
estimates lie in between the $O(a)$ and  $O(a^2)$ values. 
Nevertheless, we model average the three 
cases and these averaged values are given in Table~\ref{tab:CCFVresults}. 
A third systematic uncertainty is taken to be half the 
difference  between the  $O(a)$ and  $O(a^2)$ values.  

Results for the CCFV fits to all 13 ensembles are summarized in 
Table~\ref{tab:CCFV_13pt_fit} and for CC fits to 11 ensembles in 
Table~\ref{tab:CC_11pt_fits}.  We note that for all three moments and for 
a given ESC strategy, renormalization method and discretization 
correction ansatz, CCFV fits to 13 ensembles, to 11 ensembles and the 
CC fits to 11 ensembles give essentially the same results. The 
differences are much smaller that the errors.  In the CCFV fits, 
keeping the finite-volume term $c_4$ improves the fit 
to the small-volume points, $a087m290$ and $a086m180$, 
but does not change the result. We choose the CCFV fit to the 13 ensembles 
for the central value and assign half the difference 
between the CCFV and CC estimates as the fourth systematic uncertainty. 

The bottom line is, we choose the $\{4,3^*\}$ strategy and the CCFV fit and model average the discretization correction and the two renormalization methods for the central value and 
the analysis error. To this, we assign the four systematic uncertainties discussed above. With these choices, our final results are 
\begin{widetext}
\begin{equation}
\begin{tabular}{>{\(}l<{\)} >{\(}l<{\)} >{\(}l<{\)} >{\(}l<{\)} >{\(}l<{\)} >{\(}l<{\)} >{\(}l<{\)} >{\(}l<{\)}}
\langle x \rangle_{u -d}                &=  0.154(10) & (8)_{ES}  & (4)_{Z} & (2)_a & (1)_{FV} &=  0.154(10)_{stat.} & (9)_{sys.} \,,  \\
\langle x \rangle_{\Delta u- \Delta d}  &=  0.177(10) & (13)_{ES} & (5)_{Z} & (4)_a & (1)_{FV} &=  0.177(10)_{stat.} & (15)_{sys.} \,, \\
\langle x \rangle_{\delta u- \delta d}  &=  0.197(12) & (16)_{ES} & (5)_{Z} & (5)_a & (1)_{FV} &=  0.197(12)_{stat.} & (18)_{sys.} \,,
\end{tabular}
\label{eq:finalresults}
\end{equation}
\end{widetext}
where, on the right, the four systematic errors have 
been combined in quadrature. Overall,  the errors in order of size to address in 
future studies are from ESC, statistical, and discretization effects.  


\section{Comparison With Other Determinations}
\label{sec:results}

Results for these three isovector moments are now being reviewed by 
the Flavor Lattice Averaging Group (FLAG) and the 2024 
report~\cite{FlavourLatticeAveragingGroupFLAG:2024oxs} lists all the calculations 
published by  30, 2024 and their lattice parameters. Only this work is new and not covered. 
With this background, we restrict the comparison to only the 2+1 (green) and 
the 2+1+1-flavor (blue) lattice results shown 
in the top  half of Table~\ref{tab:Compare} and Fig.~\ref{fig:FLAG} 
that qualify for the FLAG averages. 

For 2+1-flavor theory, the NME~26 supercedes NME~20 
and Mainz~24 supercedes Mainz~19. Under an increase in the data  
by both collaborations, the results have not changed significantly 
but there is a reduction in 
both the statistical and systematic errors. 
As summarized in Table~\ref{tab:Compare}, the results from our NME~26, the Mainz~24~\cite{Djukanovic:2024krw} 
and the $\chi$QCD~18~\cite{Yang:2018nqn}, are consistent. 
They also agree with the 2+1+1-flavor estimates from  
PNDME~20~\cite{Mondal:2020cmt} and ETM~22~\cite{Alexandrou:2022dtc} that have 
larger errors and a larger spread. 

Overall, note that 
calculations of the moments have now been done with four different 
lattice formulations, therefore, their agreement is a non-trivial check  
of the stability of the results within the quoted errors.\looseness-1

The three 2+1 flavor results for the momentum fraction, 
NME~26, Mainz~24 and $\chi$QCD~18, are in very good agreement with
estimates from phenomenological global fit values summarized in the 
bottom half of Table~\ref{tab:Compare}, and shown in Fig~\ref{fig:FLAG}. 
(Also see the review in Ref.~\cite{Lin:2017snn}). 
Our result NME~26 for the helicity moment is about $1\sigma$ smaller 
than the weighted mean of the other lattice 
calculations and of the phenomenological
values. The transversity moment, again consistent between the 
various lattice calculations, is a prediction.



\section{Conclusions}
\label{sec:summary}

This paper presents results for the isovector quark momentum fraction,
$\la x \ra_{u-d}^{{\rm \ol{MS}}}$, helicity moment, $\la
x \ra^{{\rm \ol{MS}}}_{\Delta u- \Delta d}$, and transversity moment,
$\la x \ra^{{\rm \ol{MS}}}_{\delta u- \delta d}$, all renormalized in the $\ol{MS}$ 
scheme at scale 2~GeV.  The calculation has been done on thirteen ensembles
generated with 2+1-flavors of Wilson-clover fermions by the 
 JLab/W\&M/LANL/MIT/Marseille collaborations~\cite{Yoo:2026uul}. Data with higher statistics 
 and a larger number of ensembles allowed us to make the four improvements 
 over the results  presented in Ref.~\cite{Mondal:2020ela} as specified in the Introduction~\ref{sec:intro}.  The final results are given in Eq.~\eqref{eq:finalresults}. 
 Notable features  of the data and the analysis presented here are: 
\begin{itemize}
\item
On each ensemble,  the symmetry  
about $\tau/2$ of the data has improved with higher statistics as shown in
Figs.~\ref{fig:Ratio-mom-1}--\ref{fig:Ratio-transversity-2}. Also, the data at each of the five values of source-sink separation $\tau$
simulated are well-resolved and show monotonic convergence towards 
the $\tau \to \infty$ value. As a result, 
 the fits to remove ESC have improved.
\item
A complete analysis is carried out with three inputs for the first 
excite-state mass: the strategy $\{4,3^*\}$ with $M_{\{0,1,2\}}$ taken
from the two-point function;  $\{4^{N\pi},3^*\}$ with $M_1$ equal to the 
non-interacting energy of the $N({\vec p}=1) \pi({\vec p}=-1$) state and 
$M_{\{0,2\}}$ from the two-point function; 
and $\{4,2^{\rm free}\}$ in which 
$M_1$ is left as a free parameter in the fit to the 
three-point function. High statistics allowed all fits to be made 
using the full covariance matrix.
The spread in results between these three analyses is used to assign a systematic 
uncertainty that accounts for possible residual excited-state contributions.
\item
The analysis has been done with two methods for removing
discretization errors in the calculations of renormalization
constants as discussed in Appendix~\ref{sec:renormalization}. The
results in Table~\ref{tab:CCFVresults} show that the results, 
in the continuum limit, overlap indicating that, as expected, most of the difference 
in the two sets of $Z$'s was due to discrezitation errors.
\item
The final results are obtained using a combined continuum-chiral-finite-volume 
(CCFV) fit to the data from the 13 ensembles. The CCFV ansatz used, 
Eq.~\eqref{eq:CCFV}, includes corrections due to discretization 
and finite volume artifacts and 
dependence on both the light and the strange quark mass. 
Figure~\ref{fig:CCFV_explicit} shows that the main 
variation of the results is with $M_\pi$ and $M_K$.  
\item
The two pairs of points,
$\{a087m290,a087m290L\}$ and $\{a086m180,a086m180L\}$, differ
only in the lattice volume and were intended to provide 
a handle on finite volume effects.  However, the CCFV fit with or without 
the two smaller volume ensembles gives  
estimates overlapping with the CC fit  as shown in Table~\ref{tab:CCFVresults}, 
i.e., the extra parameter $c_4$ in the CCFV fits is 
mainly fixed by the two smaller volume points.  To resolve finite-volume 
effects  in the future will require more points differing only in the lattice volume.
\end{itemize}

Our results for the various models, summarized in Tables~~\ref{tab:CCFVresults},~\ref{tab:CCFV_13pt_fit}, and~\ref{tab:CC_11pt_fits}, 
show little variation. The final results given
in Eq.~\eqref{eq:finalresults} are in good agreement with other lattice
calculations, and now have the smallest errors. They also agree with the 
phenomenological global fit estimates shown in
Table~\ref{tab:Compare} and Fig.~\ref{fig:FLAG}.  
Our estimate for the momentum fraction has a larger uncertainty 
than the global fit value, while that in  the helicity moment 
is smaller than in the three global fits values that show a large spread.
 Lattice estimates for the transversity moment are a prediction. 
 
The agreement between results for the three moments obtained by 
different lattice collaborations using four different lattice 
formulations (see Table~\ref{tab:Compare}) is a non-trivial
check of the stability of the results within the quoted errors. 
To improve precision in the future requires higher statistics data 
on more physical pion mass ensembles.  \looseness-1

\begin{acknowledgments}
The calculations used the Chroma software
suite~\cite{Edwards:2004sx}. This research used resources at (i) the
National Energy Research Scientific Computing Center, a DOE Office of
Science User Facility supported by the Office of Science of the
U.S. Department of Energy under Contract No. DE-AC02-05CH11231; (ii)
the Oak Ridge Leadership Computing Facility, which is a DOE Office of
Science User Facility supported under Contract DE-AC05-00OR22725, and
was awarded through the INCITE Program projects PHY138 and HEP133, and the ALCC projects
LGT107 and HEP145; (iii) the USQCD Collaboration, which are funded by the
Office of Science of the U.S. Department of Energy, and (iv)
Institutional Computing at Los Alamos National Laboratory.
T. Bhattacharya and R. Gupta were partly supported by the
U.S. Department of Energy, Office of Science, Office of High Energy
Physics under Contract No.~DE-AC52-06NA25396.  F. Winter is supported
by the U.S. Department of Energy, Office of Science, Office of Nuclear
Physics under contract DE-AC05-06OR23177.  B. Jo\'o is supported by
the U.S. Department of Energy, Office of Science, under contract
DE-AC05-06OR22725. We acknowledge support from the U.S. Department of
Energy, Office of Science, Office of Advanced Scientific Computing
Research and Office of Nuclear Physics, Scientific Discovery through
Advanced Computing (SciDAC) program, and of the U.S. Department of
Energy Exascale Computing Project.  T. Bhattacharya, R. Gupta,
S. Mondal, S. Park and B.Yoon were partly supported by the LANL LDRD
program, and S. Park by the Center for Nonlinear Studies.

\end{acknowledgments}


\clearpage
\bibliography{ref} 

\begin{thebibliography}{61}%
\makeatletter
\providecommand \@ifxundefined [1]{%
 \@ifx{#1\undefined}
}%
\providecommand \@ifnum [1]{%
 \ifnum #1\expandafter \@firstoftwo
 \else \expandafter \@secondoftwo
 \fi
}%
\providecommand \@ifx [1]{%
 \ifx #1\expandafter \@firstoftwo
 \else \expandafter \@secondoftwo
 \fi
}%
\providecommand \natexlab [1]{#1}%
\providecommand \enquote  [1]{``#1''}%
\providecommand \bibnamefont  [1]{#1}%
\providecommand \bibfnamefont [1]{#1}%
\providecommand \citenamefont [1]{#1}%
\providecommand \href@noop [0]{\@secondoftwo}%
\providecommand \href [0]{\begingroup \@sanitize@url \@href}%
\providecommand \@href[1]{\@@startlink{#1}\@@href}%
\providecommand \@@href[1]{\endgroup#1\@@endlink}%
\providecommand \@sanitize@url [0]{\catcode `\\12\catcode `\$12\catcode
  `\&12\catcode `\#12\catcode `\^12\catcode `\_12\catcode `\%12\relax}%
\providecommand \@@startlink[1]{}%
\providecommand \@@endlink[0]{}%
\providecommand \url  [0]{\begingroup\@sanitize@url \@url }%
\providecommand \@url [1]{\endgroup\@href {#1}{\urlprefix }}%
\providecommand \urlprefix  [0]{URL }%
\providecommand \Eprint [0]{\href }%
\providecommand \doibase [0]{http://dx.doi.org/}%
\providecommand \selectlanguage [0]{\@gobble}%
\providecommand \bibinfo  [0]{\@secondoftwo}%
\providecommand \bibfield  [0]{\@secondoftwo}%
\providecommand \translation [1]{[#1]}%
\providecommand \BibitemOpen [0]{}%
\providecommand \bibitemStop [0]{}%
\providecommand \bibitemNoStop [0]{.\EOS\space}%
\providecommand \EOS [0]{\spacefactor3000\relax}%
\providecommand \BibitemShut  [1]{\csname bibitem#1\endcsname}%
\let\auto@bib@innerbib\@empty
\bibitem [{\citenamefont {Accardi}\ \emph
  {et~al.}(2016{\natexlab{a}})\citenamefont {Accardi} \emph
  {et~al.}}]{Accardi:2012qut}%
  \BibitemOpen
  \bibfield  {author} {\bibinfo {author} {\bibfnamefont {A.}~\bibnamefont
  {Accardi}} \emph {et~al.},\ }\href {\doibase 10.1140/epja/i2016-16268-9}
  {\bibfield  {journal} {\bibinfo  {journal} {Eur. Phys. J. A}\ }\textbf
  {\bibinfo {volume} {52}},\ \bibinfo {pages} {268} (\bibinfo {year}
  {2016}{\natexlab{a}})},\ \Eprint {http://arxiv.org/abs/1212.1701}
  {arXiv:1212.1701 [nucl-ex]} \BibitemShut {NoStop}%
\bibitem [{\citenamefont {Dudek}\ \emph {et~al.}(2012)\citenamefont {Dudek}
  \emph {et~al.}}]{Dudek:2012vr}%
  \BibitemOpen
  \bibfield  {author} {\bibinfo {author} {\bibfnamefont {J.}~\bibnamefont
  {Dudek}} \emph {et~al.},\ }\href {\doibase 10.1140/epja/i2012-12187-1}
  {\bibfield  {journal} {\bibinfo  {journal} {Eur. Phys. J. A}\ }\textbf
  {\bibinfo {volume} {48}},\ \bibinfo {pages} {187} (\bibinfo {year} {2012})},\
  \Eprint {http://arxiv.org/abs/1208.1244} {arXiv:1208.1244 [hep-ex]}
  \BibitemShut {NoStop}%
\bibitem [{\citenamefont {Brock}\ \emph {et~al.}(1995)\citenamefont {Brock}
  \emph {et~al.}}]{Brock:1993sz}%
  \BibitemOpen
  \bibfield  {author} {\bibinfo {author} {\bibfnamefont {R.}~\bibnamefont
  {Brock}} \emph {et~al.} (\bibinfo {collaboration} {CTEQ}),\ }\href {\doibase
  10.1103/RevModPhys.67.157} {\bibfield  {journal} {\bibinfo  {journal} {Rev.
  Mod. Phys.}\ }\textbf {\bibinfo {volume} {67}},\ \bibinfo {pages} {157}
  (\bibinfo {year} {1995})}\BibitemShut {NoStop}%
\bibitem [{\citenamefont {Ji}\ \emph {et~al.}(2020)\citenamefont {Ji},
  \citenamefont {Liu}, \citenamefont {Liu}, \citenamefont {Zhang},\ and\
  \citenamefont {Zhao}}]{Ji:2020ect}%
  \BibitemOpen
  \bibfield  {author} {\bibinfo {author} {\bibfnamefont {X.}~\bibnamefont
  {Ji}}, \bibinfo {author} {\bibfnamefont {Y.-S.}\ \bibnamefont {Liu}},
  \bibinfo {author} {\bibfnamefont {Y.}~\bibnamefont {Liu}}, \bibinfo {author}
  {\bibfnamefont {J.-H.}\ \bibnamefont {Zhang}}, \ and\ \bibinfo {author}
  {\bibfnamefont {Y.}~\bibnamefont {Zhao}},\ }\href@noop {} {\enquote {\bibinfo
  {title} {{Large-Momentum Effective Theory}},}\ } (\bibinfo {year} {2020}),\
  \Eprint {http://arxiv.org/abs/2004.03543} {arXiv:2004.03543 [hep-ph]}
  \BibitemShut {NoStop}%
\bibitem [{\citenamefont {Yoon}\ \emph
  {et~al.}(2017{\natexlab{a}})\citenamefont {Yoon}, \citenamefont {Engelhardt},
  \citenamefont {Gupta}, \citenamefont {Bhattacharya}, \citenamefont {Green},
  \citenamefont {Musch}, \citenamefont {Negele}, \citenamefont {Pochinsky},
  \citenamefont {Schafer},\ and\ \citenamefont {Syritsyn}}]{Yoon:2017qzo}%
  \BibitemOpen
  \bibfield  {author} {\bibinfo {author} {\bibfnamefont {B.}~\bibnamefont
  {Yoon}}, \bibinfo {author} {\bibfnamefont {M.}~\bibnamefont {Engelhardt}},
  \bibinfo {author} {\bibfnamefont {R.}~\bibnamefont {Gupta}}, \bibinfo
  {author} {\bibfnamefont {T.}~\bibnamefont {Bhattacharya}}, \bibinfo {author}
  {\bibfnamefont {J.~R.}\ \bibnamefont {Green}}, \bibinfo {author}
  {\bibfnamefont {B.~U.}\ \bibnamefont {Musch}}, \bibinfo {author}
  {\bibfnamefont {J.~W.}\ \bibnamefont {Negele}}, \bibinfo {author}
  {\bibfnamefont {A.~V.}\ \bibnamefont {Pochinsky}}, \bibinfo {author}
  {\bibfnamefont {A.}~\bibnamefont {Schafer}}, \ and\ \bibinfo {author}
  {\bibfnamefont {S.~N.}\ \bibnamefont {Syritsyn}},\ }\href {\doibase
  10.1103/PhysRevD.96.094508} {\bibfield  {journal} {\bibinfo  {journal} {Phys.
  Rev. D}\ }\textbf {\bibinfo {volume} {96}},\ \bibinfo {pages} {094508}
  (\bibinfo {year} {2017}{\natexlab{a}})},\ \Eprint
  {http://arxiv.org/abs/1706.03406} {arXiv:1706.03406 [hep-lat]} \BibitemShut
  {NoStop}%
\bibitem [{\citenamefont {Diehl}(2003)}]{Diehl:2003ny}%
  \BibitemOpen
  \bibfield  {author} {\bibinfo {author} {\bibfnamefont {M.}~\bibnamefont
  {Diehl}},\ }\emph {\bibinfo {title} {{Generalized parton distributions}}},\
  \href {\doibase 10.1016/j.physrep.2003.08.002} {Ph.D. thesis},\ \bibinfo
  {school} {DESY} (\bibinfo {year} {2003}),\ \Eprint
  {http://arxiv.org/abs/hep-ph/0307382} {arXiv:hep-ph/0307382} \BibitemShut
  {NoStop}%
\bibitem [{\citenamefont {Cichy}\ and\ \citenamefont
  {Constantinou}(2019)}]{Cichy:2018mum}%
  \BibitemOpen
  \bibfield  {author} {\bibinfo {author} {\bibfnamefont {K.}~\bibnamefont
  {Cichy}}\ and\ \bibinfo {author} {\bibfnamefont {M.}~\bibnamefont
  {Constantinou}},\ }\href {\doibase 10.1155/2019/3036904} {\bibfield
  {journal} {\bibinfo  {journal} {Adv. High Energy Phys.}\ }\textbf {\bibinfo
  {volume} {2019}},\ \bibinfo {pages} {3036904} (\bibinfo {year} {2019})},\
  \Eprint {http://arxiv.org/abs/1811.07248} {arXiv:1811.07248 [hep-lat]}
  \BibitemShut {NoStop}%
\bibitem [{\citenamefont {Karthik}(2019)}]{Karthik:2019}%
  \BibitemOpen
  \bibfield  {author} {\bibinfo {author} {\bibfnamefont {N.}~\bibnamefont
  {Karthik}},\ }\href@noop {} {\enquote {\bibinfo {title} {Lattice computations
  of {PDF}: Challenges and progress},}\ }\bibinfo {howpublished}
  {\url{https://indico.cern.ch/event/764552/contributions/3420535/attachments/1864018/3064443/LatticeTalk19_NK.pdf}}
  (\bibinfo {year} {2019}),\ \bibinfo {note} {accessed: 2020-05-20}\BibitemShut
  {NoStop}%
\bibitem [{\citenamefont {Lin}\ \emph {et~al.}(2018)\citenamefont {Lin} \emph
  {et~al.}}]{Lin:2017snn}%
  \BibitemOpen
  \bibfield  {author} {\bibinfo {author} {\bibfnamefont {H.-W.}\ \bibnamefont
  {Lin}} \emph {et~al.},\ }\href {\doibase 10.1016/j.ppnp.2018.01.007}
  {\bibfield  {journal} {\bibinfo  {journal} {Prog. Part. Nucl. Phys.}\
  }\textbf {\bibinfo {volume} {100}},\ \bibinfo {pages} {107} (\bibinfo {year}
  {2018})},\ \Eprint {http://arxiv.org/abs/1711.07916} {arXiv:1711.07916
  [hep-ph]} \BibitemShut {NoStop}%
\bibitem [{\citenamefont {Constantinou}\ \emph {et~al.}(2020)\citenamefont
  {Constantinou} \emph {et~al.}}]{Lin:2020rut}%
  \BibitemOpen
  \bibfield  {author} {\bibinfo {author} {\bibfnamefont {M.}~\bibnamefont
  {Constantinou}} \emph {et~al.},\ }\href@noop {} {\enquote {\bibinfo {title}
  {{Parton distributions and lattice QCD calculations: toward 3D structure}},}\
  } (\bibinfo {year} {2020}),\ \Eprint {http://arxiv.org/abs/2006.08636}
  {arXiv:2006.08636 [hep-ph]} \BibitemShut {NoStop}%
\bibitem [{\citenamefont {Aoki}\ \emph {et~al.}(2020)\citenamefont {Aoki} \emph
  {et~al.}}]{Aoki:2019cca}%
  \BibitemOpen
  \bibfield  {author} {\bibinfo {author} {\bibfnamefont {S.}~\bibnamefont
  {Aoki}} \emph {et~al.} (\bibinfo {collaboration} {Flavour Lattice Averaging
  Group}),\ }\href {\doibase 10.1140/epjc/s10052-019-7354-7} {\bibfield
  {journal} {\bibinfo  {journal} {Eur. Phys. J. C}\ }\textbf {\bibinfo {volume}
  {80}},\ \bibinfo {pages} {113} (\bibinfo {year} {2020})},\ \Eprint
  {http://arxiv.org/abs/1902.08191} {arXiv:1902.08191 [hep-lat]} \BibitemShut
  {NoStop}%
\bibitem [{\citenamefont {Aoki}\ \emph {et~al.}(2022)\citenamefont {Aoki} \emph
  {et~al.}}]{FlavourLatticeAveragingGroupFLAG:2021npn}%
  \BibitemOpen
  \bibfield  {author} {\bibinfo {author} {\bibfnamefont {Y.}~\bibnamefont
  {Aoki}} \emph {et~al.} (\bibinfo {collaboration} {Flavour Lattice Averaging
  Group (FLAG)}),\ }\href {\doibase 10.1140/epjc/s10052-022-10536-1} {\bibfield
   {journal} {\bibinfo  {journal} {Eur. Phys. J. C}\ }\textbf {\bibinfo
  {volume} {82}},\ \bibinfo {pages} {869} (\bibinfo {year} {2022})},\ \Eprint
  {http://arxiv.org/abs/2111.09849} {arXiv:2111.09849 [hep-lat]} \BibitemShut
  {NoStop}%
\bibitem [{\citenamefont {Aoki}\ \emph {et~al.}(2026)\citenamefont {Aoki} \emph
  {et~al.}}]{FlavourLatticeAveragingGroupFLAG:2024oxs}%
  \BibitemOpen
  \bibfield  {author} {\bibinfo {author} {\bibfnamefont {Y.}~\bibnamefont
  {Aoki}} \emph {et~al.} (\bibinfo {collaboration} {Flavour Lattice Averaging
  Group (FLAG)}),\ }\href {\doibase 10.1103/nfzp-p5dn} {\bibfield  {journal}
  {\bibinfo  {journal} {Phys. Rev. D}\ }\textbf {\bibinfo {volume} {113}},\
  \bibinfo {pages} {014508} (\bibinfo {year} {2026})},\ \Eprint
  {http://arxiv.org/abs/2411.04268} {arXiv:2411.04268 [hep-lat]} \BibitemShut
  {NoStop}%
\bibitem [{\citenamefont {Gupta}\ \emph {et~al.}(2018)\citenamefont {Gupta},
  \citenamefont {Jang}, \citenamefont {Yoon}, \citenamefont {Lin},
  \citenamefont {Cirigliano},\ and\ \citenamefont
  {Bhattacharya}}]{Gupta:2018qil}%
  \BibitemOpen
  \bibfield  {author} {\bibinfo {author} {\bibfnamefont {R.}~\bibnamefont
  {Gupta}}, \bibinfo {author} {\bibfnamefont {Y.-C.}\ \bibnamefont {Jang}},
  \bibinfo {author} {\bibfnamefont {B.}~\bibnamefont {Yoon}}, \bibinfo {author}
  {\bibfnamefont {H.-W.}\ \bibnamefont {Lin}}, \bibinfo {author} {\bibfnamefont
  {V.}~\bibnamefont {Cirigliano}}, \ and\ \bibinfo {author} {\bibfnamefont
  {T.}~\bibnamefont {Bhattacharya}},\ }\href {\doibase
  10.1103/PhysRevD.98.034503} {\bibfield  {journal} {\bibinfo  {journal} {Phys.
  Rev.}\ }\textbf {\bibinfo {volume} {D98}},\ \bibinfo {pages} {034503}
  (\bibinfo {year} {2018})},\ \Eprint {http://arxiv.org/abs/1806.09006}
  {arXiv:1806.09006 [hep-lat]} \BibitemShut {NoStop}%
\bibitem [{\citenamefont {Yoo}\ \emph {et~al.}(2026)\citenamefont {Yoo},
  \citenamefont {Ee}, \citenamefont {Park}, \citenamefont {Gupta},
  \citenamefont {Bhattacharya}, \citenamefont {Mondal}, \citenamefont
  {Jo{\'o}}, \citenamefont {Edwards}, \citenamefont {Orginos},\ and\
  \citenamefont {Winter}}]{Yoo:2026uul}%
  \BibitemOpen
  \bibfield  {author} {\bibinfo {author} {\bibfnamefont {J.-S.}\ \bibnamefont
  {Yoo}}, \bibinfo {author} {\bibfnamefont {J.-H.}\ \bibnamefont {Ee}},
  \bibinfo {author} {\bibfnamefont {S.}~\bibnamefont {Park}}, \bibinfo {author}
  {\bibfnamefont {R.}~\bibnamefont {Gupta}}, \bibinfo {author} {\bibfnamefont
  {T.}~\bibnamefont {Bhattacharya}}, \bibinfo {author} {\bibfnamefont
  {S.}~\bibnamefont {Mondal}}, \bibinfo {author} {\bibfnamefont
  {B.}~\bibnamefont {Jo{\'o}}}, \bibinfo {author} {\bibfnamefont
  {R.}~\bibnamefont {Edwards}}, \bibinfo {author} {\bibfnamefont
  {K.}~\bibnamefont {Orginos}}, \ and\ \bibinfo {author} {\bibfnamefont
  {F.}~\bibnamefont {Winter}},\ }\href@noop {} {\  (\bibinfo {year} {2026})},\
  \Eprint {http://arxiv.org/abs/2601.10857} {arXiv:2601.10857 [hep-lat]}
  \BibitemShut {NoStop}%
\bibitem [{\citenamefont {Mondal}\ \emph
  {et~al.}(2020{\natexlab{a}})\citenamefont {Mondal}, \citenamefont {Gupta},
  \citenamefont {Park}, \citenamefont {Yoon}, \citenamefont {Bhattacharya},
  \citenamefont {Jo\'o},\ and\ \citenamefont {Winter}}]{Mondal:2020ela}%
  \BibitemOpen
  \bibfield  {author} {\bibinfo {author} {\bibfnamefont {S.}~\bibnamefont
  {Mondal}}, \bibinfo {author} {\bibfnamefont {R.}~\bibnamefont {Gupta}},
  \bibinfo {author} {\bibfnamefont {S.}~\bibnamefont {Park}}, \bibinfo {author}
  {\bibfnamefont {B.}~\bibnamefont {Yoon}}, \bibinfo {author} {\bibfnamefont
  {T.}~\bibnamefont {Bhattacharya}}, \bibinfo {author} {\bibfnamefont
  {B.}~\bibnamefont {Jo\'o}}, \ and\ \bibinfo {author} {\bibfnamefont
  {F.}~\bibnamefont {Winter}} (\bibinfo {collaboration} {Nucleon Matrix
  Elements (NME)}),\ }\href {\doibase 10.1007/JHEP04(2021)044} {\bibfield
  {journal} {\bibinfo  {journal} {JHEP}\ }\textbf {\bibinfo {volume} {21}},\
  \bibinfo {pages} {004} (\bibinfo {year} {2020}{\natexlab{a}})},\ \Eprint
  {http://arxiv.org/abs/2011.12787} {arXiv:2011.12787 [hep-lat]} \BibitemShut
  {NoStop}%
\bibitem [{\citenamefont {Mondal}\ \emph
  {et~al.}(2020{\natexlab{b}})\citenamefont {Mondal}, \citenamefont {Gupta},
  \citenamefont {Park}, \citenamefont {Yoon}, \citenamefont {Bhattacharya},\
  and\ \citenamefont {Lin}}]{Mondal:2020cmt}%
  \BibitemOpen
  \bibfield  {author} {\bibinfo {author} {\bibfnamefont {S.}~\bibnamefont
  {Mondal}}, \bibinfo {author} {\bibfnamefont {R.}~\bibnamefont {Gupta}},
  \bibinfo {author} {\bibfnamefont {S.}~\bibnamefont {Park}}, \bibinfo {author}
  {\bibfnamefont {B.}~\bibnamefont {Yoon}}, \bibinfo {author} {\bibfnamefont
  {T.}~\bibnamefont {Bhattacharya}}, \ and\ \bibinfo {author} {\bibfnamefont
  {H.-W.}\ \bibnamefont {Lin}},\ }\href {\doibase 10.1103/PhysRevD.102.054512}
  {\bibfield  {journal} {\bibinfo  {journal} {Phys. Rev. D}\ }\textbf {\bibinfo
  {volume} {102}},\ \bibinfo {pages} {054512} (\bibinfo {year}
  {2020}{\natexlab{b}})},\ \Eprint {http://arxiv.org/abs/2005.13779}
  {arXiv:2005.13779 [hep-lat]} \BibitemShut {NoStop}%
\bibitem [{\citenamefont {Bali}\ \emph {et~al.}(2010)\citenamefont {Bali},
  \citenamefont {Collins},\ and\ \citenamefont {Schafer}}]{Bali:2009hu}%
  \BibitemOpen
  \bibfield  {author} {\bibinfo {author} {\bibfnamefont {G.~S.}\ \bibnamefont
  {Bali}}, \bibinfo {author} {\bibfnamefont {S.}~\bibnamefont {Collins}}, \
  and\ \bibinfo {author} {\bibfnamefont {A.}~\bibnamefont {Schafer}},\ }\href
  {\doibase 10.1016/j.cpc.2010.05.008} {\bibfield  {journal} {\bibinfo
  {journal} {Comput.Phys.Commun.}\ }\textbf {\bibinfo {volume} {181}},\
  \bibinfo {pages} {1570} (\bibinfo {year} {2010})},\ \Eprint
  {http://arxiv.org/abs/0910.3970} {arXiv:0910.3970 [hep-lat]} \BibitemShut
  {NoStop}%
\bibitem [{\citenamefont {Blum}\ \emph {et~al.}(2013)\citenamefont {Blum},
  \citenamefont {Izubuchi},\ and\ \citenamefont {Shintani}}]{Blum:2012uh}%
  \BibitemOpen
  \bibfield  {author} {\bibinfo {author} {\bibfnamefont {T.}~\bibnamefont
  {Blum}}, \bibinfo {author} {\bibfnamefont {T.}~\bibnamefont {Izubuchi}}, \
  and\ \bibinfo {author} {\bibfnamefont {E.}~\bibnamefont {Shintani}},\ }\href
  {\doibase 10.1103/PhysRevD.88.094503} {\bibfield  {journal} {\bibinfo
  {journal} {Phys.Rev.}\ }\textbf {\bibinfo {volume} {D88}},\ \bibinfo {pages}
  {094503} (\bibinfo {year} {2013})},\ \Eprint {http://arxiv.org/abs/1208.4349}
  {arXiv:1208.4349 [hep-lat]} \BibitemShut {NoStop}%
\bibitem [{\citenamefont {Morningstar}\ and\ \citenamefont
  {Peardon}(2004)}]{Morningstar:2003gk}%
  \BibitemOpen
  \bibfield  {author} {\bibinfo {author} {\bibfnamefont {C.}~\bibnamefont
  {Morningstar}}\ and\ \bibinfo {author} {\bibfnamefont {M.~J.}\ \bibnamefont
  {Peardon}},\ }\href {\doibase 10.1103/PhysRevD.69.054501} {\bibfield
  {journal} {\bibinfo  {journal} {Phys. Rev. D}\ }\textbf {\bibinfo {volume}
  {69}},\ \bibinfo {pages} {054501} (\bibinfo {year} {2004})},\ \Eprint
  {http://arxiv.org/abs/hep-lat/0311018} {arXiv:hep-lat/0311018} \BibitemShut
  {NoStop}%
\bibitem [{\citenamefont {Sheikholeslami}\ and\ \citenamefont
  {Wohlert}(1985)}]{Sheikholeslami:1985ij}%
  \BibitemOpen
  \bibfield  {author} {\bibinfo {author} {\bibfnamefont {B.}~\bibnamefont
  {Sheikholeslami}}\ and\ \bibinfo {author} {\bibfnamefont {R.}~\bibnamefont
  {Wohlert}},\ }\href {\doibase 10.1016/0550-3213(85)90002-1} {\bibfield
  {journal} {\bibinfo  {journal} {Nucl. Phys.}\ }\textbf {\bibinfo {volume}
  {B259}},\ \bibinfo {pages} {572} (\bibinfo {year} {1985})}\BibitemShut
  {NoStop}%
\bibitem [{\citenamefont {Luscher}\ \emph {et~al.}(1997)\citenamefont
  {Luscher}, \citenamefont {Sint}, \citenamefont {Sommer}, \citenamefont
  {Weisz},\ and\ \citenamefont {Wolff}}]{Luscher:1996ug}%
  \BibitemOpen
  \bibfield  {author} {\bibinfo {author} {\bibfnamefont {M.}~\bibnamefont
  {Luscher}}, \bibinfo {author} {\bibfnamefont {S.}~\bibnamefont {Sint}},
  \bibinfo {author} {\bibfnamefont {R.}~\bibnamefont {Sommer}}, \bibinfo
  {author} {\bibfnamefont {P.}~\bibnamefont {Weisz}}, \ and\ \bibinfo {author}
  {\bibfnamefont {U.}~\bibnamefont {Wolff}},\ }\href {\doibase
  10.1016/S0550-3213(97)00080-1} {\bibfield  {journal} {\bibinfo  {journal}
  {Nucl. Phys. B}\ }\textbf {\bibinfo {volume} {491}},\ \bibinfo {pages} {323}
  (\bibinfo {year} {1997})},\ \Eprint {http://arxiv.org/abs/hep-lat/9609035}
  {arXiv:hep-lat/9609035} \BibitemShut {NoStop}%
\bibitem [{\citenamefont {Duane}\ \emph {et~al.}(1987)\citenamefont {Duane},
  \citenamefont {Kennedy}, \citenamefont {Pendleton},\ and\ \citenamefont
  {Roweth}}]{Duane:1987de}%
  \BibitemOpen
  \bibfield  {author} {\bibinfo {author} {\bibfnamefont {S.}~\bibnamefont
  {Duane}}, \bibinfo {author} {\bibfnamefont {A.}~\bibnamefont {Kennedy}},
  \bibinfo {author} {\bibfnamefont {B.}~\bibnamefont {Pendleton}}, \ and\
  \bibinfo {author} {\bibfnamefont {D.}~\bibnamefont {Roweth}},\ }\href
  {\doibase 10.1016/0370-2693(87)91197-X} {\bibfield  {journal} {\bibinfo
  {journal} {Phys. Lett. B}\ }\textbf {\bibinfo {volume} {195}},\ \bibinfo
  {pages} {216} (\bibinfo {year} {1987})}\BibitemShut {NoStop}%
\bibitem [{\citenamefont {Yoon}\ \emph
  {et~al.}(2017{\natexlab{b}})\citenamefont {Yoon} \emph
  {et~al.}}]{Yoon:2016jzj}%
  \BibitemOpen
  \bibfield  {author} {\bibinfo {author} {\bibfnamefont {B.}~\bibnamefont
  {Yoon}} \emph {et~al.},\ }\href {\doibase 10.1103/PhysRevD.95.074508}
  {\bibfield  {journal} {\bibinfo  {journal} {Phys. Rev. D}\ }\textbf {\bibinfo
  {volume} {95}},\ \bibinfo {pages} {074508} (\bibinfo {year}
  {2017}{\natexlab{b}})},\ \Eprint {http://arxiv.org/abs/1611.07452}
  {arXiv:1611.07452 [hep-lat]} \BibitemShut {NoStop}%
\bibitem [{\citenamefont {G{\"u}sken}\ \emph {et~al.}(1988)\citenamefont
  {G{\"u}sken}, \citenamefont {Schilling}, \citenamefont {Sommer},
  \citenamefont {M{\"u}tter},\ and\ \citenamefont {Patel}}]{Gusken:1988yi}%
  \BibitemOpen
  \bibfield  {author} {\bibinfo {author} {\bibfnamefont {S.}~\bibnamefont
  {G{\"u}sken}}, \bibinfo {author} {\bibfnamefont {K.}~\bibnamefont
  {Schilling}}, \bibinfo {author} {\bibfnamefont {R.}~\bibnamefont {Sommer}},
  \bibinfo {author} {\bibfnamefont {K.~H.}\ \bibnamefont {M{\"u}tter}}, \ and\
  \bibinfo {author} {\bibfnamefont {A.}~\bibnamefont {Patel}},\ }\href
  {\doibase 10.1016/0370-2693(88)90528-X} {\bibfield  {journal} {\bibinfo
  {journal} {Phys. Lett.}\ }\textbf {\bibinfo {volume} {B212}},\ \bibinfo
  {pages} {216} (\bibinfo {year} {1988})}\BibitemShut {NoStop}%
\bibitem [{\citenamefont {Yoon}\ \emph {et~al.}(2016)\citenamefont {Yoon} \emph
  {et~al.}}]{Yoon:2016dij}%
  \BibitemOpen
  \bibfield  {author} {\bibinfo {author} {\bibfnamefont {B.}~\bibnamefont
  {Yoon}} \emph {et~al.},\ }\href@noop {} {\bibfield  {journal} {\bibinfo
  {journal} {Phys. Rev.}\ }\textbf {\bibinfo {volume} {D93}},\ \bibinfo {pages}
  {114506} (\bibinfo {year} {2016})},\ \Eprint
  {http://arxiv.org/abs/1602.07737} {arXiv:1602.07737 [hep-lat]} \BibitemShut
  {NoStop}%
\bibitem [{\citenamefont {Babich}\ \emph {et~al.}(2010)\citenamefont {Babich},
  \citenamefont {Brannick}, \citenamefont {Brower}, \citenamefont {Clark},
  \citenamefont {Manteuffel} \emph {et~al.}}]{Babich:2010qb}%
  \BibitemOpen
  \bibfield  {author} {\bibinfo {author} {\bibfnamefont {R.}~\bibnamefont
  {Babich}}, \bibinfo {author} {\bibfnamefont {J.}~\bibnamefont {Brannick}},
  \bibinfo {author} {\bibfnamefont {R.}~\bibnamefont {Brower}}, \bibinfo
  {author} {\bibfnamefont {M.}~\bibnamefont {Clark}}, \bibinfo {author}
  {\bibfnamefont {T.}~\bibnamefont {Manteuffel}},  \emph {et~al.},\ }\href
  {\doibase 10.1103/PhysRevLett.105.201602} {\bibfield  {journal} {\bibinfo
  {journal} {Phys.Rev.Lett.}\ }\textbf {\bibinfo {volume} {105}},\ \bibinfo
  {pages} {201602} (\bibinfo {year} {2010})},\ \Eprint
  {http://arxiv.org/abs/1005.3043} {arXiv:1005.3043 [hep-lat]} \BibitemShut
  {NoStop}%
\bibitem [{\citenamefont {Clark}\ \emph {et~al.}(2010)\citenamefont {Clark},
  \citenamefont {Babich}, \citenamefont {Barros}, \citenamefont {Brower},\ and\
  \citenamefont {Rebbi}}]{Clark:2009wm}%
  \BibitemOpen
  \bibfield  {author} {\bibinfo {author} {\bibfnamefont {M.}~\bibnamefont
  {Clark}}, \bibinfo {author} {\bibfnamefont {R.}~\bibnamefont {Babich}},
  \bibinfo {author} {\bibfnamefont {K.}~\bibnamefont {Barros}}, \bibinfo
  {author} {\bibfnamefont {R.}~\bibnamefont {Brower}}, \ and\ \bibinfo {author}
  {\bibfnamefont {C.}~\bibnamefont {Rebbi}},\ }\href {\doibase
  10.1016/j.cpc.2010.05.002} {\bibfield  {journal} {\bibinfo  {journal}
  {Comput.Phys.Commun.}\ }\textbf {\bibinfo {volume} {181}},\ \bibinfo {pages}
  {1517} (\bibinfo {year} {2010})},\ \Eprint {http://arxiv.org/abs/0911.3191}
  {arXiv:0911.3191 [hep-lat]} \BibitemShut {NoStop}%
\bibitem [{\citenamefont {Clark}\ \emph {et~al.}(2016)\citenamefont {Clark},
  \citenamefont {Jo\'{o}}, \citenamefont {Strelchenko}, \citenamefont {Cheng},
  \citenamefont {Gambhir},\ and\ \citenamefont
  {Brower}}]{10.5555/3014904.3014995}%
  \BibitemOpen
  \bibfield  {author} {\bibinfo {author} {\bibfnamefont {M.~A.}\ \bibnamefont
  {Clark}}, \bibinfo {author} {\bibfnamefont {B.}~\bibnamefont {Jo\'{o}}},
  \bibinfo {author} {\bibfnamefont {A.}~\bibnamefont {Strelchenko}}, \bibinfo
  {author} {\bibfnamefont {M.}~\bibnamefont {Cheng}}, \bibinfo {author}
  {\bibfnamefont {A.}~\bibnamefont {Gambhir}}, \ and\ \bibinfo {author}
  {\bibfnamefont {R.~C.}\ \bibnamefont {Brower}},\ }in\ \href@noop {} {\emph
  {\bibinfo {booktitle} {Proceedings of the International Conference for High
  Performance Computing, Networking, Storage and Analysis}}},\ \bibinfo {series
  and number} {SC '16}\ (\bibinfo  {publisher} {IEEE Press},\ \bibinfo {year}
  {2016})\ p.~\bibinfo {pages} {68}\BibitemShut {NoStop}%
\bibitem [{\citenamefont {Akaike}(1974)}]{1100705}%
  \BibitemOpen
  \bibfield  {author} {\bibinfo {author} {\bibfnamefont {H.}~\bibnamefont
  {Akaike}},\ }\href {\doibase 10.1109/TAC.1974.1100705} {\bibfield  {journal}
  {\bibinfo  {journal} {IEEE Transactions on Automatic Control}\ }\textbf
  {\bibinfo {volume} {19}},\ \bibinfo {pages} {716} (\bibinfo {year}
  {1974})}\BibitemShut {NoStop}%
\bibitem [{\citenamefont {Edwards}\ and\ \citenamefont
  {Joo}(2005)}]{Edwards:2004sx}%
  \BibitemOpen
  \bibfield  {author} {\bibinfo {author} {\bibfnamefont {R.~G.}\ \bibnamefont
  {Edwards}}\ and\ \bibinfo {author} {\bibfnamefont {B.}~\bibnamefont {Joo}}
  (\bibinfo {collaboration} {SciDAC, LHPC, UKQCD}),\ }\href {\doibase
  10.1016/j.nuclphysbps.2004.11.254} {\bibfield  {journal} {\bibinfo  {journal}
  {Nucl. Phys. B Proc. Suppl.}\ }\textbf {\bibinfo {volume} {140}},\ \bibinfo
  {pages} {832} (\bibinfo {year} {2005})},\ \Eprint
  {http://arxiv.org/abs/hep-lat/0409003} {arXiv:hep-lat/0409003} \BibitemShut
  {NoStop}%
\bibitem [{\citenamefont {Gusken}\ \emph {et~al.}(1989)\citenamefont {Gusken},
  \citenamefont {Low}, \citenamefont {Mutter}, \citenamefont {Sommer},
  \citenamefont {Patel},\ and\ \citenamefont {Schilling}}]{Gusken:1989ad}%
  \BibitemOpen
  \bibfield  {author} {\bibinfo {author} {\bibfnamefont {S.}~\bibnamefont
  {Gusken}}, \bibinfo {author} {\bibfnamefont {U.}~\bibnamefont {Low}},
  \bibinfo {author} {\bibfnamefont {K.~H.}\ \bibnamefont {Mutter}}, \bibinfo
  {author} {\bibfnamefont {R.}~\bibnamefont {Sommer}}, \bibinfo {author}
  {\bibfnamefont {A.}~\bibnamefont {Patel}}, \ and\ \bibinfo {author}
  {\bibfnamefont {K.}~\bibnamefont {Schilling}},\ }\href {\doibase
  10.1016/S0370-2693(89)80034-6} {\bibfield  {journal} {\bibinfo  {journal}
  {Phys. Lett.}\ }\textbf {\bibinfo {volume} {B227}},\ \bibinfo {pages} {266}
  (\bibinfo {year} {1989})}\BibitemShut {NoStop}%
\bibitem [{\citenamefont {Gockeler}\ \emph {et~al.}(1996)\citenamefont
  {Gockeler}, \citenamefont {Horsley}, \citenamefont {Ilgenfritz},
  \citenamefont {Perlt}, \citenamefont {Rakow}, \citenamefont {Schierholz},\
  and\ \citenamefont {Schiller}}]{Gockeler:1995wg}%
  \BibitemOpen
  \bibfield  {author} {\bibinfo {author} {\bibfnamefont {M.}~\bibnamefont
  {Gockeler}}, \bibinfo {author} {\bibfnamefont {R.}~\bibnamefont {Horsley}},
  \bibinfo {author} {\bibfnamefont {E.-M.}\ \bibnamefont {Ilgenfritz}},
  \bibinfo {author} {\bibfnamefont {H.}~\bibnamefont {Perlt}}, \bibinfo
  {author} {\bibfnamefont {P.~E.~L.}\ \bibnamefont {Rakow}}, \bibinfo {author}
  {\bibfnamefont {G.}~\bibnamefont {Schierholz}}, \ and\ \bibinfo {author}
  {\bibfnamefont {A.}~\bibnamefont {Schiller}},\ }\href {\doibase
  10.1103/PhysRevD.53.2317} {\bibfield  {journal} {\bibinfo  {journal} {Phys.
  Rev.}\ }\textbf {\bibinfo {volume} {D53}},\ \bibinfo {pages} {2317} (\bibinfo
  {year} {1996})},\ \Eprint {http://arxiv.org/abs/hep-lat/9508004}
  {arXiv:hep-lat/9508004 [hep-lat]} \BibitemShut {NoStop}%
\bibitem [{\citenamefont {Harris}\ \emph {et~al.}(2019)\citenamefont {Harris},
  \citenamefont {von Hippel}, \citenamefont {Junnarkar}, \citenamefont {Meyer},
  \citenamefont {Ottnad}, \citenamefont {Wilhelm}, \citenamefont {Wittig},\
  and\ \citenamefont {Wrang}}]{Harris:2019bih}%
  \BibitemOpen
  \bibfield  {author} {\bibinfo {author} {\bibfnamefont {T.}~\bibnamefont
  {Harris}}, \bibinfo {author} {\bibfnamefont {G.}~\bibnamefont {von Hippel}},
  \bibinfo {author} {\bibfnamefont {P.}~\bibnamefont {Junnarkar}}, \bibinfo
  {author} {\bibfnamefont {H.~B.}\ \bibnamefont {Meyer}}, \bibinfo {author}
  {\bibfnamefont {K.}~\bibnamefont {Ottnad}}, \bibinfo {author} {\bibfnamefont
  {J.}~\bibnamefont {Wilhelm}}, \bibinfo {author} {\bibfnamefont
  {H.}~\bibnamefont {Wittig}}, \ and\ \bibinfo {author} {\bibfnamefont
  {L.}~\bibnamefont {Wrang}},\ }\href {\doibase 10.1103/PhysRevD.100.034513}
  {\bibfield  {journal} {\bibinfo  {journal} {Phys. Rev.}\ }\textbf {\bibinfo
  {volume} {D100}},\ \bibinfo {pages} {034513} (\bibinfo {year} {2019})},\
  \Eprint {http://arxiv.org/abs/1905.01291} {arXiv:1905.01291 [hep-lat]}
  \BibitemShut {NoStop}%
\bibitem [{\citenamefont {Bhattacharya}\ \emph {et~al.}(2014)\citenamefont
  {Bhattacharya}, \citenamefont {Cohen}, \citenamefont {Gupta}, \citenamefont
  {Joseph}, \citenamefont {Lin},\ and\ \citenamefont
  {Yoon}}]{Bhattacharya:2013ehc}%
  \BibitemOpen
  \bibfield  {author} {\bibinfo {author} {\bibfnamefont {T.}~\bibnamefont
  {Bhattacharya}}, \bibinfo {author} {\bibfnamefont {S.~D.}\ \bibnamefont
  {Cohen}}, \bibinfo {author} {\bibfnamefont {R.}~\bibnamefont {Gupta}},
  \bibinfo {author} {\bibfnamefont {A.}~\bibnamefont {Joseph}}, \bibinfo
  {author} {\bibfnamefont {H.-W.}\ \bibnamefont {Lin}}, \ and\ \bibinfo
  {author} {\bibfnamefont {B.}~\bibnamefont {Yoon}},\ }\href {\doibase
  10.1103/PhysRevD.89.094502} {\bibfield  {journal} {\bibinfo  {journal} {Phys.
  Rev.}\ }\textbf {\bibinfo {volume} {D89}},\ \bibinfo {pages} {094502}
  (\bibinfo {year} {2014})},\ \Eprint {http://arxiv.org/abs/1306.5435}
  {arXiv:1306.5435 [hep-lat]} \BibitemShut {NoStop}%
\bibitem [{\citenamefont {Bali}\ \emph {et~al.}(2014)\citenamefont {Bali},
  \citenamefont {Collins}, \citenamefont {Gläßle}, \citenamefont {Göckeler},
  \citenamefont {Najjar}, \citenamefont {Rödl}, \citenamefont {Schäfer},
  \citenamefont {Schiel}, \citenamefont {Sternbeck},\ and\ \citenamefont
  {Söldner}}]{Bali:2014gha}%
  \BibitemOpen
  \bibfield  {author} {\bibinfo {author} {\bibfnamefont {G.~S.}\ \bibnamefont
  {Bali}}, \bibinfo {author} {\bibfnamefont {S.}~\bibnamefont {Collins}},
  \bibinfo {author} {\bibfnamefont {B.}~\bibnamefont {Gläßle}}, \bibinfo
  {author} {\bibfnamefont {M.}~\bibnamefont {Göckeler}}, \bibinfo {author}
  {\bibfnamefont {J.}~\bibnamefont {Najjar}}, \bibinfo {author} {\bibfnamefont
  {R.~H.}\ \bibnamefont {Rödl}}, \bibinfo {author} {\bibfnamefont
  {A.}~\bibnamefont {Schäfer}}, \bibinfo {author} {\bibfnamefont {R.~W.}\
  \bibnamefont {Schiel}}, \bibinfo {author} {\bibfnamefont {A.}~\bibnamefont
  {Sternbeck}}, \ and\ \bibinfo {author} {\bibfnamefont {W.}~\bibnamefont
  {Söldner}},\ }\href {\doibase 10.1103/PhysRevD.90.074510} {\bibfield
  {journal} {\bibinfo  {journal} {Phys. Rev.}\ }\textbf {\bibinfo {volume}
  {D90}},\ \bibinfo {pages} {074510} (\bibinfo {year} {2014})},\ \Eprint
  {http://arxiv.org/abs/1408.6850} {arXiv:1408.6850 [hep-lat]} \BibitemShut
  {NoStop}%
\bibitem [{\citenamefont {Bali}\ \emph {et~al.}(2012)\citenamefont {Bali},
  \citenamefont {Collins}, \citenamefont {Deka}, \citenamefont {Glassle},
  \citenamefont {Gockeler}, \citenamefont {Najjar}, \citenamefont {Nobile},
  \citenamefont {Pleiter}, \citenamefont {Schafer},\ and\ \citenamefont
  {Sternbeck}}]{Bali:2012av}%
  \BibitemOpen
  \bibfield  {author} {\bibinfo {author} {\bibfnamefont {G.~S.}\ \bibnamefont
  {Bali}}, \bibinfo {author} {\bibfnamefont {S.}~\bibnamefont {Collins}},
  \bibinfo {author} {\bibfnamefont {M.}~\bibnamefont {Deka}}, \bibinfo {author}
  {\bibfnamefont {B.}~\bibnamefont {Glassle}}, \bibinfo {author} {\bibfnamefont
  {M.}~\bibnamefont {Gockeler}}, \bibinfo {author} {\bibfnamefont
  {J.}~\bibnamefont {Najjar}}, \bibinfo {author} {\bibfnamefont
  {A.}~\bibnamefont {Nobile}}, \bibinfo {author} {\bibfnamefont
  {D.}~\bibnamefont {Pleiter}}, \bibinfo {author} {\bibfnamefont
  {A.}~\bibnamefont {Schafer}}, \ and\ \bibinfo {author} {\bibfnamefont
  {A.}~\bibnamefont {Sternbeck}},\ }\href {\doibase 10.1103/PhysRevD.86.054504}
  {\bibfield  {journal} {\bibinfo  {journal} {Phys. Rev.}\ }\textbf {\bibinfo
  {volume} {D86}},\ \bibinfo {pages} {054504} (\bibinfo {year} {2012})},\
  \Eprint {http://arxiv.org/abs/1207.1110} {arXiv:1207.1110 [hep-lat]}
  \BibitemShut {NoStop}%
\bibitem [{\citenamefont {Bar}(2019)}]{Bar:2018xyi}%
  \BibitemOpen
  \bibfield  {author} {\bibinfo {author} {\bibfnamefont {O.}~\bibnamefont
  {Bar}},\ }\href {\doibase 10.1103/PhysRevD.99.054506} {\bibfield  {journal}
  {\bibinfo  {journal} {Phys. Rev. D}\ }\textbf {\bibinfo {volume} {99}},\
  \bibinfo {pages} {054506} (\bibinfo {year} {2019})},\ \Eprint
  {http://arxiv.org/abs/1812.09191} {arXiv:1812.09191 [hep-lat]} \BibitemShut
  {NoStop}%
\bibitem [{\citenamefont {Jang}\ \emph {et~al.}(2020)\citenamefont {Jang},
  \citenamefont {Gupta}, \citenamefont {Yoon},\ and\ \citenamefont
  {Bhattacharya}}]{Jang:2019vkm}%
  \BibitemOpen
  \bibfield  {author} {\bibinfo {author} {\bibfnamefont {Y.-C.}\ \bibnamefont
  {Jang}}, \bibinfo {author} {\bibfnamefont {R.}~\bibnamefont {Gupta}},
  \bibinfo {author} {\bibfnamefont {B.}~\bibnamefont {Yoon}}, \ and\ \bibinfo
  {author} {\bibfnamefont {T.}~\bibnamefont {Bhattacharya}},\ }\href {\doibase
  10.1103/PhysRevLett.124.072002} {\bibfield  {journal} {\bibinfo  {journal}
  {Phys. Rev. Lett.}\ }\textbf {\bibinfo {volume} {124}},\ \bibinfo {pages}
  {072002} (\bibinfo {year} {2020})},\ \Eprint
  {http://arxiv.org/abs/1905.06470} {arXiv:1905.06470 [hep-lat]} \BibitemShut
  {NoStop}%
\bibitem [{\citenamefont {Djukanovic}\ \emph {et~al.}(2024)\citenamefont
  {Djukanovic}, \citenamefont {von Hippel}, \citenamefont {Meyer},
  \citenamefont {Ottnad},\ and\ \citenamefont {Wittig}}]{Djukanovic:2024krw}%
  \BibitemOpen
  \bibfield  {author} {\bibinfo {author} {\bibfnamefont {D.}~\bibnamefont
  {Djukanovic}}, \bibinfo {author} {\bibfnamefont {G.}~\bibnamefont {von
  Hippel}}, \bibinfo {author} {\bibfnamefont {H.~B.}\ \bibnamefont {Meyer}},
  \bibinfo {author} {\bibfnamefont {K.}~\bibnamefont {Ottnad}}, \ and\ \bibinfo
  {author} {\bibfnamefont {H.}~\bibnamefont {Wittig}},\ }\href {\doibase
  10.1103/PhysRevD.109.074507} {\bibfield  {journal} {\bibinfo  {journal}
  {Phys. Rev. D}\ }\textbf {\bibinfo {volume} {109}},\ \bibinfo {pages}
  {074507} (\bibinfo {year} {2024})},\ \Eprint
  {http://arxiv.org/abs/2402.03024} {arXiv:2402.03024 [hep-lat]} \BibitemShut
  {NoStop}%
\bibitem [{\citenamefont {Yang}\ \emph {et~al.}(2018)\citenamefont {Yang},
  \citenamefont {Liang}, \citenamefont {Bi}, \citenamefont {Chen},
  \citenamefont {Draper}, \citenamefont {Liu},\ and\ \citenamefont
  {Liu}}]{Yang:2018nqn}%
  \BibitemOpen
  \bibfield  {author} {\bibinfo {author} {\bibfnamefont {Y.-B.}\ \bibnamefont
  {Yang}}, \bibinfo {author} {\bibfnamefont {J.}~\bibnamefont {Liang}},
  \bibinfo {author} {\bibfnamefont {Y.-J.}\ \bibnamefont {Bi}}, \bibinfo
  {author} {\bibfnamefont {Y.}~\bibnamefont {Chen}}, \bibinfo {author}
  {\bibfnamefont {T.}~\bibnamefont {Draper}}, \bibinfo {author} {\bibfnamefont
  {K.-F.}\ \bibnamefont {Liu}}, \ and\ \bibinfo {author} {\bibfnamefont
  {Z.}~\bibnamefont {Liu}},\ }\href {\doibase 10.1103/PhysRevLett.121.212001}
  {\bibfield  {journal} {\bibinfo  {journal} {Phys. Rev. Lett.}\ }\textbf
  {\bibinfo {volume} {121}},\ \bibinfo {pages} {212001} (\bibinfo {year}
  {2018})},\ \Eprint {http://arxiv.org/abs/1808.08677} {arXiv:1808.08677
  [hep-lat]} \BibitemShut {NoStop}%
\bibitem [{\citenamefont {Alexandrou}\ \emph {et~al.}(2023)\citenamefont
  {Alexandrou} \emph {et~al.}}]{Alexandrou:2022dtc}%
  \BibitemOpen
  \bibfield  {author} {\bibinfo {author} {\bibfnamefont {C.}~\bibnamefont
  {Alexandrou}} \emph {et~al.},\ }\href {\doibase 10.1103/PhysRevD.107.054504}
  {\bibfield  {journal} {\bibinfo  {journal} {Phys. Rev. D}\ }\textbf {\bibinfo
  {volume} {107}},\ \bibinfo {pages} {054504} (\bibinfo {year} {2023})},\
  \Eprint {http://arxiv.org/abs/2202.09871} {arXiv:2202.09871 [hep-lat]}
  \BibitemShut {NoStop}%
\bibitem [{\citenamefont {Li}\ \emph {et~al.}(2024)\citenamefont {Li},
  \citenamefont {Accardi}, \citenamefont {Cerutti}, \citenamefont {Fernando},
  \citenamefont {Keppel}, \citenamefont {Melnitchouk}, \citenamefont
  {Monaghan}, \citenamefont {Niculescu}, \citenamefont {Niculescu},\ and\
  \citenamefont {Owens}}]{Li:2023yda}%
  \BibitemOpen
  \bibfield  {author} {\bibinfo {author} {\bibfnamefont {S.}~\bibnamefont
  {Li}}, \bibinfo {author} {\bibfnamefont {A.}~\bibnamefont {Accardi}},
  \bibinfo {author} {\bibfnamefont {M.}~\bibnamefont {Cerutti}}, \bibinfo
  {author} {\bibfnamefont {I.~P.}\ \bibnamefont {Fernando}}, \bibinfo {author}
  {\bibfnamefont {C.~E.}\ \bibnamefont {Keppel}}, \bibinfo {author}
  {\bibfnamefont {W.}~\bibnamefont {Melnitchouk}}, \bibinfo {author}
  {\bibfnamefont {P.}~\bibnamefont {Monaghan}}, \bibinfo {author}
  {\bibfnamefont {G.}~\bibnamefont {Niculescu}}, \bibinfo {author}
  {\bibfnamefont {M.~I.}\ \bibnamefont {Niculescu}}, \ and\ \bibinfo {author}
  {\bibfnamefont {J.~F.}\ \bibnamefont {Owens}},\ }\href {\doibase
  10.1103/PhysRevD.109.074036} {\bibfield  {journal} {\bibinfo  {journal}
  {Phys. Rev. D}\ }\textbf {\bibinfo {volume} {109}},\ \bibinfo {pages}
  {074036} (\bibinfo {year} {2024})},\ \Eprint
  {http://arxiv.org/abs/2309.16851} {arXiv:2309.16851 [hep-ph]} \BibitemShut
  {NoStop}%
\bibitem [{\citenamefont {Cocuzza}\ \emph {et~al.}(2021)\citenamefont
  {Cocuzza}, \citenamefont {Keppel}, \citenamefont {Liu}, \citenamefont
  {Melnitchouk}, \citenamefont {Metz}, \citenamefont {Sato},\ and\
  \citenamefont {Thomas}}]{Cocuzza:2021rfn}%
  \BibitemOpen
  \bibfield  {author} {\bibinfo {author} {\bibfnamefont {C.}~\bibnamefont
  {Cocuzza}}, \bibinfo {author} {\bibfnamefont {C.~E.}\ \bibnamefont {Keppel}},
  \bibinfo {author} {\bibfnamefont {H.}~\bibnamefont {Liu}}, \bibinfo {author}
  {\bibfnamefont {W.}~\bibnamefont {Melnitchouk}}, \bibinfo {author}
  {\bibfnamefont {A.}~\bibnamefont {Metz}}, \bibinfo {author} {\bibfnamefont
  {N.}~\bibnamefont {Sato}}, \ and\ \bibinfo {author} {\bibfnamefont {A.~W.}\
  \bibnamefont {Thomas}} (\bibinfo {collaboration} {Jefferson Lab Angular
  Momentum (JAM)}),\ }\href {\doibase 10.1103/PhysRevLett.127.242001}
  {\bibfield  {journal} {\bibinfo  {journal} {Phys. Rev. Lett.}\ }\textbf
  {\bibinfo {volume} {127}},\ \bibinfo {pages} {242001} (\bibinfo {year}
  {2021})},\ \Eprint {http://arxiv.org/abs/2104.06946} {arXiv:2104.06946
  [hep-ph]} \BibitemShut {NoStop}%
\bibitem [{\citenamefont {Hou}\ \emph {et~al.}(2019)\citenamefont {Hou} \emph
  {et~al.}}]{Hou:2019efy}%
  \BibitemOpen
  \bibfield  {author} {\bibinfo {author} {\bibfnamefont {T.-J.}\ \bibnamefont
  {Hou}} \emph {et~al.},\ }\href@noop {} {\enquote {\bibinfo {title} {{New CTEQ
  global analysis of quantum chromodynamics with high-precision data from the
  LHC}},}\ } (\bibinfo {year} {2019}),\ \Eprint
  {http://arxiv.org/abs/1912.10053} {arXiv:1912.10053 [hep-ph]} \BibitemShut
  {NoStop}%
\bibitem [{\citenamefont {Ethier}\ \emph {et~al.}(2017)\citenamefont {Ethier},
  \citenamefont {Sato},\ and\ \citenamefont {Melnitchouk}}]{Ethier:2017zbq}%
  \BibitemOpen
  \bibfield  {author} {\bibinfo {author} {\bibfnamefont {J.}~\bibnamefont
  {Ethier}}, \bibinfo {author} {\bibfnamefont {N.}~\bibnamefont {Sato}}, \ and\
  \bibinfo {author} {\bibfnamefont {W.}~\bibnamefont {Melnitchouk}},\ }\href
  {\doibase 10.1103/PhysRevLett.119.132001} {\bibfield  {journal} {\bibinfo
  {journal} {Phys. Rev. Lett.}\ }\textbf {\bibinfo {volume} {119}},\ \bibinfo
  {pages} {132001} (\bibinfo {year} {2017})},\ \Eprint
  {http://arxiv.org/abs/1705.05889} {arXiv:1705.05889 [hep-ph]} \BibitemShut
  {NoStop}%
\bibitem [{\citenamefont {Ball}\ \emph {et~al.}(2017)\citenamefont {Ball} \emph
  {et~al.}}]{Ball:2017nwa}%
  \BibitemOpen
  \bibfield  {author} {\bibinfo {author} {\bibfnamefont {R.~D.}\ \bibnamefont
  {Ball}} \emph {et~al.} (\bibinfo {collaboration} {NNPDF}),\ }\href {\doibase
  10.1140/epjc/s10052-017-5199-5} {\bibfield  {journal} {\bibinfo  {journal}
  {Eur. Phys. J.}\ }\textbf {\bibinfo {volume} {C77}},\ \bibinfo {pages} {663}
  (\bibinfo {year} {2017})},\ \Eprint {http://arxiv.org/abs/1706.00428}
  {arXiv:1706.00428 [hep-ph]} \BibitemShut {NoStop}%
\bibitem [{\citenamefont {Alekhin}\ \emph {et~al.}(2017)\citenamefont
  {Alekhin}, \citenamefont {Blümlein}, \citenamefont {Moch},\ and\
  \citenamefont {Placakyte}}]{Alekhin:2017kpj}%
  \BibitemOpen
  \bibfield  {author} {\bibinfo {author} {\bibfnamefont {S.}~\bibnamefont
  {Alekhin}}, \bibinfo {author} {\bibfnamefont {J.}~\bibnamefont {Blümlein}},
  \bibinfo {author} {\bibfnamefont {S.}~\bibnamefont {Moch}}, \ and\ \bibinfo
  {author} {\bibfnamefont {R.}~\bibnamefont {Placakyte}},\ }\href {\doibase
  10.1103/PhysRevD.96.014011} {\bibfield  {journal} {\bibinfo  {journal} {Phys.
  Rev.}\ }\textbf {\bibinfo {volume} {D96}},\ \bibinfo {pages} {014011}
  (\bibinfo {year} {2017})},\ \Eprint {http://arxiv.org/abs/1701.05838}
  {arXiv:1701.05838 [hep-ph]} \BibitemShut {NoStop}%
\bibitem [{\citenamefont {Accardi}\ \emph
  {et~al.}(2016{\natexlab{b}})\citenamefont {Accardi}, \citenamefont {Brady},
  \citenamefont {Melnitchouk}, \citenamefont {Owens},\ and\ \citenamefont
  {Sato}}]{Accardi:2016qay}%
  \BibitemOpen
  \bibfield  {author} {\bibinfo {author} {\bibfnamefont {A.}~\bibnamefont
  {Accardi}}, \bibinfo {author} {\bibfnamefont {L.~T.}\ \bibnamefont {Brady}},
  \bibinfo {author} {\bibfnamefont {W.}~\bibnamefont {Melnitchouk}}, \bibinfo
  {author} {\bibfnamefont {J.~F.}\ \bibnamefont {Owens}}, \ and\ \bibinfo
  {author} {\bibfnamefont {N.}~\bibnamefont {Sato}},\ }\href {\doibase
  10.1103/PhysRevD.93.114017} {\bibfield  {journal} {\bibinfo  {journal} {Phys.
  Rev.}\ }\textbf {\bibinfo {volume} {D93}},\ \bibinfo {pages} {114017}
  (\bibinfo {year} {2016}{\natexlab{b}})},\ \Eprint
  {http://arxiv.org/abs/1602.03154} {arXiv:1602.03154 [hep-ph]} \BibitemShut
  {NoStop}%
\bibitem [{\citenamefont {Abramowicz}\ \emph {et~al.}(2015)\citenamefont
  {Abramowicz} \emph {et~al.}}]{Abramowicz:2015mha}%
  \BibitemOpen
  \bibfield  {author} {\bibinfo {author} {\bibfnamefont {H.}~\bibnamefont
  {Abramowicz}} \emph {et~al.} (\bibinfo {collaboration} {H1, ZEUS}),\ }\href
  {\doibase 10.1140/epjc/s10052-015-3710-4} {\bibfield  {journal} {\bibinfo
  {journal} {Eur. Phys. J.}\ }\textbf {\bibinfo {volume} {C75}},\ \bibinfo
  {pages} {580} (\bibinfo {year} {2015})},\ \Eprint
  {http://arxiv.org/abs/1506.06042} {arXiv:1506.06042 [hep-ex]} \BibitemShut
  {NoStop}%
\bibitem [{\citenamefont {Dulat}\ \emph {et~al.}(2016)\citenamefont {Dulat},
  \citenamefont {Hou}, \citenamefont {Gao}, \citenamefont {Guzzi},
  \citenamefont {Huston}, \citenamefont {Nadolsky}, \citenamefont {Pumplin},
  \citenamefont {Schmidt}, \citenamefont {Stump},\ and\ \citenamefont
  {Yuan}}]{Dulat:2015mca}%
  \BibitemOpen
  \bibfield  {author} {\bibinfo {author} {\bibfnamefont {S.}~\bibnamefont
  {Dulat}}, \bibinfo {author} {\bibfnamefont {T.-J.}\ \bibnamefont {Hou}},
  \bibinfo {author} {\bibfnamefont {J.}~\bibnamefont {Gao}}, \bibinfo {author}
  {\bibfnamefont {M.}~\bibnamefont {Guzzi}}, \bibinfo {author} {\bibfnamefont
  {J.}~\bibnamefont {Huston}}, \bibinfo {author} {\bibfnamefont
  {P.}~\bibnamefont {Nadolsky}}, \bibinfo {author} {\bibfnamefont
  {J.}~\bibnamefont {Pumplin}}, \bibinfo {author} {\bibfnamefont
  {C.}~\bibnamefont {Schmidt}}, \bibinfo {author} {\bibfnamefont
  {D.}~\bibnamefont {Stump}}, \ and\ \bibinfo {author} {\bibfnamefont {C.~P.}\
  \bibnamefont {Yuan}},\ }\href {\doibase 10.1103/PhysRevD.93.033006}
  {\bibfield  {journal} {\bibinfo  {journal} {Phys. Rev.}\ }\textbf {\bibinfo
  {volume} {D93}},\ \bibinfo {pages} {033006} (\bibinfo {year} {2016})},\
  \Eprint {http://arxiv.org/abs/1506.07443} {arXiv:1506.07443 [hep-ph]}
  \BibitemShut {NoStop}%
\bibitem [{\citenamefont {Harland-Lang}\ \emph {et~al.}(2015)\citenamefont
  {Harland-Lang}, \citenamefont {Martin}, \citenamefont {Motylinski},\ and\
  \citenamefont {Thorne}}]{Harland-Lang:2014zoa}%
  \BibitemOpen
  \bibfield  {author} {\bibinfo {author} {\bibfnamefont {L.~A.}\ \bibnamefont
  {Harland-Lang}}, \bibinfo {author} {\bibfnamefont {A.~D.}\ \bibnamefont
  {Martin}}, \bibinfo {author} {\bibfnamefont {P.}~\bibnamefont {Motylinski}},
  \ and\ \bibinfo {author} {\bibfnamefont {R.~S.}\ \bibnamefont {Thorne}},\
  }\href {\doibase 10.1140/epjc/s10052-015-3397-6} {\bibfield  {journal}
  {\bibinfo  {journal} {Eur. Phys. J.}\ }\textbf {\bibinfo {volume} {C75}},\
  \bibinfo {pages} {204} (\bibinfo {year} {2015})},\ \Eprint
  {http://arxiv.org/abs/1412.3989} {arXiv:1412.3989 [hep-ph]} \BibitemShut
  {NoStop}%
\bibitem [{\citenamefont {Nocera}\ \emph {et~al.}(2014)\citenamefont {Nocera},
  \citenamefont {Ball}, \citenamefont {Forte}, \citenamefont {Ridolfi},\ and\
  \citenamefont {Rojo}}]{Nocera:2014gqa}%
  \BibitemOpen
  \bibfield  {author} {\bibinfo {author} {\bibfnamefont {E.~R.}\ \bibnamefont
  {Nocera}}, \bibinfo {author} {\bibfnamefont {R.~D.}\ \bibnamefont {Ball}},
  \bibinfo {author} {\bibfnamefont {S.}~\bibnamefont {Forte}}, \bibinfo
  {author} {\bibfnamefont {G.}~\bibnamefont {Ridolfi}}, \ and\ \bibinfo
  {author} {\bibfnamefont {J.}~\bibnamefont {Rojo}} (\bibinfo {collaboration}
  {NNPDF}),\ }\href {\doibase 10.1016/j.nuclphysb.2014.08.008} {\bibfield
  {journal} {\bibinfo  {journal} {Nucl. Phys.}\ }\textbf {\bibinfo {volume}
  {B887}},\ \bibinfo {pages} {276} (\bibinfo {year} {2014})},\ \Eprint
  {http://arxiv.org/abs/1406.5539} {arXiv:1406.5539 [hep-ph]} \BibitemShut
  {NoStop}%
\bibitem [{\citenamefont {de~Florian}\ \emph {et~al.}(2009)\citenamefont
  {de~Florian}, \citenamefont {Sassot}, \citenamefont {Stratmann},\ and\
  \citenamefont {Vogelsang}}]{deFlorian:2009vb}%
  \BibitemOpen
  \bibfield  {author} {\bibinfo {author} {\bibfnamefont {D.}~\bibnamefont
  {de~Florian}}, \bibinfo {author} {\bibfnamefont {R.}~\bibnamefont {Sassot}},
  \bibinfo {author} {\bibfnamefont {M.}~\bibnamefont {Stratmann}}, \ and\
  \bibinfo {author} {\bibfnamefont {W.}~\bibnamefont {Vogelsang}},\ }\href
  {\doibase 10.1103/PhysRevD.80.034030} {\bibfield  {journal} {\bibinfo
  {journal} {Phys. Rev.}\ }\textbf {\bibinfo {volume} {D80}},\ \bibinfo {pages}
  {034030} (\bibinfo {year} {2009})},\ \Eprint {http://arxiv.org/abs/0904.3821}
  {arXiv:0904.3821 [hep-ph]} \BibitemShut {NoStop}%
\bibitem [{\citenamefont {de~Florian}\ \emph {et~al.}(2008)\citenamefont
  {de~Florian}, \citenamefont {Sassot}, \citenamefont {Stratmann},\ and\
  \citenamefont {Vogelsang}}]{deFlorian:2008mr}%
  \BibitemOpen
  \bibfield  {author} {\bibinfo {author} {\bibfnamefont {D.}~\bibnamefont
  {de~Florian}}, \bibinfo {author} {\bibfnamefont {R.}~\bibnamefont {Sassot}},
  \bibinfo {author} {\bibfnamefont {M.}~\bibnamefont {Stratmann}}, \ and\
  \bibinfo {author} {\bibfnamefont {W.}~\bibnamefont {Vogelsang}},\ }\href
  {\doibase 10.1103/PhysRevLett.101.072001} {\bibfield  {journal} {\bibinfo
  {journal} {Phys. Rev. Lett.}\ }\textbf {\bibinfo {volume} {101}},\ \bibinfo
  {pages} {072001} (\bibinfo {year} {2008})},\ \Eprint
  {http://arxiv.org/abs/0804.0422} {arXiv:0804.0422 [hep-ph]} \BibitemShut
  {NoStop}%
\bibitem [{\citenamefont {Gockeler}\ \emph {et~al.}(2010)\citenamefont
  {Gockeler} \emph {et~al.}}]{Gockeler:2010yr}%
  \BibitemOpen
  \bibfield  {author} {\bibinfo {author} {\bibfnamefont {M.}~\bibnamefont
  {Gockeler}} \emph {et~al.},\ }\href {\doibase 10.1103/PhysRevD.82.114511,
  10.1103/PhysRevD.86.099903} {\bibfield  {journal} {\bibinfo  {journal} {Phys.
  Rev.}\ }\textbf {\bibinfo {volume} {D82}},\ \bibinfo {pages} {114511}
  (\bibinfo {year} {2010})},\ \bibinfo {note} {[Erratum: Phys.
  Rev.D86,099903(2012)]},\ \Eprint {http://arxiv.org/abs/1003.5756}
  {arXiv:1003.5756 [hep-lat]} \BibitemShut {NoStop}%
\bibitem [{\citenamefont {Constantinou}\ \emph {et~al.}(2013)\citenamefont
  {Constantinou}, \citenamefont {Costa}, \citenamefont {Göckeler},
  \citenamefont {Horsley}, \citenamefont {Panagopoulos}, \citenamefont {Perlt},
  \citenamefont {Rakow}, \citenamefont {Schierholz},\ and\ \citenamefont
  {Schiller}}]{Constantinou:2013ada}%
  \BibitemOpen
  \bibfield  {author} {\bibinfo {author} {\bibfnamefont {M.}~\bibnamefont
  {Constantinou}}, \bibinfo {author} {\bibfnamefont {M.}~\bibnamefont {Costa}},
  \bibinfo {author} {\bibfnamefont {M.}~\bibnamefont {Göckeler}}, \bibinfo
  {author} {\bibfnamefont {R.}~\bibnamefont {Horsley}}, \bibinfo {author}
  {\bibfnamefont {H.}~\bibnamefont {Panagopoulos}}, \bibinfo {author}
  {\bibfnamefont {H.}~\bibnamefont {Perlt}}, \bibinfo {author} {\bibfnamefont
  {P.~E.~L.}\ \bibnamefont {Rakow}}, \bibinfo {author} {\bibfnamefont
  {G.}~\bibnamefont {Schierholz}}, \ and\ \bibinfo {author} {\bibfnamefont
  {A.}~\bibnamefont {Schiller}},\ }\href {\doibase 10.1103/PhysRevD.87.096019}
  {\bibfield  {journal} {\bibinfo  {journal} {Phys. Rev.}\ }\textbf {\bibinfo
  {volume} {D87}},\ \bibinfo {pages} {096019} (\bibinfo {year} {2013})},\
  \Eprint {http://arxiv.org/abs/1303.6776} {arXiv:1303.6776 [hep-lat]}
  \BibitemShut {NoStop}%
\bibitem [{\citenamefont {Gracey}(2003)}]{Gracey:2003mr}%
  \BibitemOpen
  \bibfield  {author} {\bibinfo {author} {\bibfnamefont {J.~A.}\ \bibnamefont
  {Gracey}},\ }\href {\doibase 10.1016/S0550-3213(03)00543-1} {\bibfield
  {journal} {\bibinfo  {journal} {Nucl. Phys.}\ }\textbf {\bibinfo {volume}
  {B667}},\ \bibinfo {pages} {242} (\bibinfo {year} {2003})},\ \Eprint
  {http://arxiv.org/abs/hep-ph/0306163} {arXiv:hep-ph/0306163 [hep-ph]}
  \BibitemShut {NoStop}%
\bibitem [{\citenamefont {Constantinou}\ \emph {et~al.}(2015)\citenamefont
  {Constantinou}, \citenamefont {Horsley}, \citenamefont {Panagopoulos},
  \citenamefont {Perlt}, \citenamefont {Rakow}, \citenamefont {Schierholz},
  \citenamefont {Schiller},\ and\ \citenamefont
  {Zanotti}}]{Constantinou:2014fka}%
  \BibitemOpen
  \bibfield  {author} {\bibinfo {author} {\bibfnamefont {M.}~\bibnamefont
  {Constantinou}}, \bibinfo {author} {\bibfnamefont {R.}~\bibnamefont
  {Horsley}}, \bibinfo {author} {\bibfnamefont {H.}~\bibnamefont
  {Panagopoulos}}, \bibinfo {author} {\bibfnamefont {H.}~\bibnamefont {Perlt}},
  \bibinfo {author} {\bibfnamefont {P.}~\bibnamefont {Rakow}}, \bibinfo
  {author} {\bibfnamefont {G.}~\bibnamefont {Schierholz}}, \bibinfo {author}
  {\bibfnamefont {A.}~\bibnamefont {Schiller}}, \ and\ \bibinfo {author}
  {\bibfnamefont {J.}~\bibnamefont {Zanotti}},\ }\href {\doibase
  10.1103/PhysRevD.91.014502} {\bibfield  {journal} {\bibinfo  {journal} {Phys.
  Rev. D}\ }\textbf {\bibinfo {volume} {91}},\ \bibinfo {pages} {014502}
  (\bibinfo {year} {2015})},\ \Eprint {http://arxiv.org/abs/1408.6047}
  {arXiv:1408.6047 [hep-lat]} \BibitemShut {NoStop}%
\bibitem [{\citenamefont {Alexandrou}\ \emph {et~al.}(2020)\citenamefont
  {Alexandrou}, \citenamefont {Bacchio}, \citenamefont {Constantinou},
  \citenamefont {Finkenrath}, \citenamefont {Hadjiyiannakou}, \citenamefont
  {Jansen}, \citenamefont {Koutsou}, \citenamefont {Panagopoulos},\ and\
  \citenamefont {Spanoudes}}]{Alexandrou:2020sml}%
  \BibitemOpen
  \bibfield  {author} {\bibinfo {author} {\bibfnamefont {C.}~\bibnamefont
  {Alexandrou}}, \bibinfo {author} {\bibfnamefont {S.}~\bibnamefont {Bacchio}},
  \bibinfo {author} {\bibfnamefont {M.}~\bibnamefont {Constantinou}}, \bibinfo
  {author} {\bibfnamefont {J.}~\bibnamefont {Finkenrath}}, \bibinfo {author}
  {\bibfnamefont {K.}~\bibnamefont {Hadjiyiannakou}}, \bibinfo {author}
  {\bibfnamefont {K.}~\bibnamefont {Jansen}}, \bibinfo {author} {\bibfnamefont
  {G.}~\bibnamefont {Koutsou}}, \bibinfo {author} {\bibfnamefont
  {H.}~\bibnamefont {Panagopoulos}}, \ and\ \bibinfo {author} {\bibfnamefont
  {G.}~\bibnamefont {Spanoudes}},\ }\href {\doibase
  10.1103/PhysRevD.101.094513} {\bibfield  {journal} {\bibinfo  {journal}
  {Phys. Rev. D}\ }\textbf {\bibinfo {volume} {101}},\ \bibinfo {pages}
  {094513} (\bibinfo {year} {2020})},\ \Eprint
  {http://arxiv.org/abs/2003.08486} {arXiv:2003.08486 [hep-lat]} \BibitemShut
  {NoStop}%
\bibitem [{\citenamefont {Bhattacharya}\ \emph {et~al.}(2016)\citenamefont
  {Bhattacharya}, \citenamefont {Cirigliano}, \citenamefont {Cohen},
  \citenamefont {Gupta}, \citenamefont {Lin},\ and\ \citenamefont
  {Yoon}}]{Bhattacharya:2016zcn}%
  \BibitemOpen
  \bibfield  {author} {\bibinfo {author} {\bibfnamefont {T.}~\bibnamefont
  {Bhattacharya}}, \bibinfo {author} {\bibfnamefont {V.}~\bibnamefont
  {Cirigliano}}, \bibinfo {author} {\bibfnamefont {S.}~\bibnamefont {Cohen}},
  \bibinfo {author} {\bibfnamefont {R.}~\bibnamefont {Gupta}}, \bibinfo
  {author} {\bibfnamefont {H.-W.}\ \bibnamefont {Lin}}, \ and\ \bibinfo
  {author} {\bibfnamefont {B.}~\bibnamefont {Yoon}},\ }\href {\doibase
  10.1103/PhysRevD.94.054508} {\bibfield  {journal} {\bibinfo  {journal} {Phys.
  Rev.}\ }\textbf {\bibinfo {volume} {D94}},\ \bibinfo {pages} {054508}
  (\bibinfo {year} {2016})},\ \Eprint {http://arxiv.org/abs/1606.07049}
  {arXiv:1606.07049 [hep-lat]} \BibitemShut {NoStop}%
\end{thebibliography}%
\clearpage
\appendix
\section{Plots of the Ratio \texorpdfstring{$C_{\mathcal{O}}^{3\text{pt}}(\tau;t)/C^{2\text{pt}}(\tau)$}{C(3pt)/C(2pt)}}
\label{sec:ratios}


\begin{figure*}[p]  
\centering
\begin{subfigure}
\centering
\includegraphics[angle=0,width=0.32\textwidth]{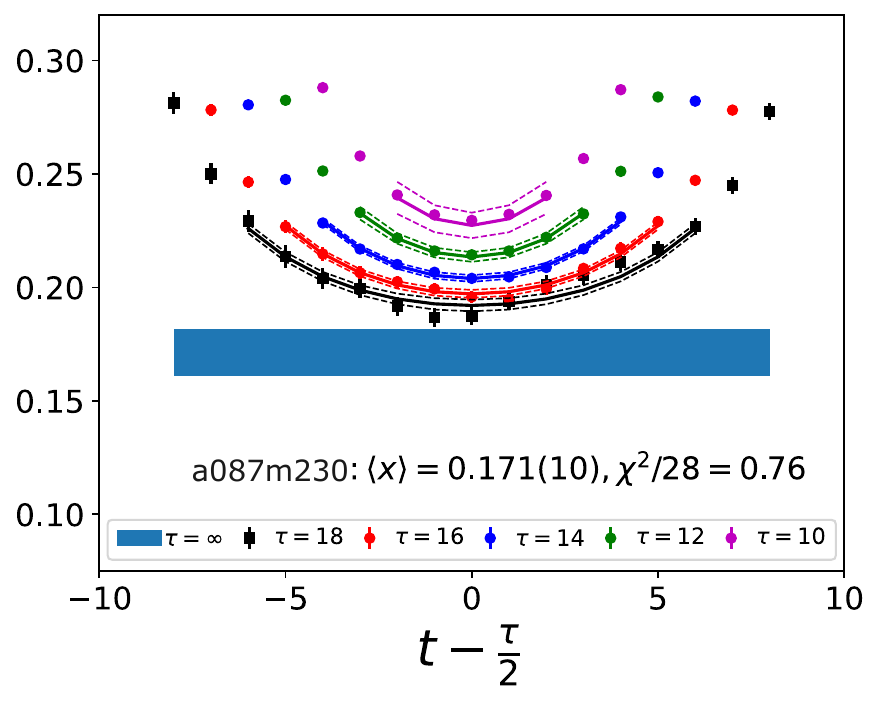}
\includegraphics[angle=0,width=0.32\textwidth]{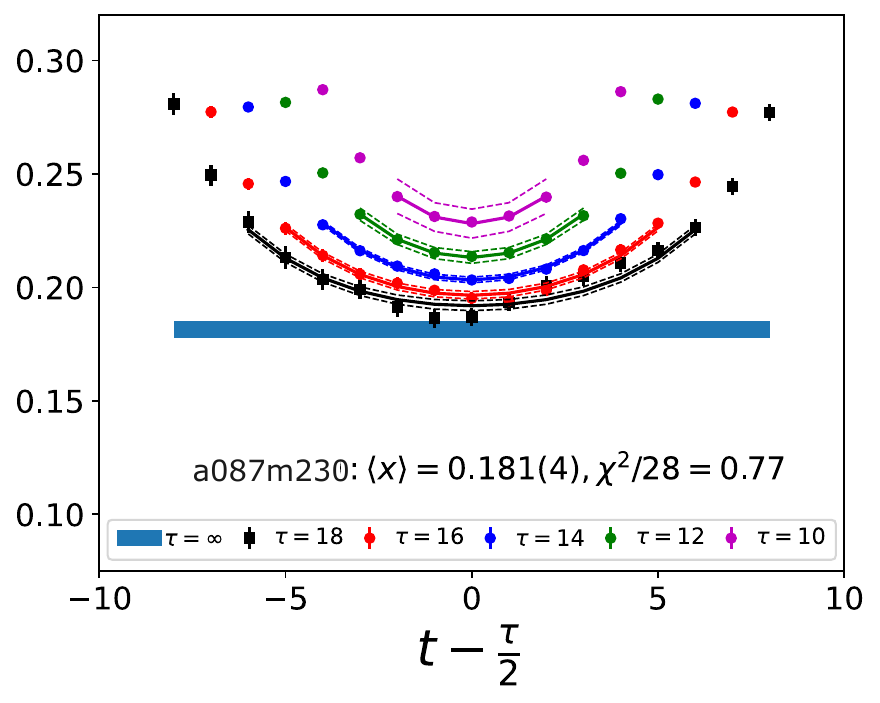}
\includegraphics[angle=0,width=0.32\textwidth]{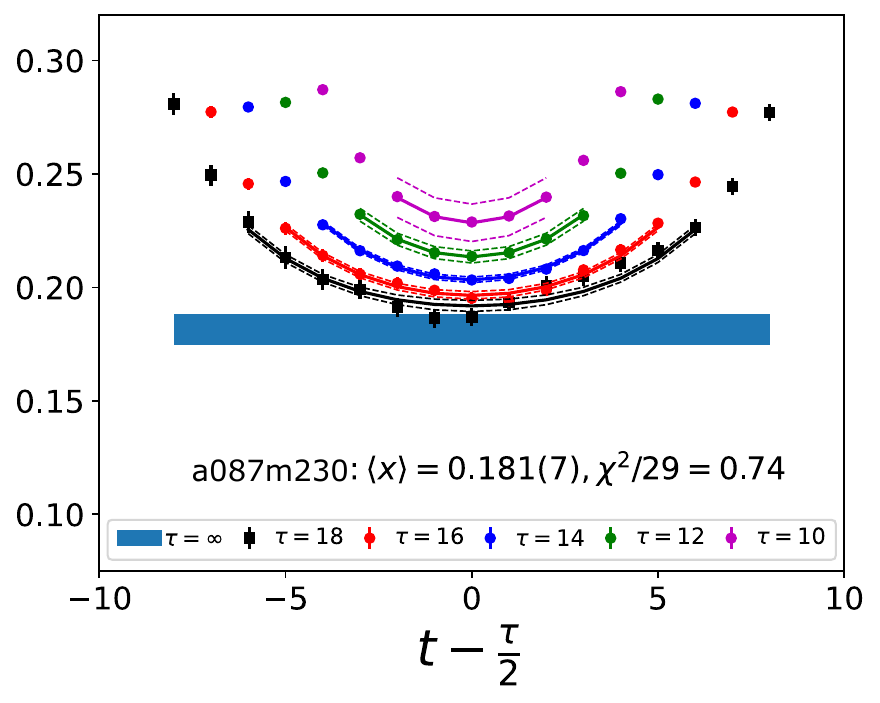}

\includegraphics[angle=0,width=0.32\textwidth]{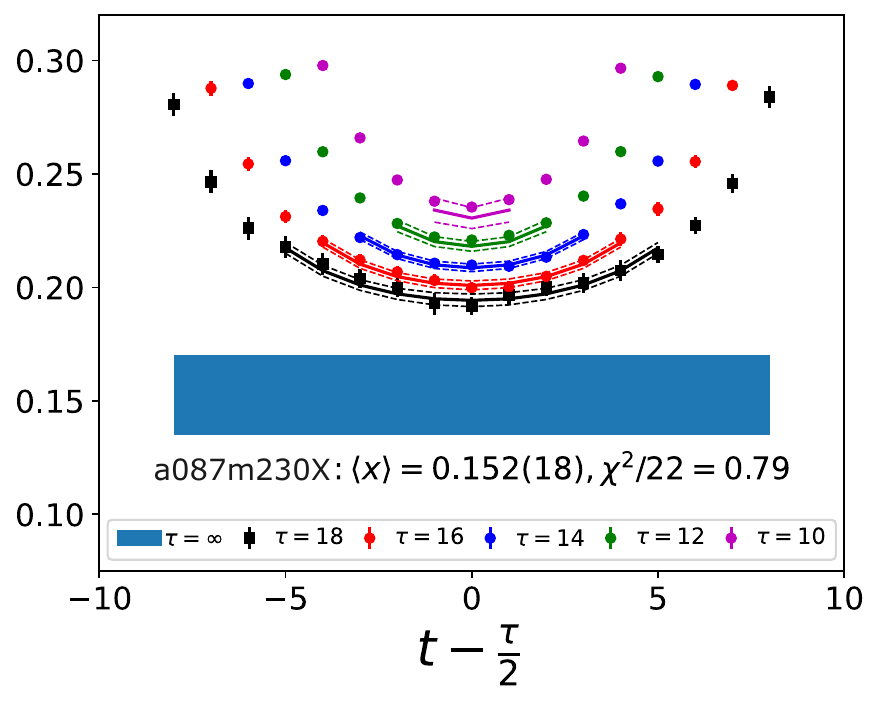}
\includegraphics[angle=0,width=0.32\textwidth]{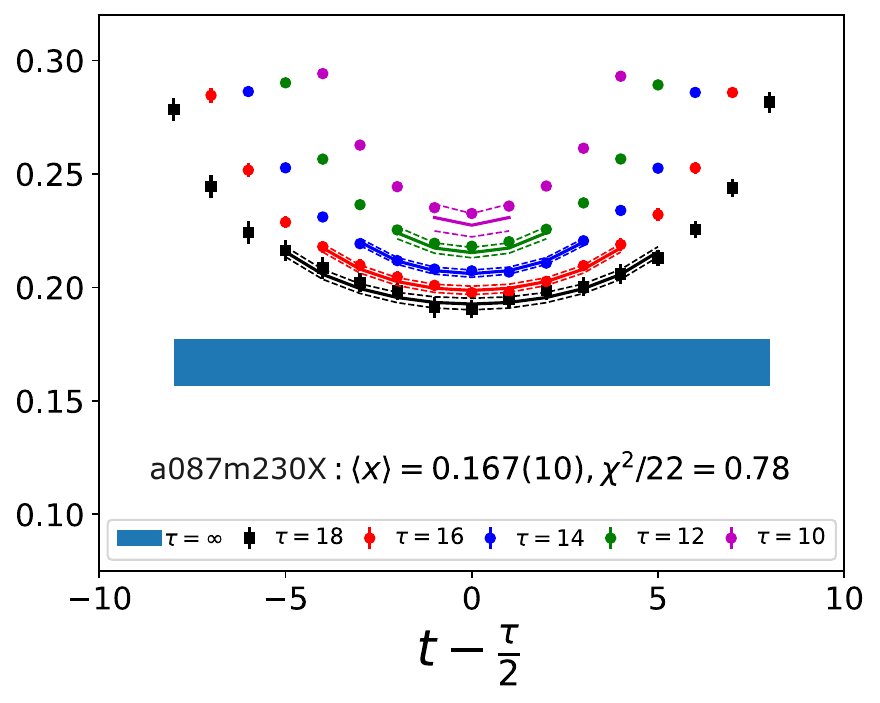}
\includegraphics[angle=0,width=0.32\textwidth]{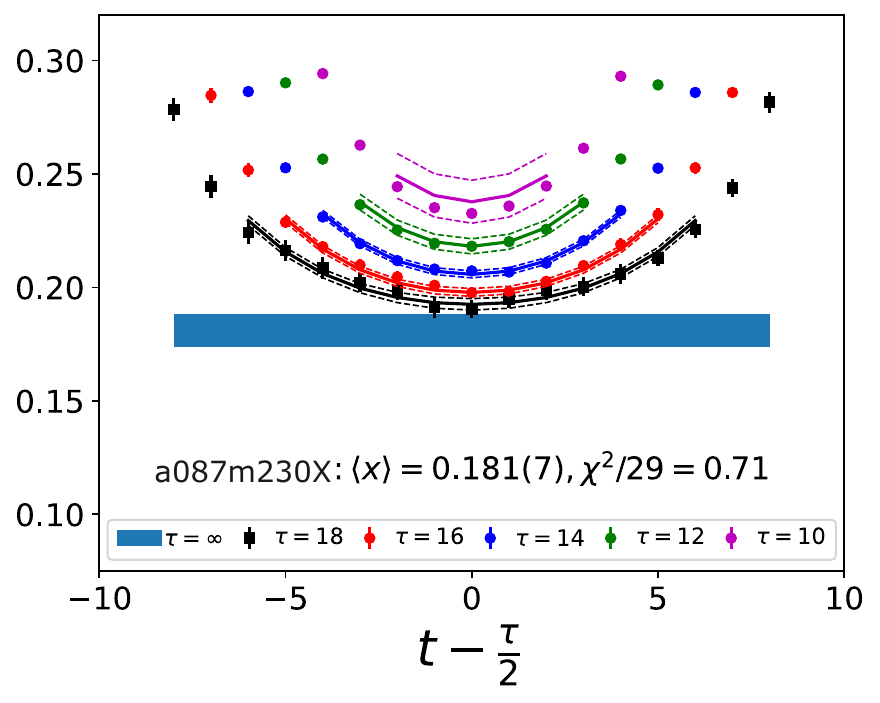}

\includegraphics[angle=0,width=0.32\textwidth]{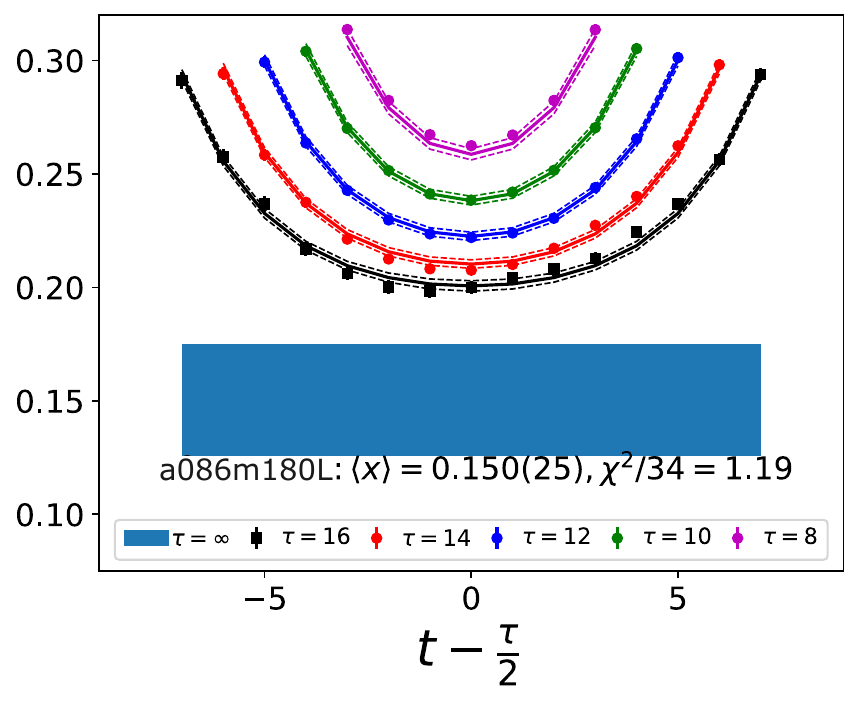}
\includegraphics[angle=0,width=0.32\textwidth]{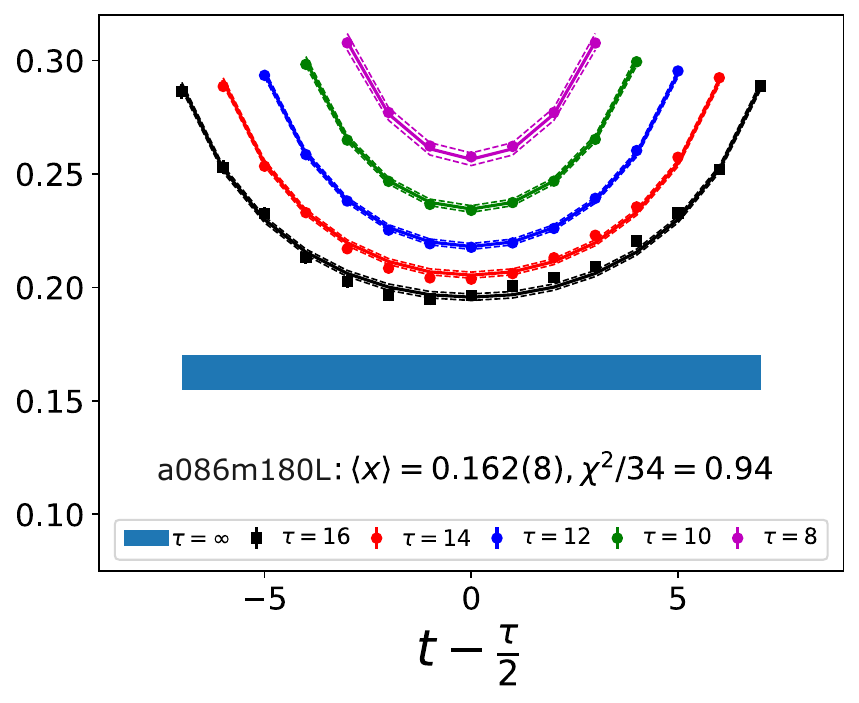}
\includegraphics[angle=0,width=0.32\textwidth]{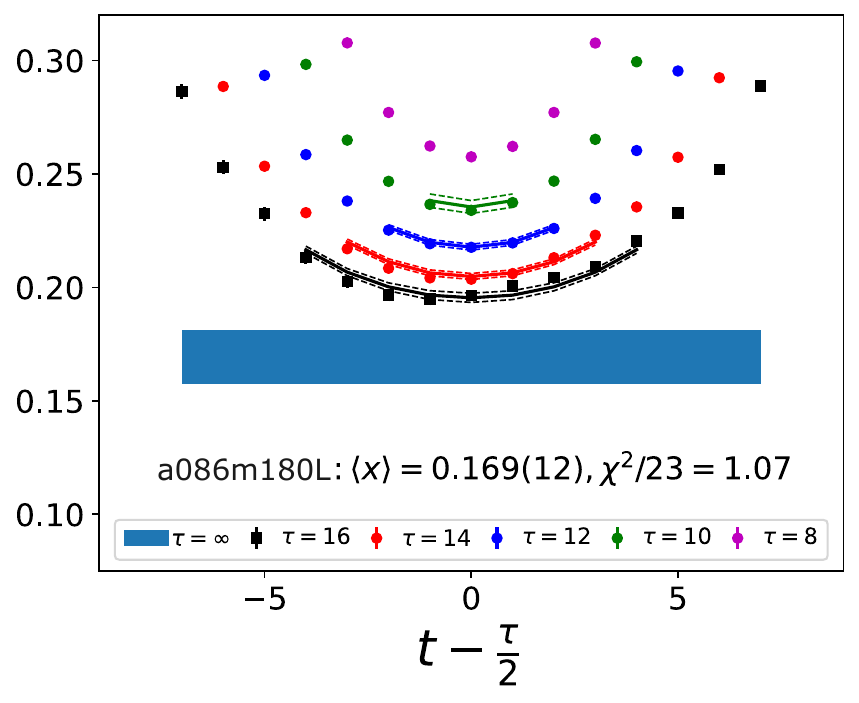}

\includegraphics[angle=0,width=0.32\textwidth]{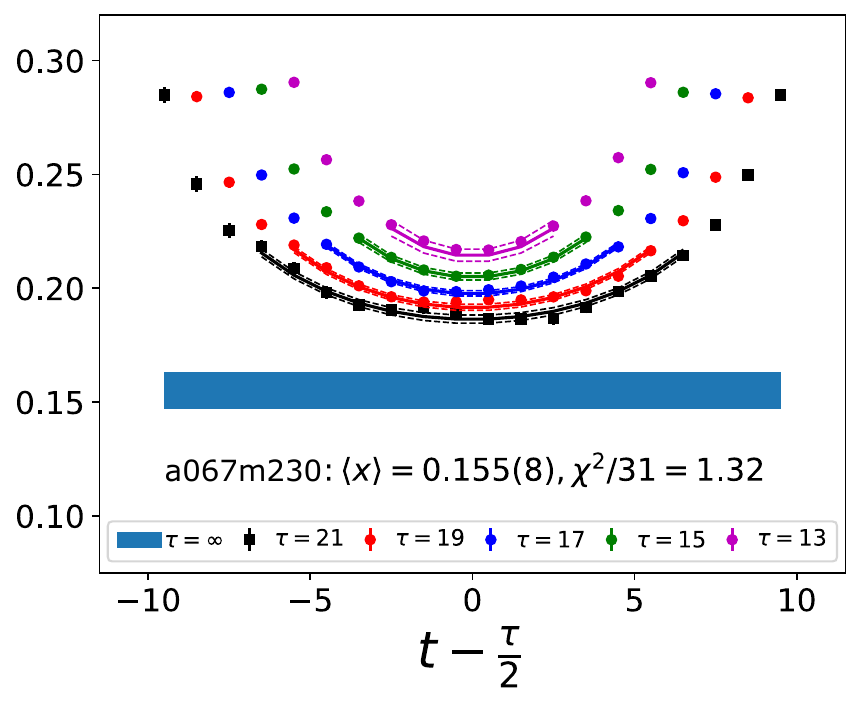}
\includegraphics[angle=0,width=0.32\textwidth]{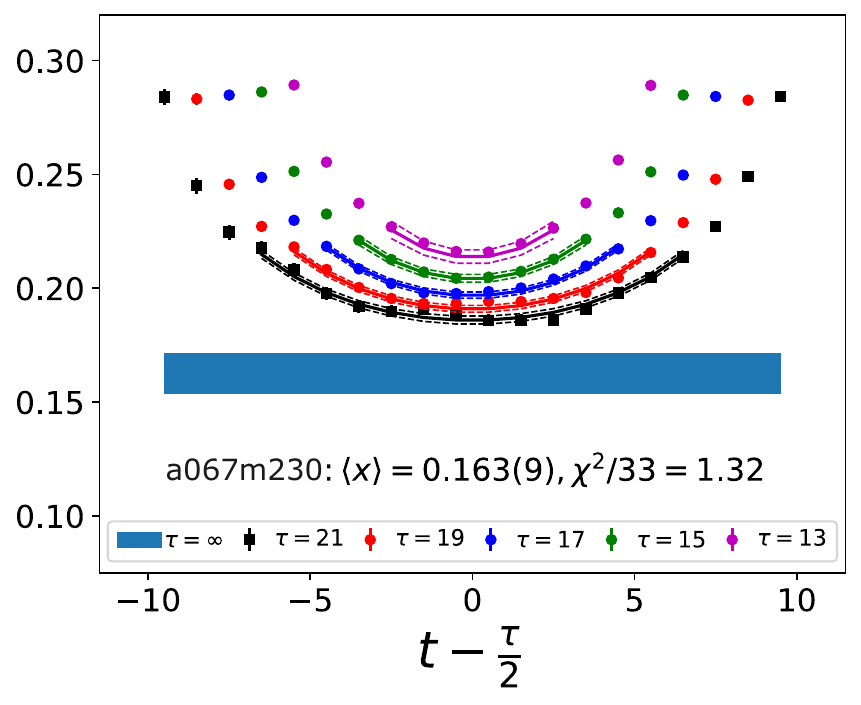}
\includegraphics[angle=0,width=0.32\textwidth]{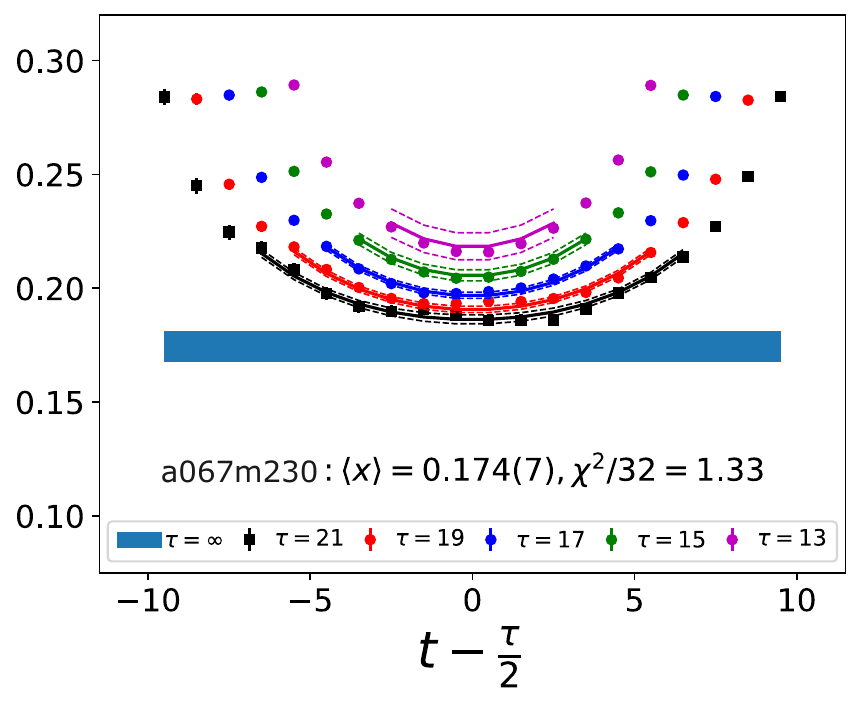}

\end{subfigure}

\caption{Data for the ratio   $C_\mathcal{O}^{3\text{pt}}(\tau;t)/C^{2\text{pt}}(\tau)$, 
  scaled using Eq.~\protect\eqref{eq:me2momentV} to give the momentum
  fraction $\langle x \rangle_{u-d}$, are shown for the $a087m230$
  (top row), $a087m230X$ (second row), $a086m180L$ (third row), and
  $a067m230$ (bottom row) ensembles.  The three panels in each row
  show fits to the same data but using the three
  strategies: $\{4^{N\pi},3^*\}$ (left), $\{4^{},3^*\}$ (middle) and
  $\{4^{},2^{\rm free}\}$ (right). See Sec.~\protect\ref{sec:ESC} for
  details.  The values of $\tau$ used in the fits are listed in 
  Table~\ref{tab:ensembles}, and the output parameters of these fits are
  summarized in Tables~\protect\ref{tab:5strategy-fits-momfrac}.  For each
  $\tau$, the line with the same color as the data shows the fit result, and connects the points used in
  the fit.  The resulting unrenormalized ground-state
  ($\tau \to \infty$) value of the moment is shown by the blue band
  and given in the label along with the $\chi^2/dof$ of the fit. The
  y-interval is selected to be the same for all the panels on this page to
  facilitate comparison. }
\label{fig:Ratio-mom-1}
\end{figure*}

\begin{figure*}[tp] 

\begin{subfigure}
\centering
\includegraphics[angle=0,width=0.32\textwidth]{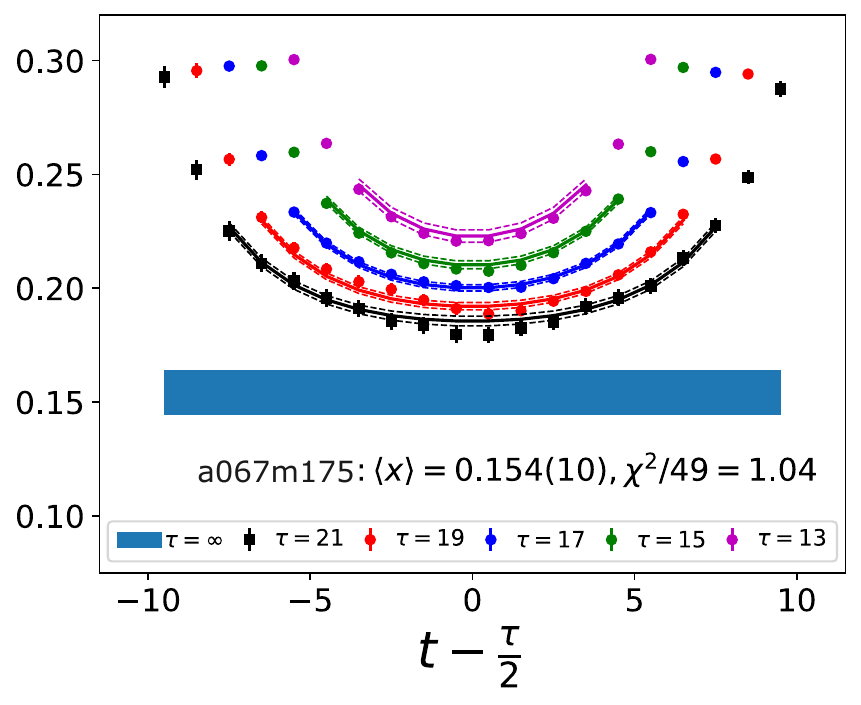}
\includegraphics[angle=0,width=0.32\textwidth]{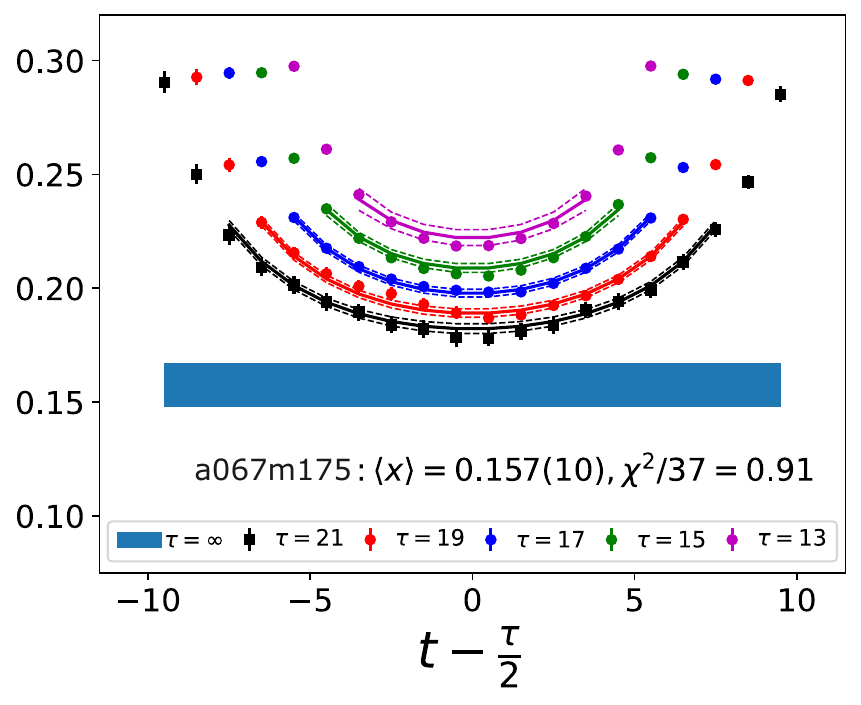}
\includegraphics[angle=0,width=0.32\textwidth]{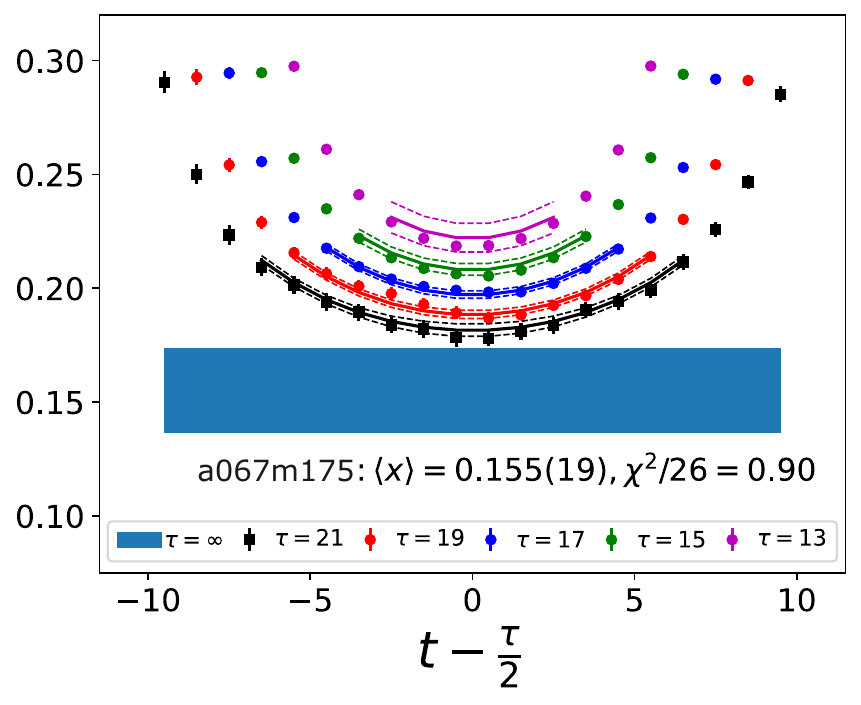}
\end{subfigure}

\begin{subfigure}
\centering
\includegraphics[angle=0,width=0.32\textwidth]{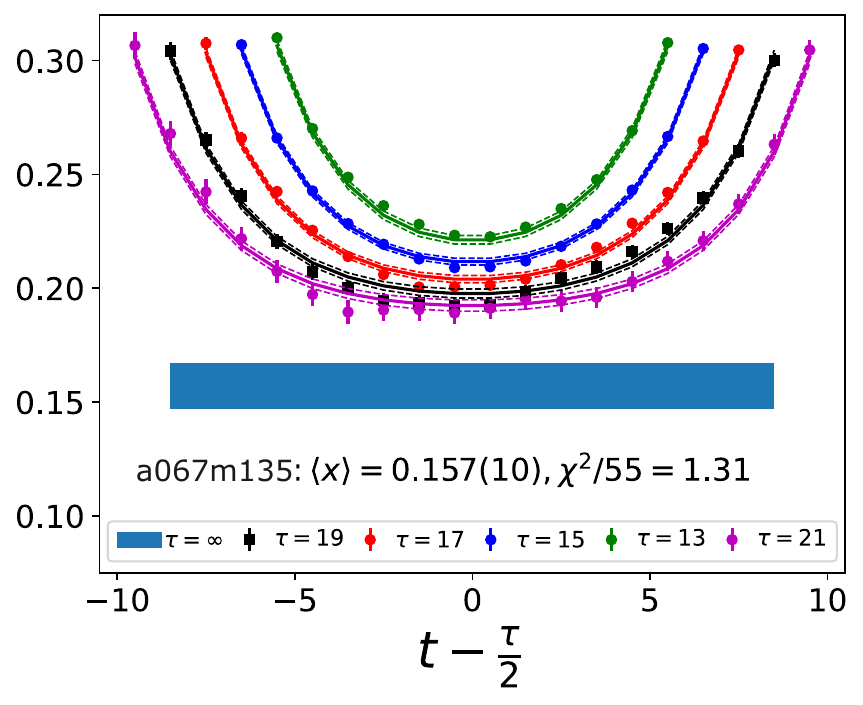}
\includegraphics[angle=0,width=0.32\textwidth]{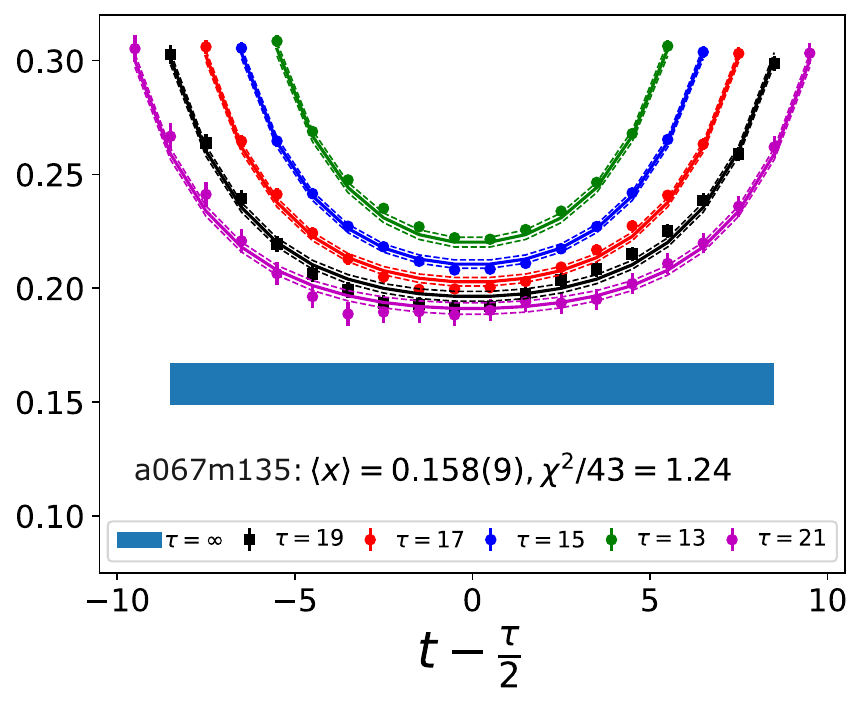}
\includegraphics[angle=0,width=0.32\textwidth]{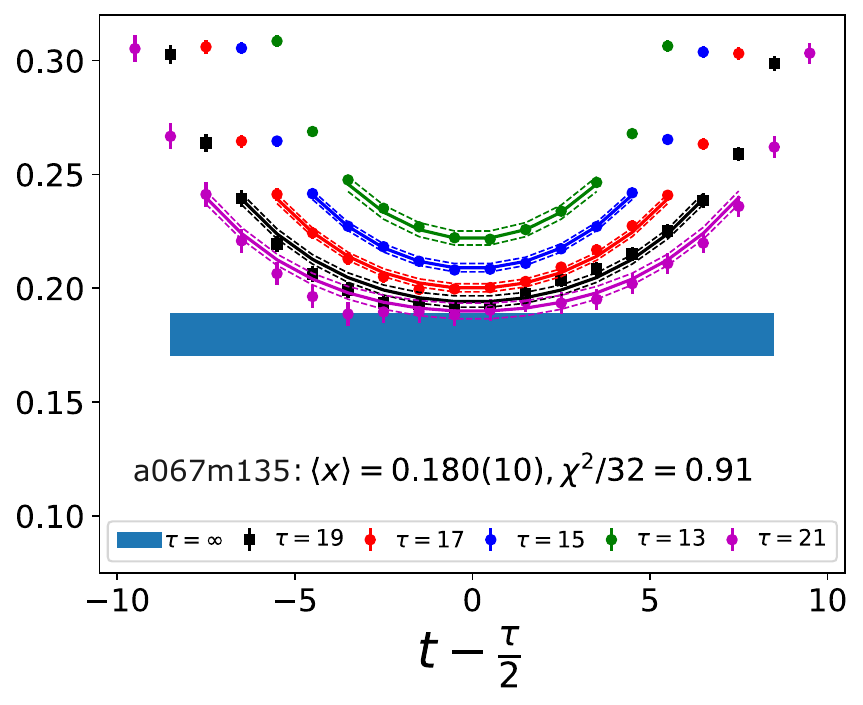}
\end{subfigure}

\begin{subfigure}
\centering
\includegraphics[angle=0,width=0.32\textwidth]{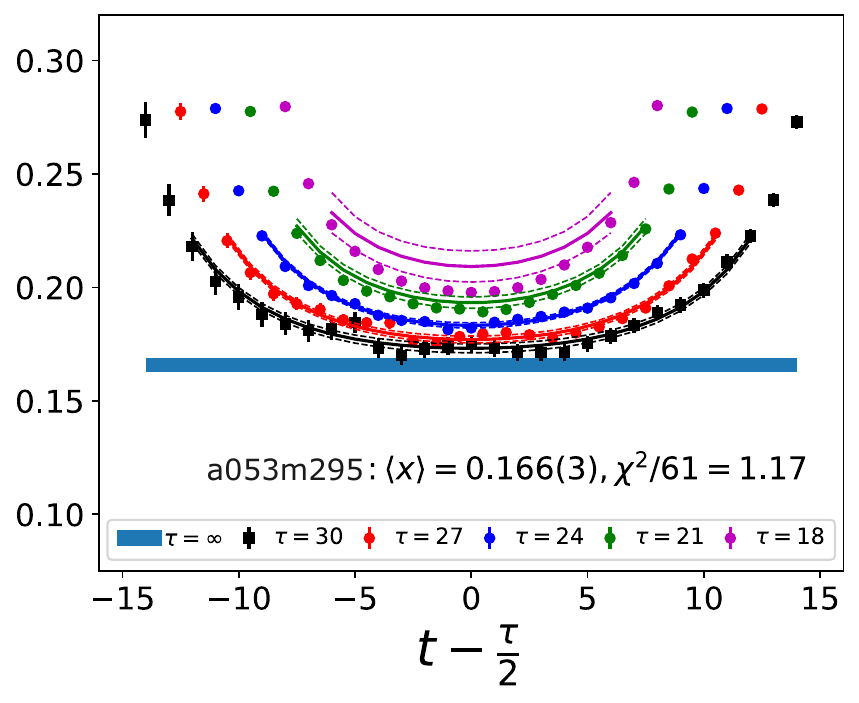}                                   
\includegraphics[angle=0,width=0.32\textwidth]{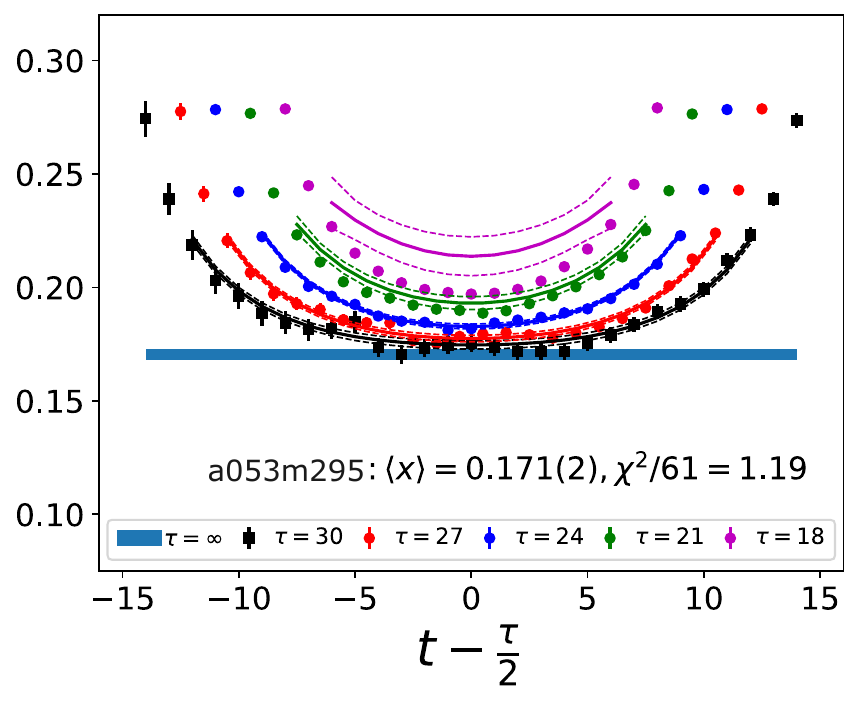}                                      
\includegraphics[angle=0,width=0.32\textwidth]{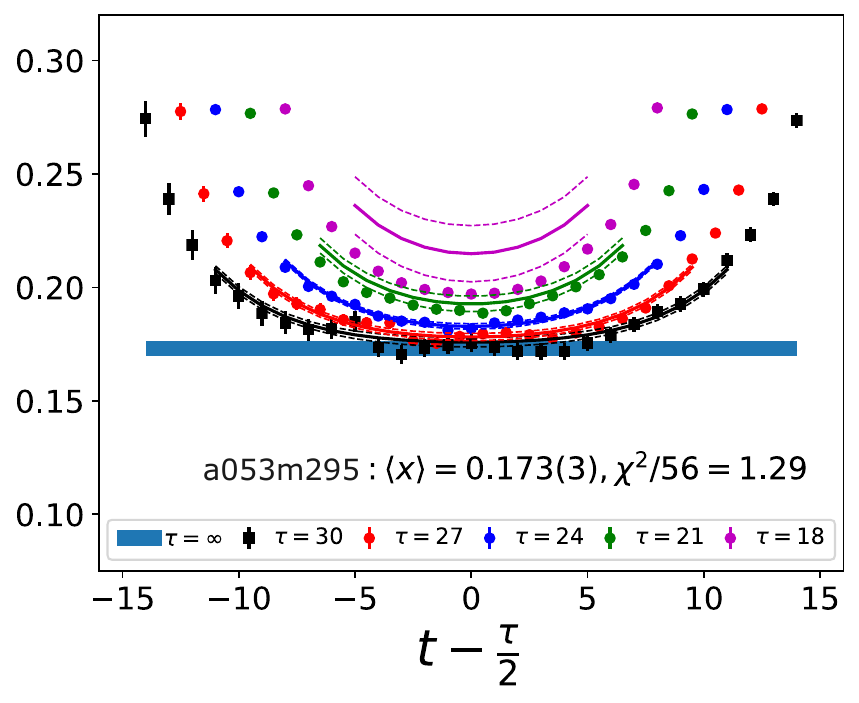}                                   
\end{subfigure}

\begin{subfigure}
\centering
\includegraphics[angle=0,width=0.32\textwidth]{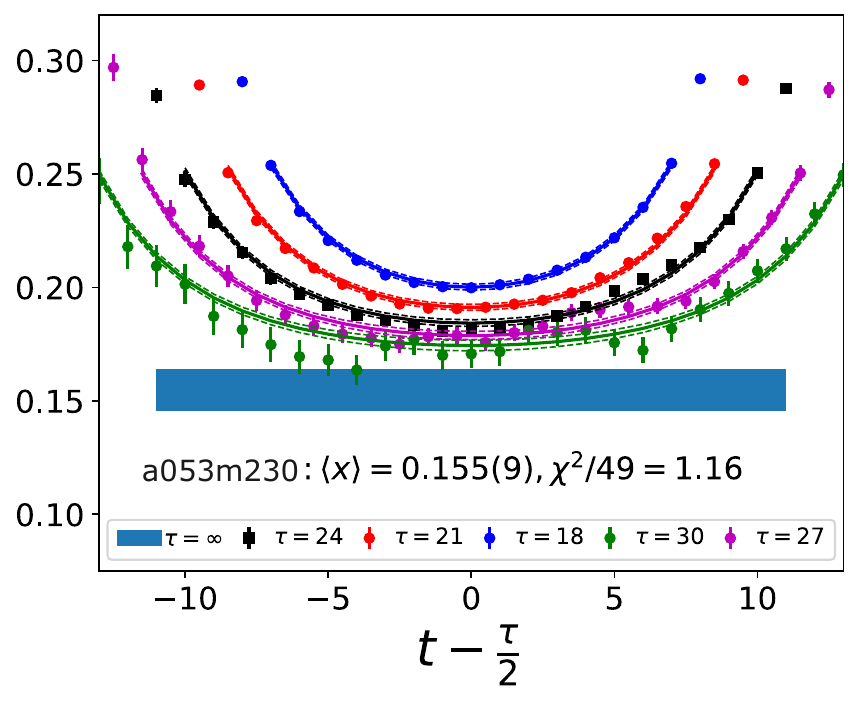}
\includegraphics[angle=0,width=0.32\textwidth]{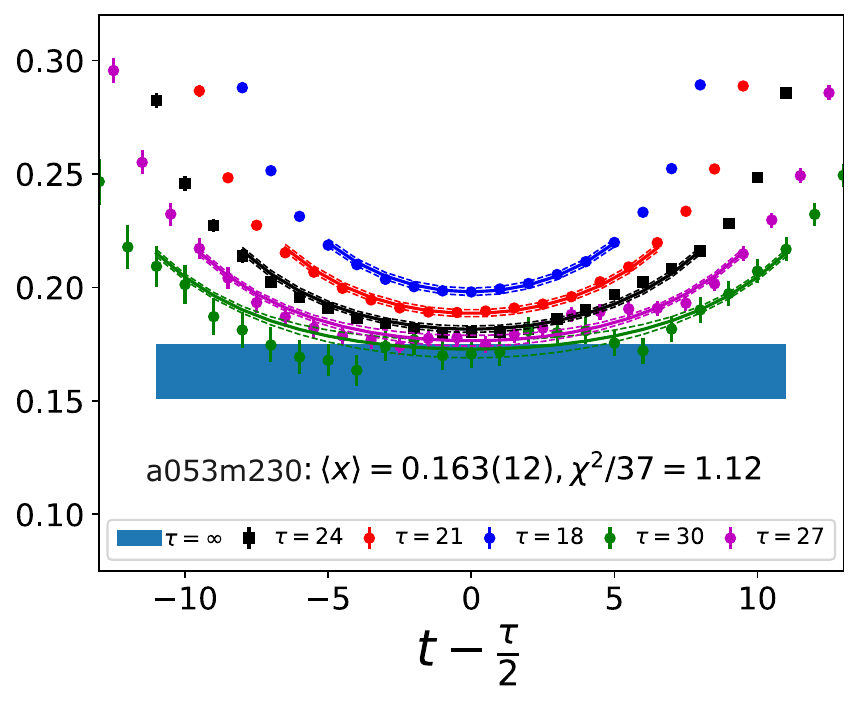}
\includegraphics[angle=0,width=0.32\textwidth]{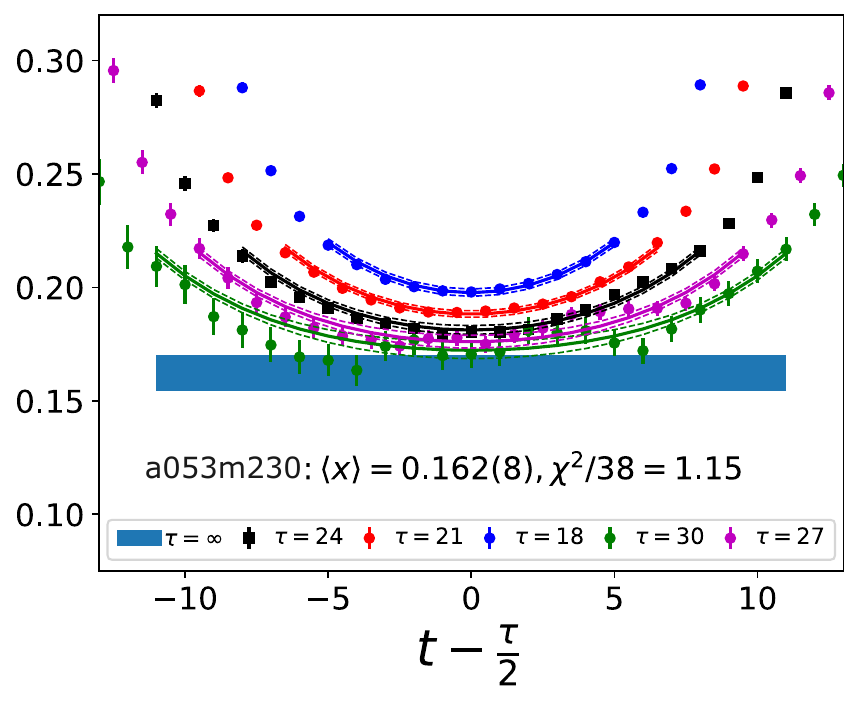}
\end{subfigure}

\caption{Continuation of the data for the ratio
  $C_\mathcal{O}^{3\text{pt}}(\tau;t)/C^{2\text{pt}}(\tau)$, scaled
  using Eq.~\protect\eqref{eq:me2momentV} to give $\langle
  x \rangle_{u-d}$, and fits to remove excited-state
  contamination for the $a067m175$
  (top row), $a067m135$ (second row), $a053m290$ (third row), and
  $a053m230$ (bottom row) ensembles.    The rest is the
  same as in Fig.~\protect\ref{fig:Ratio-mom-1}.}
\label{fig:Ratio-mom-2} 
\end{figure*}

\begin{figure*}[tp]  
\begin{subfigure}
\centering
\includegraphics[angle=0,width=0.32\textwidth]{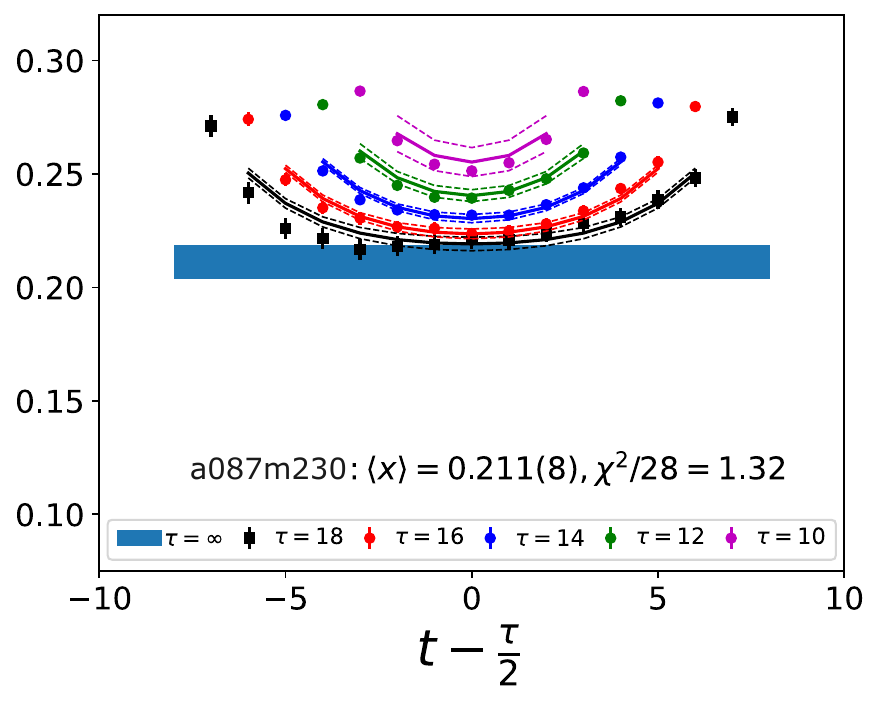}
\includegraphics[angle=0,width=0.32\textwidth]{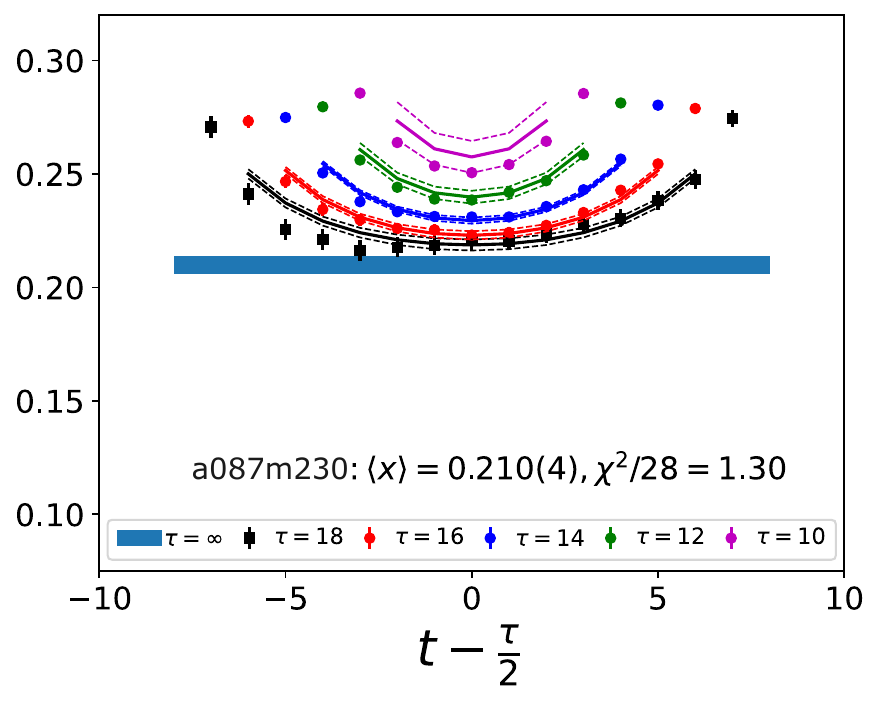}
\includegraphics[angle=0,width=0.32\textwidth]{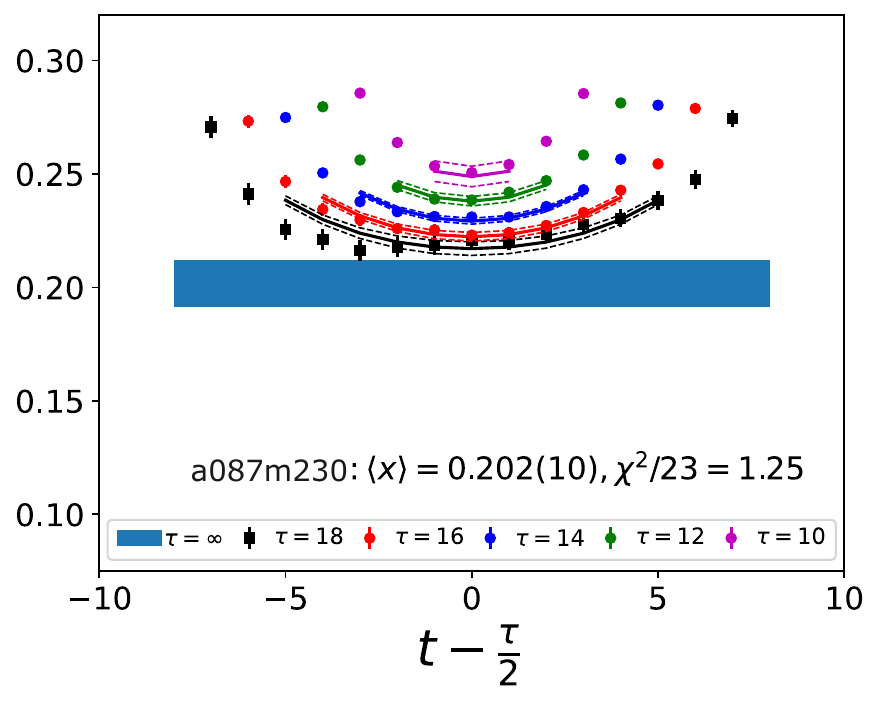}

\includegraphics[angle=0,width=0.32\textwidth]{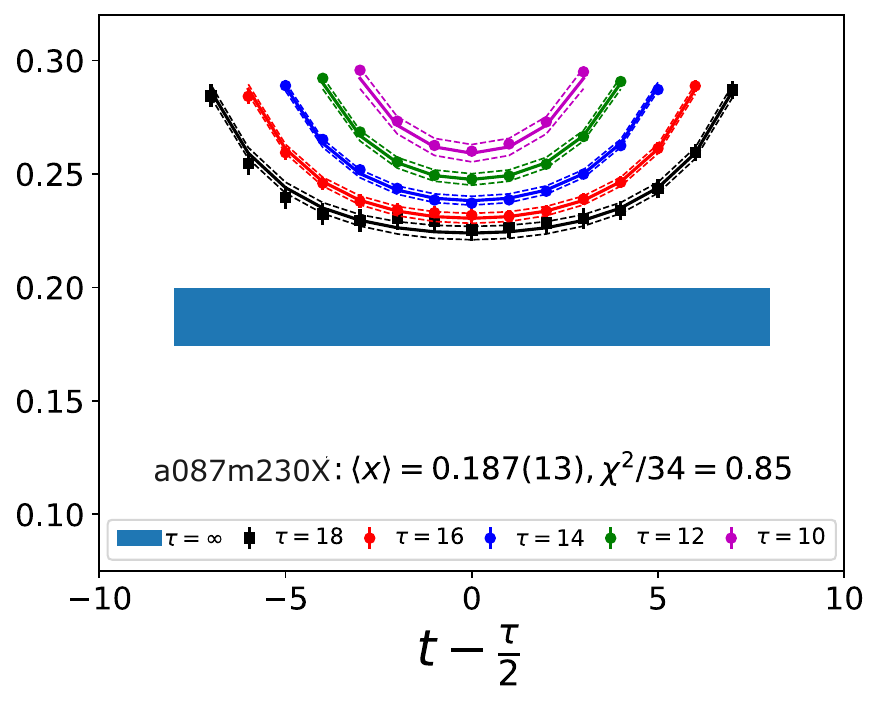}
\includegraphics[angle=0,width=0.32\textwidth]{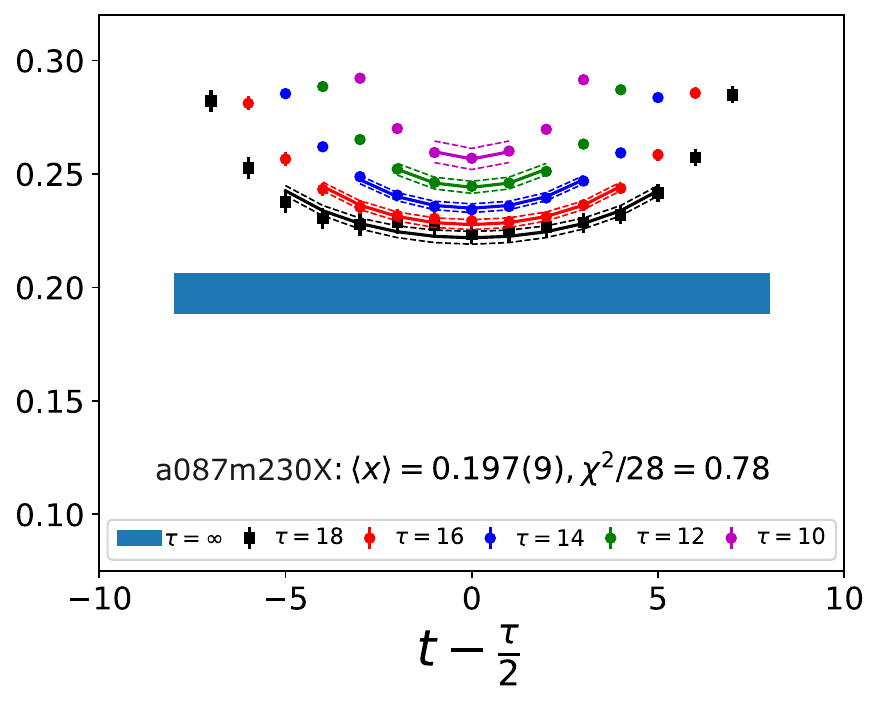}
\includegraphics[angle=0,width=0.32\textwidth]{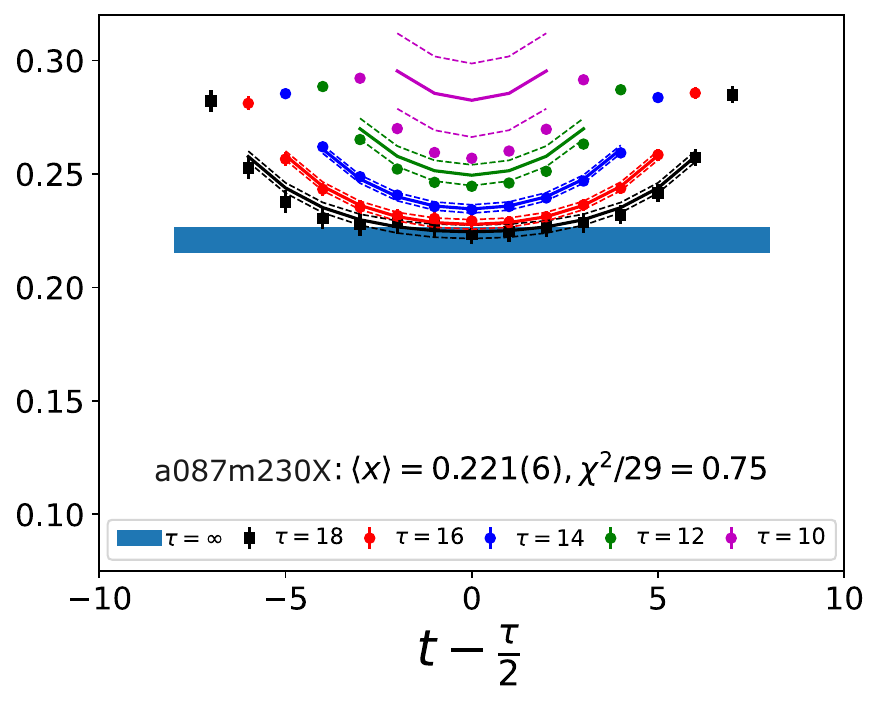}

\includegraphics[angle=0,width=0.32\textwidth]{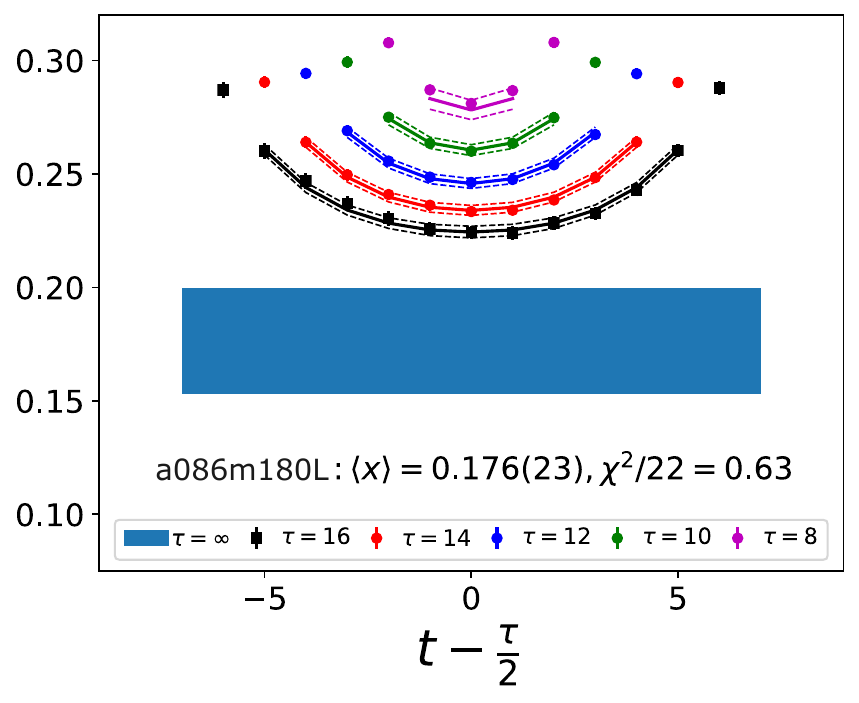}
\includegraphics[angle=0,width=0.32\textwidth]{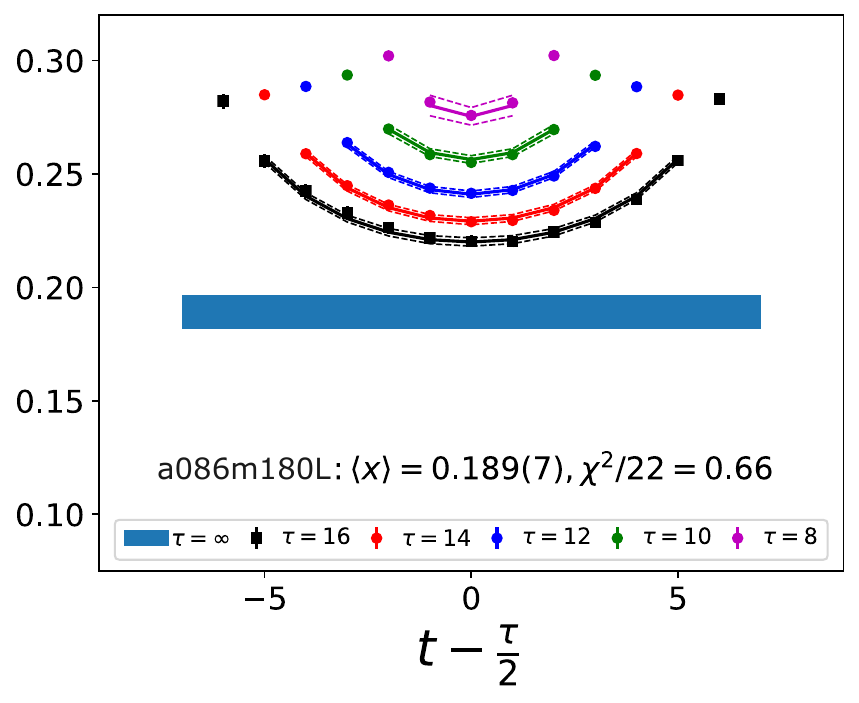}
\includegraphics[angle=0,width=0.32\textwidth]{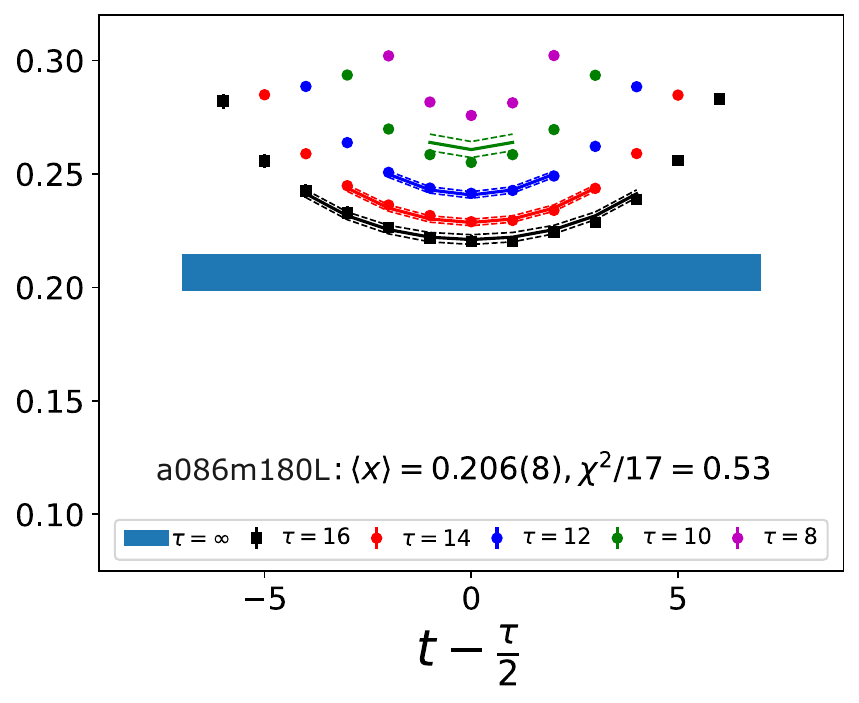}

\includegraphics[angle=0,width=0.32\textwidth]{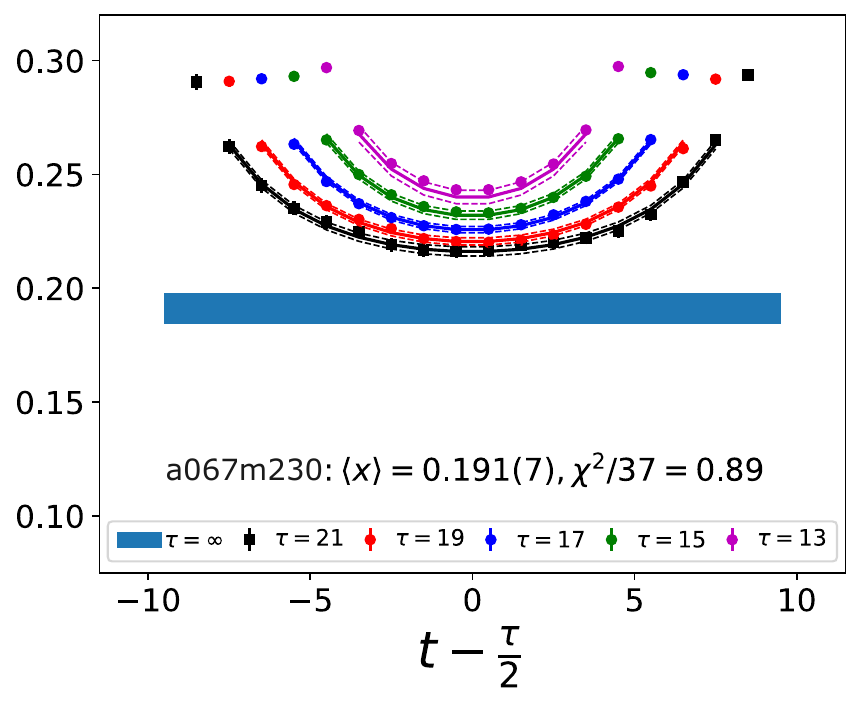}
\includegraphics[angle=0,width=0.32\textwidth]{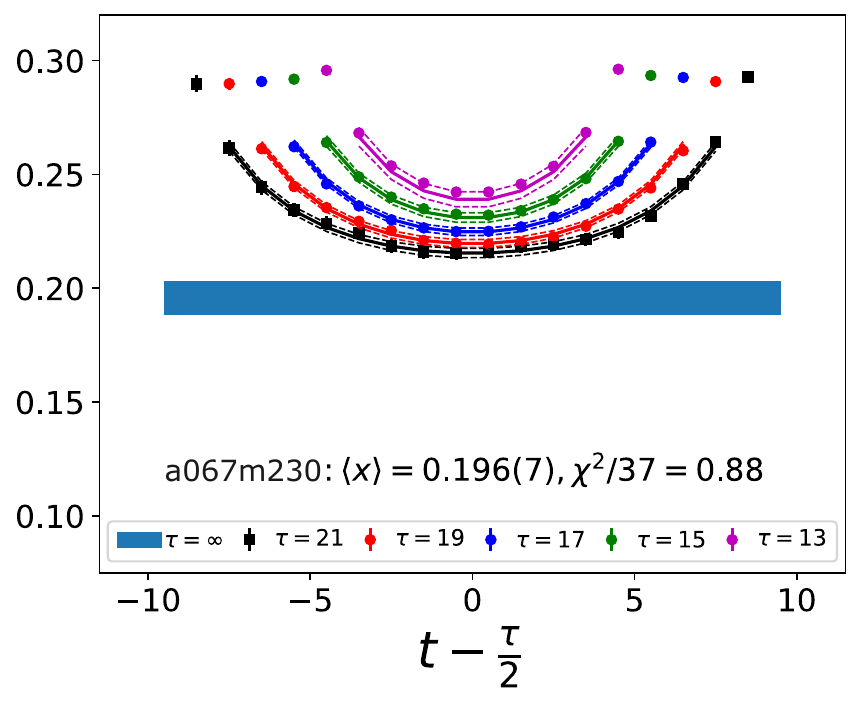}
\includegraphics[angle=0,width=0.32\textwidth]{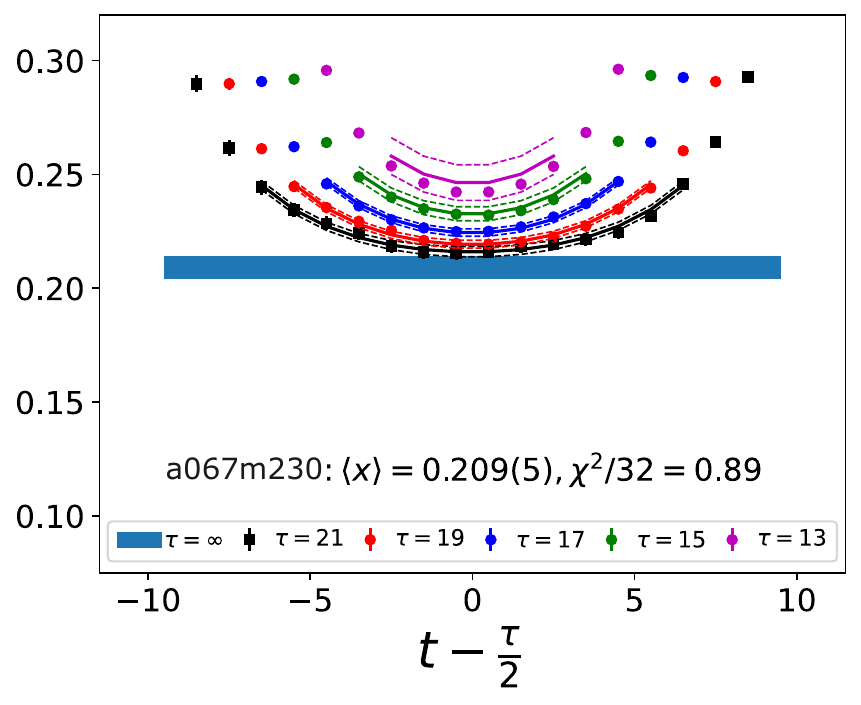}
\end{subfigure}
\caption{Data for the ratio
  $C_\mathcal{O}^{3\text{pt}}(\tau;t)/C^{2\text{pt}}(\tau)$, scaled
  using Eq.~\protect\eqref{eq:me2momentA} to give the helicity moment
  $\langle x \rangle_{\Delta u-\Delta d}$, and fits to remove
  excited-state contamination for the $a087m230$
  (top row), $a087m230X$ (second row), $a086m180L$ (third row), and
  $a067m230$ (bottom row) ensembles.
  The fit parameters are summarized in
  Table~\protect\ref{tab:5strategy-fits-helfrac}.  The rest is the
  same as in Fig.~\protect\ref{fig:Ratio-mom-1}.}
\label{fig:Ratio-helicity-1}
\end{figure*}

\begin{figure*}[tp]   
\centering

\begin{subfigure}
\centering
\includegraphics[angle=0,width=0.32\textwidth]{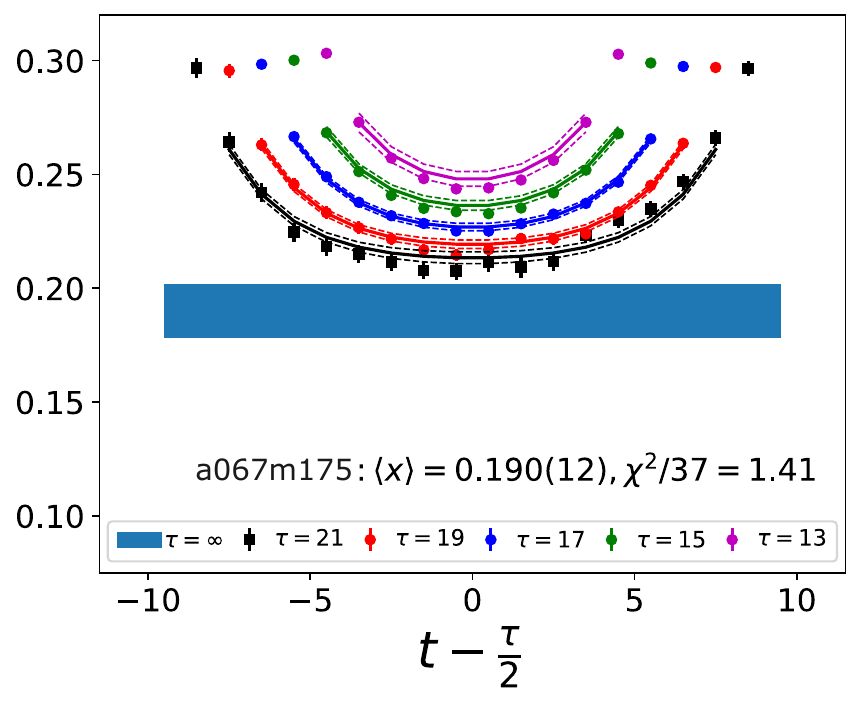}
\includegraphics[angle=0,width=0.32\textwidth]{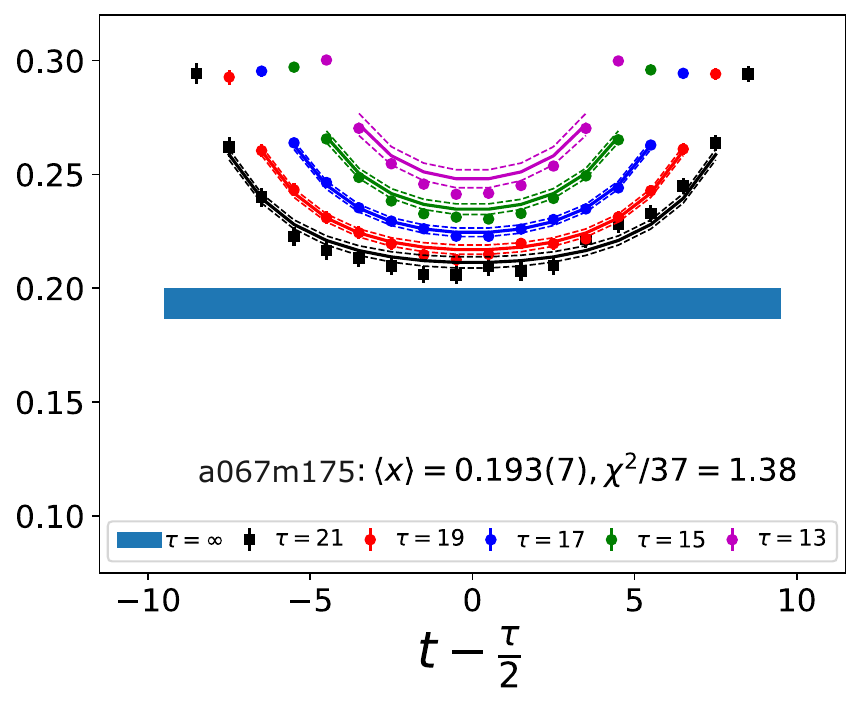}
\includegraphics[angle=0,width=0.32\textwidth]{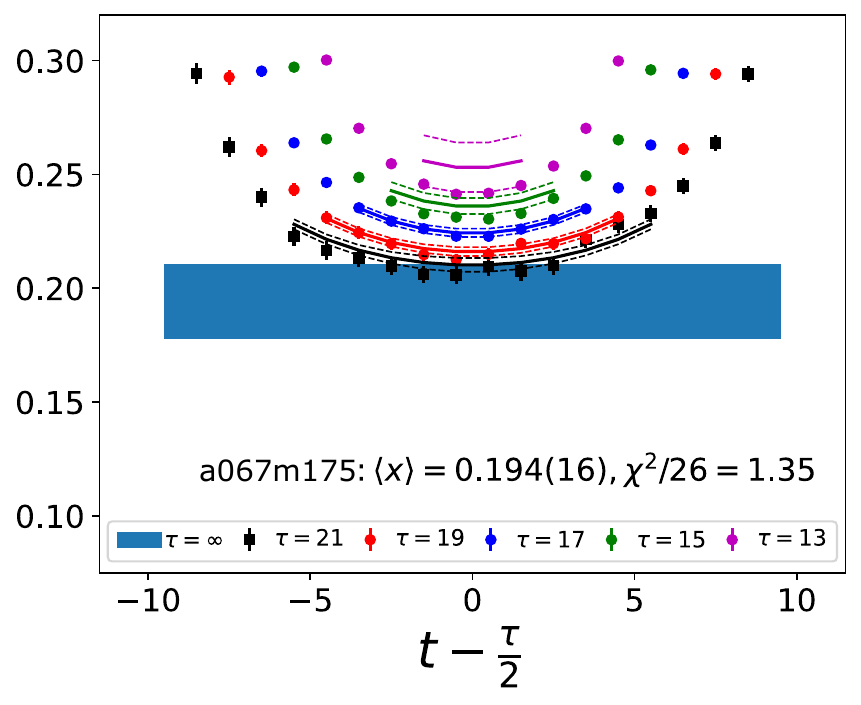}
\end{subfigure}

\begin{subfigure}
\centering
\includegraphics[angle=0,width=0.32\textwidth]{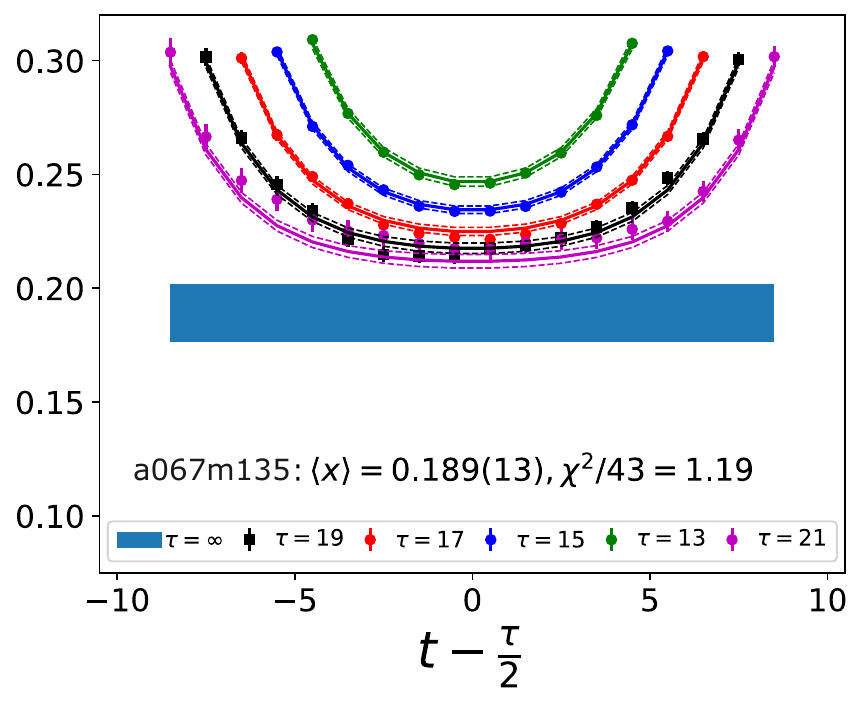}
\includegraphics[angle=0,width=0.32\textwidth]{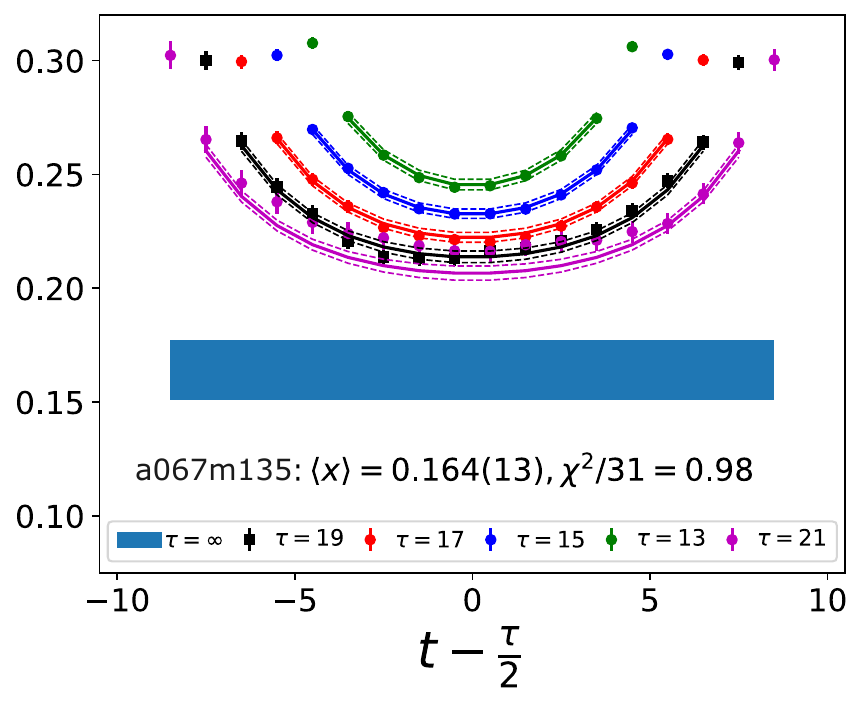}
\includegraphics[angle=0,width=0.32\textwidth]{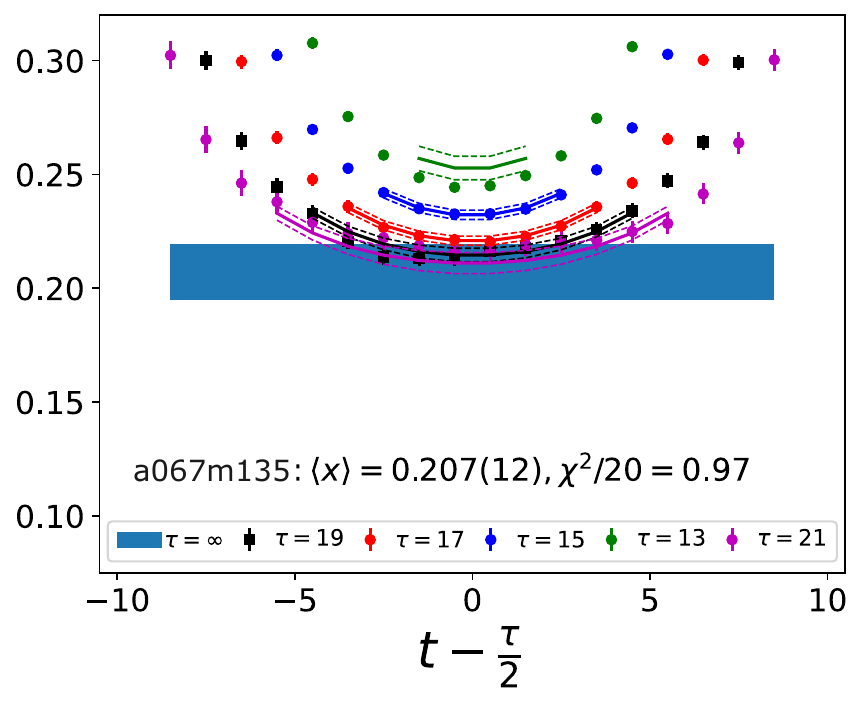}
\end{subfigure}

\begin{subfigure}
\centering
\includegraphics[angle=0,width=0.32\textwidth]{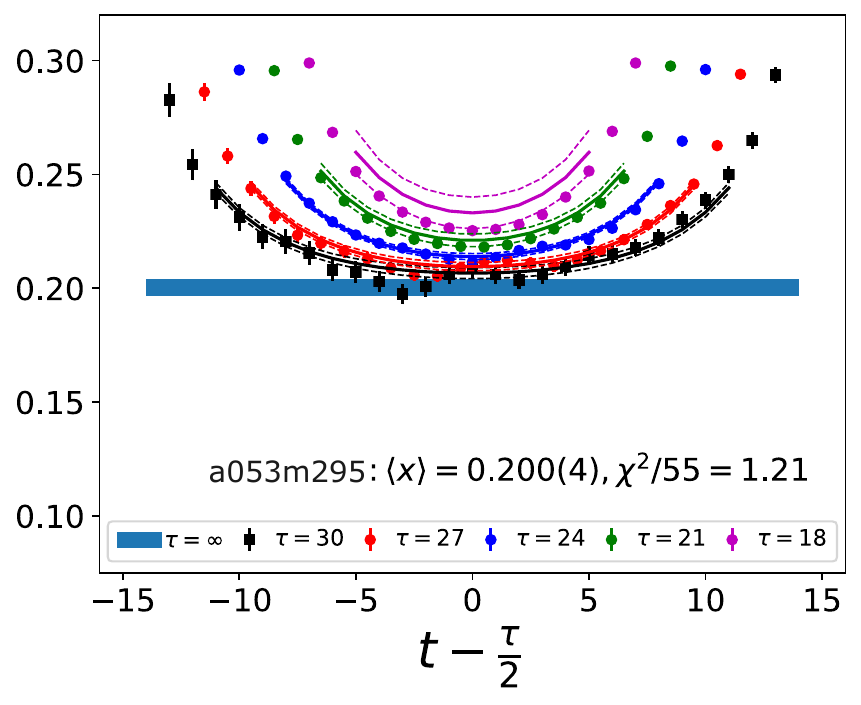}
\includegraphics[angle=0,width=0.32\textwidth]{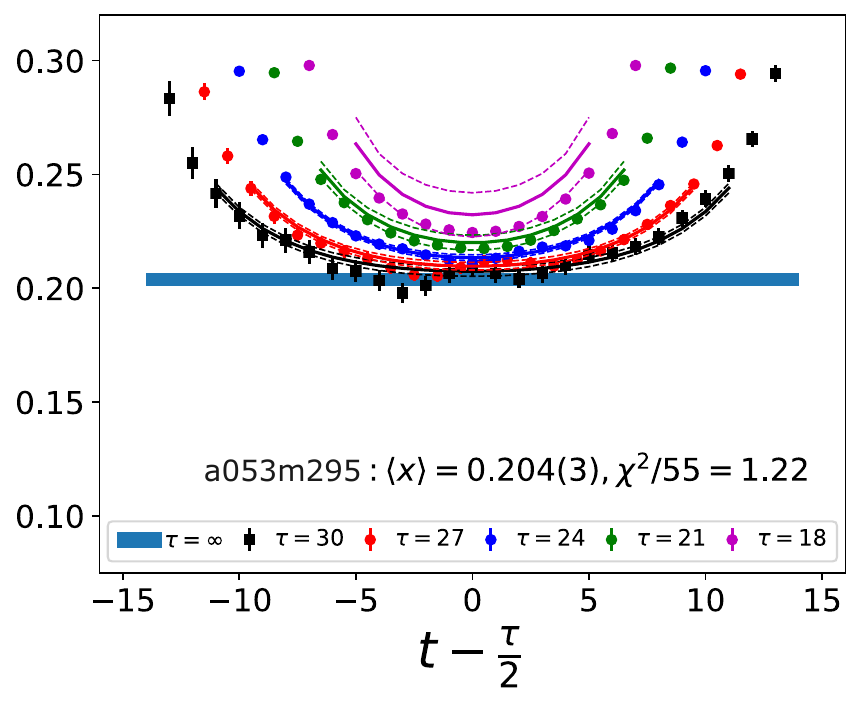}
\includegraphics[angle=0,width=0.32\textwidth]{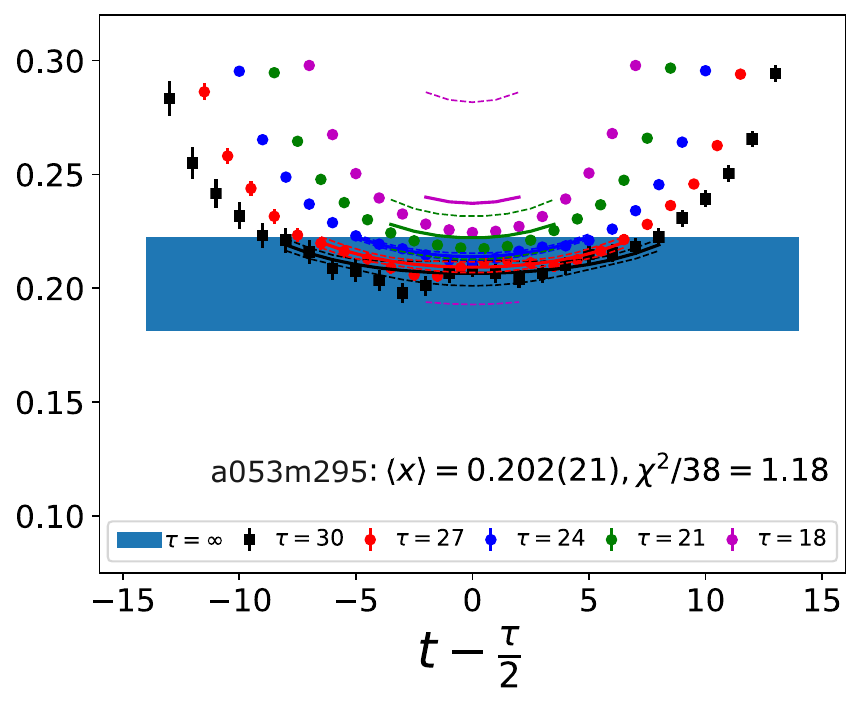}
\end{subfigure}

\begin{subfigure}
\centering

\includegraphics[angle=0,width=0.32\textwidth]{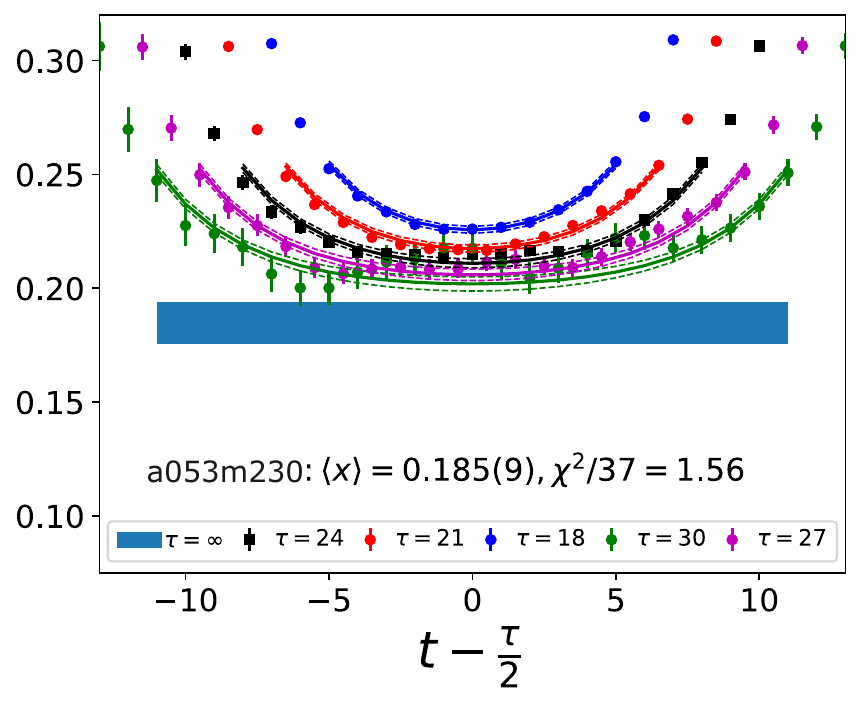}
\includegraphics[angle=0,width=0.32\textwidth]{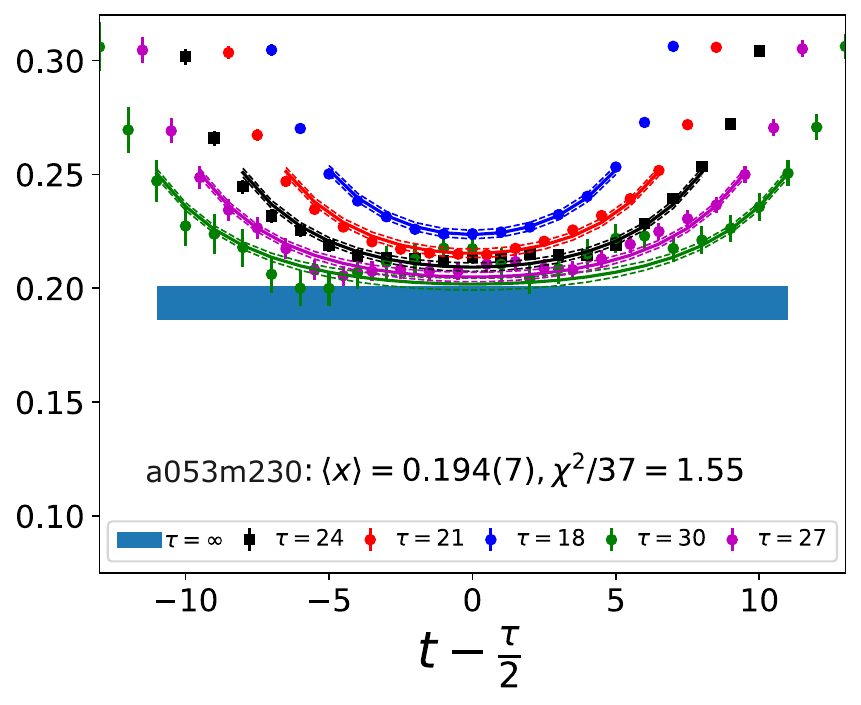}
\includegraphics[angle=0,width=0.32\textwidth]{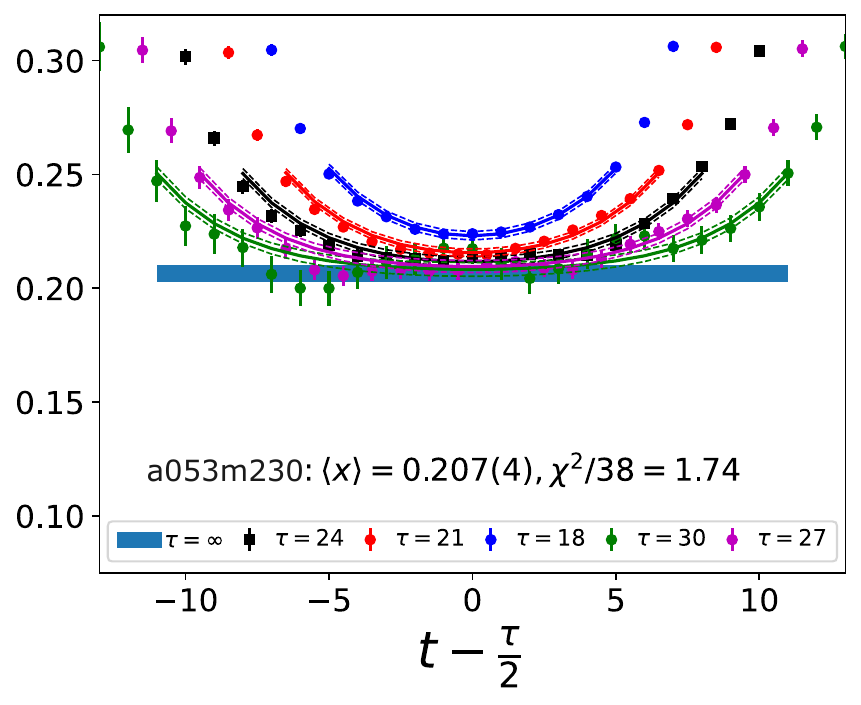}
\end{subfigure}

\caption{Continuation of the data for the ratio
  $C_\mathcal{O}^{3\text{pt}}(\tau;t)/C^{2\text{pt}}(\tau)$, scaled
  using Eq.~\protect\eqref{eq:me2momentA} to give the helicity moment
  $\langle x \rangle_{\Delta u-\Delta d}$, and fits to remove
  excited-state contamination for the  $a067m175$
  (top row), $a067m135$ (second row), $a053m290$ (third row), and
  $a053m230$ (bottom row) ensembles.
  The fit parameters are summarized in
  Table~\protect\ref{tab:5strategy-fits-helfrac}.  The rest is the
  same as in Fig.~\protect\ref{fig:Ratio-mom-1}
}
\label{fig:Ratio-helicity-2} 
\end{figure*}

\begin{figure*}[tp]  
\begin{subfigure}
\centering
\includegraphics[angle=0,width=0.32\textwidth]{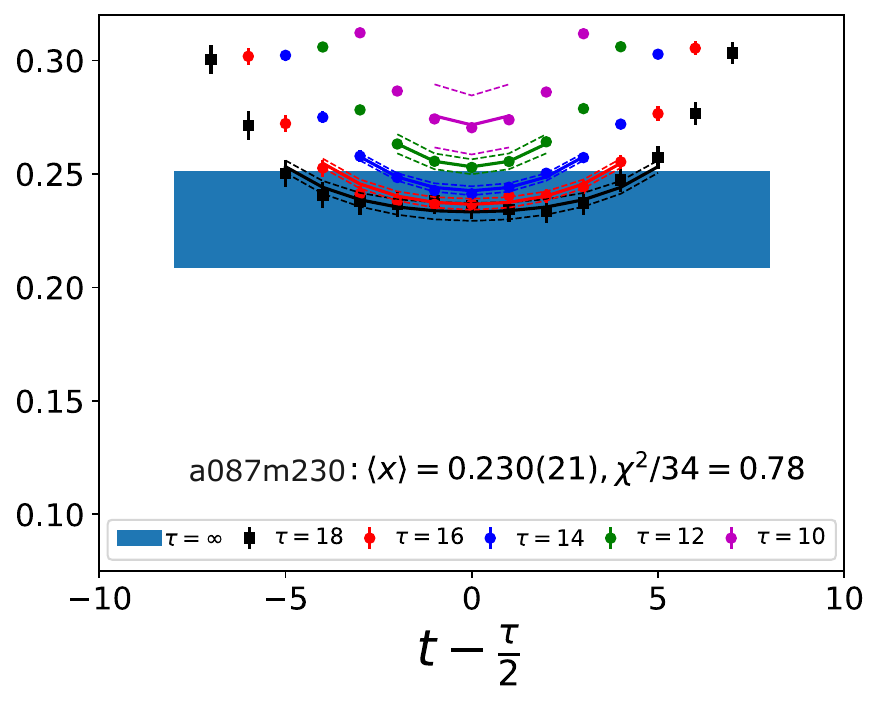}
\includegraphics[angle=0,width=0.32\textwidth]{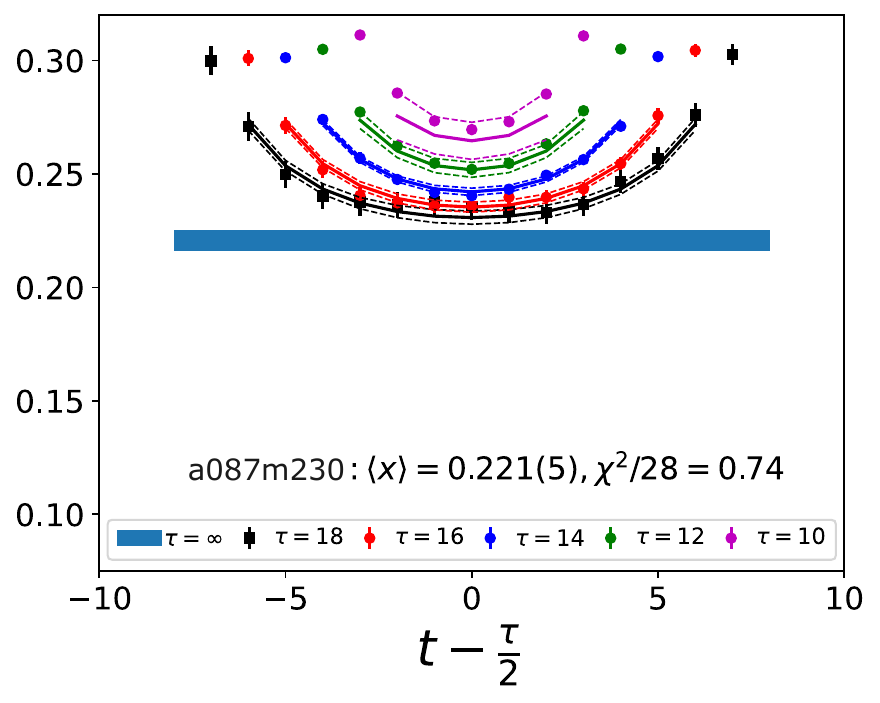}
\includegraphics[angle=0,width=0.32\textwidth]{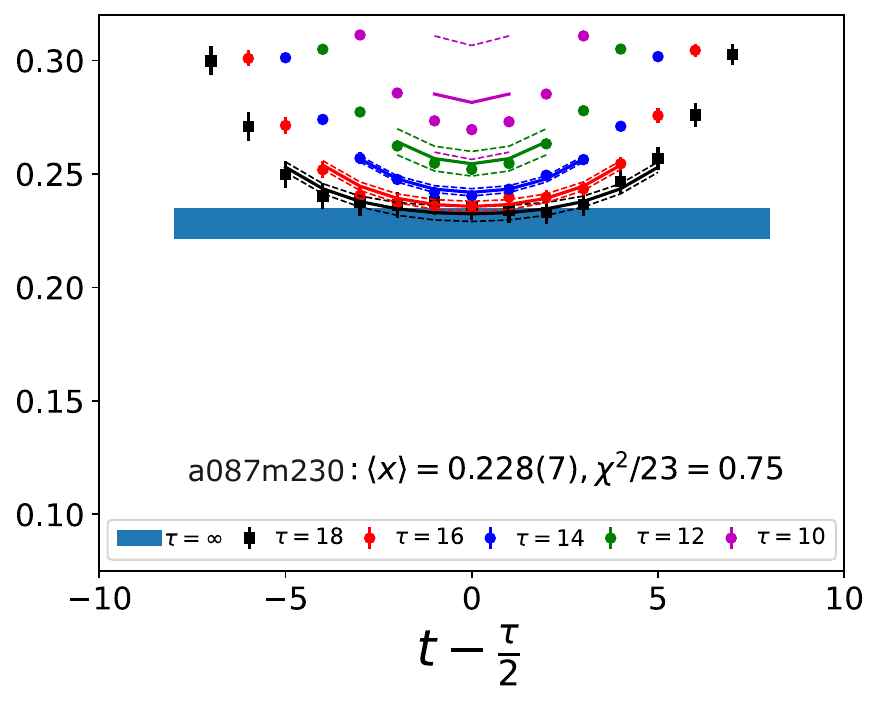}

\includegraphics[angle=0,width=0.32\textwidth]{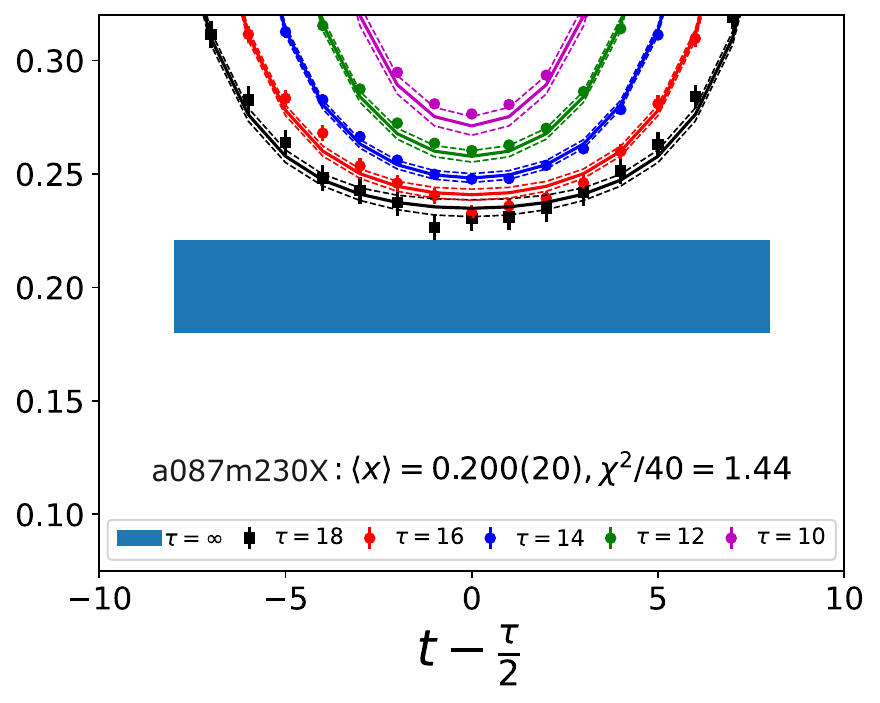}
\includegraphics[angle=0,width=0.32\textwidth]{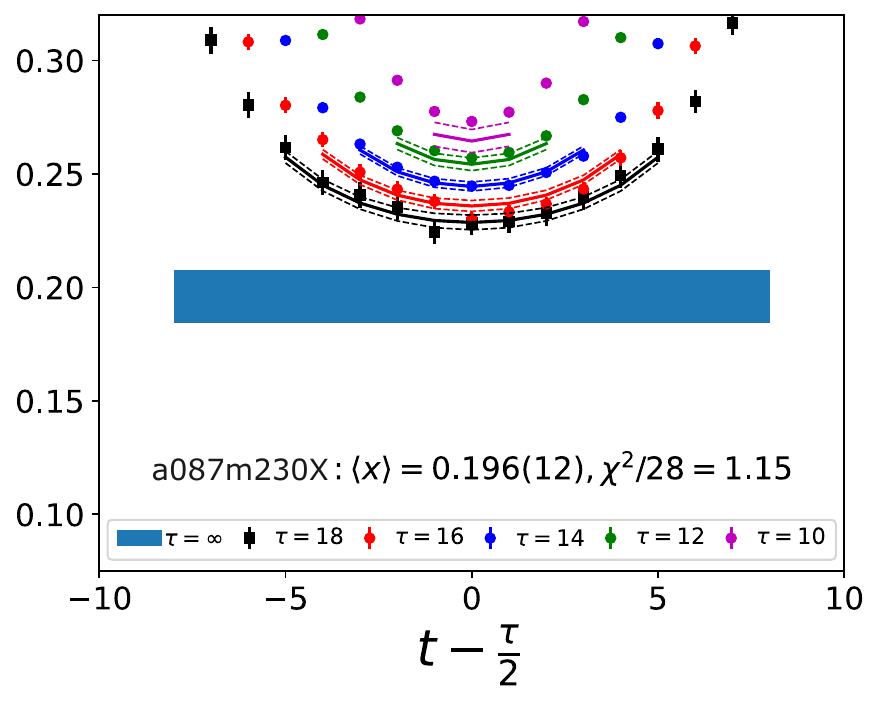}
\includegraphics[angle=0,width=0.32\textwidth]{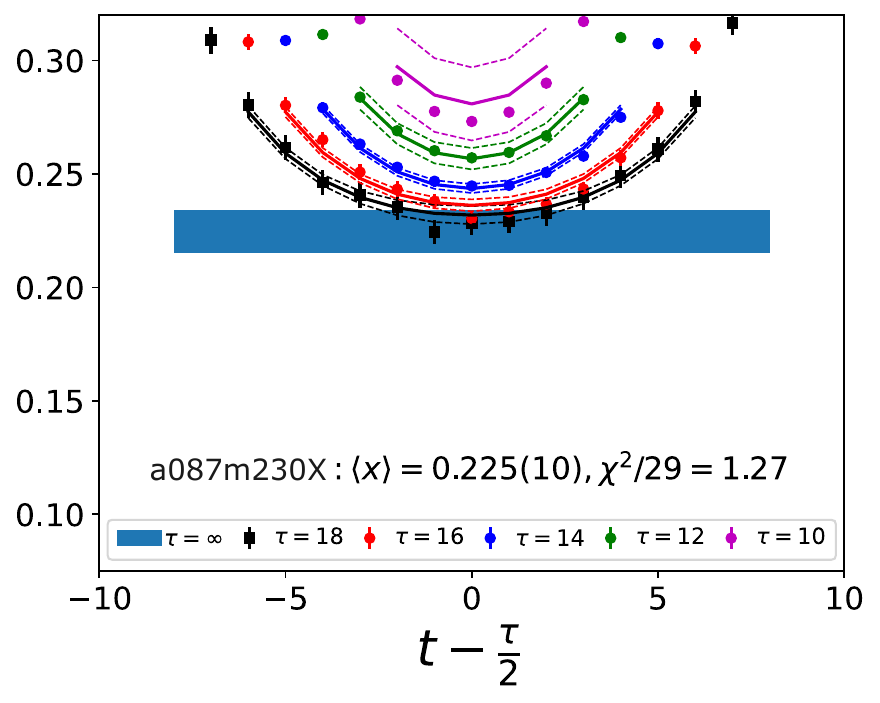}

\includegraphics[angle=0,width=0.32\textwidth]{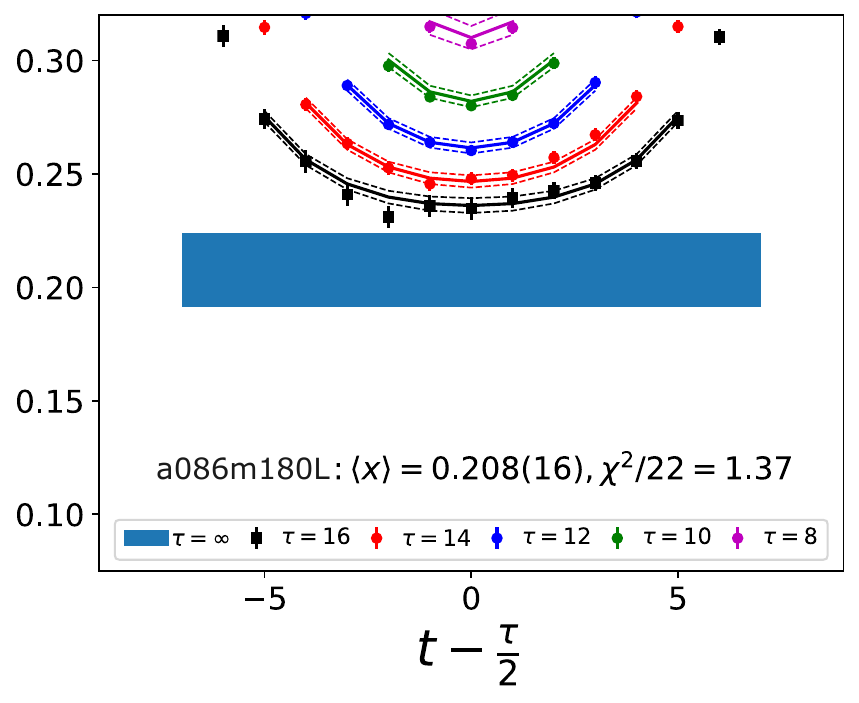}
\includegraphics[angle=0,width=0.32\textwidth]{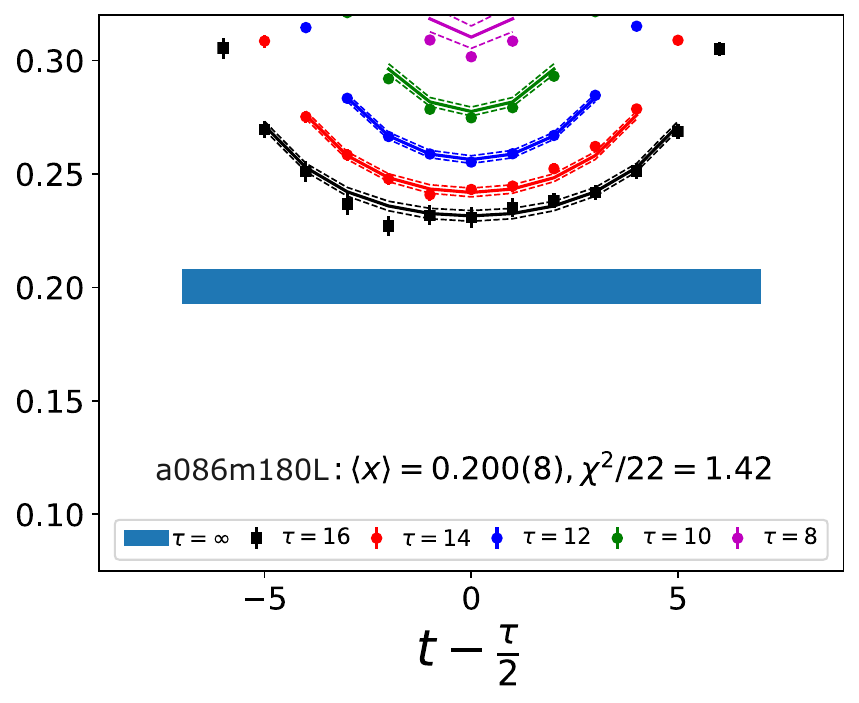}
\includegraphics[angle=0,width=0.32\textwidth]{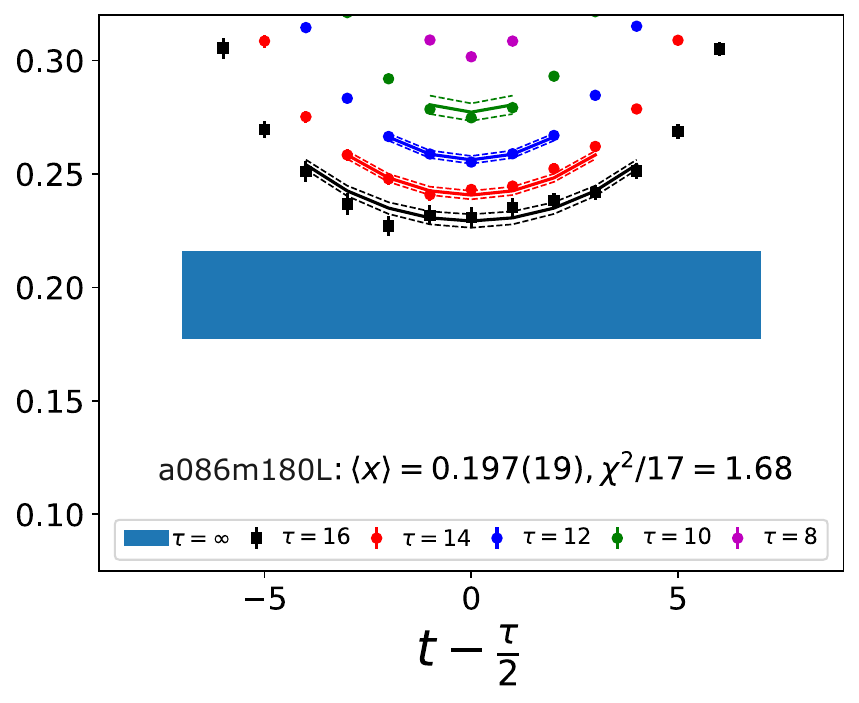}

\includegraphics[angle=0,width=0.32\textwidth]{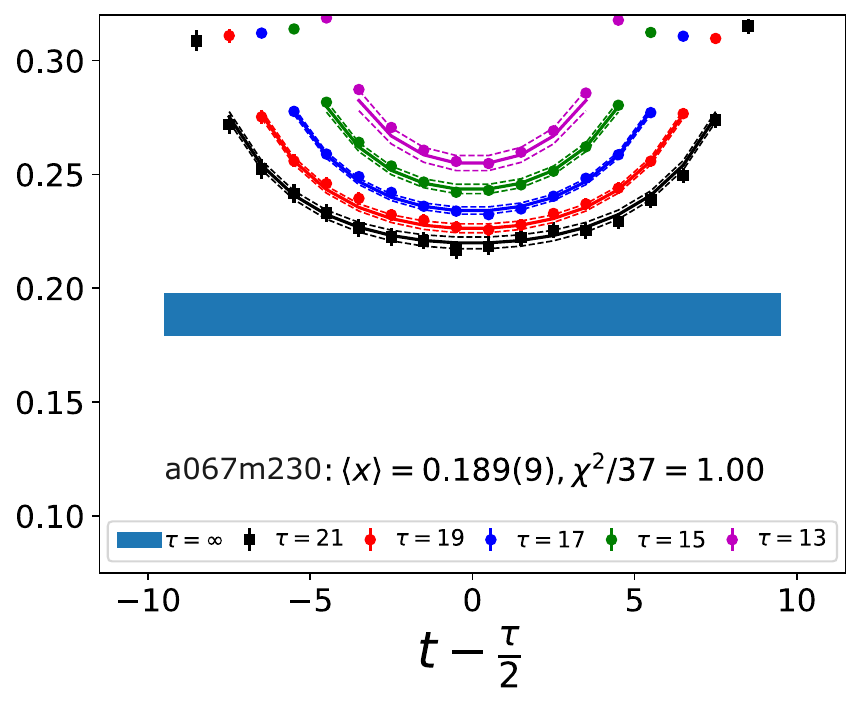}
\includegraphics[angle=0,width=0.32\textwidth]{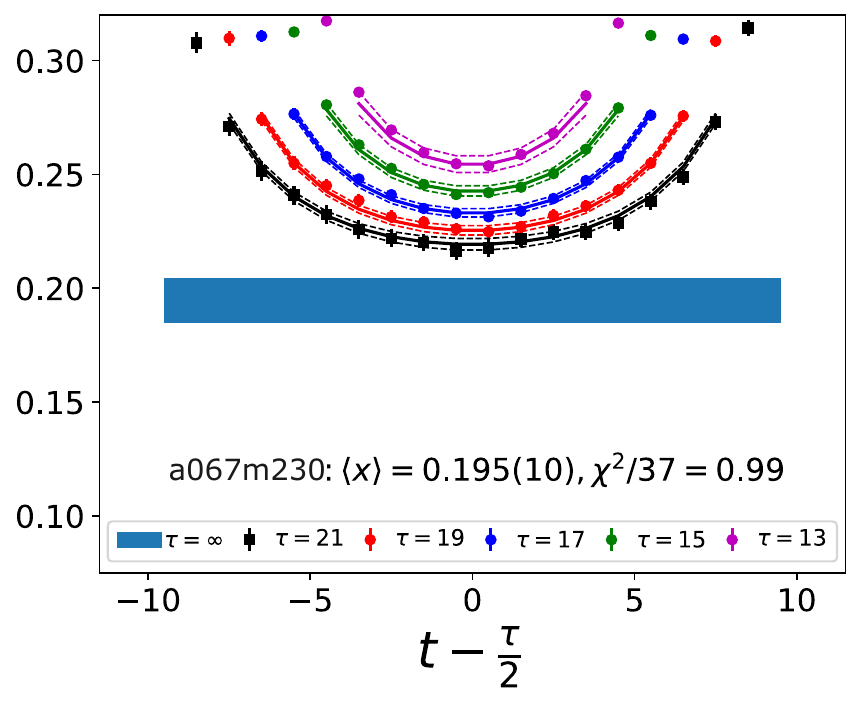}
\includegraphics[angle=0,width=0.32\textwidth]{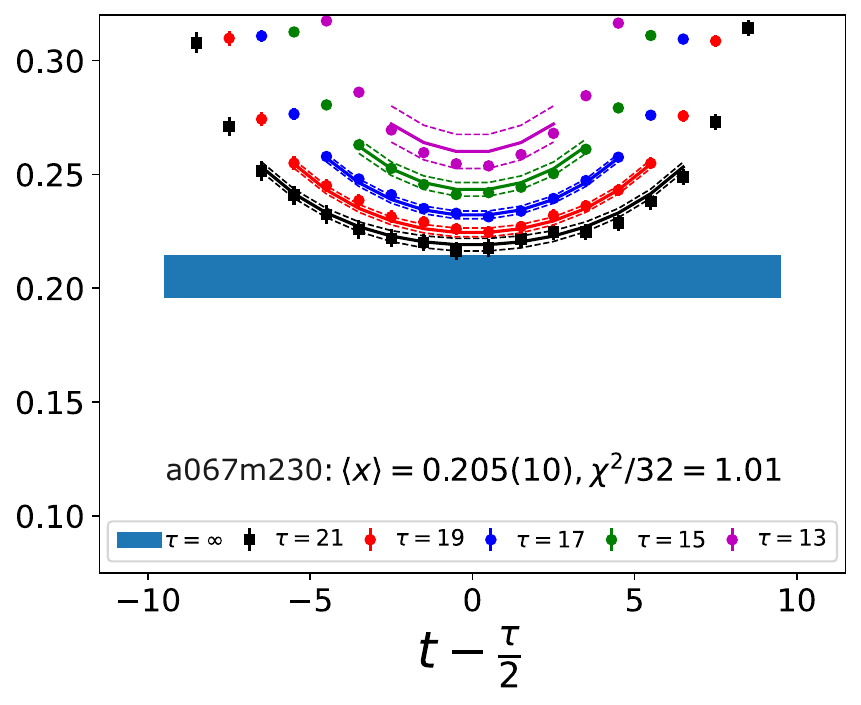}
\end{subfigure}

\caption{Data  for the ratio
  $C_\mathcal{O}^{3\text{pt}}(\tau;t)/C^{2\text{pt}}(\tau)$, scaled
 using Eq.~\protect\eqref{eq:me2momentT} to give $\langle
 x \rangle_{\delta u-\delta d}$, and fits to remove excited-state
 contamination for the $a087m230$ (top row), $a087m230X$ (second row),
 $a086m180L$ (third row), and $a067m230$ (bottom row) ensembles.  The
 fit parameters are listed in
 Table~\protect\ref{tab:5strategy-fits-transvmom}. The rest is the
 same as in Fig.~\protect\ref{fig:Ratio-mom-1}. }
\label{fig:Ratio1-transversity-1}
\end{figure*}

\begin{figure*}[tp]  

\centering

\begin{subfigure}
\centering
\includegraphics[angle=0,width=0.32\textwidth]{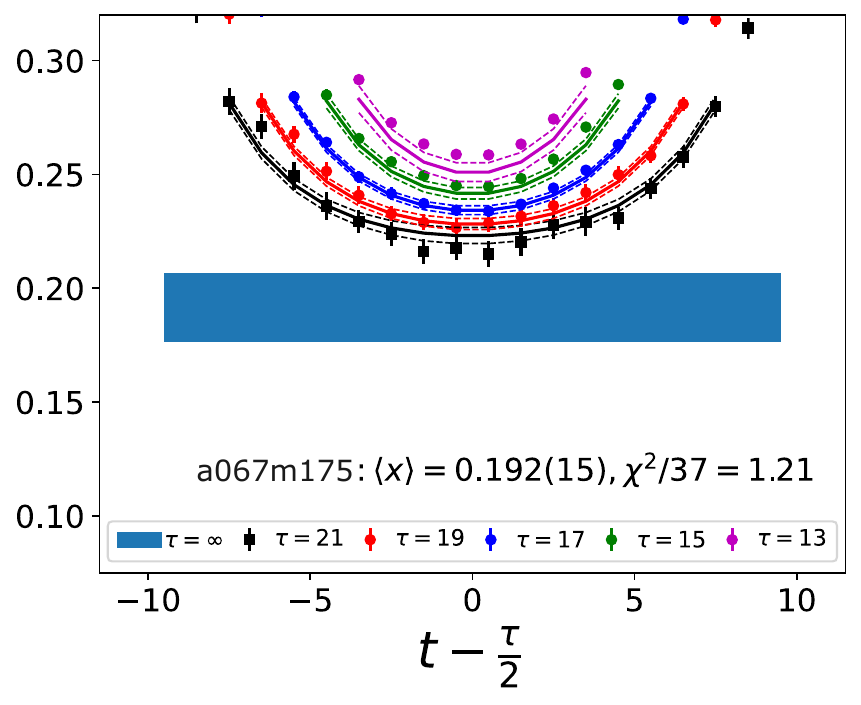}
\includegraphics[angle=0,width=0.32\textwidth]{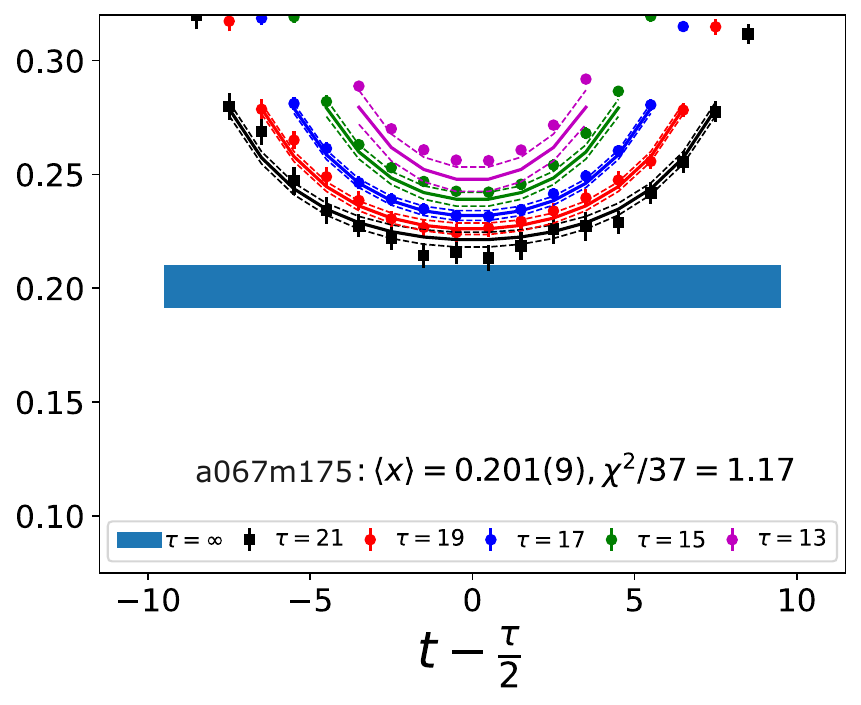}
\includegraphics[angle=0,width=0.32\textwidth]{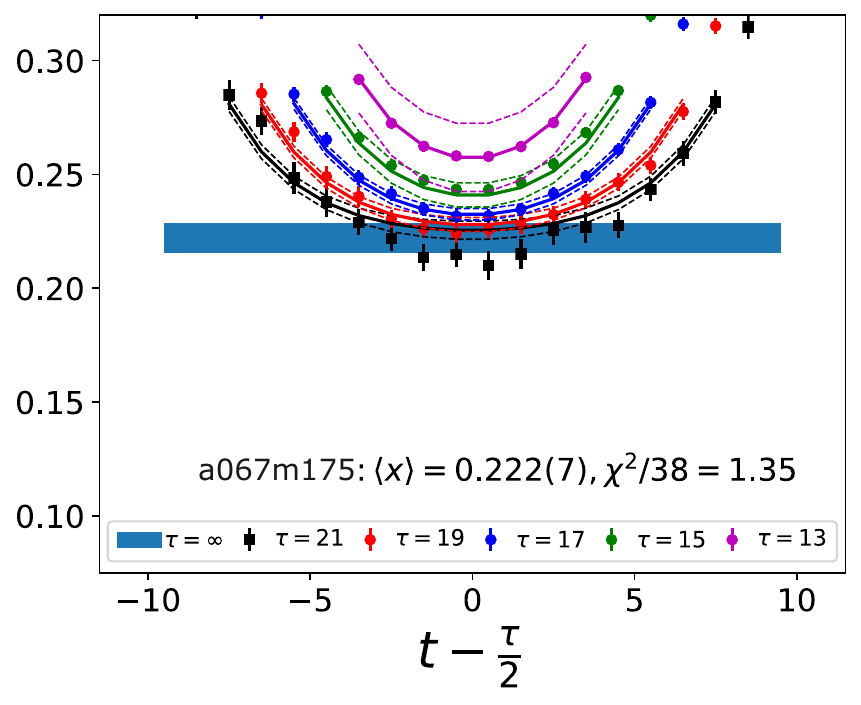}
\end{subfigure}

\begin{subfigure}
\centering
\includegraphics[angle=0,width=0.32\textwidth]{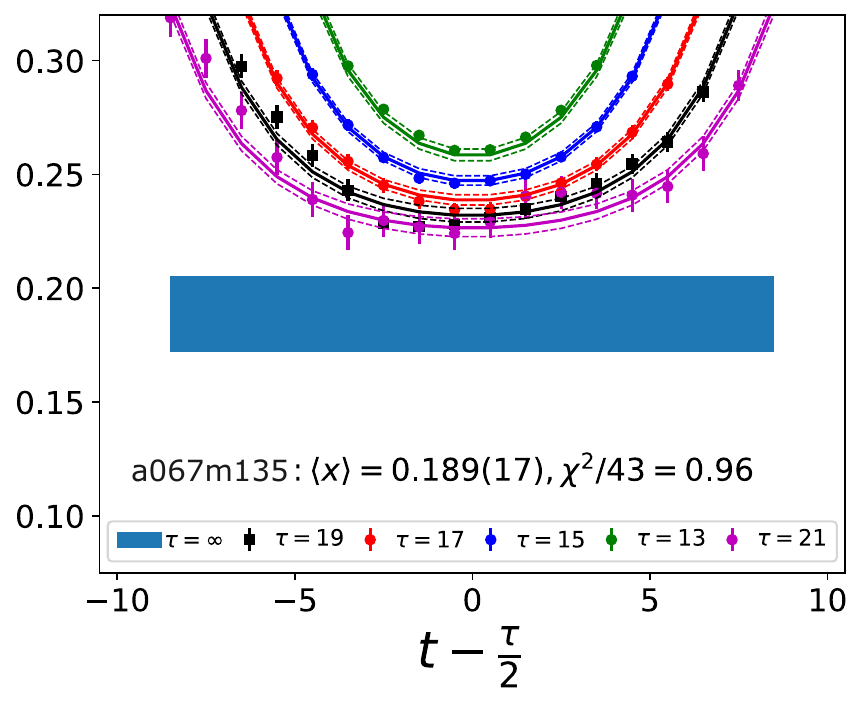}
\includegraphics[angle=0,width=0.32\textwidth]{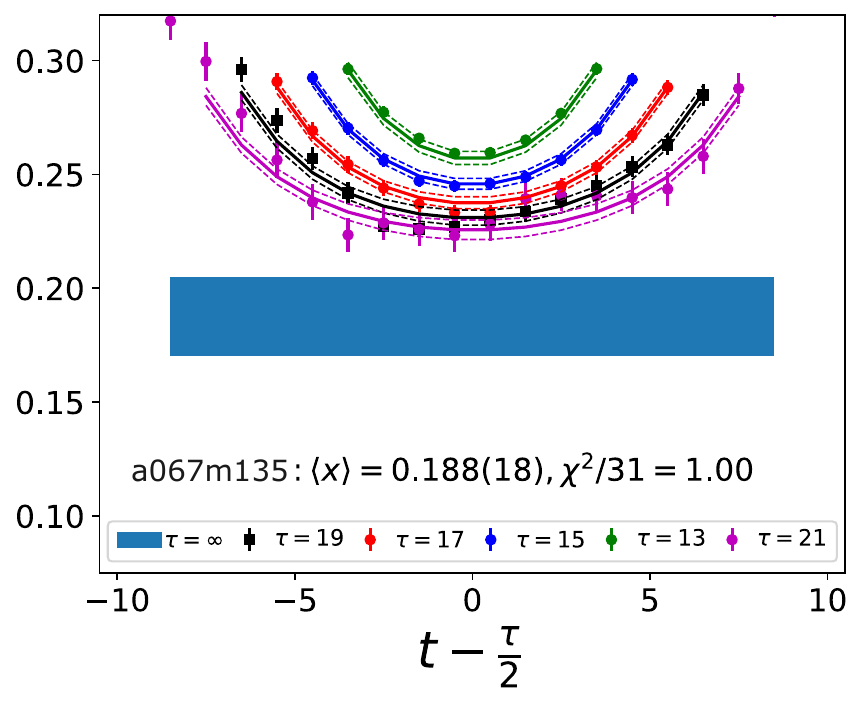}
\includegraphics[angle=0,width=0.32\textwidth]{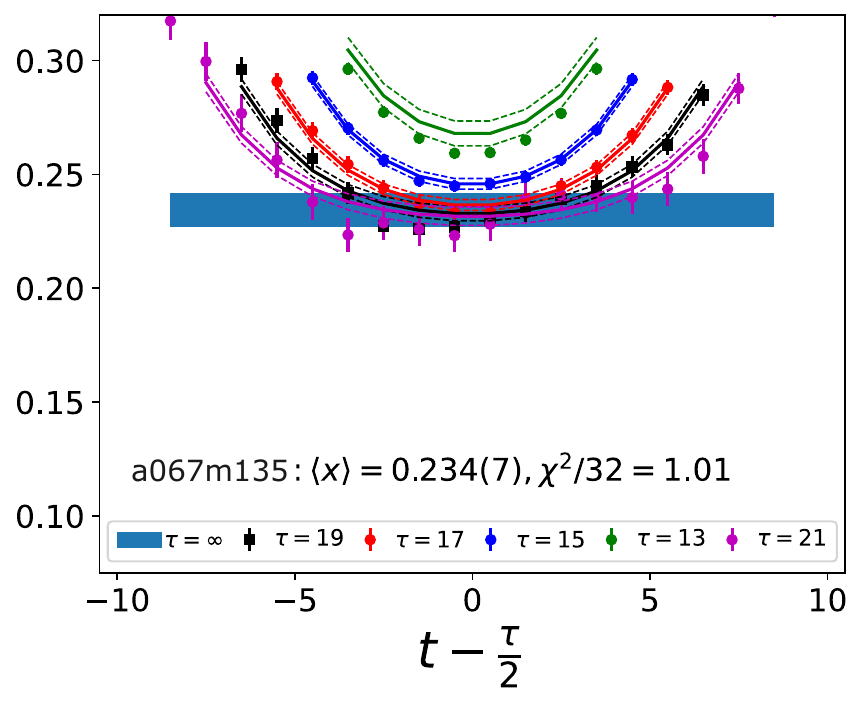}
\end{subfigure}

\begin{subfigure}
\centering
\includegraphics[angle=0,width=0.32\textwidth]{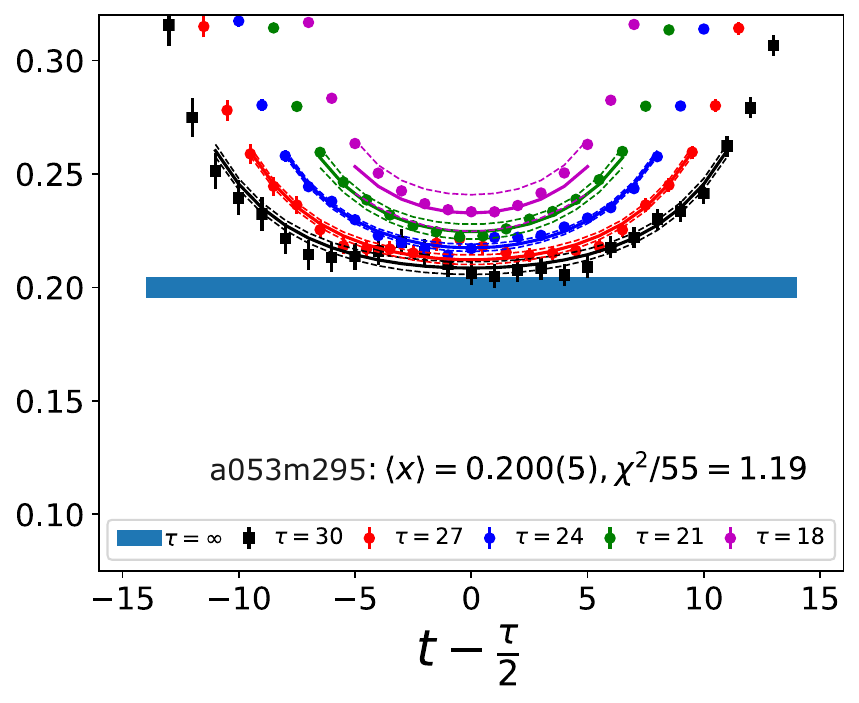}
\includegraphics[angle=0,width=0.32\textwidth]{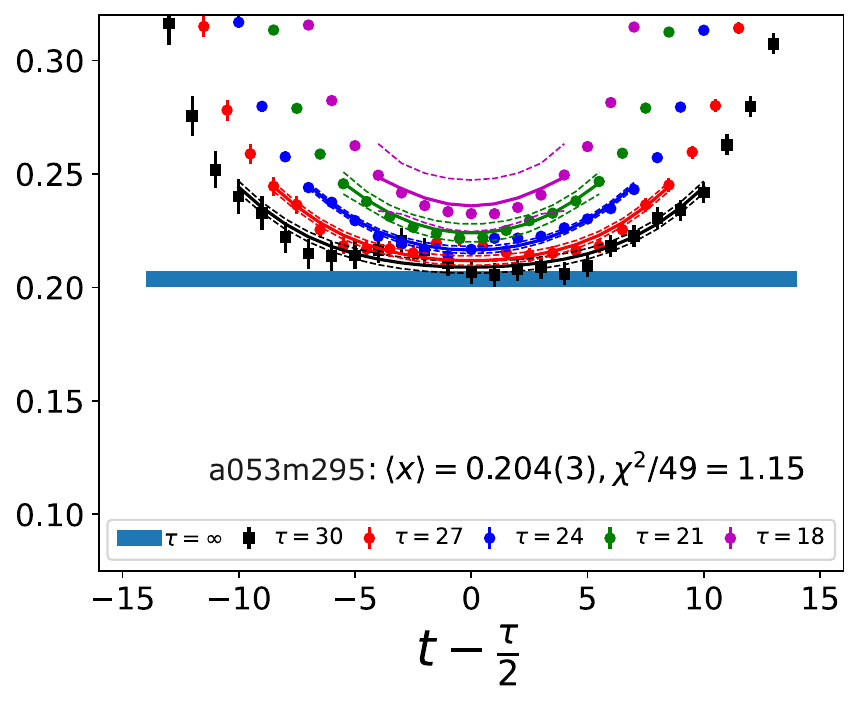}
\includegraphics[angle=0,width=0.32\textwidth]{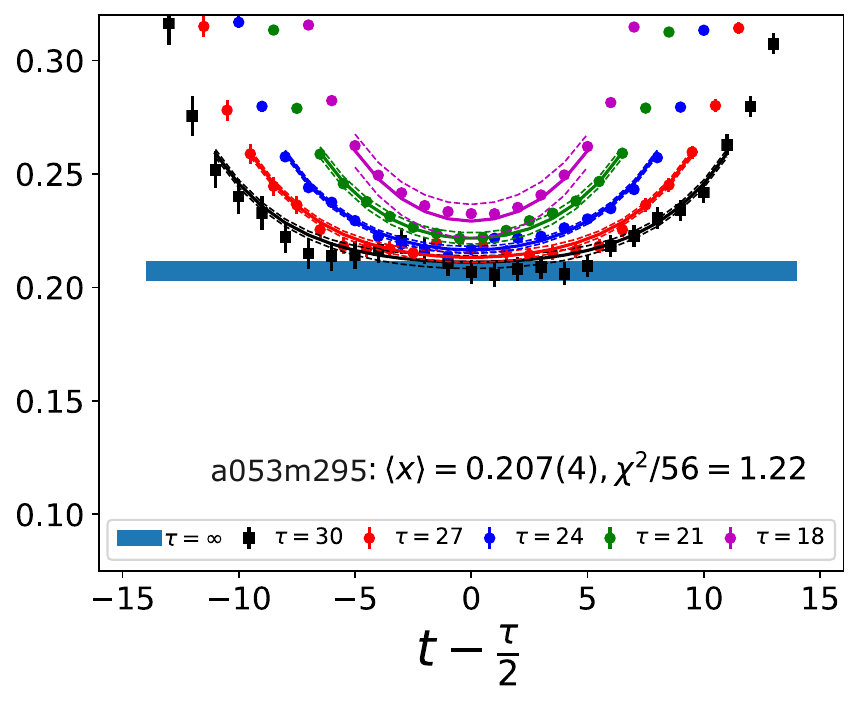}
\end{subfigure}

\begin{subfigure}
\centering
\includegraphics[angle=0,width=0.32\textwidth]{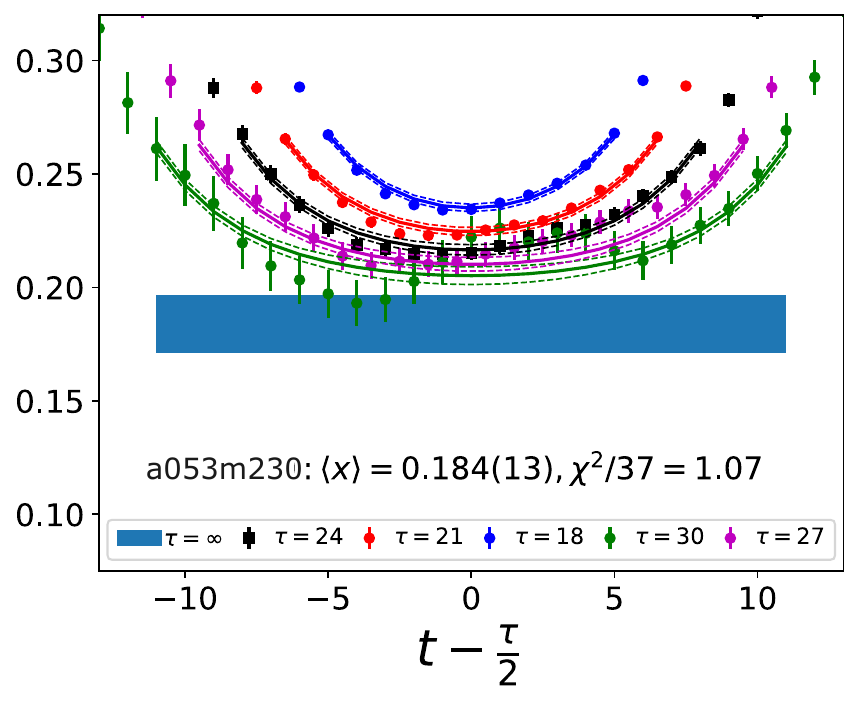}
\includegraphics[angle=0,width=0.32\textwidth]{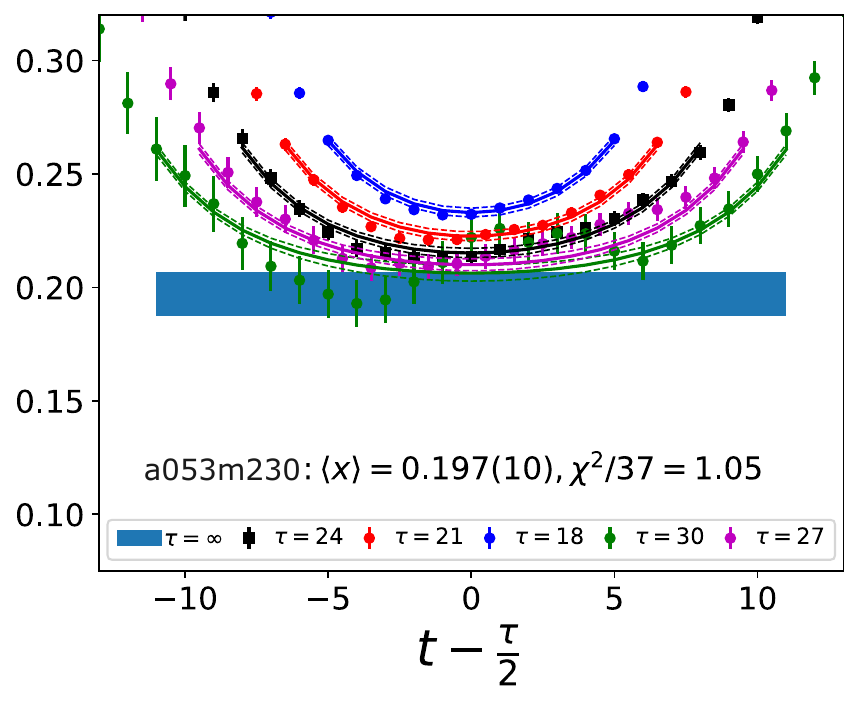}
\includegraphics[angle=0,width=0.32\textwidth]{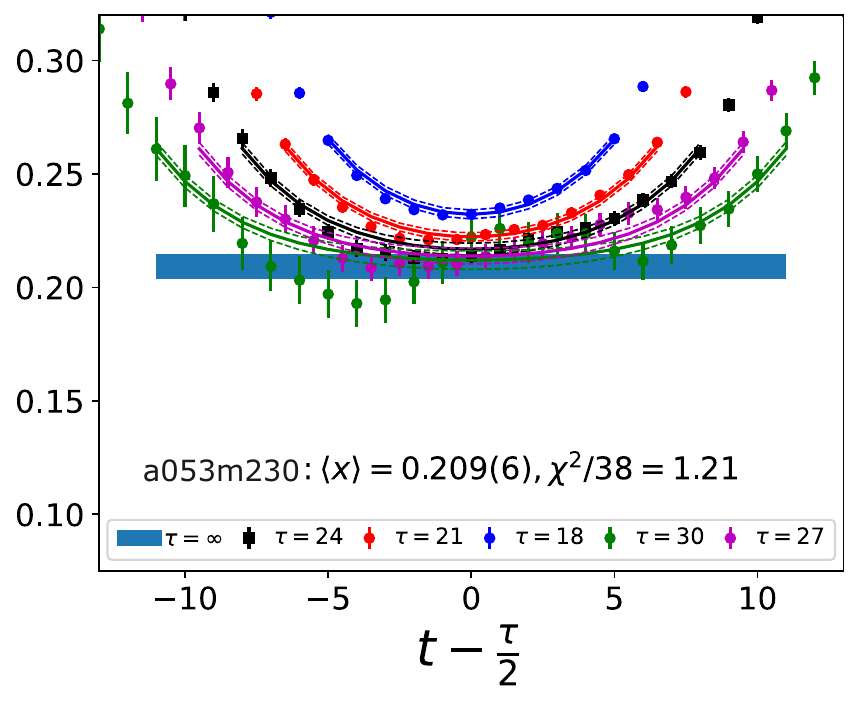}
\end{subfigure}

\caption{Continuation of the data for the ratio
  $C_\mathcal{O}^{3\text{pt}}(\tau;t)/C^{2\text{pt}}(\tau)$, scaled
 using Eq.~\protect\eqref{eq:me2momentT} to give $\langle
 x \rangle_{\delta u-\delta d}$, and fits to remove excited-state
 contamination for the $a067m175$ (top row), $a067m135$ (second row),
 $a053m290$ (third row), and $a053m230$ (bottom row) ensembles.  The
 fit parameters are summarized in
 Table~\protect\ref{tab:5strategy-fits-transvmom}. The rest is the
 same as in Fig.~\protect\ref{fig:Ratio-mom-1}. }
\label{fig:Ratio-transversity-2} 
\end{figure*}

This appendix gives, in Figs.~\ref{fig:Ratio-mom-1}--\ref{fig:Ratio-transversity-2},
plots for the ratio, 
$C_\mathcal{O}^{3\text{pt}}(\tau;t)/C^{2\text{pt}}(\tau)$, multiplied
by the appropriate factor given in
Eqs.~\eqref{eq:me2momentV}--\protect\eqref{eq:me2momentT} to get the
three moments: 
$\la x \ra_{u-d}$, $\la x \ra_{\Delta u-\Delta d}$, and $\la
x\ra_{\delta u-\delta d}$. The figures show data for  
the eight "new" ensembles defined in Table~\ref{tab:ensembles} while those for 
the remaining five are unchanged and given in Ref.~\cite{Mondal:2020ela}.

The three panels in each row have the same data for 
the ratio but show fits with
the three ESC strategies: $\{4^{N\pi},3^*\}$ (left), $\{4^{},3^*\}$
(middle) and $\{4^{},2^{\rm free}\}$ (right).  In all but two cases, 
$a067m135$ and $a053m230$, the fits used to get the final 
results using Eq.~\eqref{eq:3pt},
are made keeping data at the largest three values of $\tau$ 
as listed in Table~\ref{tab:statistics} and 
discussed in Sec.~\ref{sec:ESC} to minimize ESC.  The ground state matrix 
element (blue band) is 
obtained from a simultaneous fit to the $C^{2\text{pt}}(\tau)$ and
$C_\mathcal{O}^{3\text{pt}}(\tau;t)$ that is carried out 
within a single-elimination jackknife loop. The results of these fits 
are shown,  for each $\tau$,  by bands with the same color 
as the data. The unrenormalized value of the moment, the $\chi^2/$dof of the fit, and the values of $\tau$ at which data were 
collected are given in each figure's labels.

The data show three expected trends for these charges: (i) a monotonic convergence in $\tau$ towards the
$\tau \to \infty$ estimate. (ii) The data are, to a good degree, symmetric about $t-\tau/2$
for all values of $\tau$, except for the largest $\tau$ on the $a053m230$ ensemble. 
Our checks confirm that this expected symmetry improves with increased statistics. 
(iii) The size of the 
extrapolation, i.e., the difference between the data at
$t=\tau/2$ with the largest $\tau$ simulated and the $\tau = \infty$ value depends on the mass gap $\Delta M_1$ listed in Table~\ref{tab:massgap}. It is the largest 
for the $\{4^{N\pi},3^\ast\}$ fits that uses the smallest mass gap,  and the smallest 
for the $\{4^{},2^{\rm free}\}$ strategy which gives the largest mass gap.

\clearpage


\section{Results of the excited-state fits}
\label{sec:rawfits}
Tables~\ref{tab:5strategy-fits-momfrac},~\ref{tab:5strategy-fits-helfrac},
and~\ref{tab:5strategy-fits-transvmom} give the results of the fit
parameters defined in Eq.~\eqref{eq:3pt} for the three strategies used
for removing the excited-state contamination in 
the extraction of $\la x \ra_{u-d}$, $\la x \ra_{\Delta u-\Delta d}$, 
and $\la x\ra_{\delta u-\delta d}$.   
The values of $\tau$ used in the fits are given in Table~\ref{tab:ensembles}.  

The results of the fits for the five 
unchanged ensembles are the same as in Ref.~\cite{Mondal:2020ela} and reproduced 
here to make this paper self-contained. The resulting values of the
unrenormalized moments are given in Table~\ref{tab:ESC-fits} 
for each ensemble and ESC fit type.


\begin{table*}[htbp]   
\scriptsize
\centering
\setlength{\tabcolsep}{0.6pt}
\renewcommand{\arraystretch}{1.4}
\begin{tabular}{|c|c|c|c|c|c|c|c|c|c| }
\hline

\multicolumn{10}{|c|}{$\langle x \rangle_{ u- d}$} \\
\hline
Ensemble & fit-type&$a \Delta M_1$&$a \Delta M_2$
&$\la 0|{\cal O}|0 \ra$
&$\frac{\la 1|{\cal O}| 1\ra |A_1|^2 }{\la 0|{\cal O}|0 \ra |A_0|^2}$
&$\frac{\la 1|{\cal O}| 0\ra |A_1|}{\la 0|{\cal O}|0 \ra |A_0|}$ 
&$\frac{\la 2|{\cal O}| 0\ra |A_2|}{\la 0|{\cal O}|0 \ra |A_0|}$ 
&$\frac{\la 2|{\cal O}| 1\ra |A_2||A_1|}{\la 0|{\cal O}|0 \ra |A_0|^2}$
&$\chi^2$/dof\\
\hline
\hline

$a117m310$& $\{4,2\}$&$0.381(41)$ && $0.1166(54)$& $0.24(62)$&$0.629(63)$ & & &$1.74$\\
$a117m310$& $\{4^{N\pi},2\}$&$0.326(20)$ && $0.1114(28)$& $-0.46(31)$&$0.708(46)$ & & &$1.72$\\ 
$a117m310$& $\{4,3^*\}$&$0.381(41)$ &$0.783(54)$& $0.1176(41)$& $0.29(60)$&$0.584(31)$ &$0.048(45)$ &$1.17(78)$ &$1.72$\\ 
$a117m310$& $\{4^{N\pi},3^*\}$&$0.326(20)$ &$0.688(61)$& $0.1118(28)$& $-0.59(42)$&$0.696(68)$ &$-0.12(12)$ &$2.6(1.2)$ &$1.73$\\
$a117m310$& $\{4,2^{free}\}$&$0.359(35)$ && $0.1148(41)$& $-0.01(59)$&$0.651(52)$ & & &$1.81$\\

\hline

$a087m290$& $\{4,2\}$&$0.264(75)$ && $0.0796(88)$& $-0.19(52)$&$0.77(18)$ & & &$1.22$\\
$a087m290$& $\{4^{N\pi},2\}$&$0.263(11)$ && $0.0784(27)$& $-0.24(37)$&$0.80(10)$ & & &$1.25$\\
$a087m290$& $\{4,3^*\}$&$0.264(75)$ &$0.556(86)$& $0.0855(53)$& $0.44(49)$&$0.473(98)$ &$0.28(13)$ &$1.02(89)$ &$1.23$\\
$a087m290$& $\{4^{N\pi},3^*\}$&$0.263(11)$ &$0.55(15)$& $0.0850(33)$& $0.40(38)$&$0.47(13)$ &$0.29(12)$ &$1.00(88)$ &$1.23$\\
$a087m290$& $\{4,2^{free}\}$&$0.485(42)$ && $0.0951(22)$& $6.8(3.6)$&$0.586(20)$ & & &$1.06$\\

\hline

$a087m290L$& $\{4,2\}$&$0.303(42)$ && $0.0822(35)$& $0.74(95)$&$0.680(47)$ & & &$1.08$\\
$a087m290L$& $\{4^{N\pi},2\}$&$0.249(27)$ && $0.0767(36)$& $-0.18(45)$&$0.775(75)$ & & &$1.57$\\
$a087m290L$& $\{4,3^*\}$&$0.303(42)$ &$0.561(83)$& $0.0834(23)$& $1.29(99)$&$0.588(53)$ &$0.23(13)$ &$-1.2(3.4)$ &$0.95$\\
$a087m290L$& $\{4^{N\pi},3^*\}$&$0.249(27)$ &$0.440(61)$& $0.0806(21)$& $0.55(36)$&$0.534(61)$ &$0.316(93)$ &$-0.0012134(32)$ &$0.93$\\
$a087m290L$& $\{4,2^{free}\}$&$0.344(26)$ && $0.0849(16)$& $2.4(1.6)$&$0.654(16)$ & & &$1.02$\\

\hline

$a087m230$& $\{4,3^*\}$&$0.357(31)$ &$0.77(18)$& $0.0813(18)$& $1.5(2.4)$&$0.699(88)$ &$-0.07(45)$ &$4(21)$ &$0.77$\\
$a087m230$& $\{4^{N\pi},3^*\}$&$0.2071(52)$ &$0.272(69)$& $0.0764(47)$& $-0.7(1.5)$&$0.41(26)$ &$0.36(28)$ &$1.6(2.5)$ &$0.76$\\
$a087m230$& $\{4,2^{ free}\}$&$0.359(61)$ && $0.0813(31)$& $1.9(3.7)$&$0.698(34)$ & & &$0.74$\\

\hline

$a087m230X$& $\{4,3^*\}$&$0.220(32)$ &$0.649(33)$& $0.0723(48)$& $0.33(99)$&$0.58(21)$ &$1.1(1.0)$ &$0.7(8.1)$ &$0.78$\\
$a087m230X$& $\{4^{N\pi},3^*\}$&$0.160(18)$ &$0.577(87)$& $0.0652(77)$& $0.4(1.0)$&$0.63(39)$ &$1.0(1.1)$ &$1.2(5.3)$ &$0.79$\\
$a087m230X$& $\{4,2^{free}\}$&$0.345(50)$ && $0.0784(32)$& $4.2(4.4)$&$0.761(41)$ & & &$0.71$\\

\hline
$a086m180$& $\{4,2\}$&$0.293(45)$ && $0.0682(57)$& $0.89(95)$&$0.94(14)$ & & &$0.98$\\
$a086m180$& $\{4^{N\pi},2\}$&$0.195(19)$ && $0.0473(71)$& $-0.92(67)$&$1.78(43)$ & & &$1.34$\\
$a086m180$& $\{4,3^*\}$&$0.293(45)$ &$0.578(81)$& $0.0696(39)$& $1.4(1.2)$&$0.81(15)$ &$0.28(41)$ &$-1.2(4.9)$ &$1.04$\\
$a086m180$& $\{4^{N\pi},3^*\}$&$0.195(19)$ &$0.413(66)$& $0.0640(41)$& $1.27(39)$&$0.56(22)$ &$0.74(13)$ &$-0.55(61)$ &$1.00$\\
$a086m180$& $\{4,2^{ free}\}$&$0.311(48)$ && $0.0702(50)$& $1.5(1.7)$&$0.89(11)$ & & &$1.02$\\

\hline

$a086m180L$& $\{4,3^*\}$&$0.252(30)$ &$0.615(45)$& $0.0672(34)$& $1.04(41)$&$0.754(65)$ &$0.408(63)$ &$0.72(59)$ &$0.94$\\
$a086m180L$& $\{4^{N\pi},3^*\}$&$0.141(24)$ &$0.420(77)$& $0.061(10)$& $1.23(47)$&$0.43(50)$ &$0.90(17)$ &$0.35(65)$ &$1.19$\\
$a086m180L$& $\{4,2^{free}\}$&$0.303(50)$ && $0.0701(49)$& $1.8(1.8)$&$0.862(92)$ & & &$1.07$\\

\hline

$a068m290$& $\{4,2\}$&$0.229(41)$ && $0.0612(45)$& $0.17(77)$&$0.76(11)$ & & &$0.98$\\
$a068m290$& $\{4^{N\pi},2\}$&$0.217(24)$ && $0.0598(31)$& $-0.05(47)$&$0.792(82)$ & & &$1.02$\\
$a068m290$& $\{4,3^*\}$&$0.229(41)$ &$0.386(79)$& $0.0633(25)$& $0.97(71)$&$0.563(64)$ &$0.35(11)$ &$-0.9(1.2)$ &$1.03$\\
$a068m290$& $\{4^{N\pi},3^*\}$&$0.217(24)$ &$0.363(79)$& $0.0626(16)$& $0.83(44)$&$0.545(75)$ &$0.383(86)$ &$-0.75(87)$ &$1.03$\\
$a068m290$& $\{4,2^{free}\}$&$0.311(16)$ && $0.06699(96)$& $3.3(1.3)$&$0.695(15)$ & & &$1.56$\\

\hline

$a067m230$& $\{4,3^*\}$&$0.204(40)$ &$0.495(33)$& $0.0564(34)$& $-0.19(78)$&$0.65(11)$ &$0.28(36)$ &$4.0(4.8)$ &$1.32$\\
$a067m230$& $\{4^{N\pi},3^*\}$&$0.172(19)$ &$0.471(40)$& $0.0536(30)$& $-0.33(62)$&$0.68(15)$ &$0.31(36)$ &$3.4(3.2)$ &$1.32$\\
$a067m230$& $\{4,2^{free}\}$&$0.295(47)$ && $0.0605(24)$& $3.5(4.5)$&$0.751(32)$ & & &$1.33$\\

\hline

$a067m175$& $\{4,3^*\}$&$0.212(45)$ &$0.444(26)$& $0.0517(35)$& $2.8(1.8)$&$0.62(10)$ &$0.73(23)$ &$-5.7(5.8)$ &$0.91$\\
$a067m175$& $\{4^{N\pi},3^*\}$&$0.141(11)$ &$0.383(42)$& $0.0501(33)$& $1.66(45)$&$0.32(19)$ &$0.978(99)$ &$-0.51(75)$ &$1.04$\\
$a067m175$& $\{4,2^{free}\}$&$0.222(60)$ && $0.0509(61)$& $1.7(3.0)$&$0.83(17)$ & & &$0.90$\\

\hline



$a067m135$& $\{4,3^*\}$&$0.141(17)$ &$0.435(22)$& $0.0494(31)$& $0.89(29)$&$0.44(12)$ &$0.890(90)$ &$0.32(47)$ &$1.24$\\
$a067m135$& $\{4^{N\pi},3^*\}$&$0.1254(77)$ &$0.426(23)$& $0.0489(33)$& $1.01(29)$&$0.39(15)$ &$0.967(81)$ &$0.30(44)$ &$1.31$\\
$a067m135$& $\{4,2^{ free}\}$&$0.289(36)$ && $0.0561(29)$& $4.3(3.2)$&$0.874(67)$ & & &$0.91$\\

\hline

$a053m295$& $\{4,3^*\}$&$0.284(17)$ &$0.67(16)$& $0.050(43)$& $26(999)$&$0.6(4.5)$ &$0.3(5.5)$ &$-29(1217)$ &$1.14$\\
$a053m295$& $\{4^{N\pi},3^*\}$&$0.224(24)$ &$0.313(65)$& $0.04858(97)$& $9.4(6.1)$&$0.486(95)$ &$0.41(12)$ &$-13(14)$ &$1.17$\\
$a053m295$& $\{4,2^{ free}\}$&$0.35(28)$ && $0.051(26)$& $135(1254)$&$0.74(51)$ & & &$1.20$\\

\hline

$a053m230$& $\{4,3^*\}$&$0.208(70)$ &$0.341(44)$& $0.0448(37)$& $1.3(2.1)$&$0.665(94)$ &$0.25(25)$ &$-2.7(5.6)$ &$1.12$\\
$a053m230$& $\{4^{N\pi},3^*\}$&$0.129(21)$ &$0.313(31)$& $0.0422(27)$& $0.59(32)$&$0.42(11)$ &$0.666(77)$ &$-0.06(52)$ &$1.16$\\
$a053m230$& $\{4,2^{free}\}$&$0.212(28)$ && $0.0445(22)$& $0.6(1.3)$&$0.756(67)$ & & &$1.15$\\

\hline

\end{tabular}
\caption{Comparison of results of the fits to remove the excited-state contamination 
for the momentum fraction $\langle x \rangle_{ u- d}$ using the five strategies, $\{4,2\}$,
  $\{4^{N\pi},2\}$, $\{4,3^*\}$, $\{4^{N\pi},3^*\}$ and $\{4,2^{\rm
    free}\}$. The fit parameters, defined in Eq.~\protect\eqref{eq:3pt}, are given for all seven ensembles
 along with the $\chi^2/$dof of the fit.
}
\label{tab:5strategy-fits-momfrac}
\end{table*}


\begin{table*}[htbp]   
\scriptsize
\centering
\setlength{\tabcolsep}{0.6pt}
\renewcommand{\arraystretch}{1.4}
\begin{tabular}{|c|c|c|c|c|c|c|c|c|c| }
\hline

\multicolumn{10}{|c|}{$\langle x \rangle_{\Delta u-\Delta d}$} \\
\hline
Ensemble & fit-type&$a \Delta M_1$&$a \Delta M_2$
&$\la 0|{\cal O}|0 \ra$
&$\frac{\la 1|{\cal O}| 1\ra |A_1|^2 }{\la 0|{\cal O}|0 \ra |A_0|^2}$
&$\frac{\la 1|{\cal O}| 0\ra |A_1|}{\la 0|{\cal O}|0 \ra |A_0|}$ 
&$\frac{\la 2|{\cal O}| 0\ra |A_2|}{\la 0|{\cal O}|0 \ra |A_0|}$ 
&$\frac{\la 2|{\cal O}| 1\ra |A_2||A_1|}{\la 0|{\cal O}|0 \ra |A_0|^2}$
&$\chi^2$/dof\\
\hline
\hline

$a117m310$& $\{4,2\}$&$0.381(41)$ && $0.1404(38)$& $-0.19(35)$&$0.404(30)$ & & &$1.10$\\
$a117m310$& $\{4^{N\pi},2\}$&$0.326(20)$ && $0.1363(24)$& $-0.25(25)$&$0.431(30)$ & & &$1.30$\\
$a117m310$& $\{4,3^*\}$&$0.381(41)$ &$0.783(54)$& $0.1443(24)$& $0.40(41)$&$0.252(51)$ &$0.529(94)$ &$-0.6(1.7)$ &$1.30$\\
$a117m310$& $\{4^{N\pi},3^*\}$&$0.326(20)$ &$0.688(61)$& $0.1445(33)$& $0.59(49)$&$0.162(96)$ &$0.62(30)$ &$-0.04(2.14)$ &$0.83$\\
$a117m310$& $\{4,2^{free}\}$&$0.706(58)$ && $0.1503(14)$& $15(12)$&$0.532(28)$ & & &$1.20$\\

\hline

$a087m290$& $\{4,2\}$&$0.264(75)$ && $0.0960(80)$& $-0.01(41)$&$0.56(12)$ & & &$1.30$\\
$a087m290$& $\{4^{N\pi},2\}$&$0.263(11)$ && $0.0949(27)$& $-0.05(31)$&$0.574(74)$ & & &$1.33$\\
$a087m290$& $\{4,3^*\}$&$0.264(75)$ &$0.556(86)$& $0.1048(40)$& $0.80(43)$&$0.20(11)$ &$0.69(11)$ &$-0.29(80)$ &$1.09$\\
$a087m290$& $\{4^{N\pi},3^*\}$&$0.263(11)$ &$0.55(15)$& $0.1048(52)$& $0.80(48)$&$0.19(18)$ &$0.703(96)$ &$-0.30(85)$ &$1.09$\\
$a087m290$& $\{4,2^{free}\}$&$0.689(50)$ && $0.1136(18)$& $39(22)$&$0.692(28)$ & & &$0.94$\\

\hline

$a087m290L$& $\{4,2\}$&$0.303(42)$ && $0.0971(30)$& $0.19(59)$&$0.543(25)$ & & &$1.26$\\
$a087m290L$& $\{4^{N\pi},2\}$&$0.249(27)$ && $0.0940(26)$& $0.18(39)$&$0.499(39)$ & & &$1.30$\\
$a087m290L$& $\{4,3^*\}$&$0.303(42)$ &$0.561(83)$& $0.0986(19)$& $1.14(65)$&$0.418(63)$ &$0.497(96)$ &$-3.1(3.5)$ &$0.86$\\
$a087m290L$& $\{4^{N\pi},3^*\}$&$0.249(27)$ &$0.440(61)$& $0.0970(17)$& $0.83(41)$&$0.324(68)$ &$0.540(87)$ &$-1.4(1.3)$ &$0.85$\\
$a087m290L$& $\{4,2^{ free}\}$&$0.400(53)$ && $0.1015(18)$& $4.4(4.5)$&$0.559(35)$ & & &$1.11$\\

\hline

$a087m230$& $\{4,3^*\}$&$0.357(31)$ &$0.77(18)$& $0.0939(19)$& $2.4(2.1)$&$0.499(86)$ &$0.44(38)$ &$12(20)$ &$1.30$\\
$a087m230$& $\{4^{N\pi},3^*\}$&$0.2071(52)$ &$0.272(69)$& $0.0942(34)$& $1.2(1.0)$&$-0.004(107)$ &$0.72(16)$ &$-0.3(2.0)$ &$1.32$\\
$a087m230$& $\{4,2^{ free}\}$&$0.300(61)$ && $0.0903(47)$& $-0.006(925)$&$0.583(49)$ & & &$1.25$\\

\hline

$a087m230X$& $\{4,3^*\}$&$0.220(32)$ &$0.649(33)$& $0.0855(43)$& $0.45(54)$&$0.46(11)$ &$0.62(31)$ &$1.8(3.1)$ &$0.80$\\
$a087m230X$& $\{4^{N\pi},3^*\}$&$0.160(18)$ &$0.577(87)$& $0.0801(58)$& $0.86(39)$&$0.40(19)$ &$0.98(14)$ &$-0.47(94)$ &$0.85$\\
$a087m230X$& $\{4,2^{free}\}$&$0.458(68)$ && $0.0957(24)$& $22(23)$&$0.685(54)$ & & &$0.75$\\

\hline

$a086m180$& $\{4,2\}$&$0.293(45)$ && $0.0794(57)$& $-0.12(56)$&$0.83(12)$ & & &$1.36$\\
$a086m180$& $\{4^{N\pi},2\}$&$0.195(19)$ && $0.0646(59)$& $-0.74(44)$&$1.15(21)$ & & &$1.39$\\
$a086m180$& $\{4,3^*\}$&$0.293(45)$ &$0.578(81)$& $0.0848(32)$& $0.69(78)$&$0.52(11)$ &$0.48(27)$ &$1.3(3.2)$ &$1.02$\\
$a086m180$& $\{4^{N\pi},3^*\}$&$0.195(19)$ &$0.413(66)$& $0.0802(53)$& $0.58(71)$&$0.36(23)$ &$0.65(27)$ &$0.3(1.5)$ &$1.03$\\
$a086m180$& $\{4,2^{ free}\}$&$0.343(96)$ && $0.0856(63)$& $1.3(3.0)$&$0.681(58)$ & & &$1.23$\\

\hline

$a086m180L$& $\{4,3^*\}$&$0.252(30)$ &$0.615(45)$& $0.0783(33)$& $1.06(48)$&$0.595(94)$ &$0.76(23)$ &$-1.2(2.2)$ &$0.66$\\
$a086m180L$& $\{4^{N\pi},3^*\}$&$0.141(24)$ &$0.420(77)$& $0.133(41)$& $2.63(36)$&$-0.80(40)$ &$1.276(90)$ &$-0.99(45)$ &$1.16$\\
$a086m180L$& $\{4,2^{free}\}$&$0.377(60)$ && $0.0855(35)$& $4.8(4.0)$&$0.738(37)$ & & &$0.53$\\

\hline

$a068m290$& $\{4,2\}$&$0.229(41)$ && $0.0747(33)$& $-0.13(42)$&$0.542(50)$ & & &$1.56$\\
$a068m290$& $\{4^{N\pi},2\}$&$0.217(24)$ && $0.0737(23)$& $-0.23(31)$&$0.555(47)$ & & &$1.63$\\
$a068m290$& $\{4,3^*\}$&$0.229(41)$ &$0.386(79)$& $0.0780(13)$& $0.57(30)$&$0.30(14)$ &$0.630(96)$ &$0.61(97)$ &$1.12$\\
$a068m290$& $\{4^{N\pi},3^*\}$&$0.217(24)$ &$0.363(79)$& $0.0780(17)$& $0.59(33)$&$0.25(14)$ &$0.660(84)$ &$0.39(87)$ &$1.12$\\
$a068m290$& $\{4,2^{free}\}$&$0.457(14)$ && $0.08270(52)$& $24.0(7.1)$&$0.758(13)$ & & &$1.28$\\

\hline

$a067m230$& $\{4,3^*\}$&$0.204(40)$ &$0.495(33)$& $0.0679(29)$& $0.05(50)$&$0.451(67)$ &$0.72(13)$ &$0.9(2.2)$ &$0.88$\\
$a067m230$& $\{4^{N\pi},3^*\}$&$0.172(19)$ &$0.471(40)$& $0.0660(26)$& $0.07(39)$&$0.431(89)$ &$0.76(12)$ &$0.8(1.5)$ &$0.89$\\
$a067m230$& $\{4,2^{ free}\}$&$0.353(48)$ && $0.0726(17)$& $7.5(8.9)$&$0.715(42)$ & & &$0.89$\\

\hline

$a067m175$& $\{4,3^*\}$&$0.212(45)$ &$0.444(26)$& $0.0635(26)$& $2.7(1.4)$&$0.32(10)$ &$1.10(17)$ &$-3.8(3.7)$ &$1.38$\\
$a067m175$& $\{4^{N\pi},3^*\}$&$0.141(11)$ &$0.383(42)$& $0.0618(39)$& $1.84(50)$&$0.11(17)$ &$1.28(16)$ &$-2.0(1.3)$ &$1.41$\\
$a067m175$& $\{4,2^{ free}\}$&$0.264(95)$ && $0.0637(55)$& $4.4(7.9)$&$0.594(58)$ & & &$1.35$\\

\hline



$a067m135$& $\{4,3^*\}$&$0.141(17)$ &$0.435(22)$& $0.0513(44)$& $1.17(38)$&$0.52(18)$ &$1.13(18)$ &$-0.04(72)$ &$0.98$\\
$a067m135$& $\{4^{N\pi},3^*\}$&$0.1254(77)$ &$0.426(23)$& $0.0588(41)$& $1.86(31)$&$0.07(15)$ &$1.47(11)$ &$-0.85(55)$ &$1.19$\\
$a067m135$& $\{4,2^{ free}\}$&$0.350(84)$ && $0.0647(38)$& $16(20)$&$0.867(90)$ & & &$0.97$\\

\hline

$a053m295$& $\{4,3^*\}$&$0.284(17)$ &$0.67(16)$& $0.05988(87)$& $8(10)$&$0.587(64)$ &$0.24(46)$ &$102(179)$ &$1.22$\\
$a053m295$& $\{4^{N\pi},3^*\}$&$0.224(24)$ &$0.313(65)$& $0.0588(11)$& $2.8(3.8)$&$0.40(11)$ &$0.42(16)$ &$6(15)$ &$1.21$\\
$a053m295$& $\{4,2^{ free}\}$&$0.47(11)$ && $0.0629(69)$& $0.02299827(56)$&$0.969(53)$ & & &$1.73$\\

\hline

$a053m230$& $\{4,3^*\}$&$0.208(70)$ &$0.341(44)$& $0.0532(24)$& $1.0(1.1)$&$0.47(17)$ &$0.81(22)$ &$-3.6(5.0)$ &$1.55$\\
$a053m230$& $\{4^{N\pi},3^*\}$&$0.129(21)$ &$0.313(31)$& $0.0503(27)$& $0.57(41)$&$0.31(13)$ &$1.04(16)$ &$-1.16(97)$ &$1.56$\\
$a053m230$& $\{4,2^{free}\}$&$0.347(28)$ && $0.05674(91)$& $10.4(8.8)$&$0.858(31)$ & & &$1.74$\\

\hline
\end{tabular}
\caption{Comparison of results of the fits to remove the excited-state contamination for the helicity  moment  
$\langle x \rangle_{\Delta u - \Delta d}$. The rest is the same as in
Table~\protect\ref{tab:5strategy-fits-momfrac}.  }
\label{tab:5strategy-fits-helfrac}
\end{table*}


\begin{table*}[htbp]   
\scriptsize
\centering
\setlength{\tabcolsep}{0.4pt}
\renewcommand{\arraystretch}{1.4}
\begin{tabular}{|c|c|c|c|c|c|c|c|c|c| }
\hline

\multicolumn{10}{|c|}{$\langle x \rangle_{\delta u-\delta d}$} \\
\hline
Ensemble & fit-type&$a \Delta M_1$&$a \Delta M_2$
&$\la 0|{\cal O}|0 \ra$
&$\frac{\la 1|{\cal O}| 1\ra |A_1|^2 }{\la 0|{\cal O}|0 \ra |A_0|^2}$
&$\frac{\la 1|{\cal O}| 0\ra |A_1|}{\la 0|{\cal O}|0 \ra |A_0|}$ 
&$\frac{\la 2|{\cal O}| 0\ra |A_2|}{\la 0|{\cal O}|0 \ra |A_0|}$ 
&$\frac{\la 2|{\cal O}| 1\ra |A_2||A_1|}{\la 0|{\cal O}|0 \ra |A_0|^2}$
&$\chi^2$/dof\\
\hline
\hline
$a117m310$& $\{4,2\}$&$0.381(41)$ && $0.1406(55)$& $0.10(57)$&$0.610(50)$ & & &$1.28$\\
$a117m310$& $\{4^{N\pi},2\}$&$0.326(20)$ && $0.1344(33)$& $-0.22(32)$&$0.661(50)$ & & &$1.59$\\ 
$a117m310$& $\{4,3^*\}$&$0.381(41)$ &$0.783(54)$& $0.1448(36)$& $0.51(57)$&$0.450(64)$ &$0.36(14)$ &$2.5(2.2)$ &$1.59$\\ 
$a117m310$& $\{4^{N\pi},3^*\}$&$0.326(20)$ &$0.688(61)$& $0.1466(44)$& $0.30(89)$&$0.30(15)$ &$0.42(55)$ &$5.4(4.5)$ &$0.57$\\
$a117m310$& $\{4,2^{free}\}$&$0.64(11)$ && $0.1539(32)$& $17(21)$&$0.68(10)$ & & &$0.81$\\

\hline

$a087m290$& $\{4,2\}$&$0.264(75)$ && $0.096(10)$& $-0.19(53)$&$0.74(17)$ & & &$1.63$\\
$a087m290$& $\{4^{N\pi},2\}$&$0.263(11)$ && $0.0944(32)$& $-0.24(42)$&$0.76(10)$ & & &$1.66$\\
$a087m290$& $\{4,3^*\}$&$0.264(75)$ &$0.556(86)$& $0.1056(53)$& $0.54(51)$&$0.34(13)$ &$0.66(15)$ &$1.1(1.2)$ &$1.66$\\
$a087m290$& $\{4^{N\pi},3^*\}$&$0.263(11)$ &$0.55(15)$& $0.1054(58)$& $0.54(51)$&$0.32(20)$ &$0.67(14)$ &$1.1(1.2)$ &$1.08$\\
$a087m290$& $\{4,2^{free}\}$&$0.633(45)$ && $0.1166(20)$& $29(15)$&$0.804(25)$ & & &$0.86$\\

\hline

$a087m290L$& $\{4,2\}$&$0.303(42)$ && $0.0997(35)$& $-0.30(54)$&$0.644(31)$ & & &$0.80$\\
$a087m290L$& $\{4^{N\pi},2\}$&$0.249(27)$ && $0.0943(36)$& $-0.69(33)$&$0.700(55)$ & & &$1.06$\\
$a087m290L$& $\{4,3^*\}$&$0.303(42)$ &$0.561(83)$& $0.1024(20)$& $0.39(59)$&$0.460(77)$ &$0.53(13)$ &$3.3(3.3)$ &$0.62$\\
$a087m290L$& $\{4^{N\pi},3^*\}$&$0.249(27)$ &$0.440(61)$& $0.1009(20)$& $0.31(42)$&$0.348(85)$ &$0.59(11)$ &$1.4(1.6)$ &$0.64$\\
$a087m290L$& $\{4,2^{free}\}$&$0.466(29)$ && $0.1066(12)$& $12.3(7.0)$&$0.746(25)$ & & &$0.82$\\

\hline

$a087m230$& $\{4,3^*\}$&$0.357(31)$ &$0.77(18)$& $0.0988(22)$& $1.9(2.2)$&$0.55(11)$ &$1.27(66)$ &$-19(29)$ &$0.74$\\
$a087m230$& $\{4^{N\pi},3^*\}$&$0.2071(52)$ &$0.272(69)$& $0.1027(97)$& $0.3(2.7)$&$-0.03(42)$ &$0.72(61)$ &$1.3(4.9)$ &$0.78$\\
$a087m230$& $\{4,2^{ free}\}$&$0.50(14)$ && $0.1021(31)$& $19(43)$&$0.79(22)$ & & &$0.75$\\

\hline 

$a087m230X$& $\{4,3^*\}$&$0.220(32)$ &$0.649(33)$& $0.0849(54)$& $0.15(70)$&$0.64(15)$ &$1.18(40)$ &$-3.5(4.8)$ &$1.15$\\
$a087m230X$& $\{4^{N\pi},3^*\}$&$0.160(18)$ &$0.577(87)$& $0.0858(90)$& $0.65(53)$&$0.37(30)$ &$1.11(12)$ &$0.8(1.0)$ &$1.44$\\
$a087m230X$& $\{4,2^{free}\}$&$0.418(79)$ && $0.0973(40)$& $8(13)$&$0.850(43)$ & & &$1.27$\\

\hline

$a086m180$& $\{4,2\}$&$0.293(45)$ && $0.0780(72)$& $0.59(93)$&$1.05(17)$ & & &$1.17$\\
$a086m180$& $\{4^{N\pi},2\}$&$0.195(19)$ && $0.0517(88)$& $-1.29(78)$&$2.07(54)$ & & &$1.89$\\
$a086m180$& $\{4,3^*\}$&$0.293(45)$ &$0.578(81)$& $0.0834(47)$& $-0.2(1.1)$&$0.80(16)$ &$-0.06(39)$ &$9.9(5.9)$ &$0.87$\\
$a086m180$& $\{4^{N\pi},3^*\}$&$0.195(19)$ &$0.413(66)$& $0.0816(69)$& $1.48(62)$&$0.30(29)$ &$1.10(17)$ &$-0.8(1.1)$ &$1.08$\\
$a086m180$& $\{4,2 ^ {free}\}$&$0.424(61)$ && $0.0904(39)$& $8.5(7.0)$&$0.893(39)$ & & &$0.95$\\

\hline

$a086m180L$& $\{4,3^*\}$&$0.252(30)$ &$0.615(45)$& $0.0829(35)$& $1.61(61)$&$0.52(12)$ &$0.97(30)$ &$1.8(2.7)$ &$1.42$\\
$a086m180L$& $\{4^{N\pi},3^*\}$&$0.141(24)$ &$0.420(77)$& $0.0843(68)$& $1.86(51)$&$-0.003(184)$ &$1.27(30)$ &$-0.2(1.1)$ &$1.37$\\
$a086m180L$& $\{4,2^{free}\}$&$0.294(66)$ && $0.0814(81)$& $1.8(2.2)$&$0.87(12)$ & & &$1.68$\\

\hline

$a068m290$& $\{4,2\}$&$0.229(41)$ && $0.0736(49)$& $-0.65(46)$&$0.774(92)$ & & &$1.01$\\
$a068m290$& $\{4^{N\pi},2\}$&$0.217(24)$ && $0.0738(29)$& $-0.43(38)$&$0.695(68)$ & & &$0.73$\\
$a068m290$& $\{4,3^*\}$&$0.229(41)$ &$0.386(79)$& $0.0786(17)$& $0.38(36)$&$0.41(14)$ &$0.66(12)$ &$1.3(1.4)$ &$0.79$\\
$a068m290$& $\{4^{N\pi},3^*\}$&$0.217(24)$ &$0.363(79)$& $0.0785(18)$& $0.40(37)$&$0.36(15)$ &$0.70(10)$ &$1.0(1.2)$ &$0.80$\\
$a068m290$& $\{4,2^{free}\}$&$0.431(16)$ && $0.08430(71)$& $17.7(5.6)$&$0.868(14)$ & & &$1.11$\\

\hline
$a067m230$& $\{4,3^*\}$&$0.204(40)$ &$0.495(33)$& $0.0675(37)$& $1.25(89)$&$0.492(98)$ &$1.09(20)$ &$-3.5(4.0)$ &$0.99$ \\
$a067m230$& $\{4^{N\pi},3^*\}$&$0.172(19)$ &$0.471(40)$& $0.0651(34)$& $0.96(55)$&$0.47(13)$ &$1.16(19)$ &$-2.1(2.2)$ &$1.00$\\ 
$a067m230$& $\{4,2^{
m free}\}$&$0.297(56)$ && $0.0712(33)$& $4.5(5.7)$&$0.756(42)$ & & &$1.01$\\ 


\hline

$a067m175$& $\{4,3^*\}$&$0.212(45)$ &$0.444(26)$& $0.0659(34)$& $0.24(98)$&$0.49(13)$ &$0.92(22)$ &$-0.8(4.5)$ &$1.17$ \\
$a067m175$& $\{4^{N\pi},3^*\}$&$0.141(11)$ &$0.383(42)$& $0.0623(50)$& $0.43(65)$&$0.36(22)$ &$1.09(24)$ &$-0.4(1.9)$ &$1.21$ \\
$a067m175$& $\{4,2^{free}\}$&$0.435(50)$ && $0.0725(18)$& $28(33)$&$0.985(59)$ & & &$1.28$ \\

\hline


$a067m135$& $\{4,3^*\}$&$0.141(17)$ &$0.435(22)$& $0.0586(57)$& $0.16(61)$&$0.50(25)$ &$0.88(28)$ &$2.4(1.6)$ &$1.00$\\
$a067m135$& $\{4^{N\pi},3^*\}$&$0.1254(77)$ &$0.426(23)$& $0.0587(53)$& $0.70(42)$&$0.38(21)$ &$1.23(15)$ &$0.73(83)$ &$0.96$\\
$a067m135$& $\{4,2^{^Mm free}\}$&$0.455(44)$ && $0.0733(20)$& $57(39)$&$1.087(54)$ & & &$1.01$\\

\hline

$a053m295$& $\{4,3^*\}$&$0.284(17)$ &$0.67(16)$& $0.0599(11)$& $6(12)$&$0.87(11)$ &$-1.4(1.9)$ &$-125(440)$ &$1.15$\\
$a053m295$& $\{4^{N\pi},3^*\}$&$0.224(24)$ &$0.313(65)$& $0.0586(14)$& $3.1(4.4)$&$0.58(13)$ &$0.59(23)$ &$-20(23)$ &$1.19$\\
$a053m295$& $\{4,2^{free}\}$&$0.300(37)$ && $0.0609(13)$& $0.03(9.87)$&$0.828(53)$ & & &$1.22$\\

\hline


$a053m230$& $\{4,3^*\}$&$0.208(70)$ &$0.341(44)$& $0.0541(30)$& $1.6(1.8)$&$0.51(18)$ &$0.89(27)$ &$-4.0(6.1)$ &$1.05$\\
$a053m230$& $\{4^{N\pi},3^*\}$&$0.129(21)$ &$0.313(31)$& $0.0501(36)$& $0.69(61)$&$0.38(18)$ &$1.12(23)$ &$-1.2(1.4)$ &$1.07$\\
$a053m230$& $\{4,2^{free}\}$&$0.326(38)$ && $0.0575(15)$& $10.0(9.6)$&$0.915(39)$ & & &$1.21$\\

\hline

\end{tabular}
\caption{Comparison of results of the fits to remove the excited-state contamination 
for the transversity moment $\langle x \rangle_{\delta u -
\delta d}$. The rest is the same as in Table~\protect\ref{tab:5strategy-fits-momfrac}. 
}
\label{tab:5strategy-fits-transvmom}
\end{table*}

\onecolumngrid

\begin{table*}[b]   
\centering
\setlength{\tabcolsep}{2.25pt}
\renewcommand{\arraystretch}{1.0}
\begin{tabular}{|l|ccc|ccc|ccc| }
 \hline

Ensemble  &  \multicolumn{3}{c|}{$\la x\ra_{u-d}$}  & \multicolumn{3}{c|}{$\la x \ra_{\Delta u-\Delta d}$}  & \multicolumn{3}{c|}{ $\la x \ra_{\delta u-\delta d}$}  \\
 & {$\{4^{N\pi},3^*\}$} & {$\{4,3^*\}$} &{$\{4,2^{\rm free}\}$}  & {$\{4^{N\pi},3^*\}$} & {$\{4,3^*\}$} &{$\{4,2^{\rm free}\}$}  & {$\{4^{N\pi},3^*\}$} & {$\{4,3^*\}$} &{$\{4,2^{\rm free}\}$}  \\

\hline
\hline
$a117m310$  & $0.181(4)$ & $0.190(6)$ & $0.186(7)$ & $0.234(5)$ & $0.233(3)$ & $0.243(2)$ & $0.238(7)$ & $0.234(5)$ & $0.249(5)$   \\
\hline
$a087m290$  & $0.190(7)$ & $0.190(8)$ & $0.203(5)$ & $0.222(9)$ & $0.222(9)$ & $0.240(5)$ & $0.222(10)$ & $0.223(11)$ & $0.243(6)$ \\
\hline
$a087m290L$ & $0.173(4)$ &  $0.179(5)$ & $0.182(3)$ & $0.208(3)$ & $0.211(4)$ & $0.217(4)$ & $0.217(4)$ & $0.219(4)$ & $0.228(2)$  \\
\hline
$a087m230$  & 0.171(10)  & 0.181(4)    & 0.181(7)   & 0.211(8)   & 0.210(4)   & 0.202(10)  & 0.230(21)  & 0.221(5)   & 0.228(7)    \\
\hline
$a087m230X$ & 0.152(18)  & 0.167(10)   & 0.181(7)   & 0.187(13)  & 0.197(9)   & 0.221(6)   & 0.200(20)  & 0.196(12)  & 0.225(10)    \\
\hline
$a086m180$  & $0.148(14)$ & $0.167(9)$ &  $0.169(12)$ & $0.195(13)$ &  $0.204(7)$ &  $0.216(7)$ & $0.183(18)$ &  $0.200(11)$& $0.217(9)$\\
\hline
$a086m180L$ & $0.150(25)$ & $0.162(8)$ &  $0.169(12)$ & $0.176(23)$ &  $0.189(7)$ &  $0.206(8)$ & $0.208(16)$ &  $0.200(8)$& $0.197(19)$\\
\hline
$a068m290$  & $0.168(4)$  &  $0.170(6)$ &  $0.180(3)$  & $0.210(4)$  &  $0.210(3)$ &  $0.222(2)$ & $0.211(5)$  &  $0.211(4)$ &  $0.227(2)$\\
\hline
$a067m230$  & 0.155(8)   & 0.163(9)    & 0.174(7)   & 0.191(7)   & 0.196(7)   & 0.209(5)   & 0.189(9)   & 0.195(10)  & 0.205(10)       \\
\hline
$a067m175$  & $0.154(10)$ &  $0.157(10)$ &  $0.155(19)$ & $0.190(12)$&  $0.193(7)$ &  $0.194(16)$ & $0.192(15)$& $0.201(9)$ &  $0.222(7)$\\
\hline
$a067m135$  & 0.157(10)  & 0.158(9)    & 0.180(10)  & 0.189(13)  & 0.164(13)  & 0.207(12)  & 0.189(19)  & 0.188(21)  & 0.234(7)      \\
\hline
$a053m295$  & 0.166(3)   & 0.171(2)    & 0.173(3)   & 0.200(4)   & 0.204(8)   & 0.210(21)  & 0.200(5)   & 0.204(3)   & 0.207(4)      \\
\hline
$a053m230$  & 0.155(9)   & 0.163(12)   & 0.162(8)   & 0.185(9)   & 0.194(7)   & 0.207(4)   & 0.184(13)  & 0.197(10)  & 0.209(6)      \\
\hline
\end{tabular}
\caption{Values of the unrenormalized moments for the thirteen 
ensembles and the three fit
  strategies, $\{4^{N\pi},3^\ast \}$, $\{4,3^\ast \}$ and $\{4,2^{\rm
  free} \}$, used to remove the ESC in the three-point functions. The values of 
  $\tau$ used in the fits are given in Table~\ref{tab:ensembles}.  }
\label{tab:ESC-fits}
\end{table*}
\twocolumngrid

\clearpage

\section{Renormalization of the Operators}
\label{sec:renormalization}

\begin{figure*}[tbhp]  
\begin{subfigure}
\centering
\includegraphics[angle=0,width=0.38\textwidth]{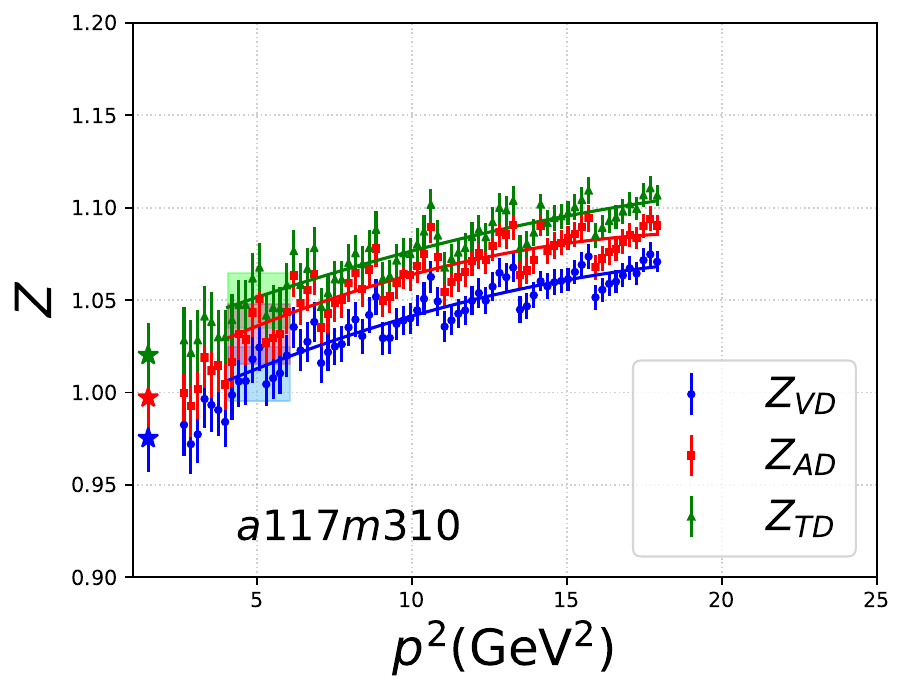}
\end{subfigure} 
\begin{subfigure}
\centering
\includegraphics[angle=0,width=0.38\textwidth]{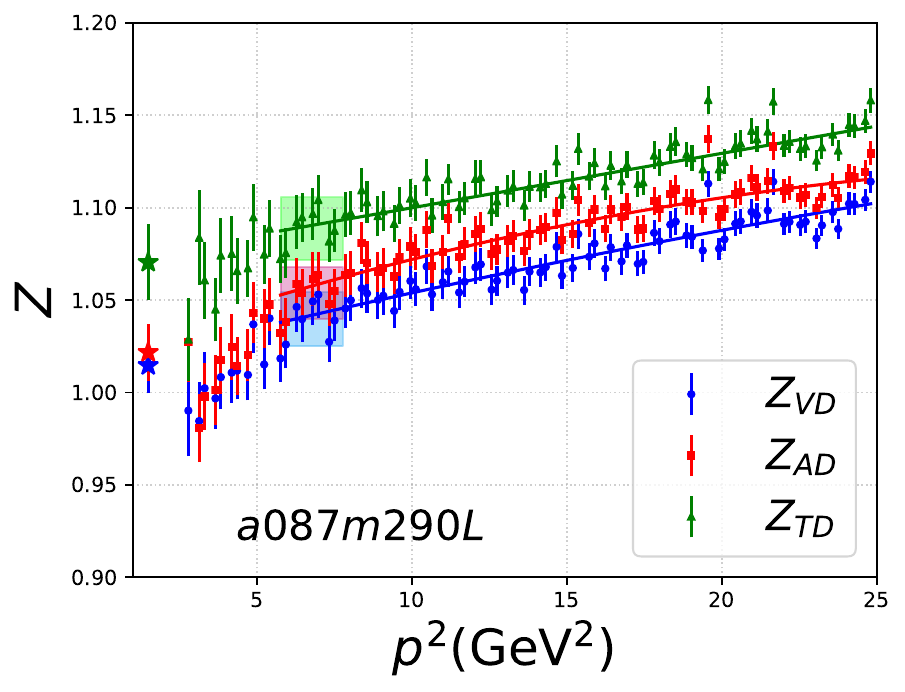}
\end{subfigure}
\begin{subfigure}
\centering
\includegraphics[angle=0,width=0.38\textwidth]{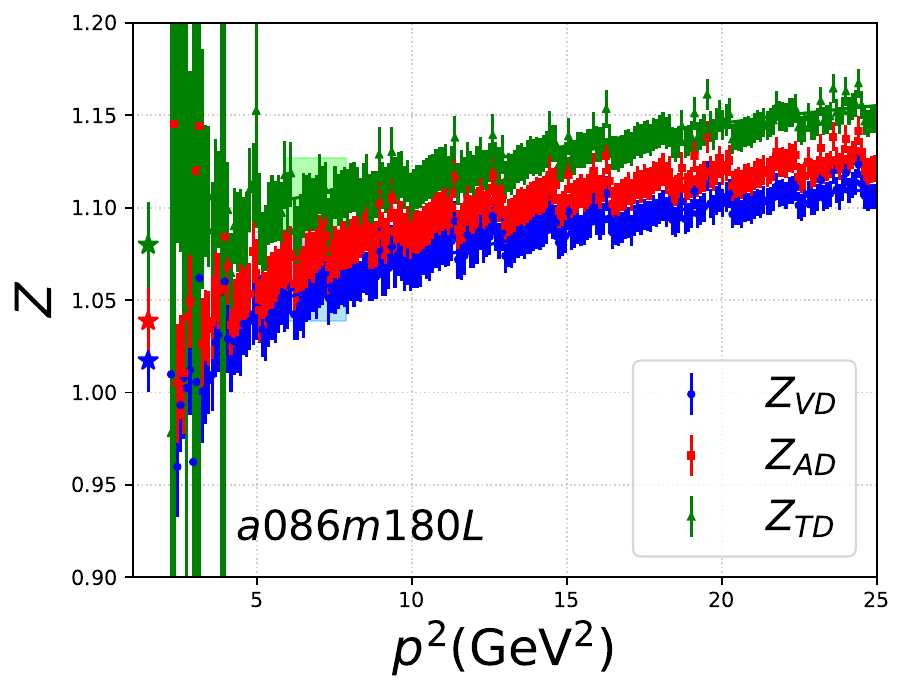}
\end{subfigure}
\begin{subfigure}
\centering
\includegraphics[angle=0,width=0.38\textwidth]{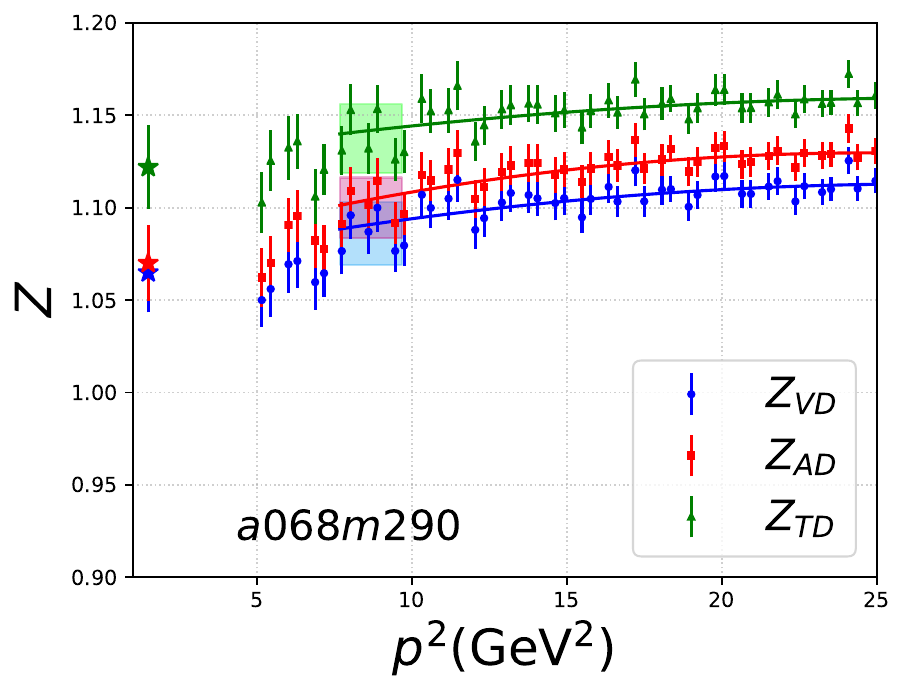}
\end{subfigure}
\begin{subfigure}
\centering
\includegraphics[angle=0,width=0.38\textwidth]{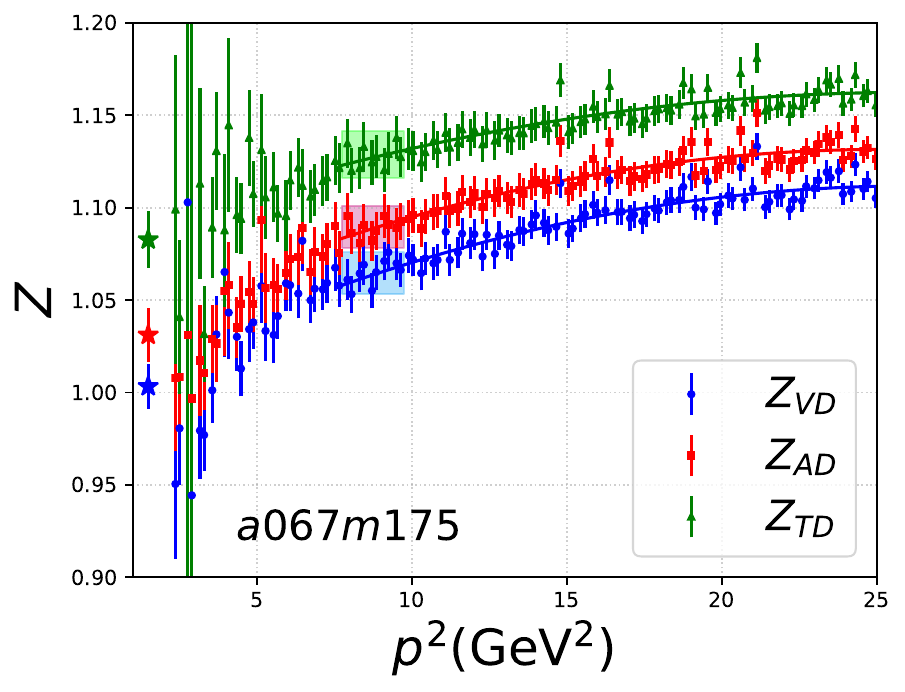}
\end{subfigure}
\begin{subfigure}
\centering
\includegraphics[angle=0,width=0.38\textwidth]{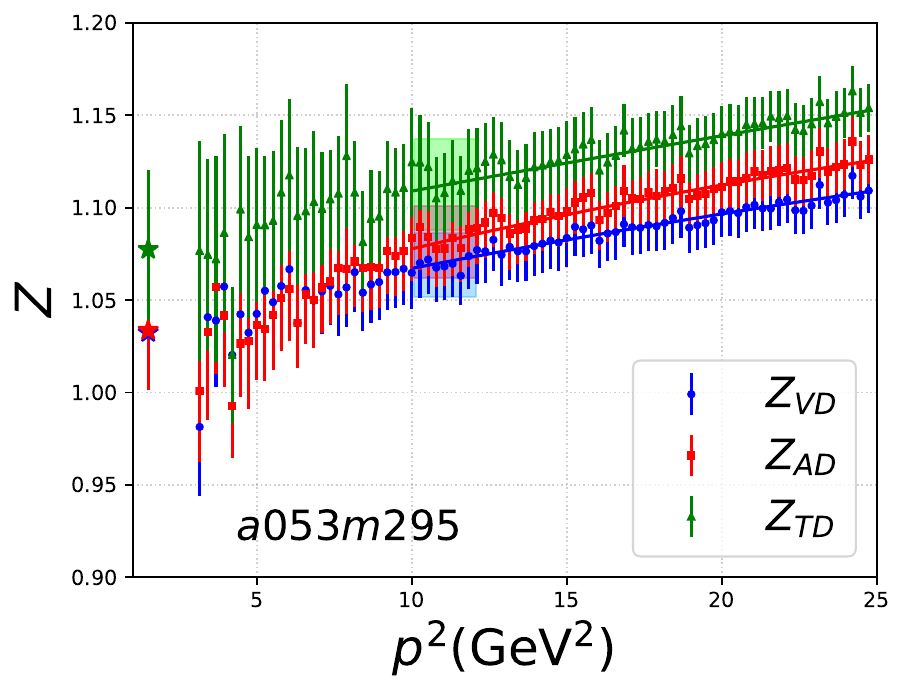}
\end{subfigure}
\caption{Nonperturbative renormalization factors for ${ \la
    x\ra_{u-d}}$, ($Z_{VD}$), ${ \la x \ra_{\Delta u - \Delta d}}$,
  ($Z_{AD}$), and ${ \la x\ra_{\delta u - \delta d}}$, ($Z_{TD}$) in
  the ${\rm \ol{MS}}$ scheme at $\mu=2\ {\rm GeV}$ for the six
  ensembles.  In all cases, the extent along the x-axis of shaded
  bands marks the region in ${p^2}$ that is averaged in Method A and
  the height is twice the error in the estimate. The points next to the
  y-axis with the star symbol give a second estimate (Method B)
  obtained from a fit with ansatz $Z(p)=Z_0+a p^2+b p^4$ to data
  starting at the left edge of the shaded band.}
\label{fig:Z-7ensembles}
\end{figure*}

In this appendix, we describe two methods used to 
calculate the renormalization
factors, $Z_{VD,AD,TD}$, for the three one-derivative operators specified in 
Eqs.~\eqref{eq:finaloperatorV},~\eqref{eq:finaloperatorA}
and~\eqref{eq:finaloperatorT}.  These $Z$'s
are first determined nonperturbatively on the lattice in the intermediate 
\ripmom\ scheme~\cite{Gockeler:2010yr,Constantinou:2013ada} as a
function of the lattice scale $p^2 = p^\mu p^\mu$, and then converted
to ${\rm \ol{MS}}$ scheme using $3$-loop perturbative factors
calculated in the continuum in Ref.~\cite{Gracey:2003mr}. For data at
each $p$, we perform horizontal matching by choosing the
${\rm \ol{MS}}$ scale $\mu=|p|$. These numbers are then run in the
continuum ${\rm \ol{MS}}$ scheme from scale $\mu$ to $2$~GeV using
three-loop anomalous dimensions~\cite{Gracey:2003mr}.  The two methods
differ in how the observed dependence of $Z^{\rm \ol{MS}}$(2GeV) on $p^2a^2$, a
lattice artifact, is removed. For details of the three operators and
their decomposition into irreducible representations, and for
alternate methods for the subtraction of discretization effects using
perturbation theory, we refer the reader to
Refs.~\cite{Gockeler:1995wg,Constantinou:2013ada,Constantinou:2014fka,Harris:2019bih}.

Two important notes.  While the following two methods are the same as
used in our previous works,
Refs.~\cite{Mondal:2020ela,Mondal:2020cmt}, the $Z_{VD,AD,TD}$
presented here in Table~\ref{tab:Z-fac} are different from those in
Ref.~\cite{Mondal:2020ela} because we have updated the values of $p^2$
used. Second, we have calculated the $Z$-factors on a subset of the
ensembles as they are found to vary less than the statistical errors
versus $M_\pi^2$ for fxed $\beta$. The interpolation/extrapolations
used to estimate the rest are given in the caption of
Fig.~\ref{tab:Z-fac}.

The data for the renormalization factors $Z_{VD, AD, TD}$ in the
${\rm \ol{MS}}$ scheme at $\mu = 2$ GeV are shown in
Fig.~\ref{fig:Z-7ensembles} for six ensembles as a function of
$p^2$---the lattice scale at which the data are collected in
the \ripmom\ scheme. These data do not show a window in $p^2$ where
the results are independent of $p^2$---indicative of lattice artifacts
including those due to the reduction of rotational invariance to
$90^o$. Many methods have been proposed to control it, see for example
Refs.~\cite{Harris:2019bih,Alexandrou:2020sml,Bhattacharya:2016zcn}.
We use the following two:
\begin{itemize}
\item
In method A, we take an average over the data points in an interval of
$2~ {\rm GeV^2}$ about ${\hat p}^2 = \Lambda/a$, where the scale
$\Lambda = 3$ GeV is chosen to be large enough to avoid
nonperturbative effects and, at the same time, 
at which perturbation theory is expected
to be reasonably well behaved. Also, this choice satisfies both 
${\hat p}a \rightarrow 0$ and $\Lambda/{\hat p} \rightarrow 0$ in the
continuum limit as desired.  The shaded bands in
Figs.~\ref{fig:Z-7ensembles} show the window over which the data are
averaged (also specified in column three of Table~\ref{tab:Z-fac}) and
the error (half the height of the band).  This method was used in our
previous work, PNDME 20, presented in Ref.~\cite{Mondal:2020cmt}.
\item
In method B, we make a fit to the data using the ansatz $Z(p)=Z_0+a
\sum_\mu p_\mu p_\mu + b (\sum_\mu p_\mu p_\mu)^2$ to remove the $p^2$ 
dependent artifacts. The fit is made with $p^2$ starting at the lower
edge of the band used in method A, which is given in column three of
Table~\ref{tab:Z-fac}, and by which a roughly linear in $p^2$ behavior
is manifest. The results of Method B are shown next to the y-axis in
Fig.~\ref{fig:Z-7ensembles} using the star symbol and 
can be compared to those from Method A.
\end{itemize}

Estimates of $Z_{VD}$, $Z_{AD}$ and $Z_{TD}$, summarized in
Table~\ref{tab:Z-fac}, are used to renormalize the bare moments
given in Table~\ref{tab:ESC-fits}. The renormalized values,  
given in Tables~\ref{tab:renormalized-moments-A}
and~\ref{tab:renormalized-moments-B} for the two methods, are 
different as the discretization errors are expected to be
different in the two methods. We, therefore, do not
average the values of the renormalization constants but perform the
full analysis, including the CCFV fits, separately for the two methods and
compare the moments only after this extrapolation.
These final results, after CCFV fits to the data given in
Tables~\ref{tab:renormalized-moments-A}
and~\ref{tab:renormalized-moments-B}, are summarized in
Table~\ref{tab:CCFVresults} and found to be consistent.

\begin{table*}[htbp]      
\centering
\setlength{\tabcolsep}{2pt}
\renewcommand{\arraystretch}{1.3}
\begin{tabular}{|c|c|c|c|c|c|c|c|c| }
\hline
        & $N_{\rm conf}$ & fit range     
        & \multicolumn{2}{|c|}{$Z_{VD} \equiv Z_{\langle x \rangle_{u -d}}$} 
        & \multicolumn{2}{|c|}{$Z_{AD} \equiv Z_{\langle x \rangle_{\Delta u- \Delta d}}$} 
        & \multicolumn{2}{|c|}{$Z_{TD} \equiv Z_{\langle x \rangle_{\delta u- \delta d}}$}  \\
Ensemble    &        &   GeV${}^2$   & Method A  & Method B   & Method A   &  Method B & Method A  & Method B \\
\hline                                                                                
\hline                                                                                
$a117m310$  & 100 & $4.06-6.06$  & 1.010(15) & 0.975(18) & 1.032(16) & 0.997(19) & 1.050(15) & 1.020(17)  \\
\hline
$a087m290L$ & 100 & $5.77-7.77$  & 1.040(15) & 1.015(15) & 1.054(14) & 1.022(15) & 1.089(17) & 1.070(21)  \\
\hline
$a087m290$  & --- & ---  & 1.040(15) & 1.015(15) & 1.054(14) & 1.022(15) & 1.089(17) & 1.070(21)  \\
\hline
$a087m230$  & --- & ---        & 1.047(15) & 1.016(17) & 1.064(15) & 1.031(18) & 1.098(20) & 1.075(23)  \\
\hline
$a087m230X$ & --- & ---        & 1.047(15) & 1.016(17) & 1.064(15) & 1.031(18) & 1.098(20) & 1.075(23)  \\
\hline
$a086m180L$ & 108 & $5.89-7.89$  & 1.053(15) & 1.017(17) & 1.073(15) & 1.039(18) & 1.107(20) & 1.080(23)  \\
\hline
$a086m180$  & --- & ---  & 1.053(15) & 1.017(17) & 1.073(15) & 1.039(18) & 1.107(20) & 1.080(23)  \\
\hline
$a068m290$  & 100 & $7.68-9.68$  & 1.086(17) & 1.065(22) & 1.100(16) & 1.070(20) & 1.137(19) & 1.122(23) \\
\hline
$a067m230$  & --- & ---        & 1.076(17) & 1.034(22) & 1.095(16) & 1.050(20) & 1.133(19) & 1.103(23)  \\
\hline
$a067m175$  & 112 & $7.74-9.74$  & 1.065(11) & 1.003(12) & 1.090(11) & 1.031(14) & 1.129(13) & 1.083(15)  \\
\hline
$a067m135$  & --- & ---        & 1.065(11) & 1.003(12) & 1.090(11) & 1.031(14) & 1.129(13) & 1.083(15)  \\
\hline
$a053m295$  & 103 & $10.08-12.08$ & 1.069(17) & 1.032(28) & 1.082(19) & 1.034(32) & 1.113(25) & 1.077(43)  \\
\hline
$a053m230$  & --- & ---        & 1.069(17) & 1.032(28) & 1.082(19) & 1.034(32) & 1.113(25) & 1.077(43)  \\
\hline
\end{tabular}
\caption{Results for the renormalization factors, $Z_{VD,AD,TD}$, in
  the $\MSbar$ scheme at $2$~GeV.  These are calculated in the RI'-MOM
  scheme as a function of scale $p=\sqrt{p_\mu p_\mu}$ on the lattice,
  matched to the $\MSbar$ scheme at the same scale $\mu=p$, and then
  run in the continuum $\MSbar$ scheme from $\mu$ to $2$~GeV. Results
  are given for two methods used to remove the $p^2$ dependent
  artifacts as described in the text.  In method A, the $Z$'s are
  obtained by averaging the data shown in
  Fig.~\protect\ref{fig:Z-7ensembles} over the range of $p^2$
  specified in the third column. Results using method B are obtained
  by fitting the data using the ansatz $Z(p)=Z_0+a p^2+b p^4$ and
  starting at the lower value of $p^2$ given in column 3.  Noting the small 
  dependence of the $Z$'s on $M_\pi$ at fixed $\beta$, we have used the following
  interpolations for the following ensembles with entry --- for $N_{\rm conf}$:   
 $a087m230 =  (a087m290L+a086m180L)/2$; 
  $a087m230X =  (a087m290L+a086m180L)/2$; 
  $a067m230 =  (a068m290+a067m175)/2$;  
 $a067m135 = a067m175$ and
  $a053m230 = a053m295$.}
\label{tab:Z-fac}
\end{table*}

\cleardoublepage
\section{Chiral, Continuum and Infinite Volume (CCFV) Extrapolation}
\label{sec:A_CCFV}

This appendix summarizes results of the CCFV and CC extrapolations 
to the physical point defined by 
$M_\pi^{\rm phy} = 135$~MeV and $M_K^{\rm phy} = 494$~MeV~\cite{Yoo:2026uul}, 
and using $\sqrt{t_0}=0.14474(57)$~fm taken from the 2024 FLAG report~\cite{FlavourLatticeAveragingGroupFLAG:2024oxs}. 

The CCFV fits use the five parameter ansatz defined in Eq.~\eqref{eq:CCFV} 
and all thirteen data points. In the CC fits, the finite volume 
parameter, $c_4$, is set to zero  and the $a087m290$ and $a086m180$ points are 
not included. Results are presented in Tables~\ref{tab:CCFV_13pt_fit} 
and~\ref{tab:CC_11pt_fits} for the three strategies for controlling ESC,  
the three options for the discretization ansatz  labeled as $c_{3_1} \sim a$, 
$c_{3_2}  \sim \alpha_s a $, and $c_{3_3} \sim a^2$, and the two methods, A and B, for 
the determination of the renormalization constants.  The parameters used in each fit 
are given in the column under the heading "Ansatz". When data are averaged, the 
model averaging is done using the AIC score corrected for finite sample 
size: $AIC_c = \chi^2 + 2k + \frac{2k(k+1)}{n-k-1}$ where $n,k$ are the 
number of data-points and parameters respectively.

\begin{figure*}[t]    
\centering
\begin{subfigure}
\centering
\includegraphics[trim=0 10 0 10,angle=0,width=1.0\textwidth]{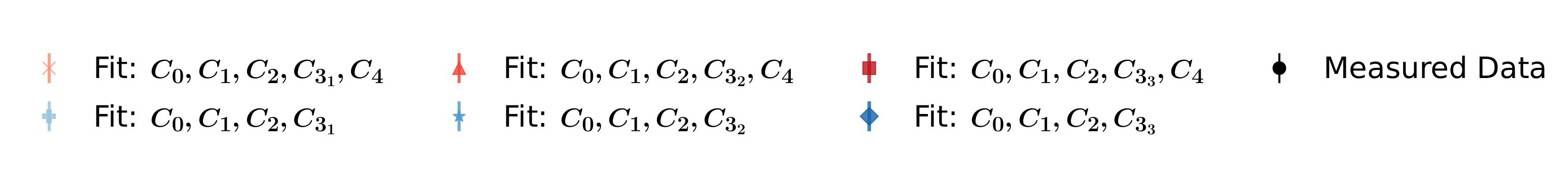}
\end{subfigure}
\vskip -0.9 cm

\begin{subfigure}
\centering
\includegraphics[trim=0 90 0 10,angle=0,width=0.48\textwidth]{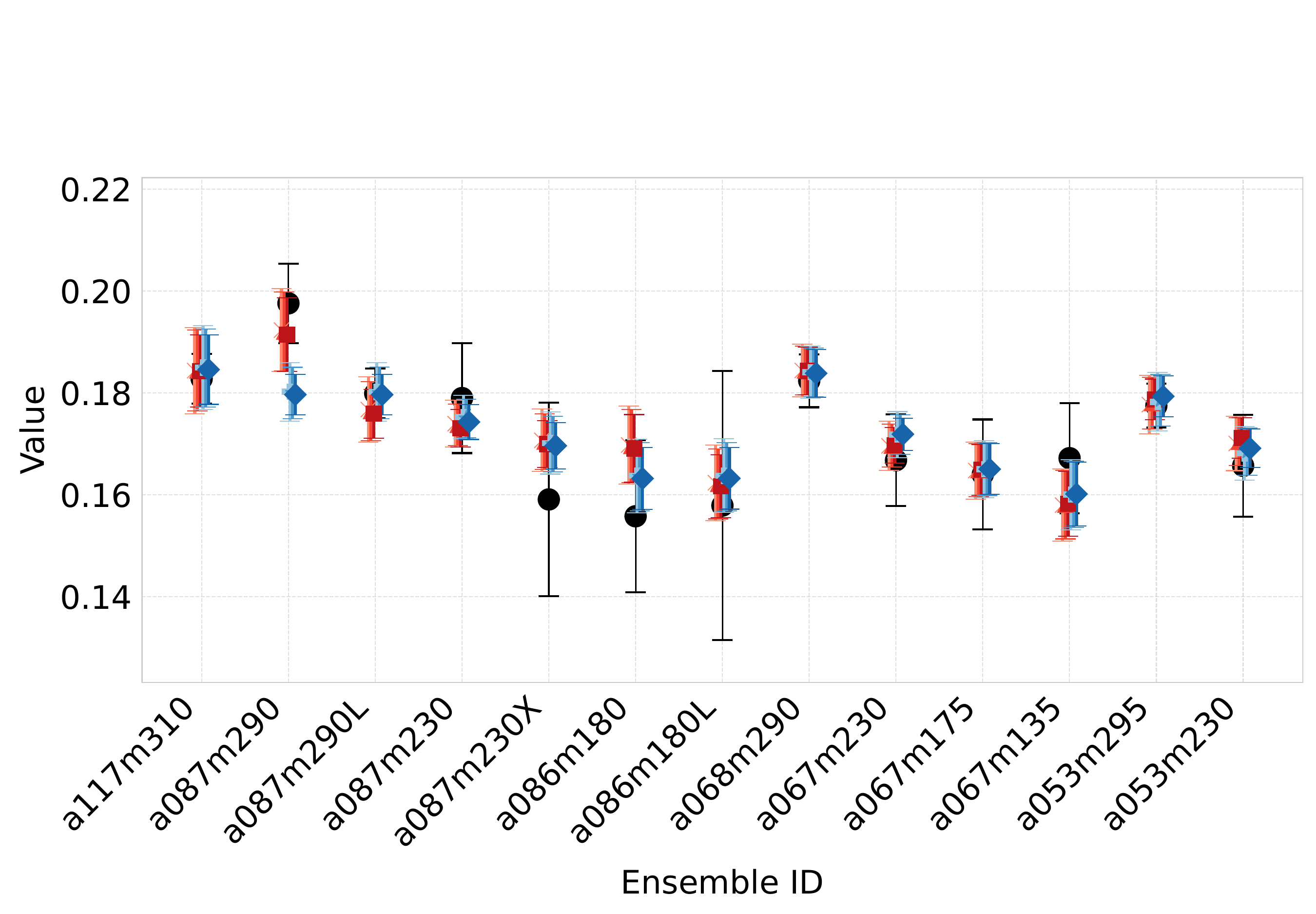}%
\hfill
\includegraphics[trim=0 90 0 10,angle=0,width=0.48\textwidth]{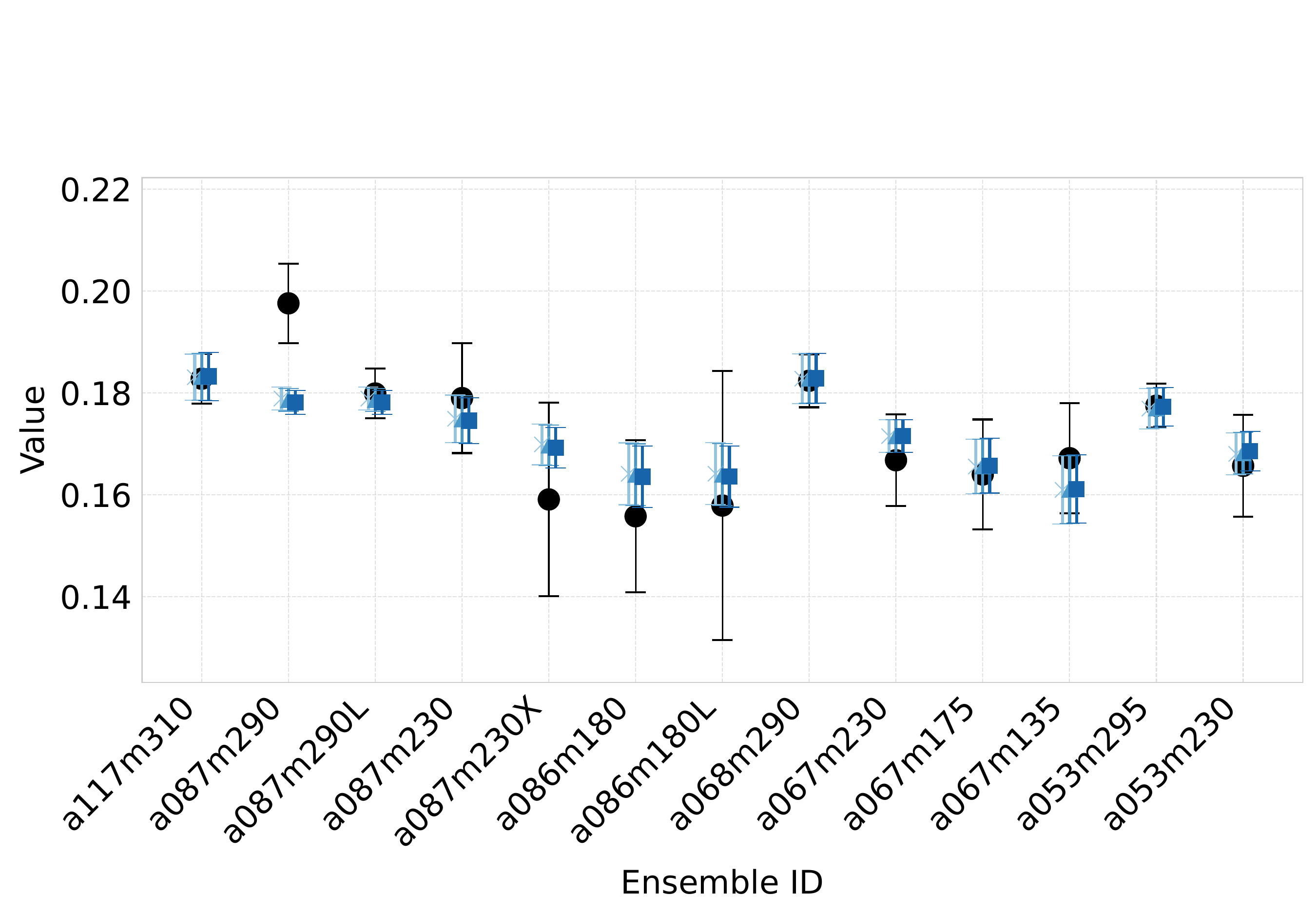}
\end{subfigure}


\begin{subfigure}
\centering
\includegraphics[trim=0 90 0 15,angle=0,width=0.48\textwidth]{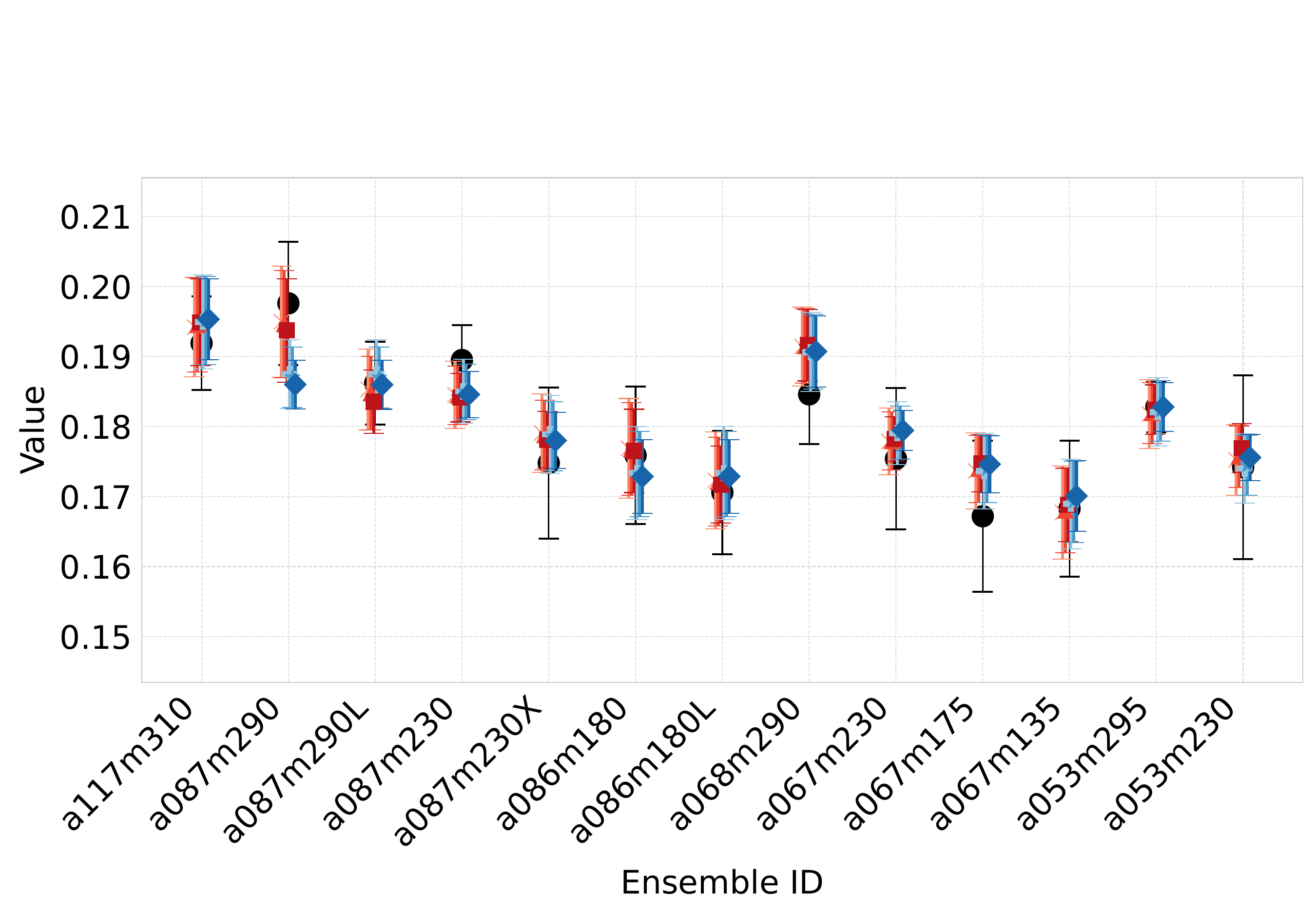}%
\hfill
\includegraphics[trim=0 90 0 15,angle=0,width=0.48\textwidth]{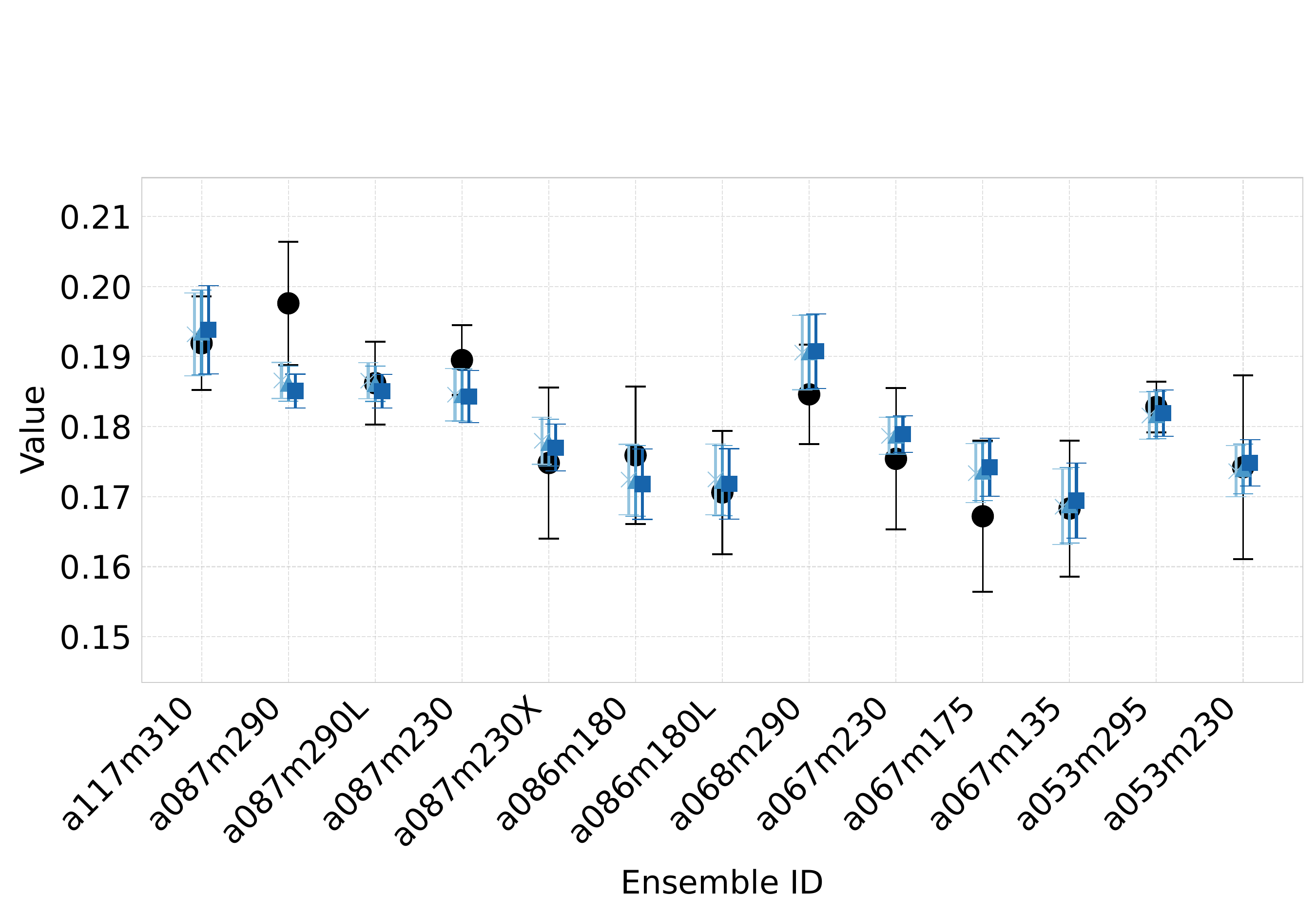}
\end{subfigure}


\begin{subfigure}
\centering
\includegraphics[trim=0 50 0 15,angle=0,width=0.48\textwidth]{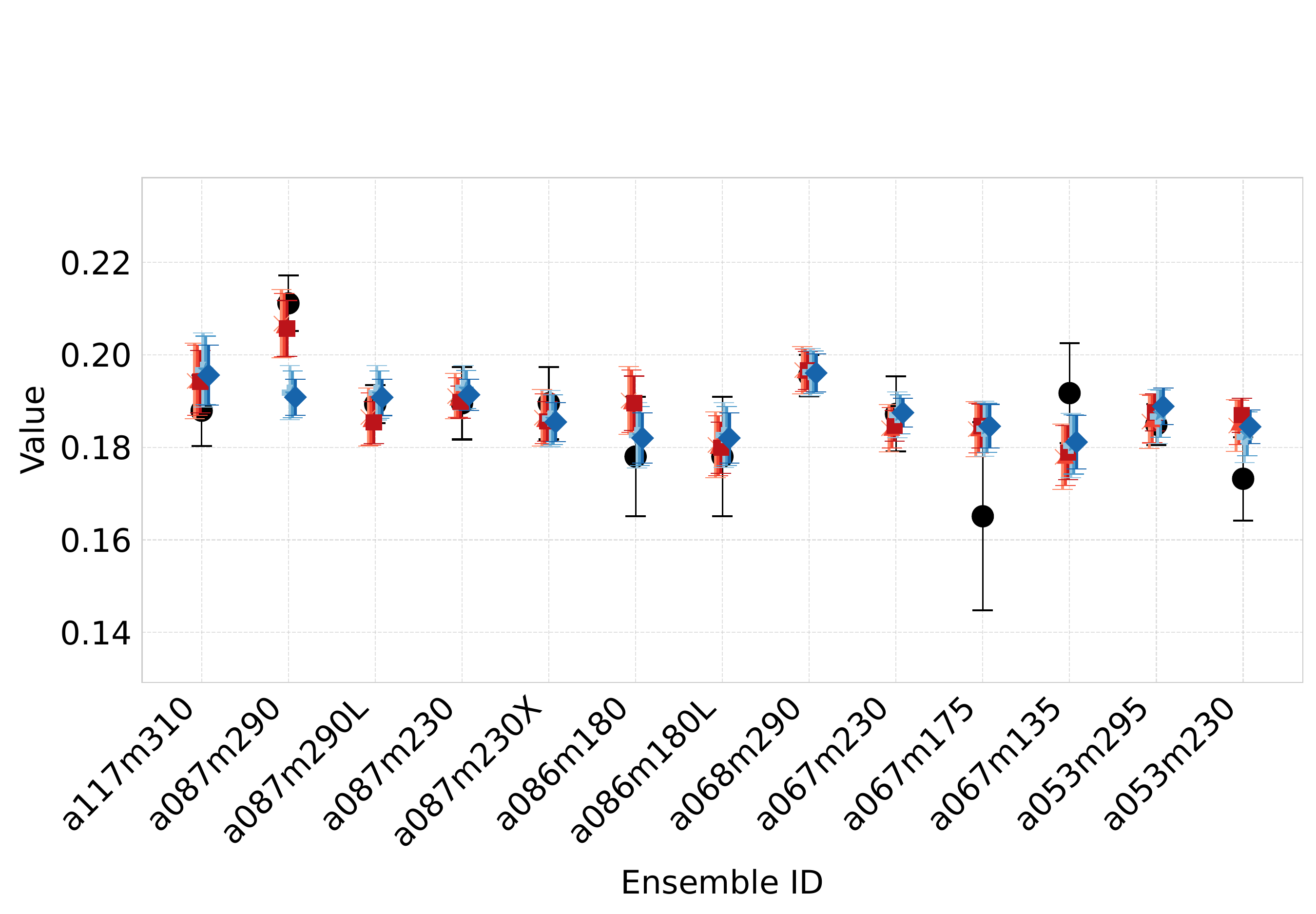}%
\hfill
\includegraphics[trim=0 50 0 15,angle=0,width=0.48\textwidth]{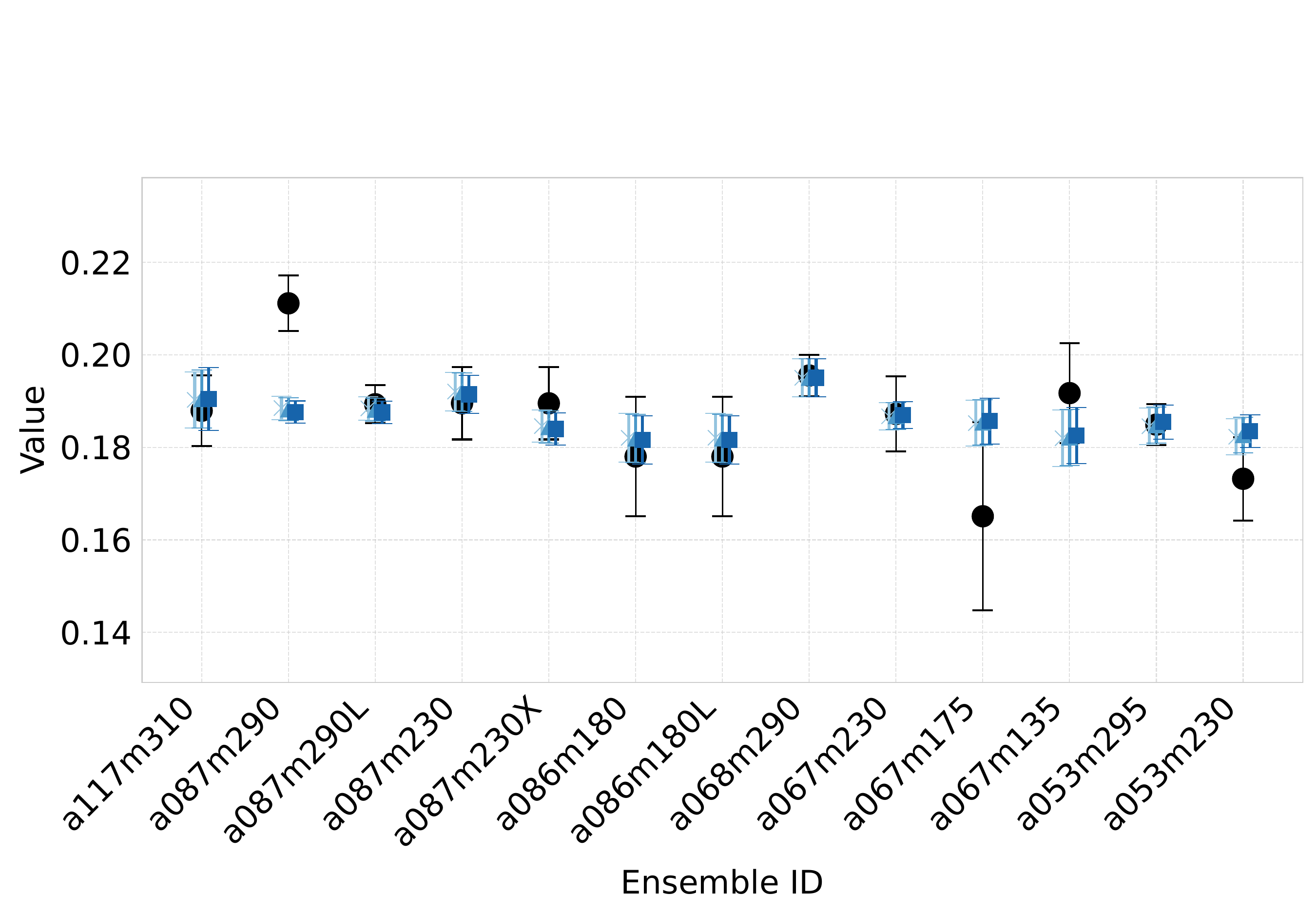}
\end{subfigure}
\vspace{0.4cm}
\caption{Comparison of the results of the CCFV fits to 13 ensembles (red symbols) 
and 11 larger volume ensembles (blue symbols) in the left panels with the CC fits to 
the 11 ensembles (blue sysbols) in the right panels for 
the momentum fraction $\langle x \rangle_{u-d}$ data (black circles). Fits with the 
three discretization ans\"atze are shown with different symbols as specified 
in the very top panel. 
The three rows give the fits to the data with the $\{4^{N\pi},3^\ast\}$ (top row), 
$\{4,3^\ast\}$ (middle row) and $\{4,2^{\rm free}\}$ (bottom row) strategies for 
handling ESC. The best fit is 
to the data obtained with the $\{4,3^\ast\}$ ansatz. }
\label{fig:momfrac_comp}
\end{figure*}

\begin{figure*}[t]     
\centering
\begin{subfigure}
\centering
\includegraphics[trim=0 10 0 10,trim=0 10 0 10,angle=0,width=1.0\textwidth]{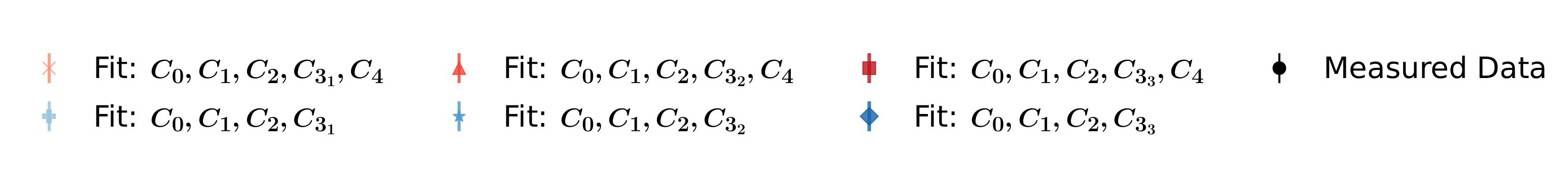}
\end{subfigure}
\vskip -0.9 cm
\begin{subfigure}
\centering
\includegraphics[trim=0 50 0 10,angle=0,width=0.48\textwidth]{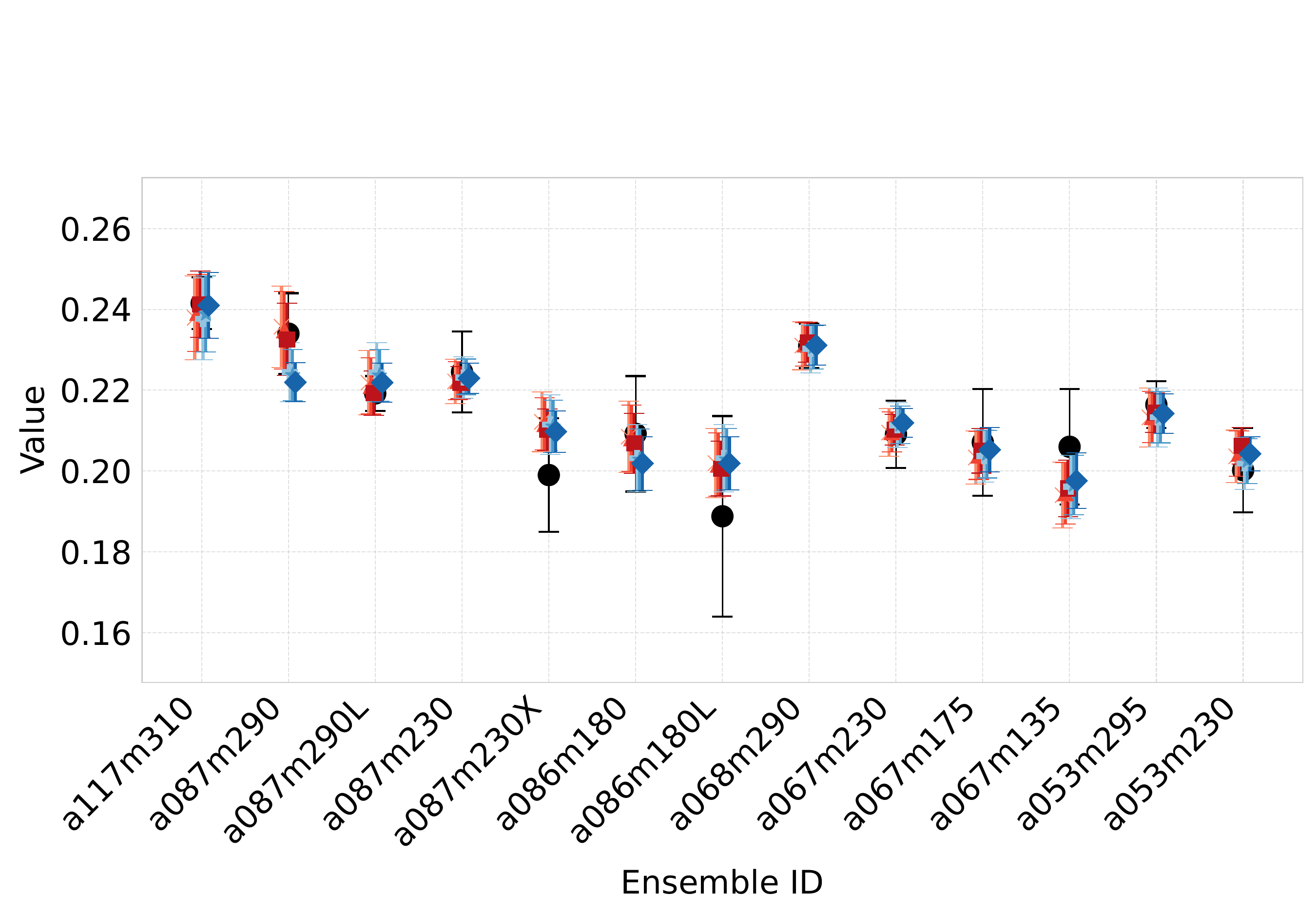}%
\hfill
\includegraphics[trim=0 50 0 10,angle=0,width=0.48\textwidth]{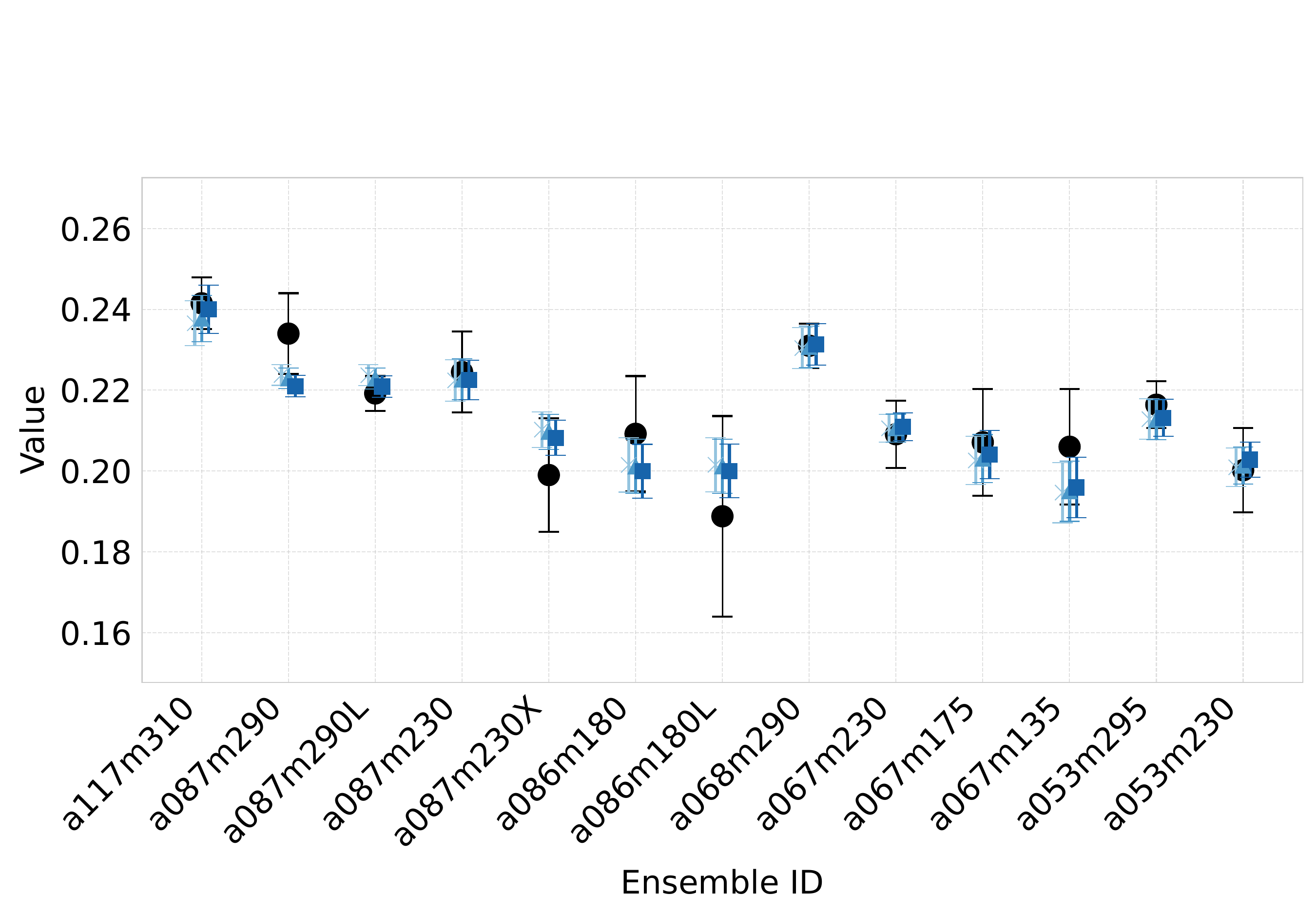}
\end{subfigure}


\begin{subfigure}
\centering
\includegraphics[trim=0 90 0 30,angle=0,width=0.48\textwidth]{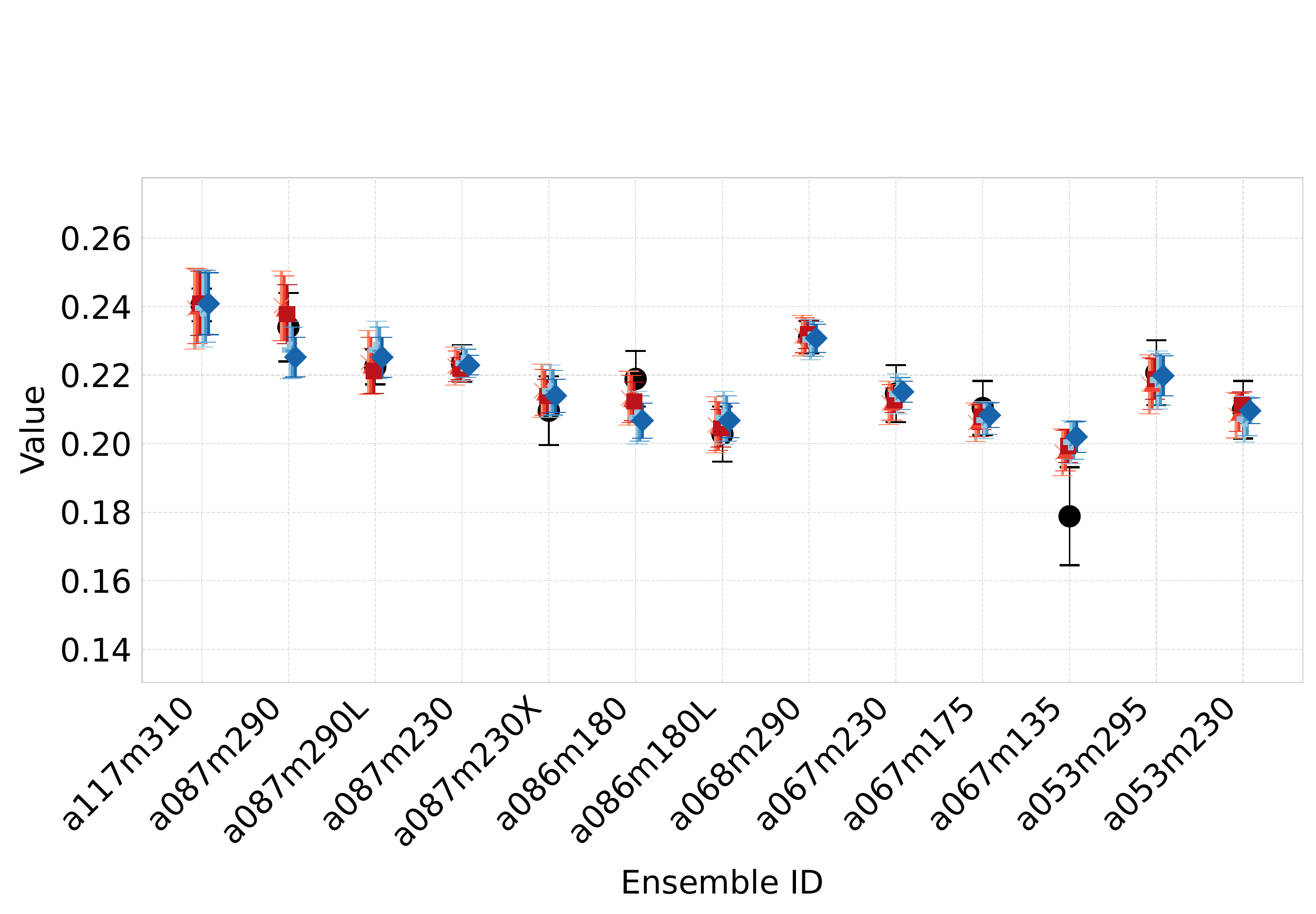}%
\hfill
\includegraphics[trim=0 90 0 30,angle=0,width=0.48\textwidth]{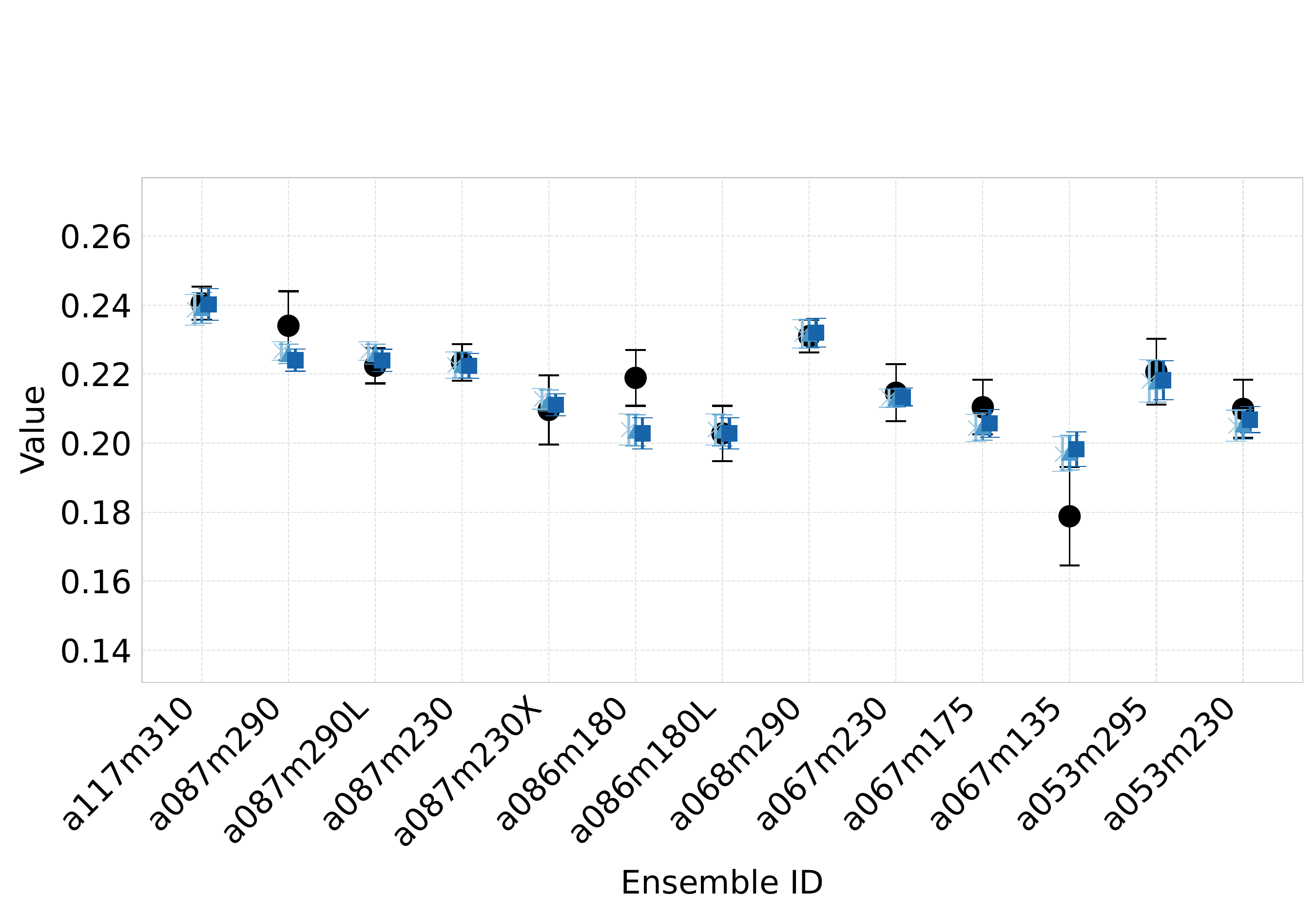}
\end{subfigure}


\begin{subfigure}
\centering
\includegraphics[trim=0 90 0 10,angle=0,width=0.48\textwidth]{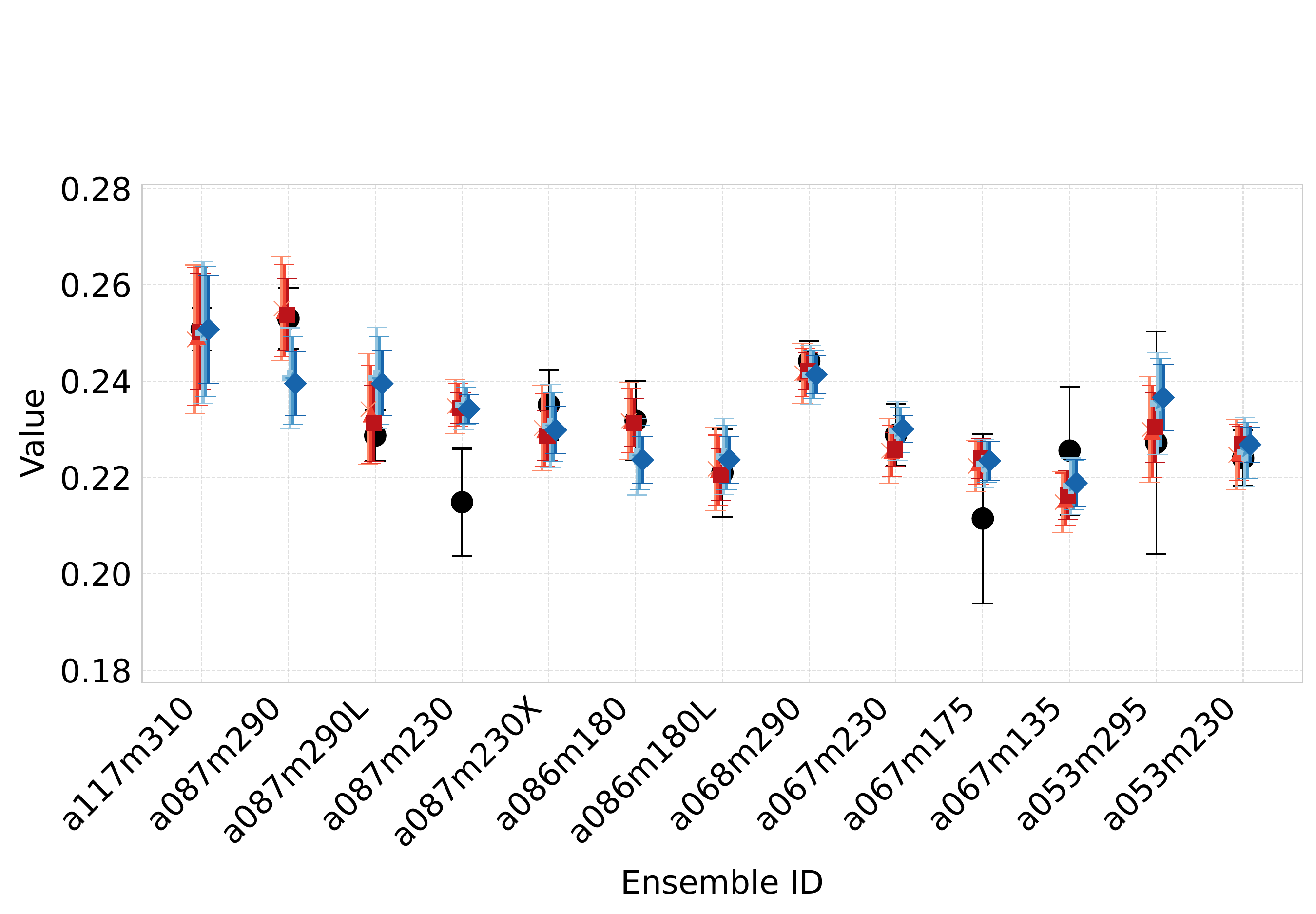}%
\hfill
\includegraphics[trim=0 90 0 10,angle=0,width=0.48\textwidth]{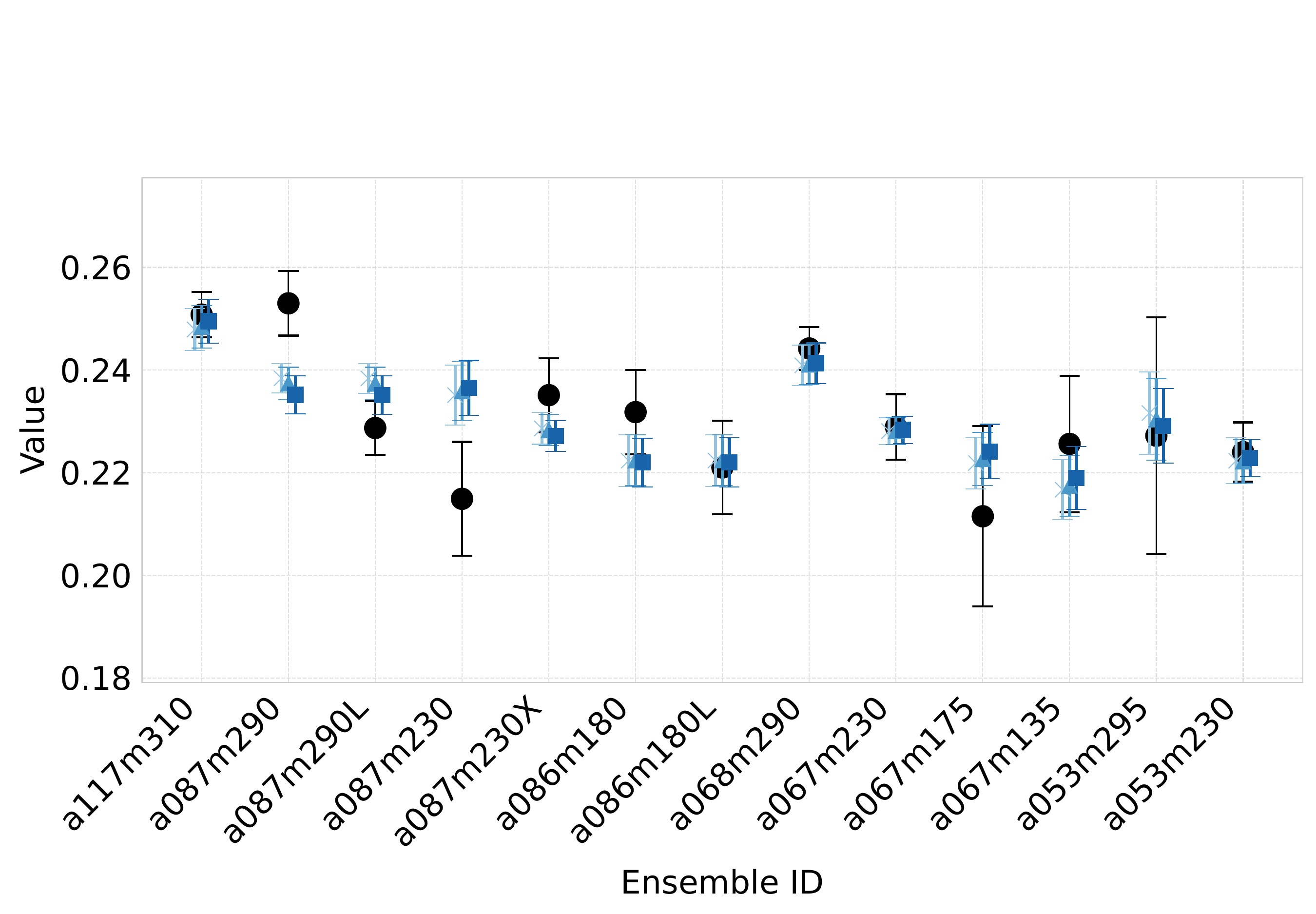}
\end{subfigure}
\vspace{0.4cm}
\caption{Comparison of the CCFV (left panels) with the CC (right panels) fits 
for the helicity moment $\langle x \rangle_{\Delta u-\Delta d}$ data. 
The best fit is again to the $\{4,3^\ast\}$ data. The 
rest is the same as in Fig.~\protect\ref{fig:momfrac_comp}. 
 }
\label{fig:helfrac_comp}
\end{figure*}

\begin{figure*}[h]   
\centering
\begin{subfigure}
\centering
\includegraphics[trim=0 10 0 10,angle=0,width=1.0\textwidth]{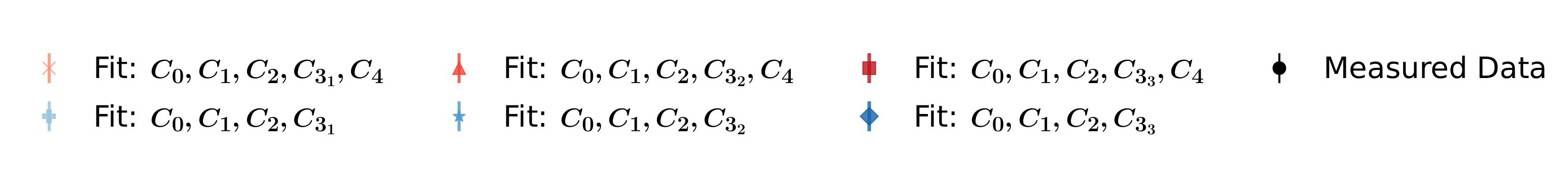}
\end{subfigure}
\vskip -0.9cm

\begin{subfigure}
\centering
\includegraphics[angle=0,width=0.48\textwidth]{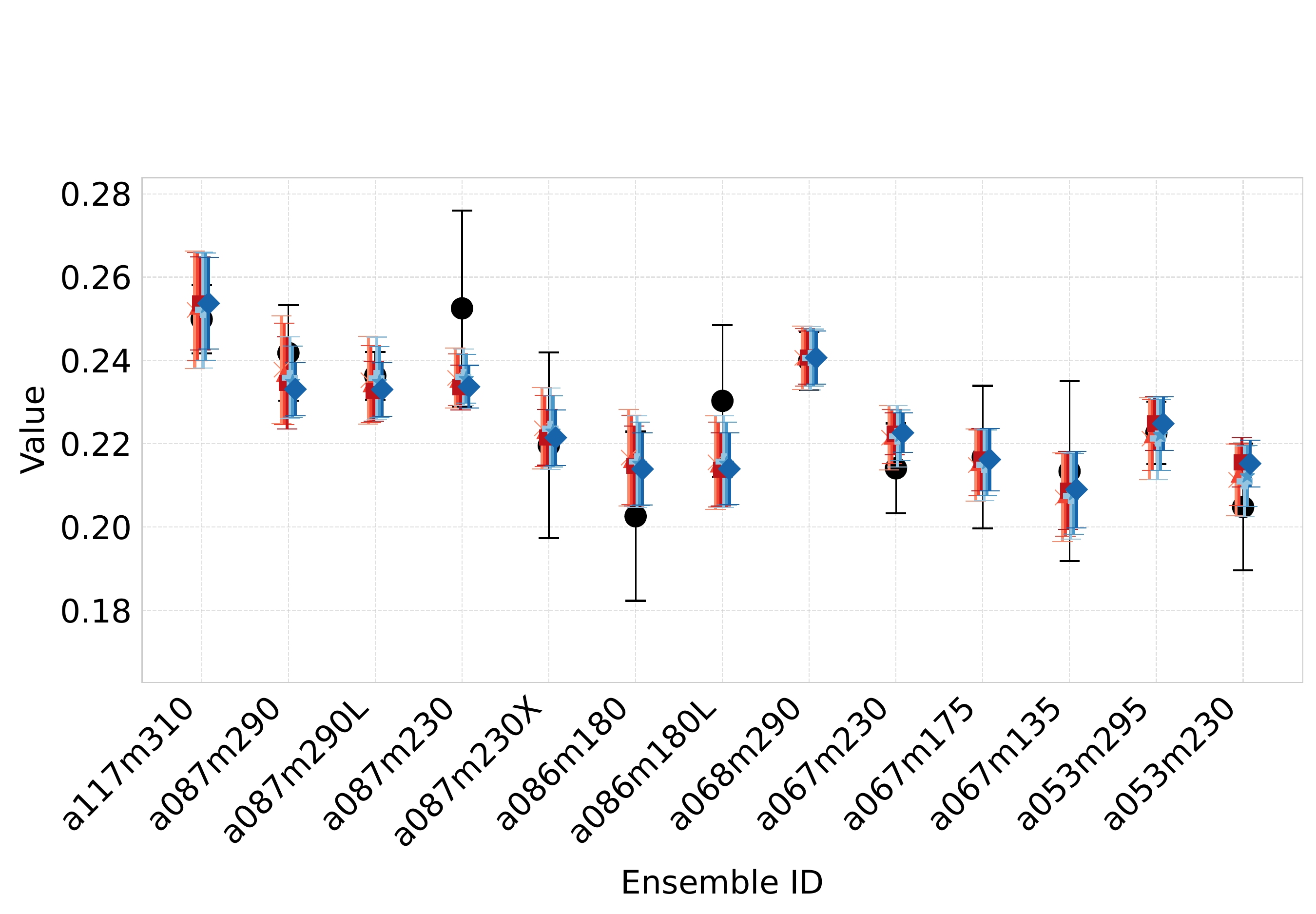}%
\hfill
\includegraphics[angle=0,width=0.48\textwidth]{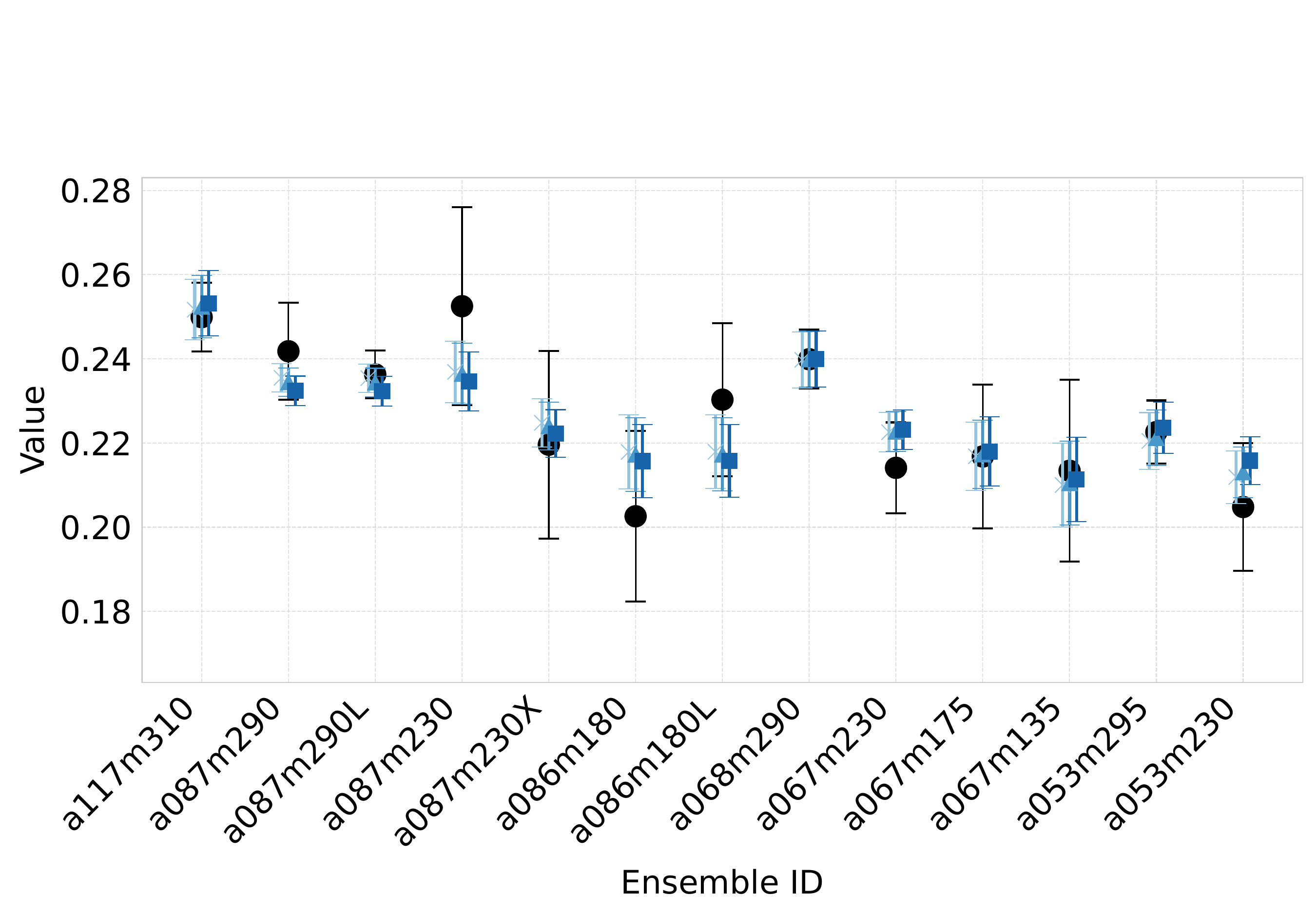}
\end{subfigure}


\begin{subfigure}
\centering
\includegraphics[trim=0 90 0 90,angle=0,width=0.48\textwidth]{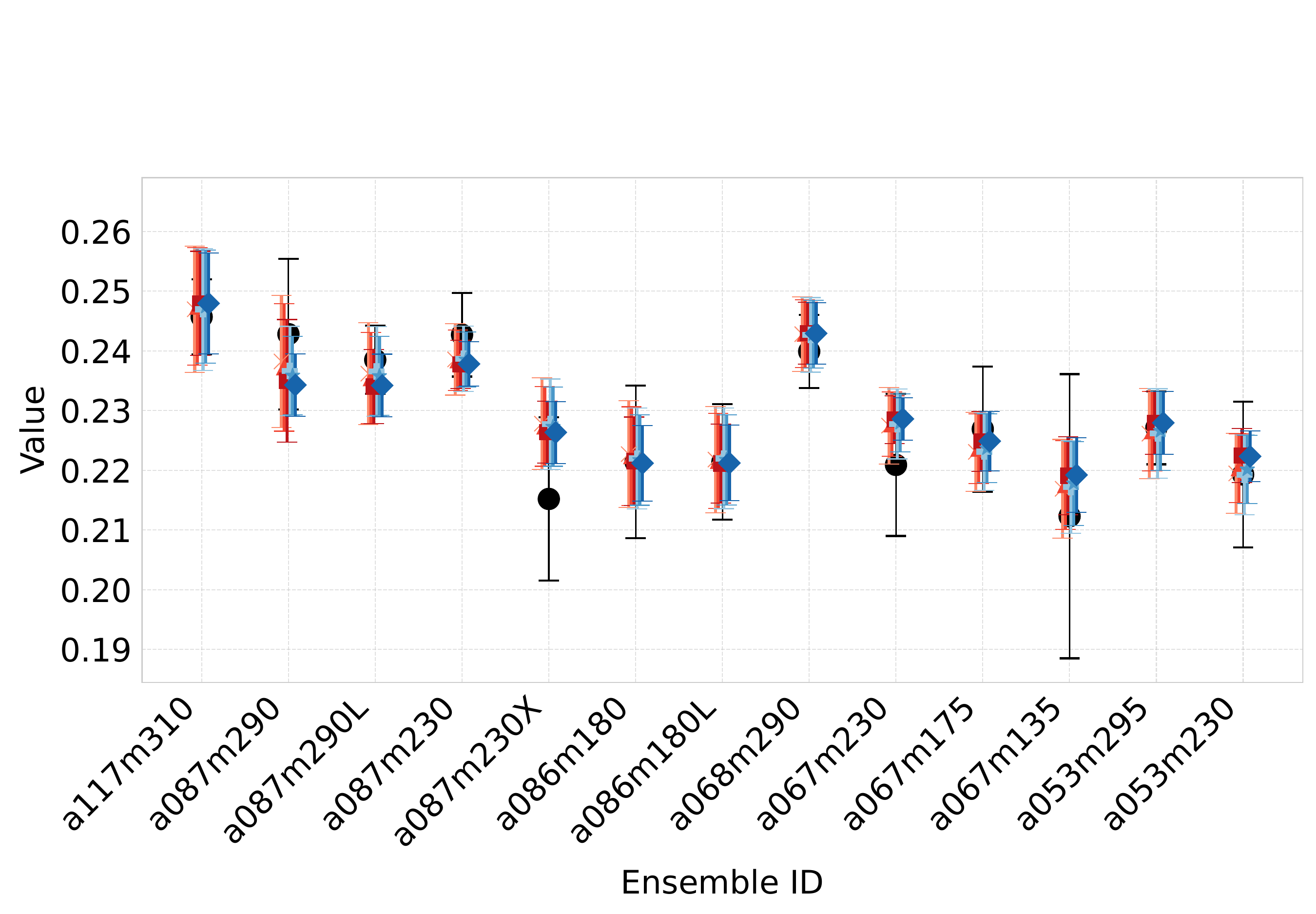}%
\hfill
\includegraphics[trim=0 90 0 90,angle=0,width=0.48\textwidth]{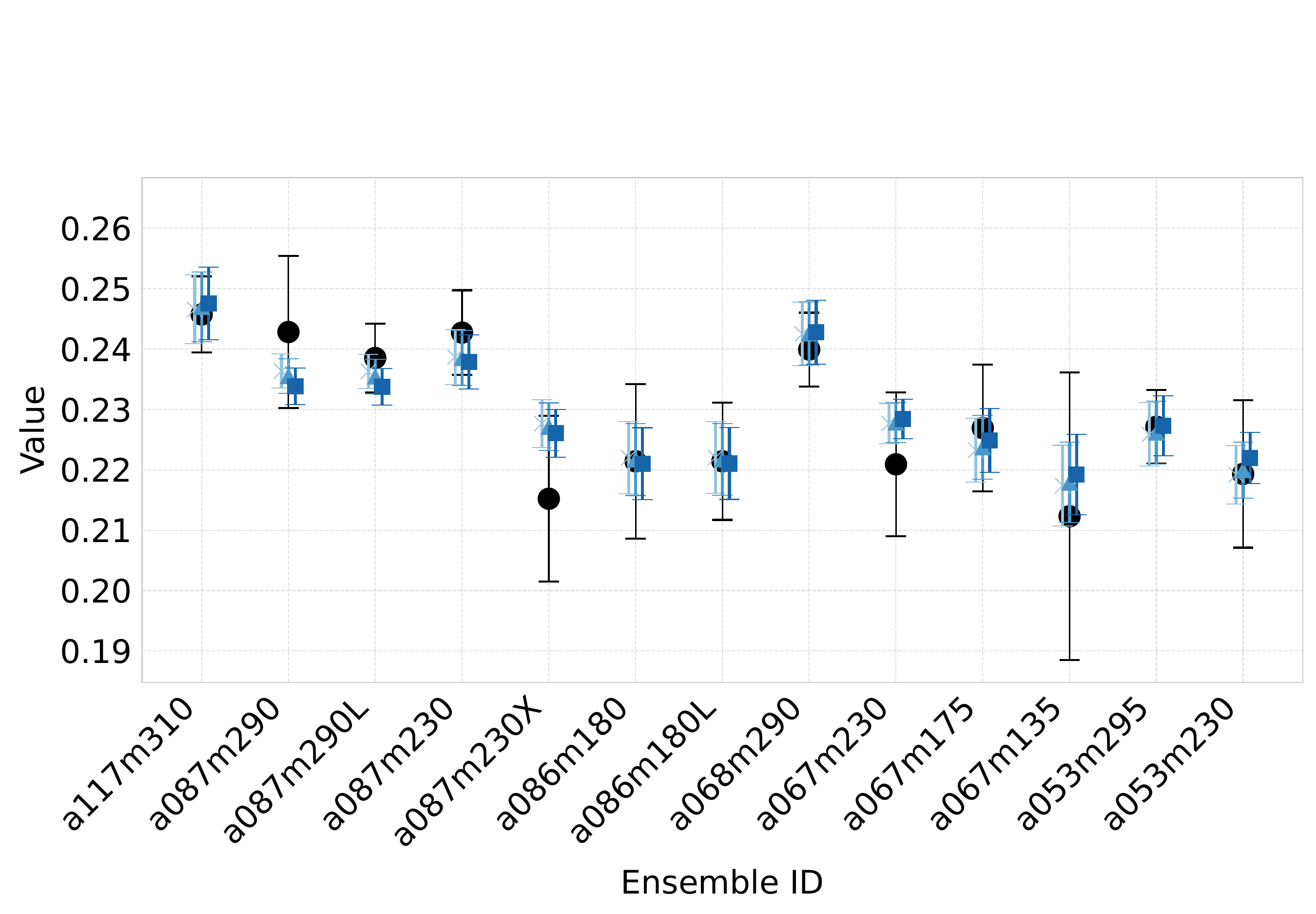}
\end{subfigure}


\begin{subfigure}
\centering
\includegraphics[trim=0 50 0 15,angle=0,width=0.48\textwidth]{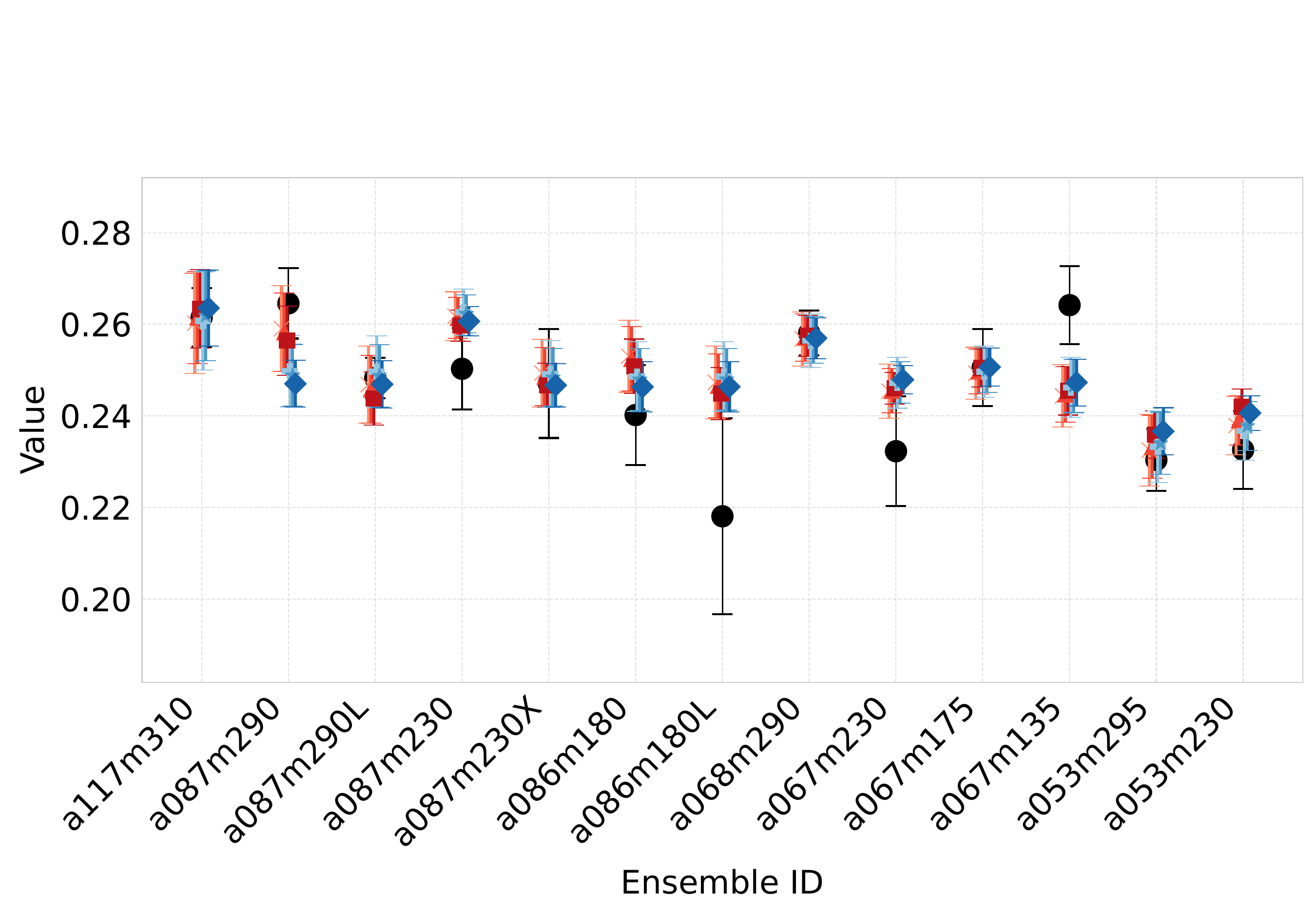}%
\hfill
\includegraphics[trim=0 50 0 15,angle=0,width=0.48\textwidth]{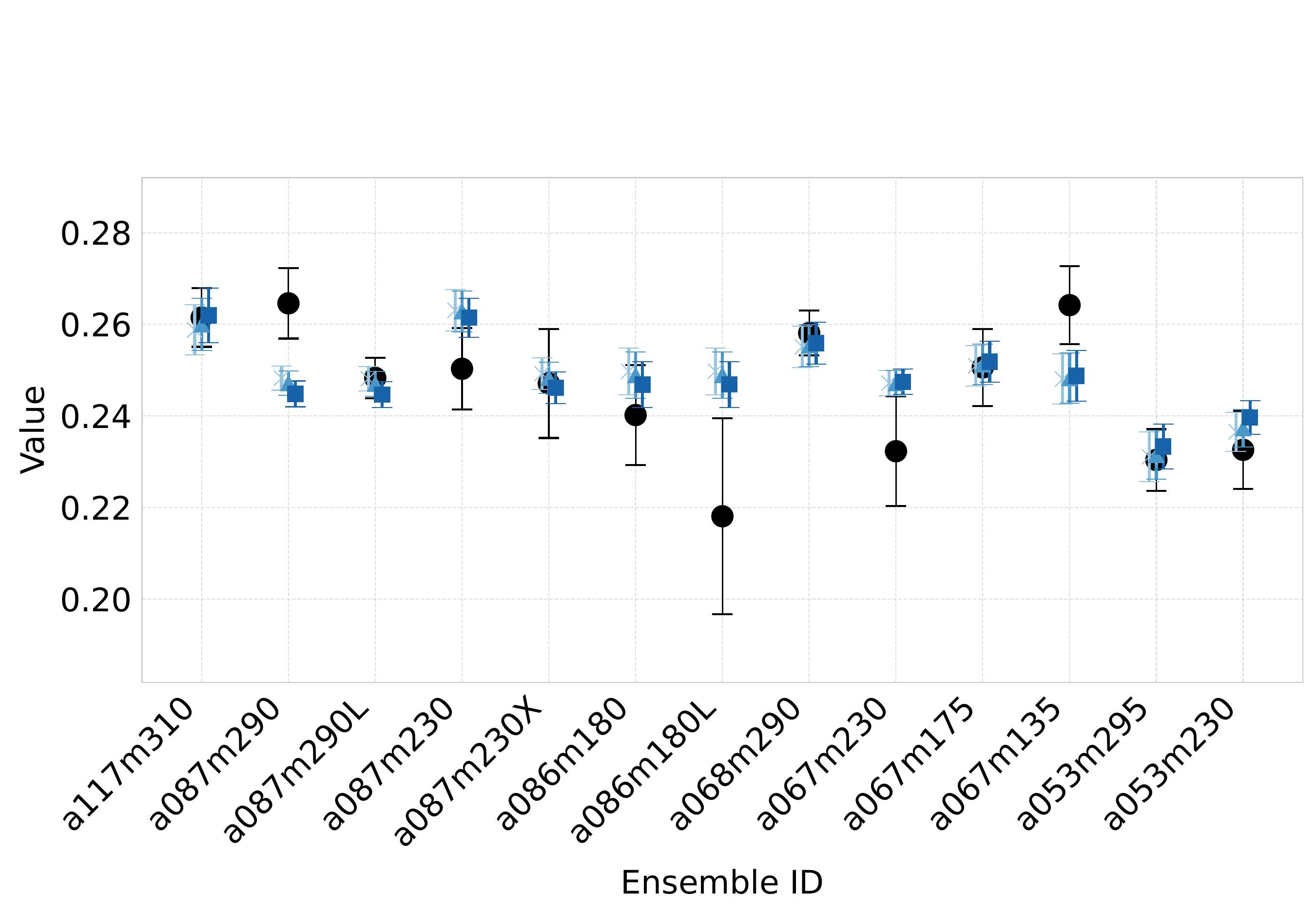}
\end{subfigure}
\vspace{0.4cm}
\caption{Comparison of the CCFV (left panels) with the CC (right panels) fits 
for the transversity moment $\langle x \rangle_{\delta u-\delta d}$ data. 
The rest is the same as in Fig.~\protect\ref{fig:momfrac_comp}. }
\label{fig:trans_comp}
\end{figure*}

A comparison of the results from the CCFV and CC fits is presented 
in Figs.~\ref{fig:momfrac_comp}, ~\ref{fig:helfrac_comp}, and~\ref{fig:trans_comp}. 
The main difference between the two fits is the $a087m290$ point in the 
momentum fraction---it is not captured by 
the CC fit. Similarly, the difference in the CCFV fit 
with and without the $c_4$ term is again $a087m290$. In contrast, 
the other small volume point $a086m180$ does not stand out since 
the $a086m180$ and $a086m180L$ values overlap. Based on these features, 
we choose the CCFV fit for the central values.


\begin{table*}[htbp]   
\fontsize{5pt}{8pt}\selectfont
\centering
\setlength{\tabcolsep}{4pt}
\renewcommand{\arraystretch}{1.3}
\begin{tabular}{|l|c|c|c|c|c|c|c|l|}
\hline
\textbf{Fit Strategy} & \textbf{Method A} & \textbf{Method B} & \textbf{Average} & $(\chi^2/dof)_{\text{A}}$ & $(\chi^2/dof)_{\text{B}}$ & $AICc_A$ & $AICc_B$ & \textbf{Ansatz} \\ \hline \hline
\multicolumn{9}{|c|}{\cellcolor[gray]{0.9}\textbf{Moment: $\langle x \rangle_{u-d}$}} \\ \hline
\multirow{6}{*}{$\{4^{N\pi},3^*\}$} & 0.1488(93) & 0.1390(98) & 0.1439(98) & 0.456 & 0.461 & 22.22 & 22.26 & $C_0, C_1, C_2, C_{3_1}, C_4$ \\
 & 0.1501(88) & 0.1406(92) & 0.1454(92) & 0.474 & 0.487 & 22.37 & 22.47 & $C_0, C_1, C_2, C_{3_2}, C_4$ \\
 & 0.1522(82) & 0.1431(84) & 0.1476(84) & 0.517 & 0.539 & 22.70 & 22.88 & $C_0, C_1, C_2, C_{3_3}, C_4$ \\
 & 0.1536(89) & 0.1441(94) & 0.1488(94) & 0.768 & 0.805 & 19.91 & 20.24 & $C_0, C_1, C_2, C_{3_1}$ \\
 & 0.1547(85) & 0.1455(88) & 0.1501(88) & 0.783 & 0.824 & 20.05 & 20.42 & $C_0, C_1, C_2, C_{3_2}$ \\
 & 0.1564(79) & 0.1475(80) & 0.1520(80) & 0.815 & 0.857 & 20.34 & 20.71 & $C_0, C_1, C_2, C_{3_3}$ \\
\hline
\multirow{6}{*}{$\{4,3^*\}$} & 0.1553(88) & 0.1486(99) & 0.1519(99) & 0.357 & 0.191 & 21.42 & 20.10 & $C_0, C_1, C_2, C_{3_1}, C_4$ \\
 & 0.1571(83) & 0.1504(90) & 0.1538(90) & 0.374 & 0.214 & 21.56 & 20.28 & $C_0, C_1, C_2, C_{3_2}, C_4$ \\
 & 0.1604(75) & 0.1536(78) & 0.1570(78) & 0.433 & 0.269 & 22.04 & 20.72 & $C_0, C_1, C_2, C_{3_3}, C_4$ \\
 & 0.1585(82) & 0.1525(94) & 0.1555(94) & 0.437 & 0.368 & 16.93 & 16.31 & $C_0, C_1, C_2, C_{3_1}$ \\
 & 0.1603(77) & 0.1543(86) & 0.1573(86) & 0.456 & 0.391 & 17.11 & 16.51 & $C_0, C_1, C_2, C_{3_2}$ \\
 & 0.1635(69) & 0.1574(73) & 0.1605(73) & 0.516 & 0.440 & 17.64 & 16.96 & $C_0, C_1, C_2, C_{3_3}$ \\
\hline
\multirow{6}{*}{$\{4,2^{\text{free}}\}\}$} & 0.1638(86) & 0.1545(92) & 0.1592(92) & 0.896 & 0.900 & 25.74 & 25.77 & $C_0, C_1, C_2, C_{3_1}, C_4$ \\
 & 0.1661(80) & 0.1572(85) & 0.1617(85) & 0.953 & 0.967 & 26.20 & 26.31 & $C_0, C_1, C_2, C_{3_2}, C_4$ \\
 & 0.1699(72) & 0.1616(74) & 0.1658(74) & 1.086 & 1.106 & 27.26 & 27.42 & $C_0, C_1, C_2, C_{3_3}, C_4$ \\
 & 0.1697(84) & 0.1611(90) & 0.1654(90) & 1.758 & 1.825 & 28.82 & 29.43 & $C_0, C_1, C_2, C_{3_1}$ \\
 & 0.1719(78) & 0.1637(82) & 0.1678(82) & 1.813 & 1.884 & 29.32 & 29.96 & $C_0, C_1, C_2, C_{3_2}$ \\
 & 0.1755(70) & 0.1676(71) & 0.1716(71) & 1.928 & 1.986 & 30.35 & 30.88 & $C_0, C_1, C_2, C_{3_3}$ \\
\hline
\multicolumn{9}{|c|}{\cellcolor[gray]{0.9}\textbf{Moment: $\langle x \rangle_{\Delta u-\Delta d}$}} \\ \hline
\multirow{6}{*}{$\{4^{N\pi},3^*\}$} & 0.1710(106) & 0.1586(113) & 0.1648(113) & 0.390 & 0.233 & 21.69 & 20.43 & $C_0, C_1, C_2, C_{3_1}, C_4$ \\
 & 0.1742(99) & 0.1626(104) & 0.1684(104) & 0.324 & 0.196 & 21.17 & 20.14 & $C_0, C_1, C_2, C_{3_2}, C_4$ \\
 & 0.1802(90) & 0.1700(92) & 0.1751(92) & 0.252 & 0.202 & 20.58 & 20.18 & $C_0, C_1, C_2, C_{3_3}, C_4$ \\
 & 0.1755(101) & 0.1633(108) & 0.1694(108) & 0.557 & 0.420 & 18.01 & 16.78 & $C_0, C_1, C_2, C_{3_1}$ \\
 & 0.1784(94) & 0.1669(100) & 0.1726(100) & 0.496 & 0.386 & 17.46 & 16.47 & $C_0, C_1, C_2, C_{3_2}$ \\
 & 0.1838(86) & 0.1736(88) & 0.1787(88) & 0.418 & 0.378 & 16.76 & 16.40 & $C_0, C_1, C_2, C_{3_3}$ \\
\hline
\multirow{6}{*}{$\{4,3^*\}$} & 0.1778(96) & 0.1665(102) & 0.1721(102) & 0.465 & 0.444 & 22.29 & 22.13 & $C_0, C_1, C_2, C_{3_1}, C_4$ \\
 & 0.1809(86) & 0.1704(91) & 0.1757(91) & 0.436 & 0.448 & 22.06 & 22.16 & $C_0, C_1, C_2, C_{3_2}, C_4$ \\
 & 0.1869(72) & 0.1774(74) & 0.1821(74) & 0.436 & 0.523 & 22.06 & 22.75 & $C_0, C_1, C_2, C_{3_3}, C_4$ \\
 & 0.1850(87) & 0.1742(93) & 0.1796(93) & 0.747 & 0.764 & 19.72 & 19.88 & $C_0, C_1, C_2, C_{3_1}$ \\
 & 0.1877(77) & 0.1776(82) & 0.1827(82) & 0.726 & 0.771 & 19.53 & 19.94 & $C_0, C_1, C_2, C_{3_2}$ \\
 & 0.1927(63) & 0.1836(66) & 0.1881(66) & 0.722 & 0.830 & 19.50 & 20.47 & $C_0, C_1, C_2, C_{3_3}$ \\
\hline
\multirow{6}{*}{$\{4,2^{\text{free}}\}\}$} & 0.1976(97) & 0.1882(112) & 0.1929(112) & 0.855 & 0.777 & 25.41 & 24.79 & $C_0, C_1, C_2, C_{3_1}, C_4$ \\
 & 0.1998(85) & 0.1910(96) & 0.1954(96) & 0.801 & 0.749 & 24.98 & 24.56 & $C_0, C_1, C_2, C_{3_2}, C_4$ \\
 & 0.2045(68) & 0.1965(74) & 0.2005(74) & 0.737 & 0.732 & 24.47 & 24.43 & $C_0, C_1, C_2, C_{3_3}, C_4$ \\
 & 0.2095(88) & 0.2007(103) & 0.2051(103) & 1.686 & 1.659 & 28.17 & 27.93 & $C_0, C_1, C_2, C_{3_1}$ \\
 & 0.2112(76) & 0.2030(87) & 0.2071(87) & 1.680 & 1.665 & 28.12 & 27.98 & $C_0, C_1, C_2, C_{3_2}$ \\
 & 0.2144(60) & 0.2066(66) & 0.2105(66) & 1.684 & 1.687 & 28.15 & 28.18 & $C_0, C_1, C_2, C_{3_3}$ \\
\hline
\multicolumn{9}{|c|}{\cellcolor[gray]{0.9}\textbf{Moment: $\langle x \rangle_{\delta u-\delta d}$}} \\ \hline
\multirow{6}{*}{$\{4^{N\pi},3^*\}$} & 0.1824(137) & 0.1727(149) & 0.1776(149) & 0.337 & 0.360 & 21.27 & 21.45 & $C_0, C_1, C_2, C_{3_1}, C_4$ \\
 & 0.1872(128) & 0.1784(136) & 0.1828(136) & 0.375 & 0.420 & 21.57 & 21.93 & $C_0, C_1, C_2, C_{3_2}, C_4$ \\
 & 0.1954(115) & 0.1877(118) & 0.1916(118) & 0.512 & 0.585 & 22.67 & 23.25 & $C_0, C_1, C_2, C_{3_3}, C_4$ \\
 & 0.1831(133) & 0.1736(144) & 0.1784(144) & 0.304 & 0.327 & 15.74 & 15.94 & $C_0, C_1, C_2, C_{3_1}$ \\
 & 0.1878(123) & 0.1793(131) & 0.1836(131) & 0.338 & 0.380 & 16.04 & 16.42 & $C_0, C_1, C_2, C_{3_2}$ \\
 & 0.1959(111) & 0.1884(114) & 0.1921(114) & 0.458 & 0.525 & 17.12 & 17.72 & $C_0, C_1, C_2, C_{3_3}$ \\
\hline
\multirow{6}{*}{$\{4,3^*\}$} & 0.1972(110) & 0.1871(124) & 0.1922(124) & 0.284 & 0.263 & 20.84 & 20.68 & $C_0, C_1, C_2, C_{3_1}, C_4$ \\
 & 0.2007(101) & 0.1915(112) & 0.1961(112) & 0.306 & 0.307 & 21.02 & 21.03 & $C_0, C_1, C_2, C_{3_2}, C_4$ \\
 & 0.2070(89) & 0.1988(94) & 0.2029(94) & 0.396 & 0.432 & 21.74 & 22.03 & $C_0, C_1, C_2, C_{3_3}, C_4$ \\
 & 0.1979(100) & 0.1884(114) & 0.1932(114) & 0.255 & 0.242 & 15.30 & 15.18 & $C_0, C_1, C_2, C_{3_1}$ \\
 & 0.2013(93) & 0.1926(103) & 0.1969(103) & 0.275 & 0.280 & 15.47 & 15.52 & $C_0, C_1, C_2, C_{3_2}$ \\
 & 0.2072(81) & 0.1995(86) & 0.2034(86) & 0.352 & 0.388 & 16.17 & 16.49 & $C_0, C_1, C_2, C_{3_3}$ \\
\hline
\multirow{6}{*}{$\{4,2^{\text{free}}\}\}$} & 0.2149(94) & 0.2067(112) & 0.2108(112) & 1.602 & 1.337 & 31.38 & 29.26 & $C_0, C_1, C_2, C_{3_1}, C_4$ \\
 & 0.2194(86) & 0.2115(100) & 0.2154(100) & 1.610 & 1.362 & 31.45 & 29.47 & $C_0, C_1, C_2, C_{3_2}, C_4$ \\
 & 0.2274(74) & 0.2197(83) & 0.2236(83) & 1.744 & 1.481 & 32.52 & 30.42 & $C_0, C_1, C_2, C_{3_3}, C_4$ \\
 & 0.2193(89) & 0.2118(106) & 0.2156(106) & 1.663 & 1.424 & 27.97 & 25.82 & $C_0, C_1, C_2, C_{3_1}$ \\
 & 0.2237(81) & 0.2165(94) & 0.2201(94) & 1.680 & 1.459 & 28.12 & 26.13 & $C_0, C_1, C_2, C_{3_2}$ \\
 & 0.2314(69) & 0.2243(78) & 0.2279(78) & 1.806 & 1.577 & 29.25 & 27.20 & $C_0, C_1, C_2, C_{3_3}$ \\
\hline
\end{tabular}
\vspace*{-4pt}
\caption{Results for the three moments given by the CCFV fits using data from the 13 ensembles are given for the three strategies for controlling ESC, the two renormalization methods A and B, and 
three choices of the parameter $c_3$, each with and without the FV parameter $c_4$ defined in Eq.~\eqref{eq:CCFV} and specified in the last column labeled "Ansatz".}
\label{tab:CCFV_13pt_fit}
\end{table*}


\begin{table*}[htbp]         
\fontsize{7pt}{9pt}\selectfont
\centering
\setlength{\tabcolsep}{4pt}
\renewcommand{\arraystretch}{1.3}
\begin{tabular}{|l|c|c|c|c|c|c|c|l|}
\hline
\textbf{Fit Strategy} & \textbf{Method A} & \textbf{Method B} & \textbf{Average} & $(\chi^2/dof)_{\text{A}}$ & $(\chi^2/dof)_{\text{B}}$ & $AICc_A$ & $AICc_B$ & \textbf{Ansatz} \\ \hline \hline
\multicolumn{9}{|c|}{\cellcolor[gray]{0.9}\textbf{Moment: $\langle x \rangle_{u-d}$}} \\ \hline
\multirow{3}{*}{$\{4^{N\pi},3^*\}$} & 0.1546(93) & 0.1440(98) & 0.1493(98) & 0.182 & 0.188 & 15.94 & 15.98 & $C_{0}, C_{1}, C_{2}, C_{3_1}$ \\
 & 0.1554(89) & 0.1450(93) & 0.1502(93) & 0.185 & 0.197 & 15.96 & 16.05 & $C_{0}, C_{1}, C_{2}, C_{3_2}$ \\
 & 0.1566(84) & 0.1466(85) & 0.1516(85) & 0.193 & 0.219 & 16.02 & 16.20 & $C_{0}, C_{1}, C_{2}, C_{3_3}$ \\
\hline
\multirow{3}{*}{$\{4,3^*\}$} & 0.1581(88) & 0.1513(98) & 0.1547(98) & 0.343 & 0.139 & 17.07 & 15.64 & $C_{0}, C_{1}, C_{2}, C_{3_1}$ \\
 & 0.1596(83) & 0.1527(90) & 0.1562(90) & 0.348 & 0.150 & 17.10 & 15.72 & $C_{0}, C_{1}, C_{2}, C_{3_2}$ \\
 & 0.1623(76) & 0.1550(79) & 0.1586(79) & 0.377 & 0.180 & 17.30 & 15.93 & $C_{0}, C_{1}, C_{2}, C_{3_3}$ \\
\hline
\multirow{3}{*}{$\{4,2^{free}\}$} & 0.1717(87) & 0.1620(92) & 0.1669(92) & 0.511 & 0.541 & 18.25 & 18.45 & $C_{0}, C_{1}, C_{2}, C_{3_1}$ \\
 & 0.1731(81) & 0.1638(85) & 0.1684(85) & 0.529 & 0.570 & 18.37 & 18.66 & $C_{0}, C_{1}, C_{2}, C_{3_2}$ \\
 & 0.1754(74) & 0.1666(76) & 0.1710(76) & 0.575 & 0.635 & 18.69 & 19.11 & $C_{0}, C_{1}, C_{2}, C_{3_3}$ \\
\hline
\multicolumn{9}{|c|}{\cellcolor[gray]{0.9}\textbf{Moment: $\langle x \rangle_{\Delta u-\Delta d}$}} \\ \hline
\multirow{3}{*}{$\{4^{N\pi},3^*\}$} & 0.1738(106) & 0.1608(113) & 0.1673(113) & 0.544 & 0.312 & 18.47 & 16.85 & $C_{0}, C_{1}, C_{2}, C_{3_1}$ \\
 & 0.1763(100) & 0.1640(105) & 0.1702(105) & 0.440 & 0.232 & 17.74 & 16.29 & $C_{0}, C_{1}, C_{2}, C_{3_2}$ \\
 & 0.1811(93) & 0.1700(96) & 0.1755(96) & 0.266 & 0.136 & 16.53 & 15.62 & $C_{0}, C_{1}, C_{2}, C_{3_3}$ \\
\hline
\multirow{3}{*}{$\{4,3^*\}$} & 0.1802(91) & 0.1684(97) & 0.1743(97) & 0.521 & 0.403 & 18.31 & 17.49 & $C_{0}, C_{1}, C_{2}, C_{3_1}$ \\
 & 0.1825(82) & 0.1713(87) & 0.1769(87) & 0.461 & 0.373 & 17.89 & 17.28 & $C_{0}, C_{1}, C_{2}, C_{3_2}$ \\
 & 0.1869(70) & 0.1766(73) & 0.1817(73) & 0.376 & 0.363 & 17.30 & 17.21 & $C_{0}, C_{1}, C_{2}, C_{3_3}$ \\
\hline
\multirow{3}{*}{$\{4,2^{free}\}$} & 0.2047(91) & 0.1938(107) & 0.1992(107) & 1.368 & 1.223 & 24.24 & 23.23 & $C_{0}, C_{1}, C_{2}, C_{3_1}$ \\
 & 0.2058(80) & 0.1954(92) & 0.2006(92) & 1.289 & 1.169 & 23.69 & 22.85 & $C_{0}, C_{1}, C_{2}, C_{3_2}$ \\
 & 0.2089(66) & 0.1994(73) & 0.2042(73) & 1.131 & 1.079 & 22.58 & 22.22 & $C_{0}, C_{1}, C_{2}, C_{3_3}$ \\
\hline
\multicolumn{9}{|c|}{\cellcolor[gray]{0.9}\textbf{Moment: $\langle x \rangle_{\delta u-\delta d}$}} \\ \hline
\multirow{3}{*}{$\{4^{N\pi},3^*\}$} & 0.1856(138) & 0.1753(149) & 0.1805(149) & 0.281 & 0.308 & 16.63 & 16.82 & $C_{0}, C_{1}, C_{2}, C_{3_1}$ \\
 & 0.1902(130) & 0.1808(137) & 0.1855(137) & 0.316 & 0.366 & 16.88 & 17.23 & $C_{0}, C_{1}, C_{2}, C_{3_2}$ \\
 & 0.1978(119) & 0.1896(122) & 0.1937(122) & 0.449 & 0.535 & 17.81 & 18.41 & $C_{0}, C_{1}, C_{2}, C_{3_3}$ \\
\hline
\multirow{3}{*}{$\{4,3^*\}$} & 0.1980(106) & 0.1876(119) & 0.1928(119) & 0.292 & 0.253 & 16.71 & 16.44 & $C_{0}, C_{1}, C_{2}, C_{3_1}$ \\
 & 0.2012(98) & 0.1915(109) & 0.1964(109) & 0.307 & 0.290 & 16.82 & 16.70 & $C_{0}, C_{1}, C_{2}, C_{3_2}$ \\
 & 0.2069(88) & 0.1982(94) & 0.2026(94) & 0.384 & 0.405 & 17.35 & 17.50 & $C_{0}, C_{1}, C_{2}, C_{3_3}$ \\
\hline
\multirow{3}{*}{$\{4,2^{free}\}$} & 0.2203(91) & 0.2111(110) & 0.2157(110) & 1.458 & 1.222 & 24.88 & 23.22 & $C_{0}, C_{1}, C_{2}, C_{3_1}$ \\
 & 0.2243(84) & 0.2154(99) & 0.2198(99) & 1.417 & 1.212 & 24.59 & 23.15 & $C_{0}, C_{1}, C_{2}, C_{3_2}$ \\
 & 0.2312(75) & 0.2227(85) & 0.2269(85) & 1.436 & 1.261 & 24.72 & 23.49 & $C_{0}, C_{1}, C_{2}, C_{3_3}$ \\
\hline
\end{tabular}
\caption{Results for the three moments from the CC fits using data from 
the 11 ensembles, i.e., excluding the smaller volume ensembles $a087m290$ 
and $a086m180$, are given for the three strategies for controlling ESC, 
the two renormalization methods A and B, and 
three choices of the parameter $c_3$, each with and without the FV parameter $c_4$ defined in Eq.~\eqref{eq:CCFV} and specified in the last column labeled "Ansatz".}
\label{tab:CC_11pt_fits}
\end{table*}


\clearpage


\end{document}